\documentclass[
    11pt,
    headings=normal,
    bibliography=totoc,
    index=totoc,
    headsepline,
    BCOR=10mm,
    DIV=11,
    numbers=noenddot,
    twoside=semi,
    open=any
]{scrbook}

\usepackage[T1]{fontenc}
\usepackage[utf8]{inputenc}
\usepackage[english]{babel}
\usepackage{silence}
\usepackage{libertinus} 
\usepackage[final]{microtype}
\usepackage{hyphenat}

\usepackage{amsmath, amssymb, amsthm}

\usepackage[automark]{scrlayer-scrpage}
\clearpairofpagestyles
\ohead{\headmark}
\ofoot*{\pagemark}
\addtokomafont{pagehead}{\small\itshape}
\addtokomafont{pagenumber}{\small\sffamily}

\usepackage[dvipsnames]{xcolor}
\usepackage{soul}
\definecolor{DarkBlue}{rgb}{0,0,0.5}
\definecolor{DeepRed}{rgb}{0.6,0,0}

\usepackage{caption}
\usepackage[autostyle]{csquotes}

\usepackage[
    hidelinks,
    breaklinks=true,
    colorlinks=true,
    linkcolor=DarkBlue,
    citecolor=DeepRed,
    urlcolor=DarkBlue,
    bookmarksopen=true,
    bookmarksopenlevel=1
]{hyperref}

\usepackage{enumitem}
\setlist[enumerate,1]{label=\arabic*., font=\bfseries, leftmargin=2em}
\setlist[itemize,1]{label=\textbullet, leftmargin=2em}
\usepackage{array,tabularx,makecell}
\usepackage{booktabs}
\newcolumntype{L}{>{\raggedright\arraybackslash}X}

\theoremstyle{plain}
\newtheorem{theorem}{Theorem}[chapter]
\newtheorem{lemma}[theorem]{Lemma}
\newtheorem{proposition}[theorem]{Proposition}
\newtheorem{corollary}[theorem]{Corollary}

\theoremstyle{definition}
\newtheorem{definition}{Definition}[chapter]
\newtheorem{example}{Example}[chapter]
\newtheorem{remark}{Remark}[chapter]
\newtheorem{assumption}{Assumption}[chapter]

\DeclareMathOperator{\Deck}{Deck}

\title{Geometric Quantization on Orbifolds}
\author{Peiyuan Teng}

\begin{document}
\setcounter{tocdepth}{1}

\frontmatter

\begin{titlepage}
    \centering
    \vspace*{5cm}
    {\huge\bfseries Geometric Quantization on Orbifolds (Draft) \par}
    \vspace{2cm}
    {\Large\itshape Peiyuan Teng\par}
\end{titlepage}

\cleardoublepage
\tableofcontents
\cleardoublepage

\mainmatter

\addchap{Introduction}
\label{sec:introduction}

The interface between classical and quantum mechanics has been a profound source of inspiration in both physics and mathematics. Geometric quantization, as pioneered by Kostant \cite{Kostant1970} and Souriau \cite{Souriau1970}, offers a rigorous mathematical framework to construct quantum theories from classical counterparts defined on symplectic manifolds. Further developed in standard texts such as Woodhouse \cite{Woodhouse1992} and \'Sniatycki \cite{Sniatycki1980}, this framework systematically addresses prequantization, polarization, and metaplectic correction, providing deep insights into the geometry of quantum states. 

While classical mechanics on smooth manifolds is well-understood, physical systems often exhibit singularities. Phase spaces modeled as orbifolds naturally arise in many physical scenarios, such as systems with discrete gauge symmetries or indistinguishable particles. An orbifold, introduced abstractly by Satake \cite{Satake1957} and further studied topologically by many authors (e.g., Adem, Leida, and Ruan \cite{AdemLeidaRuan}, Moerdijk and Pronk \cite{MoerdijkPronk1997}), provides a generalized manifold structure that locally looks like the quotient of an open subset of a Euclidean space by a finite group action. The geometry and topology of orbifolds extend standard manifold theory to these mild singularities, allowing generalizations of de Rham cohomology, line bundles, and characteristic classes \cite{Kawasaki1979}.

Applying geometric quantization to orbifolds is a challenging yet mathematically rich endeavor, motivated by quantum mechanical models with quotient singularities. Foundational aspects of associating quantization spaces to symplectic orbifolds with Hamiltonian group actions were established by Cannas da Silva and Guillemin \cite{SilvaGuillemin1999}. The closely related article \textit{Geometric quantization for proper actions} \cite{MathaiZhang2010} places quotient-type phenomena in the broader setting of proper cocompact group actions and index theory. These works locate orbifolds and proper quotient spaces within the geometric quantization program, while the analysis of quantum dynamics on singular configuration spaces often emphasizes self-adjoint extensions, boundary conditions, or spectral theory for particular models rather than the full prequantum-bundle and polarization package. The contribution of this text is more specific: it gives a unified, example-driven treatment of how stack/classifying-space line-bundle data, orbifold holonomy, polarization choices, Bohr--Sommerfeld conditions, and half-form obstructions interact in concrete quotient models. In doing so, it develops these ingredients directly in the local group-quotient setting and keeps track of the sector data that are often hidden when one works only on the coarse space or only on a smooth cover.

This text rigorously examines and extends the geometric quantization program over orbifold phase spaces. We begin in Chapter 1 by reviewing the standard foundations of geometric quantization, covering prequantization, polarization, and metaplectic correction. Chapter 2 develops the foundational theory of orbifolds, including covering theory, de Rham cohomology, and the classification of orbifold line bundles. In Chapter 3, we analyze key geometric examples such as the orbifold cone, orbispheres, and dihedral quotients, while Chapter 4 investigates the corresponding quantum mechanical problems, deriving spectra and wavefunctions for these prototypical systems. Chapter 5 establishes the theoretical core of this work by extending the geometric quantization program to symplectic orbifolds, addressing the specific requirements for connections, prequantum bundles, and half-form corrections in the singular setting. Finally, Chapter 6 provides systematic example-driven applications of this framework to representative orbifold phase spaces---ranging from free circle quotients and planar cones to football orbifolds, weighted orbispheres, and cotangent dihedral quotients. These examples do not claim to quantize every observable directly in every chosen polarization; rather, they show precisely where geometric quantization applies, where Schr\"odinger or spectral methods must be used instead, and how the two descriptions agree on sector structure, holonomy, and spectral shifts. By synthesizing topological properties, symplectic geometry, and the representation of observable operators, this work contributes to a structured understanding of geometric quantization over orbifolds with mild singularities.

\chapter{Geometric Quantization}

\section{Motivation and Overview}

Canonical quantization provides the foundational bridge between classical and quantum mechanics. However, it faces a fundamental challenge: operator-ordering ambiguities~\cite{Groenewold1946}. For a general classical observable \(f(q,p)\), there is no unique prescription for the corresponding operator \(\hat f(\hat q,\hat p)\). On nontrivial phase spaces, different ordering choices can lead to inequivalent quantum theories, highlighting the need for a more robust, geometrically grounded approach.

Geometric quantization addresses this by providing a systematic, symplectic-geometric framework~\cite{Souriau1970,Kostant1970}. It seeks to represent the Poisson algebra of classical observables by operators on a Hilbert space in a way that respects the underlying symplectic structure. By utilizing complex line bundles with connection and the choice of a polarization, the theory yields a rigorous map from classical dynamics to quantum states~\cite{Woodhouse1992}.

The procedure of geometric quantization is traditionally structured into three refining stages:

\subsection*{Stage 1: Prequantization}

Given a symplectic manifold \((M,\omega)\) satisfying the prequantization integrality condition, we first construct a complex line bundle \(L\to M\) equipped with a Hermitian metric and a connection \(\nabla\). The connection is chosen such that its curvature is proportional to the symplectic form, \(F_\nabla = -\frac{i}{\hbar}\omega\). This geometric data allows us to define a representation of the full Poisson algebra on a dense subspace of smooth sections, whose Hilbert completion is \(\mathcal{H}_{\mathrm{pre}} = L^2(M, L)\). While this prequantum Hilbert space is mathematically well-defined, it is physically too large, as wavefunctions depend on all phase space variables~\cite{Kostant1970}.

\subsection*{Stage 2: Polarization}

To recover the correct number of degrees of freedom, we must restrict the wavefunctions to depend on only half of the phase space variables (e.g., positions only or momenta only). This is achieved by choosing a \emph{polarization} \(P\)---a maximal integrable isotropic distribution in the complexified tangent bundle~\cite{Woodhouse1992}. Standard choices include the vertical polarization, which leads to the Schr\"odinger position representation, and complex polarizations, which yield the Bargmann--Fock representation.

\subsection*{Stage 3: Metaplectic (Half-Form) Correction}

The final refinement addresses subtle spectral and measure-theoretic issues. Even with a polarization, the resulting theory may yield incorrect energy spectra. The metaplectic (or half-form) correction rectifies this by ensuring that wavefunctions transform as half-densities rather than scalars~\cite{Woodhouse1992}. This correction is responsible for the zero-point energy \(E_0 = \tfrac{1}{2}\hbar\omega\) in the harmonic oscillator and ensures consistency with semiclassical analysis.

In the following sections, we formalize these stages, beginning with the analytic and topological foundations of prequantization.

\section{Prequantization}
\label{sec:prequantization-manifold}

\subsection{Prequantization Requirements}

Prequantization provides a map from the Poisson algebra of classical observables to an algebra of operators on a prequantum Hilbert space. The construction, formulated by Kostant~\cite{Kostant1970} and Souriau~\cite{Souriau1970}, formalizes the structural requirements first outlined by Dirac.

\begin{enumerate}
    \item \textbf{Linearity}: For any two classical observables \(f, g\) and constants \(a, b \in \mathbb{R}\), the quantization map \(Q\) must satisfy:
    \[
    Q(a f + b g) = a Q(f) + b Q(g).
    \]
    This ensures that the quantum theory respects the linear structure of classical observables.

    \item \textbf{Constant Function}: The constant function \(f=1\) must map to the identity operator \(\hat{I}\):
    \[
    Q(1) = \hat{I}.
    \]

    \item \textbf{Commutation Relations}: The map must preserve the algebraic structure by relating the commutator to the Poisson bracket (with our convention \(\{f,g\}=\omega(X_g,X_f)\)):
    \[
    {}[Q(f), Q(g)] = i \hbar Q(\{f, g\}),
    \]
    where \( [\cdot, \cdot] \) is the commutator, \( \hbar \) is the reduced Planck constant, and \( \{f, g\} \) is the Poisson bracket.

    This requirement preserves the fundamental classical structure: the Poisson bracket encodes the system's dynamics, and the commutator mirrors it.
\end{enumerate}

To implement these requirements concretely, we realize states as sections of a line bundle whose curvature reproduces the symplectic form.

\subsubsection{The Fundamental Lemma of Prequantization}

The next lemma packages the curvature condition and the commutator identity into one precise statement.

\begin{lemma}[Kostant~\cite{Kostant1970}]\label{lem:prequantization}
	Let \((M,\omega)\) be a symplectic manifold and let \(L\) be a complex line bundle over \(M\) equipped with a connection \(\nabla\) whose curvature is denoted by
	\[
	F_\nabla = \operatorname{Curv}(\nabla).
	\]
	Define, for any smooth function \(f\in C^\infty(M)\), the prequantum operator
	\[
	Q(f) = -i\hbar\, \nabla_{X_f} + f,
	\]
	where \(X_f\) is the Hamiltonian vector field associated with \(f\) (defined via \(\iota_{X_f}\omega = -df\)). Then the commutator of the operators satisfies
	\[
	{}[Q(f), Q(g)] = i\hbar\, Q(\{f, g\})
	\]
	for all \(f, g \in C^\infty(M)\) if and only if the curvature of the connection satisfies
	\[
	F_\nabla = -\frac{i}{\hbar}\,\omega.
	\]
\end{lemma}

\begin{proof}
	Let \(s\) be a smooth section of the prequantum line bundle (or compactly supported if \(M\) is non-compact). Using \(Q(f) = -i\hbar\nabla_{X_f} + f\), we expand:
	\begin{align*}
		[Q(f),Q(g)]s
		=& (-i\hbar\nabla_{X_f}+f)(-i\hbar\nabla_{X_g}s+gs) \\
		&\quad - (-i\hbar\nabla_{X_g}+g)(-i\hbar\nabla_{X_f}s+fs) \\
		=& -\hbar^2\big(\nabla_{X_f}\nabla_{X_g}-\nabla_{X_g}\nabla_{X_f}\big)s \\
		&\quad - i\hbar\big(\nabla_{X_f}(gs)-g\nabla_{X_f}s-\nabla_{X_g}(fs)+f\nabla_{X_g}s\big).
	\end{align*}
	The scalar term \(fgs-gfs\) vanishes, and the Leibniz rule gives
	\[
	[Q(f),Q(g)]s
	=
	-\hbar^2\big(\nabla_{X_f}\nabla_{X_g}-\nabla_{X_g}\nabla_{X_f}\big)s
	-i\hbar\big(X_f(g)-X_g(f)\big)s.
	\]
	Using
	\[
	F_\nabla(X_f,X_g)=[\nabla_{X_f},\nabla_{X_g}]-\nabla_{[X_f,X_g]},
	\]
	we obtain
	\[
	[Q(f),Q(g)]s
	=
	-\hbar^2\nabla_{[X_f,X_g]}s
	-\hbar^2F_\nabla(X_f,X_g)s
	-i\hbar\big(X_f(g)-X_g(f)\big)s.
	\]
	Now apply the conventions \(\iota_{X_f}\omega=-df\) and \(\{f,g\}=\omega(X_g,X_f)\):
	\begin{align*}
		X_f(g) &= -\{f,g\}, \\
		X_g(f) &= \{f,g\}, \\
		[X_f,X_g] &= -X_{\{f,g\}}.
	\end{align*}
	Therefore,
	\[
	[Q(f),Q(g)]s
	=
	\hbar^2\nabla_{X_{\{f,g\}}}s
	+2i\hbar\{f,g\}s
	-\hbar^2F_\nabla(X_f,X_g)s.
	\]
	On the other hand,
	\[
	i\hbar Q(\{f,g\})s
	=
	\hbar^2\nabla_{X_{\{f,g\}}}s+i\hbar\{f,g\}s.
	\]
	Subtracting,
	\[
	\big([Q(f),Q(g)]-i\hbar Q(\{f,g\})\big)s
	=
	\big(i\hbar\{f,g\}-\hbar^2F_\nabla(X_f,X_g)\big)s.
	\]
	Hence \([Q(f),Q(g)] = i\hbar Q(\{f,g\})\) for all \(f,g\) if and only if
	\[
	F_\nabla(X_f,X_g)=\frac{i}{\hbar}\{f,g\}.
	\]
	Since \(\{f,g\}=\omega(X_g,X_f)=-\omega(X_f,X_g)\), this is
	\[
	F_\nabla(X_f,X_g)=-\frac{i}{\hbar}\omega(X_f,X_g).
	\]
	As this holds for all Hamiltonian vector fields (hence all tangent directions), we conclude
	\[
	F_\nabla=-\frac{i}{\hbar}\omega. \qedhere
	\]
\end{proof}

\subsubsection{Equivalence of Abstract and Local Curvature.}
To illuminate the step utilizing \(F_\nabla(X,Y)=[\nabla_X,\nabla_Y]-\nabla_{[X,Y]}\), it is instructive to verify that this abstract operator definition coincides with the local 2-form. Let \(s\) be a local section over a trivializing chart where the covariant derivative takes the form \(\nabla_X s = X(s) - i A(X)s\), with \(A\) being a real-valued connection 1-form (the gauge potential). Evaluating the abstract commutator on \(s\) yields:
\begin{align*}
	\nabla_X (\nabla_Y s) &= \nabla_X \big( Y(s) - i A(Y)s \big) \\
	&= X\big(Y(s)\big) - i X\big(A(Y)s\big) - i A(X)\big(Y(s) - i A(Y)s\big) \\
	&= X\big(Y(s)\big) - i X(A(Y))s - i A(Y)X(s) \\
	&\quad - i A(X)Y(s) - A(X)A(Y)s.
\end{align*}
When we subtract \(\nabla_Y (\nabla_X s)\) to compute the commutator, the symmetric cross-terms \(-iA(Y)X(s)\) and \(-iA(X)Y(s)\) cancel out. Furthermore, because the complex line bundle is abelian, the connection 1-forms map to commuting scalars (complex numbers), meaning \(A(X)A(Y) = A(Y)A(X)\), leading to the cancellation of the quadratic terms. The commutator reduces to:
\[
	[\nabla_X, \nabla_Y]s = [X,Y](s) - i \big( X(A(Y)) - Y(A(X)) \big) s.
\]
Next, we subtract the covariant derivative along the Lie bracket vector field, \(\nabla_{[X,Y]}s = [X,Y](s) - i A([X,Y])s\), obtaining:
\begin{align*}
	\big([\nabla_X, \nabla_Y] - \nabla_{[X,Y]}\big)s &= -i \Big( X(A(Y)) - Y(A(X)) \\
	&\quad - A([X,Y]) \Big) s \\
	&= -i dA(X,Y) s.
\end{align*}
The resulting factor inside the parentheses is precisely Cartan's invariant formula for the exterior derivative \(dA\), acting by scalar multiplication on the section. Denoting the real curvature 2-form by \(R = dA\), this algebraically confirms the equivalence \(F_\nabla(X,Y) = -i R(X,Y)\).

\subsubsection{Convention Note.}\label{sec:conventions}
Throughout this text, we employ the sign conventions: \(\iota_{X_f}\omega = -df\), \(\{f,g\} = \omega(X_g, X_f)\), and \(F_\nabla = -\frac{i}{\hbar}\omega\). This choice explicitly ensures that for the canonical symplectic form \(\omega = dp\wedge dq\), we obtain \(X_q = -\partial_p\), \(X_p = \partial_q\), and \(\{q,p\}=1\). Consequently, the quantization of coordinates faithfully reproduces the standard quantum commutation relation \([Q(q),Q(p)] = +i\hbar\hat{I}\).

\subparagraph{Verification of Quantization Requirements}
With the prequantum operator \(Q(f) = -i\hbar\, \nabla_{X_f} + f\) defined as above, we can explicitly verify that it satisfies the three structural requirements of quantization outlined at the beginning of this section:
\begin{enumerate}
	\item \textbf{Linearity}: The assignment of a Hamiltonian vector field, \(f \mapsto X_f\), is linear. Since the covariant derivative \(\nabla_X\) is linear with respect to the vector field \(X\) (i.e., \(\nabla_{aX+bY} = a\nabla_X + b\nabla_Y\)), the operator \(Q(f)\) is manifestly linear: \(Q(af + bg) = aQ(f) + bQ(g)\).
	\item \textbf{Constant Function}: For the constant function \(f = 1\), the exterior derivative is zero (\(d(1) = 0\)), yielding a vanishing Hamiltonian vector field (\(X_1 = 0\)). The derivative part of the operator disappears (\(-i\hbar\nabla_{X_1} = 0\)), leaving only multiplication by the function itself. Thus, \(Q(1)\) acts purely by multiplying by \(1\), meaning \(Q(1) = \hat{I}\).
	\item \textbf{Commutation Relations}: As demonstrated by the fundamental lemma, the relation \([Q(f), Q(g)] = i\hbar Q(\{f,g\})\) holds exactly. The operator is carefully engineered so that its commutator generates the curvature \(F_\nabla(X_f, X_g)\). The prequantization condition \(F_\nabla = -\frac{i}{\hbar}\omega\) ensures that this curvature term precisely absorbs the residual scalar contribution, reproducing the classical Poisson bracket \(\{f, g\} = \omega(X_g, X_f)\).
\end{enumerate}

\subsubsection{The Prequantum Line Bundle and Hilbert Space}

To realize this map, quantum states are identified with sections of a complex line bundle \(L\) over the classical phase space \((M,\omega)\), equipped with a Hermitian metric and a connection \(\nabla\)~\cite{Kostant1970,Woodhouse1992}.

The objective is to construct a pair \((L, \nabla)\) such that the curvature two-form \(F_\nabla = \operatorname{Curv}(\nabla)\) is proportional to the symplectic form \(\omega\)~\cite{Woodhouse1992}. This geometric condition is necessary to satisfy the commutation relations. The natural prequantum Hilbert space is
\[
\mathcal{H}_{\mathrm{pre}} = L^2\big(M, L; \mu_{\!L}\big),
\]
the space of square-integrable sections with respect to the Liouville volume \(\mu_{\!L}=\omega^n/(n!)\) and the fiber metric.

\subsection{Complex Line Bundles and the Integrality Condition}

\subsubsection{Classification of Complex Line Bundles}
The curvature condition forces a topological quantization: admissible symplectic forms must satisfy an integrality constraint. The construction of the prequantum line bundle requires understanding the classification of complex line bundles over manifolds. This classification is intimately connected to the integrality condition that determines when prequantization is possible. In the context of geometric quantization, the prequantum line bundle is not merely an auxiliary structure but the geometric carrier of quantum states. This classification is topological and relies on the correspondence between line bundles and the second integral cohomology group \(H^2(M,\mathbb{Z})\)~\cite{Woodhouse1992}.

The group \(H^2(M, \mathbb{Z})\) is the second singular cohomology group of the manifold \(M\) with coefficients in the abelian group of integers. Its free part pairs with closed two-dimensional cycles, while torsion classes are invisible to real de Rham periods. For example, the 2-sphere \(S^2\) has \(H^2(S^2, \mathbb{Z}) \cong \mathbb{Z}\), reflecting its fundamental 2-cycle, whereas a contractible space such as \(\mathbb{R}^n\) has a trivial second cohomology group. For a non-torsion class, its pairing with a 2-cycle evaluates to an integer. The relationship between these cohomology classes and complex line bundles is one of scaffolding: a non-trivial line bundle carries an integral Chern class that can pair nontrivially with such cycles. Thus, when applied to complex line bundles, the evaluated integer intuitively counts the discrete number of times the fiber twists over that 2-cycle. Geometrically, for a generic smooth global section, this integer corresponds to the net number of topological vortices, or in higher dimensions the codimension-two zero locus, where the section is geometrically forced to vanish. In this theory, the cohomology group contains the first Chern class, a topological invariant that quantifies this exact integer twist and represents the primary obstruction to the existence of a non-vanishing global section; such a section would trivialize the line bundle. Because this obstruction completely characterizes the bundle's topology, the distinct isomorphism classes of complex line bundles over \(M\) are in bijective correspondence with the elements of \(H^2(M, \mathbb{Z})\). In the context of geometric quantization, this means that the classical symplectic area (flux) through any closed 2-dimensional surface must correspond to an integer number of twists for the quantum line bundle to be globally well-defined; flat torsion line bundles, when present, carry additional discrete holonomy data without changing the real curvature periods.

\subparagraph{Curvature and Holonomy}
A complex line bundle \(L \to M\) with connection \(\nabla\) is characterized locally by a connection 1-form \(A\). To see this, choose a local non-vanishing section (a frame) \(s\) over an open set \(U\). Any section \(\psi\) over \(U\) can be written as \(\psi = f s\) for a smooth function \(f\). The connection \(\nabla\) satisfies the Leibniz rule \(\nabla(fs) = df \otimes s + f \nabla s\). Since \(\nabla s\) must be a section of \(L \otimes T^*M\), we can write \(\nabla s = -i A \otimes s\) for some 1-form \(A\) on \(U\). This 1-form \(A\) acts as the gauge potential (analogous to the electromagnetic potential). The curvature of this connection is the globally defined closed 2-form \(R = dA\). While \(A\) is only locally defined and depends on the choice of frame \(s\) (transforming as \(A' = A - d\chi\) under a change of frame \(s' = e^{i\chi}s\)), the curvature \(R\) is global and gauge-invariant.

	The obstruction to the existence of such a bundle is encoded in the holonomy of the connection. While we can always define \(A\) and \(R\) locally, constructing a global bundle requires these local descriptions to glue together consistently. The holonomy provides the test for this consistency. For a closed loop \(\gamma\) in \(M\), the holonomy is the phase factor by which a section is multiplied after parallel transport around \(\gamma\):
	\[
	\mathrm{hol}(\gamma) = \exp\left( i \oint_\gamma A \right).
	\]
For the bundle to be well-defined globally, the holonomy around any contractible loop must be determined solely by the curvature enclosed by the loop (via Stokes' theorem). If the holonomy depended on the choice of local potential \(A\) or the specific surface spanning the loop in a way that contradicted the curvature integral, a global bundle could not exist. This consistency requirement leads directly to the quantization of flux.

\subparagraph{The Integrality Condition}
Consider a closed oriented surface \(\Sigma \subset M\). By Stokes' theorem, the integral of the curvature \(R\) over \(\Sigma\) can be related to the holonomy around the boundary of patches covering \(\Sigma\). If we decompose \(\Sigma\) into two patches \(U_1\) and \(U_2\) with intersection \(C = U_1 \cap U_2\) (a circle), and let \(A_1, A_2\) be the connection forms on these patches, they are related on \(C\) by a gauge transformation. If the transition function is \(g_{12} = e^{i\chi_{12}}\), meaning the local frames are related by \(s_2 = e^{i\chi_{12}} s_1\), then by the same Leibniz calculation as the gauge transformation above:
\[
A_2 = A_1 - d\chi_{12}.
\]
The integral of the curvature over the entire closed surface is:
\[
\int_\Sigma R = \int_{U_1} dA_1 + \int_{U_2} dA_2 = \oint_{\partial U_1} A_1 + \oint_{\partial U_2} A_2.
\]
Noting that \(\partial U_1 = C\) and \(\partial U_2 = -C\) (with opposite orientation), we have:
\[
\int_\Sigma R = \oint_C (A_1 - A_2) = \oint_C d\chi_{12} = \Delta_C\chi_{12}.
\]
Since the transition function \(g_{12}\) must be single-valued, the change in phase \(\chi_{12}\) around the closed loop \(C\) must be an integer multiple of \(2\pi\):
\[
\Delta_C\chi_{12} = 2\pi k, \quad k \in \mathbb{Z}.
\]
Thus, we arrive at the fundamental integrality condition:
\[
\frac{1}{2\pi} \int_\Sigma R \in \mathbb{Z}.
\]
This implies that the cohomology class of the form \(R/(2\pi)\) is an \emph{integral cohomology class}.

\begin{definition}[Integral Cohomology Class]
In algebraic topology, cohomology can be computed using different coefficients, such as integers \(\mathbb{Z}\) or real numbers \(\mathbb{R}\). The standard inclusion of integers into the reals, \( \mathbb{Z} \hookrightarrow \mathbb{R} \), induces a natural homomorphism between their respective cohomology groups:
\[
\iota: H^2(M, \mathbb{Z}) \to H^2(M, \mathbb{R}).
\]
By de Rham's theorem, the real cohomology group \( H^2(M, \mathbb{R}) \) is isomorphic to the de Rham cohomology group composed of closed 2-forms modulo exact 2-forms. A de Rham cohomology class \( [\Omega] \) is called \textbf{integral} if it lies in the image of this natural map \(\iota\), meaning there exists an underlying geometric integer class that maps to it.
\end{definition}

A real cohomology class represented by a closed 2-form \(\Omega\) is integral if and only if its period over \emph{every} integral \(2\)-cycle \(z \in H_2(M,\mathbb{Z})\) is an integer:
\[
\int_z \Omega \in \mathbb{Z}.
\]
This theorem forms the rigid topological backbone of geometric quantization. While a symplectic form \(\omega\) is a purely continuous object describing phase-space geometry, requiring that the scaled class \([\omega / 2\pi\hbar]\) be integral directly enforces that the macroscopic classical flux (phase-space area) over any closed 2-cycle is strictly quantized in discrete units of Planck's constant.

\subparagraph{Relating Real and Formal Curvatures}
Throughout this discussion, different notations for curvature naturally arise depending on whether the context is topological integration or algebraic operators. The real-valued closed 2-form introduced earlier as \( R = dA \) (where \( A \) is the real gauge potential) represents the physical field strength. It is used during the discussion of holonomy and the integrality condition because it is strictly real-valued, making the integrality condition \( \frac{1}{2\pi} \int_\Sigma R \in \mathbb{Z} \) intuitive as a quantization of macroscopic physical flux.

When it satisfies this analytical integrality condition, its de Rham class represents the real image of the first Chern class:
\[
\iota\!\big(c_1(L)\big) = \left[ \frac{1}{2\pi} R \right] \in H^2(M,\mathbb{R}),
\]
where \(\iota:H^2(M,\mathbb{Z})\to H^2(M,\mathbb{R})\) is the natural map.

Conversely, in formal proofs and operator algebras (such as the commutator proof in Lemma~\ref{lem:prequantization}), the curvature is defined intrinsically via the covariant derivative: \( F_\nabla(X, Y) = [\nabla_X, \nabla_Y] - \nabla_{[X,Y]} \). Because the prequantum connection is unitary and constructed locally as \( \nabla = d - iA \) to preserve probability on a complex Hilbert space, applying two covariant derivatives mathematically forces the imaginary unit into the curvature expression. As verified locally above, this yields the exact relation:
\[
F_\nabla = -i R.
\]
This aligns perfectly with the strict mathematical convention utilized in the formal proofs, where the formal curvature \(F_{\nabla}\) is inherently purely imaginary (anti-Hermitian). This imaginary nature is not arbitrary; it is a direct geometric consequence of unitarity. Because the prequantum line bundle is equipped with a Hermitian metric (to ensure quantum probabilities are preserved), the connection must be metrically compatible, meaning that probability does not ``leak'' during parallel transport. This rigid compatibility restricts the intrinsic connection 1-form \(-iA\) to take values exclusively in the purely imaginary Lie algebra \(\mathfrak{u}(1)\) of the structure group \(U(1)\). Since the intrinsic connection is purely imaginary, its exterior derivative, the formal curvature \(F_\nabla = d(-iA) = -i dA = -iR\), must also be purely imaginary. By cleanly absorbing this imaginary geometry, \(F_\nabla\) matches the quantum operator algebra directly, leading to the ubiquitous topological formula \(\iota\!\big(c_1(L)\big) = \big[\frac{i}{2\pi} F_{\nabla}\big]\) in \(H^2(M,\mathbb{R})\).

Having established the operator algebra, we now restrict the prequantum Hilbert space via a choice of polarization.

\section{Polarization}
\label{sec:polarization-manifold}

The prequantum Hilbert space constructed above, while geometrically natural, is physically unsatisfactory: it is far too large. For a system with $n$ degrees of freedom, the prequantum wavefunctions depend on all $2n$ phase space variables (both positions and momenta), whereas physical wavefunctions should depend on only $n$ variables (e.g., positions alone, or momenta alone). In representation theoretic terms, the prequantum representation is highly reducible.

To obtain a physical quantum theory, we must cut the number of variables in half. Geometrically, this is achieved by choosing a \emph{polarization}---a specific selection of directions along which the wavefunctions must be constant~\cite{Woodhouse1992,Sniatycki1980}.

\subsection{Geometric Definition and the Polarized Hilbert Space}

A polarization selects a maximal set of commuting directions at each point in phase space. Formally, a \emph{polarization} \(P\) on a symplectic manifold \((M,\omega)\) is a Lagrangian subbundle of the complexified tangent bundle \(TM^{\mathbb{C}}\). That is, \(P\) is a complex subbundle of rank \(n = \tfrac{1}{2}\dim M\) such that \(\omega|_P = 0\) and \(P\) is closed under the Lie bracket (integrable)~\cite{Woodhouse1992}.

\subsubsection{Integrability vs. Integrality.}
It is essential to distinguish the \emph{integrability} of a polarization from the \emph{integrality} condition encountered in prequantization (see Section~\ref{sec:prequantization-manifold}). While the terms are linguistically similar, they address distinct requirements:
\begin{itemize}
    \item \textbf{Integrality} is a global \emph{topological} condition on the symplectic form $\omega$, ensuring the existence of the prequantum line bundle.
    \item \textbf{Integrability} is a local \emph{differential-geometric} condition on the distribution $P$, ensuring that it is involutive so that it integrates to a foliation.
\end{itemize}
In the context of real polarizations, these concepts meet in the Bohr--Sommerfeld condition (Section~\ref{sec:real-polarizations-manifold}), where the existence of the bundle (integrality) and the structure of the leaves (integrability) together determine the discrete set of physically allowed quantum states.

The condition \(\omega|_P = 0\) ensures that the variables constant along \(P\) commute with each other, while the rank condition ensures we eliminate exactly half the degrees of freedom.

The integrability condition---that \(P\) is closed under the Lie bracket---is mathematically crucial to avoid over-constraining the quantum states. In quantum mechanics, requiring a physical state \(\psi\) to be ``covariantly constant'' along a direction \(X \in P\) (i.e., \(\nabla_X \psi = 0\)) is the geometric way of stating that the wavefunction does not depend on the variable associated with \(X\). For example, if we require the state to be constant along all momentum directions (\(\nabla_{\partial/\partial p} \psi = 0\)), we are enforcing that the wavefunction depends only on position, \(\psi = \psi(q)\).

If we require physical states \(\psi\) to be covariantly constant along two such commuting directions \(X, Y \in P\) (i.e., \(\nabla_X \psi = \nabla_Y \psi = 0\)), the curvature of the connection enforces
\[
[\nabla_X, \nabla_Y]\psi - \nabla_{[X,Y]}\psi = -\frac{i}{\hbar} \omega(X,Y) \psi.
\]
Because \(P\) is a Lagrangian subbundle, \(\omega(X,Y) = 0\), and by the assumed constancy, \([\nabla_X, \nabla_Y]\psi = 0\). This leaves \(\nabla_{[X,Y]}\psi = 0\), which forces the wavefunction to additionally be constant along the bracket direction \([X,Y]\). If \([X,Y]\) were outside of \(P\), this would unintentionally impose new, independent constraints~\cite{Woodhouse1992}.

In quantum mechanical terms, choosing a polarization corresponds to selecting a \emph{representation} (such as the position or momentum representation). Because the uncertainty principle forbids simultaneous eigenstates for conjugate variables, a physical quantum state cannot depend on all \(2n\) phase space parameters. The Lie bracket integrability condition guarantees that the chosen directions that we ``ignore'' form a closed algebra, so locally the quotient by the polarization is \(n\)-dimensional. In the case of a real polarization, this corresponds to retaining \(n\) commuting variables, allowing one to build the corresponding representation without contradictory constraints.

Once a polarization \(P\) is chosen, the physical Hilbert space, or \emph{polarized Hilbert space} \(\mathcal{H}_{\mathrm{pol}}\), consists of those prequantum sections \(s\) that are covariantly constant along \(P\)~\cite{Sniatycki1980,Woodhouse1992}:
\[
\nabla_X s = 0 \qquad \text{for all } X \in P.
\]
Operators are similarly restricted: a classical observable \(f\) is \emph{quantizable} only if its Hamiltonian flow preserves the polarization (i.e., \([X_f, P] \subset P\)). For such observables, the quantum operator is defined as:
\[
Q_{\mathrm{pol}}(f) = -i\hbar\,\nabla_{X_f} + f.
\]

\subsection{Real Polarizations}

The most intuitive polarizations are \emph{real polarizations}, which are constructed from strictly real vector fields on the phase space.

\begin{definition}[Real Polarization]
\label{def:real-polarization}
A real polarization on a symplectic manifold \((M,\omega)\) is a polarization \(P \subset TM^{\mathbb{C}}\) that takes the form \(P = D \otimes \mathbb{C}\), where \(D \subset TM\) is a real distribution.
\end{definition}

In differential geometry, a distribution \(D\) is simply a smooth assignment of a specific vector subspace \(D_x\) to the tangent space \(T_x M\) at every point \(x \in M\). For \(D \otimes \mathbb{C}\) to constitute a valid polarization for geometric quantization, the underlying real distribution \(D\) must satisfy three strict rules:
\begin{enumerate}
    \item \textbf{Rank:} It must have rank \(n\) (half the dimension of the \(2n\)-dimensional phase space).
    \item \textbf{Lagrangian:} The symplectic form \(\omega\) must vanish when restricted to \(D\), ensuring that all constrained observables mutually commute.
    \item \textbf{Integrability:} It must be closed under the Lie bracket, also known as being \emph{involutive}.
\end{enumerate}

Geometrically, \(D\) serves as the blueprint defining the ``allowed'' local directions along which physical states must be constant. The integrability requirement ensures that this local blueprint can be integrated into tangible, global surfaces spanning the phase space. By Frobenius' theorem, an involutive regular distribution \(D \subset TM\) uniquely integrates to form a \emph{foliation} of the manifold.

\begin{definition}[Foliation and Leaf]
A \emph{foliation} is a partition of \(M\) into a disjoint union of connected, immersed submanifolds called \emph{leaves}. For a real polarization generated by a distribution \(D\), a leaf \(\Lambda\) is a maximal connected submanifold of \(M\) such that its tangent space \(T_x\Lambda\) coincides with the subspace \(D_x\) at every point \(x \in \Lambda\).
\end{definition}

Locally, a foliation guarantees that around any point there exist ``flat'' coordinates where the leaves are simply the slices defined by setting half the coordinates to constants. The standard physical example is the \emph{vertical polarization} on a cotangent bundle \(T^*Q\). Here, the leaves are the cotangent fibers (the momentum spaces attached to each point \(q \in Q\)). Requiring prequantum wavefunctions to be covariantly constant along these vertical leaves physically restricts them to depend strictly on the base space variables \(q\), thereby recovering the standard Schr\"odinger position representation.

\subsection{Bohr--Sommerfeld Quantization}
\label{sec:real-polarizations-manifold}

For real polarizations, the covariant constancy condition \(\nabla_X s = 0\) for all \(X \in P\) imposes severe global restrictions on the prequantum sections. By definition, the real distribution \(D\) generating the polarization restricts to the tangent bundle \(T\Lambda\) of the leaves \(\Lambda\). Since \(\Lambda\) is Lagrangian, the symplectic form vanishes on it: \(\omega|_\Lambda = 0\). Consequently, the prequantum connection \(\nabla\), whose curvature is proportional to \(\omega\), restricts to a \emph{flat connection} on the prequantum line bundle \({L|_\Lambda}\) over each leaf.

Because the connection is flat, locally on any simply connected open subset of \(\Lambda\), a non-zero covariantly constant section always exists and is determined uniquely by its value at a single point via parallel transport. However, extending this section globally over a leaf that is not simply connected requires the holonomy of the flat connection to be trivial. If the holonomy around any non-contractible loop is non-trivial, a section parallel transported around the loop will acquire a phase factor. For the section to be cleanly single-valued and continuous globally, this phase factor must be strictly the identity; otherwise, any global polarized section must vanish on that leaf. Thus, non-zero global polarized sections can occur only along those special leaves where the holonomy is trivial.

To make this explicit, consider the case where the symplectic form is globally exact, admitting a symplectic potential \(\theta\) such that \(\omega = d\theta\) (as in a cotangent bundle with \(\theta = p\,dq\)). Following our conventions, we can choose a global trivializing frame \(s_0\) such that the connection acts via \(\nabla s_0 = -iA \otimes s_0\). To recover the curvature \(F_\nabla = -\frac{i}{\hbar}\omega = -idA\), we must set the real connection 1-form to \(A = \frac{1}{\hbar}\theta\). A general section \(s = f s_0\) satisfies the covariant constancy equation \(\nabla s = 0\) along a leaf \(\Lambda\) when \(df - i A f = 0\). This implies the section's coefficient is given by \(f(x) \propto \exp\left(i\int_{x_0}^x A\right) = \exp\left(\frac{i}{\hbar}\int_{x_0}^x \theta\right)\). When we transport \(s\) around a closed loop \(\gamma \subset \Lambda\), it acquires a holonomy phase factor \(\operatorname{Hol}_\nabla(\gamma) = \exp\left(i\oint_\gamma A\right) = \exp\left(\frac{i}{\hbar} \oint_\gamma \theta\right)\). For the section to be single-valued globally, this phase factor must be 1, requiring its argument to be an integer multiple of \(2\pi\). This geometrically fundamental requirement is known as the Bohr--Sommerfeld condition.

\begin{theorem}[Bohr--Sommerfeld~\cite{Woodhouse1992}]\label{thm:BS}
A Lagrangian leaf \(\Lambda\) supports a non-zero global polarized section if and only if the holonomy of the prequantum connection restricts to the identity on \(\Lambda\). When \(\omega = d\theta\) is exact, this is equivalent to the condition
\[
\frac{1}{2\pi\hbar} \oint_\gamma \theta \in \mathbb{Z}
\]
for every closed loop \(\gamma \subset \Lambda\).
\end{theorem}

\noindent
Leaves that satisfy this condition are called \textbf{Bohr--Sommerfeld leaves}. In systems possessing completely integrable dynamics, the leaves are invariant tori (such as those constructed from action-angle variables). The Bohr--Sommerfeld condition then selects a discrete set of these tori---quantizing the classical action variables---recovering the identical rules of the ``old quantum theory''.

\subsubsection{Relation to the Integrality Condition of Prequantization}

The Bohr--Sommerfeld quantization condition and the prequantum integrality condition both involve the quantization of a geometric integral into integer units of Planck's constant.

\begin{remark}[Integrality vs. Integrability]
\label{rem:integrality-vs-integrability-manifold}
The \emph{integrality} condition discussed here is a global topological requirement on the symplectic form $\omega$, ensuring the existence of the prequantum line bundle. It should not be confused with the \emph{integrability} of a polarization (Definition~\ref{def:real-polarization}), which is a local differential-geometric condition on the involutivity of a distribution.
\end{remark}

However, they operate at different stages of the quantization process and on topological spaces of different dimensions.

The Integrality Condition of prequantization imposes a global topological requirement on the entire symplectic manifold \(M\): for every integral \(2\)-cycle \(z \in H_2(M,\mathbb{Z})\), the symplectic period must satisfy \(\frac{1}{2\pi\hbar} \int_z \omega \in \mathbb{Z}\). This condition acts as an \emph{existence theorem}, ensuring the global consistency of the prequantum line bundle by restricting its curvature's macroscopic flux such that its associated first Chern class is integral.

\subsection{Complex and K\"ahler Polarizations}

Real polarizations yield quantum representations with profound ties to classical trajectories, but their analytic behavior depends on the foliation. For instance, the vertical polarization on \(T^*Q\) gives the standard Schr\"odinger representation, with polarized states depending on the base variables \(q\). In contrast, real polarizations with compact leaves or non-trivial leaf holonomy lead to Bohr--Sommerfeld conditions and may be represented by states supported on the Bohr--Sommerfeld leaves. To obtain a holomorphic representation adapted to a complex structure, we may instead use complex polarizations. Because polarizations are formally defined within the complexified tangent bundle \(TM^{\mathbb{C}}\), they can naturally accommodate complex directions.

\begin{definition}[Complex and Purely Complex Polarization]
A \emph{complex polarization} is a polarization \(P \subset TM^{\mathbb{C}}\) that is not strictly real, meaning it differs from its complex conjugate bundle: \(P \neq \overline{P}\). The intersection \(P \cap \overline{P}\) corresponds to the real directions preserved by the polarization. If this intersection is trivial, \(P \cap \overline{P} = \{0\}\), the polarization consists entirely of independent complex directions and is called \emph{purely complex}.
\end{definition}

The inclusion of complex directions fundamentally alters the mathematical nature of the quantum states. Rather than acting as a directional derivative that forces the state to be constant along a real spatial axis, the covariant constancy condition \(\nabla_X s = 0\) (for \(X \in P\)) transforms into a system of generalized Cauchy--Riemann equations. Consequently, a complex polarization demands that physical wavefunctions be \emph{holomorphic} (or partially holomorphic along the complex directions) with respect to the complex coordinate variables defined by the polarization.

The most physically significant and mathematically elegant class of purely complex polarizations arises from K\"ahler geometry. Suppose the phase space \((M, \omega)\) admits a compatible complex structure \(J\), making it a K\"ahler manifold. We can canonically construct a purely complex polarization by choosing \(P\) to be the antiholomorphic tangent bundle, \(P = T^{0,1}M\).

In this K\"ahler polarization, the physical states must be annihilated by the covariant antiholomorphic derivatives:
\[
\nabla_{\bar{Z}} s = 0 \qquad \text{for all } \bar{Z} \in T^{0,1}M.
\]
Because the K\"ahler form \(\omega\) is of type \((1,1)\), the prequantum curvature \(F_\nabla=-\frac{i}{\hbar}\omega\) has no \((0,2)\) component. Hence \((\nabla^{0,1})^2=0\), so \(\nabla^{0,1}\) defines a holomorphic structure on \(L\). The polarization condition above is therefore precisely the Cauchy--Riemann equation for sections of \(L\), and the polarized Hilbert space \(\mathcal{H}_{\mathrm{pol}}\) consists of the \emph{holomorphic sections} of the prequantum line bundle \(L\).

By identifying quantum states with holomorphic sections, K\"ahler polarizations replace the leafwise Bohr--Sommerfeld picture by a smooth holomorphic one. The resulting sections are smooth and single-valued as global sections of the line bundle, while their zeros and normalizability are controlled by the line bundle metric and the global geometry.

\section{The Metaplectic Correction}

While polarization reduces the prequantum Hilbert space to the physically appropriate ``size,'' it overlooks a fundamental geometric requirement: quantum wavefunctions cannot transform simply as scalars if their squared modulus is to yield a well-defined probability distribution. This omission leads to significant spectral discrepancies, most notably the absence of zero-point energy in the harmonic oscillator. The \emph{metaplectic} (or \emph{half-form}) correction resolves this by tensoring the prequantum line bundle with a square root of the canonical bundle associated with the polarization~\cite{Woodhouse1992,Sniatycki1980}.

\subsubsection{Half-Forms and Densities}

In classical mechanics, we integrate functions against a volume form to compute observables. In quantum mechanics, however, the situation is different: the probability density $|\psi|^2$ must itself be a volume form, so that integrating it gives unity. This implies that the wavefunction $\psi$ should transform as a ``square root'' of a volume form---a \emph{half-density}.

For a Lagrangian polarization \(P \subset TM^{\mathbb{C}}\), let \(P^\circ \subset T^*M^{\mathbb{C}}\) denote its annihilator (covectors that vanish on \(P\)). The top exterior power
\[
K_P \;=\; \bigwedge^{n} P^\circ
\]
is the canonical bundle associated to the polarization. For a K\"ahler polarization \(P=T^{0,1}M\), one has \(P^\circ=\Omega^{1,0}M\), so \(K_P\) is the usual canonical bundle \(K_M\). For a real polarization, \(P^\circ\) is the complexified conormal bundle, so \(K_P\) encodes transverse densities. We define the \emph{half-form bundle} \(\delta_P\) as a square root of \(K_P\).

In order to construct physically meaningful wavefunctions, we form the tensor product \(L \otimes \delta_P\). Sections of the prequantum line bundle \(L\) supply the complex phase information that encodes the symplectic geometry, but they transform merely as scalars. Meanwhile, sections of \(\delta_P\) supply the transformation property of a ``half-volume''. Taking their tensor product ensures that the inner product of two wavefunctions (or the absolute square \(|\psi|^2\)) naturally yields a genuine, coordinate-independent density that can be integrated.

\subsubsection{The Corrected Hilbert Space and Operators}

With the half-form bundle in hand, we can now define the corrected quantum structures. The metaplectically corrected state space is formed from sections of \(L \otimes \delta_P\) that are covariantly constant along the polarization:
\[
\mathcal{H}_{\mathrm{meta}}=\bigl\{\psi\in\Gamma(L\otimes\delta_P)\,\bigm|\,\nabla^{L\otimes\delta_P}_X\psi=0\ \text{for all }X\in P\bigr\},
\]
where \(L\) is the prequantum line bundle over \((M,\omega)\), \(\delta_P\) is the half-form bundle, and \(\nabla^{L\otimes\delta_P}_X\) denotes the derivative along \(X\in P\) obtained from the prequantum connection on \(L\) together with the half-form Lie derivative on \(\delta_P\). In K\"ahler cases one then takes the usual \(L^2\)-completion, while for real polarizations the half-form factor supplies the correct transverse density on the leaf space.

For an observable \(f\) whose Hamiltonian flow preserves the polarization \(P\), the corrected operator on a decomposable section \(\psi=s\otimes\nu\) is
\[
Q_{\mathrm{meta}}(f)(s\otimes\nu)=\bigl(Q_{\mathrm{pre}}(f)s\bigr)\otimes\nu-i\hbar\,s\otimes\bigl(\mathcal{L}^{1/2}_{X_f}\nu\bigr),
\]
with
\[
Q_{\mathrm{pre}}(f)=-i\hbar\nabla^L_{X_f}+f.
\]
Here \(\mathcal{L}^{1/2}_{X_f}\) denotes the Lie derivative on half-forms, defined by
\[
2\bigl(\mathcal{L}^{1/2}_{X_f}\nu\bigr)\otimes\nu
\;=\;
\mathcal{L}_{X_f}(\nu\otimes\nu)
\;=\;
\bigl(\operatorname{div}_P X_f\bigr)(\nu\otimes\nu).
\]
Hence
\[
\mathcal{L}^{1/2}_{X_f}\nu=\tfrac12\bigl(\operatorname{div}_P X_f\bigr)\nu,
\]
and therefore
\[
Q_{\mathrm{meta}}(f)(s\otimes\nu)
=
\left[
Q_{\mathrm{pre}}(f)s
-\frac{i\hbar}{2}\bigl(\operatorname{div}_P X_f\bigr)s
\right]\otimes\nu.
\]
The extra term \(-\tfrac{i\hbar}{2}\operatorname{div}_P X_f\) is exactly what restores formal symmetry: in the inner-product computation, integration by parts of \(-i\hbar\nabla^L_{X_f}\) produces a divergence contribution, and the half-form term cancels it.

\subsubsection{The Maslov Index and Spectral Shifts}

Beyond ensuring the self-adjointness of quantum observables, the half-form bundle has profound physical implications for semiclassical quantization. Consider the WKB approximation: as a particle's semiclassical wavefunction propagates along a classical trajectory \(\gamma\), it accumulates a dynamical phase governed by the classical action. However, when the trajectory passes through a classical turning point---a caustic where the classical probability density formally diverges---the wavefunction experiences an abrupt phase shift of \(-\pi/2\).

This phenomenon finds a natural geometric explanation in the framework of half-forms. Parallel-transporting a half-form \(\nu\) around a closed orbit \(\gamma\) on a Lagrangian invariant torus tracks these phase jumps precisely. Each caustic crossing rotates the half-form by a phase of \(e^{-i\pi/2}\), so the total half-form holonomy around the orbit is given by \(e^{-i\mu(\gamma)\pi/2}\). Here, the integer \(\mu(\gamma)\) counts (with sign) the number of caustic crossings and is recognized as a fundamental topological invariant known as the \emph{Maslov index}~\cite{maslov}.

Crucially, the quantum single-valuedness condition applies not to \(\nu\) alone, but to the combined section of \(L\otimes\delta_P\). This means the holonomy of the prequantum connection and the half-form holonomy must together yield unity around every closed orbit. Consequently, this geometric phase requirement directly modifies the naive Bohr--Sommerfeld quantization rule. While prequantization alone leads to the primitive condition \(\frac{1}{2\pi\hbar}\oint_\gamma \theta \in \mathbb{Z}\), incorporating the half-form holonomy correctly shifts the quantized action by the Maslov phase. This yields the WKB quantization condition, often called the Bohr--Sommerfeld--Maslov rule~\cite{maslov}:
\[
\frac{1}{2\pi\hbar}\oint_\gamma \theta - \frac{\mu(\gamma)}{4} \in \mathbb{Z}.
\]
To see this in practice, consider the one-dimensional harmonic oscillator, where the action integral is \(J = \oint p\,dq\). A complete cycle in phase space encounters two classical turning points (the extremal amplitudes of oscillation), giving a Maslov index of \(\mu=2\). The corrected quantization condition then demands \(\frac{J}{2\pi\hbar} - \frac{2}{4} = n\), with \(n\in\mathbb{Z}_{\ge 0}\), or equivalently \(J = 2\pi\hbar(n+\tfrac{1}{2})\). Plugging this back into the classical Hamiltonian \(E = \omega J/(2\pi)\), we naturally recover the exact quantum energy levels:
\[
E_n = \hbar\omega\bigl(n + \tfrac{1}{2}\bigr).
\]
In this light, the emergence of the oscillator zero-point energy \(\tfrac{1}{2}\hbar\omega\) is thoroughly demystified. It is not an ad~hoc correction patched in to match experiment, but rather the contribution of the half-form geometry for this polarization and orbit. More generally, the Maslov index supplies the geometric phase term \(\mu(\gamma)/4\) in the Bohr--Sommerfeld--Maslov rule, so spectral shifts depend on the Maslov index, boundary conditions, and the chosen polarization rather than being universally equal to \(\tfrac{1}{2}\).

\section{Examples}\label{sec:examples}

Having developed the general framework, we now illustrate geometric quantization through three canonical examples of increasing topological complexity. We begin with the harmonic oscillator on the simply connected phase space \(\mathbb{R}^2\), proceed to the Dirac monopole whose non-trivial topology enforces charge quantization, and conclude with the Landau problem where magnetic flux quantization underlies the quantum Hall effect.

\subsection{The Harmonic Oscillator: A Quantum Mechanical Workhorse}

The harmonic oscillator serves as a canonical test case: it is analytically tractable and exhibits the essential structures of geometric quantization.

Consider a one-dimensional harmonic oscillator with classical Hamiltonian
\[
H = \frac{p^2}{2m} + \frac{1}{2}m\omega^2 q^2.
\]
The phase space is \(M = \mathbb{R}^2\) with canonical symplectic form \(\omega = dp \wedge dq\).

\subsubsection{Prequantization}

Since \(\mathbb{R}^2\) is contractible, the second real cohomology group \(H^2(M, \mathbb{R})\) is trivial, so the de Rham class \([\omega/(2\pi\hbar)]\) vanishes and is therefore integral. Choosing the symplectic potential \(\theta = p\,dq\) such that \(d\theta = \omega\), the prequantum line bundle \(L = M \times \mathbb{C}\) is trivial. The prequantum connection is then given by
\[
\nabla = d - \tfrac{i}{\hbar}\,\theta = d - \tfrac{i}{\hbar}\,p\,dq,
\]
with curvature 2-form \(F_\nabla = -\tfrac{i}{\hbar}\,\omega\).

For any classical observable \(f\in C^\infty(M)\), the Hamiltonian vector field \(X_f\) satisfies \(\iota_{X_f}\omega = -df\). To find the explicit form of \(X_f\), we write an arbitrary vector field as \(X_f = A\,\partial_q + B\,\partial_p\). The interior product with the symplectic form \(\omega = dp \wedge dq\) is evaluated as
\begin{multline*}
\iota_{X_f}\omega = \iota_{A\,\partial_q + B\,\partial_p}(dp \wedge dq) \\
= (\iota_{A\,\partial_q + B\,\partial_p}dp) \wedge dq - dp \wedge (\iota_{A\,\partial_q + B\,\partial_p}dq) = B\,dq - A\,dp.
\end{multline*}
Equating this to the exterior derivative \(-df = -\frac{\partial f}{\partial q}\,dq - \frac{\partial f}{\partial p}\,dp\) yields the coefficients
\[
A = \frac{\partial f}{\partial p} \qquad \text{and} \qquad B = -\frac{\partial f}{\partial q}.
\]
Thus, the general expression for the Hamiltonian vector field is
\[
X_f = \frac{\partial f}{\partial p}\,\partial_q - \frac{\partial f}{\partial q}\,\partial_p.
\]
Applying this formula to the coordinate functions (\(f=q\) and \(f=p\)), and to the classical Hamiltonian \(H\), explicitly yields
\[
X_q = -\partial_p,\qquad X_p = \partial_q,\qquad X_H = -m\omega^2 q\,\partial_p + \frac{p}{m}\,\partial_q.
\]
The Kostant--Souriau prequantum operator is \(Q(f) = -i\hbar\,\nabla_{X_f} + f\). Using the definition of the covariant derivative \(\nabla_{X_f} = X_f - \frac{i}{\hbar}\theta(X_f)\) with the symplectic potential \(\theta = p\,dq\), the operator acts as
\[
Q(f) = -i\hbar \left( X_f - \frac{i}{\hbar}\theta(X_f) \right) + f = -i\hbar X_f - \theta(X_f) + f.
\]
For the position coordinate \(f = q\), the Hamiltonian vector field is \(X_q = -\partial_p\). Evaluating the symplectic potential on \(X_q\) gives \(\theta(X_q) = p\,dq(-\partial_p) = 0\). Substituting these into the operator expression yields
\[
Q(q) = -i\hbar(-\partial_p) - 0 + q = i\hbar\,\partial_p + q.
\]
Similarly, for the momentum coordinate \(f = p\), the corresponding vector field is \(X_p = \partial_q\). Evaluating \(\theta\) on \(X_p\) gives \(\theta(X_p) = p\,dq(\partial_q) = p\). This results in
\[
Q(p) = -i\hbar(\partial_q) - p + p = -i\hbar\,\partial_q.
\]
One readily verifies the prequantum commutation relation \([Q(q), Q(p)] = i\hbar\hat{I}\). With \(\{f,g\} = \omega(X_g, X_f)\), the canonical bracket evaluates to \(\{q,p\} = \omega(X_p, X_q) = 1\). This classically positive bracket faithfully propagates through the representation to yield the standard quantum mechanical commutator \([Q(q), Q(p)] = i\hbar Q(\{q,p\}) = i\hbar\hat{I}\).

\subsubsection{The Schr\"odinger Representation (Real Polarization)}

To pass from prequantization to true quantization, we must choose a polarization. The vertical real polarization \(P = \mathrm{span}\{\partial_p\}\) is the natural choice for the Schr\"odinger representation. A section \(s\) of \(L\) is polarized if \(\nabla_{\partial_p}s = 0\), which forces
\[
\partial_p s = 0 \quad\Rightarrow\quad s(q,p) = \psi(q).
\]
Since the Liouville measure is infinite along the fibres, after restricting to polarized sections we quotient out the \(p\)-dependence and use the induced base measure \(dq\); thus \(\psi\in L^2(\mathbb{R}_q,dq)\) and \(\mathcal{H}_{\mathrm{pol}} = L^2(\mathbb{R}_q)\).

On polarized sections, the prequantum operators reduce to the standard Schr\"odinger representation. To see this, we apply the operators $Q(q) = i\hbar\,\partial_p + q$ and $Q(p) = -i\hbar\,\partial_q$ to a polarized section $\psi\in L^2(\mathbb{R}_q)$. Since a polarized section is independent of $p$ we have $\partial_p\psi = 0$. Thus,
\[
Q(q)\,\psi = (i\hbar\,\partial_p + q)\,\psi = i\hbar(0) + q\,\psi = q\,\psi,
\]
and
\[
Q(p)\,\psi = -i\hbar\,\partial_q\psi.
\]

\subsubsection{Metaplectic Correction and the Heisenberg Algebra}

To rigorously define the inner product, one must incorporate the metaplectic correction. The half-form bundle \(\delta_P\) over \(\mathbb{R}^2\) is the square root of the line bundle of 1-forms annihilating \(P\), namely \(\bigwedge^1 P^\circ \cong \mathbb{C}\,dq\). Sections of \(L\otimes\delta_P\) transform as \(\psi(q)\,|dq|^{1/2}\), and the metaplectically corrected Hilbert space is
\[
\mathcal{H}_{\mathrm{meta}} = L^2(\mathbb{R}_q, |dq|).
\]
The Kostant--Souriau operator \(Q_{\mathrm{meta}}(f)\) evaluated on half-forms acquires a correction term from the Lie derivative \(L_{X_f}|dq|^{1/2} = \frac{1}{2}(\partial_q A)|dq|^{1/2}\), where \(X_f = A\,\partial_q + B\,\partial_p\). Thus,
\[
Q_{\mathrm{meta}}(f) = Q_{\mathrm{pol}}(f) - \tfrac{i\hbar}{2}\,\partial_q A.
\]
Let us calculate the new position and momentum operators explicitly. For the position coordinate \(f = q\), we found earlier that \(X_q = -\partial_p\). Here \(A=0\), so its correction vanishes: \(\partial_q(0) = 0\). Consequently, the metaplectic correction term is zero, and the position operator remains unchanged:
\[
Q_{\mathrm{meta}}(q)\psi = Q_{\mathrm{pol}}(q)\psi - \tfrac{i\hbar}{2}(0)\psi = q\,\psi.
\]
Similarly, for the momentum coordinate \(f = p\), the Hamiltonian vector field is \(X_p = \partial_q\). Here \(A=1\), whose derivative is also trivially zero: \(\partial_q(1) = 0\). Thus, the momentum operator also remains unchanged:
\[
Q_{\mathrm{meta}}(p)\psi = Q_{\mathrm{pol}}(p)\psi - \tfrac{i\hbar}{2}(0)\psi = -i\hbar\,\partial_q\psi.
\]
Because the base-projected divergence \(\partial_q A\) is zero for both generators, the linear operators forming the Heisenberg algebra are entirely unaffected by the metaplectic correction. Note that the classical Hamiltonian \(H\) does not preserve the real polarization (\([X_H, \partial_p] \notin P\)), so it is not a directly quantizable first-order Kostant--Souriau observable in this representation. This does not remove the physical Hamiltonian: in the Schr\"odinger realization it is introduced as the second-order operator \(-\hbar^2\partial_q^2/(2m)+m\omega^2q^2/2\), while the semiclassical Bohr--Sommerfeld calculation below recovers its spectrum.

\subsubsection{Bohr--Sommerfeld Quantization}

The Bohr--Sommerfeld condition provides a semiclassical perspective on the energy spectrum. To apply it to the harmonic oscillator, one uses the real polarization by invariant energy curves (equivalently, the action-angle polarization on \(\mathbb{R}^2\setminus\{0\}\)), whose leaves are the level sets \(H=E\). The action integral around such a leaf is
\[
\oint_{H=E} \theta = \oint p\,dq = \frac{2\pi E}{\omega}.
\]
Applying the Bohr--Sommerfeld rule with Maslov correction (\(\mu=2\) for the ellipse) yields
\[
\frac{1}{2\pi\hbar}\oint p\,dq - \frac{\mu}{4} = \frac{E}{\hbar\omega} - \frac{1}{2} = n \in \mathbb{Z}_{\ge 0},
\]
whence \(E_n = \hbar\omega(n+\tfrac{1}{2})\), recovering the metaplectically corrected spectrum.

\subsubsection{The Bargmann--Fock Representation (Complex Polarization)}

The real polarization is not the only choice. An equally natural alternative is the complex (K\"ahler) polarization, which leads to the Bargmann--Fock representation. Introduce the complex coordinate
\[
z = \sqrt{\tfrac{m\omega}{2\hbar}}\,q - \frac{i}{\sqrt{2m\hbar\omega}}\,p,\qquad
\bar z = \sqrt{\tfrac{m\omega}{2\hbar}}\,q + \frac{i}{\sqrt{2m\hbar\omega}}\,p.
\]
Inverting this coordinate transformation provides the standard phase space coordinates \(q = \sqrt{\hbar/(2m\omega)}(z+\bar{z})\) and \(p = i\sqrt{m\hbar\omega/2}\,(z-\bar{z})\). A direct calculation transforms the symplectic form \(\omega = dp \wedge dq\) into its complex equivalent:
\[
\omega = i\hbar\,dz\wedge d\bar z.
\]
Selecting the complex polarization \(P_{\mathbb{C}} = \mathrm{span}\{\partial_{\bar z}\}\), physical quantum states are restricted to polarized sections defined by \(\nabla_{\partial_{\bar{z}}}s = 0\). In a holomorphic frame, use the K\"ahler primitive \(\theta_{\mathbb{C}} = -i\hbar\,\bar{z}\,dz\), for which \(d\theta_{\mathbb{C}} = \omega\). The covariant derivative then restricts to the partial derivative \(\nabla_{\partial_{\bar{z}}} = \partial_{\bar z}\). Thus, polarized sections are precisely the space of holomorphic functions \(s(z,\bar{z}) = \phi(z)\).

The physical inner product of two states requires integrating them over the phase space. In K\"ahler quantization, the symplectic form is globally generated by a K\"ahler potential \(K(z,\bar{z})\) such that \(\omega = i\hbar\,\partial\bar{\partial}K\); for the harmonic oscillator, \(K = |z|^2\). Geometric quantization mandates that the prequantum connection \(\nabla\) must be precisely compatible with the Hermitian metric \(h\) placed on the line bundle \(L\). If \(s_0\) is a local holomorphic non-vanishing frame (\(\nabla_{\partial_{\bar{z}}} s_0 = 0\)), the geometric compatibility condition \( d|s_0|_h^2 = h(\nabla s_0, s_0) + h(s_0, \nabla s_0) \) strictly links the metric to the connection potential, forcing the pointwise geometric norm of the base section to scale exponentially with the K\"ahler potential: \(|s_0|_h^2 \propto e^{-K}\).

Consequently, the norm of any polarized state \(s = \phi(z)s_0\) intrinsically acquires a Gaussian modulating factor: \(|s|_h^2 = |\phi(z)|^2 e^{-|z|^2}\). This geometric scaling mechanism introduces the Gaussian weight into the inner product measure without ad-hoc insertions. It does not make every holomorphic \(\phi\) normalizable; rather, the physical Hilbert space is precisely the subspace satisfying the corresponding square-integrability condition:
\[
\mathcal{H}_{\mathrm{BF}} = \Big\{\phi\,\text{holomorphic}\,:\,\int_{\mathbb{C}}|\phi(z)|^2\,e^{-|z|^2}\,\frac{i}{2\pi}\,dz\wedge d\bar z < \infty\Big\},
\]
equipped with the orthonormal basis \(\phi_n(z)=z^n/\sqrt{n!}\).

To explicitly derive the elemental quantum operators, we substitute the coordinate functions \(z\) and \(\bar{z}\) into the Kostant--Souriau formula \(Q(f) = -i\hbar\nabla_{X_f} + f\). First, we determine the Hamiltonian vector fields from the defining relation \(\iota_{X_f}\omega = -df\). Evaluating this geometric identity for \(\omega = i\hbar\,dz\wedge d\bar z\) directly yields the vector fields:
\[
X_z = -\frac{i}{\hbar}\partial_{\bar{z}}, \qquad X_{\bar{z}} = \frac{i}{\hbar}\partial_z.
\]
Next, evaluating the connection potential on these fields yields \(\theta_{\mathbb{C}}(X_z) = (-i\hbar\,\bar{z}\,dz)(-\tfrac{i}{\hbar}\partial_{\bar{z}}) = 0\) and \(\theta_{\mathbb{C}}(X_{\bar{z}}) = (-i\hbar\,\bar{z}\,dz)(\tfrac{i}{\hbar}\partial_z) = \bar{z}\). Thus, we can explicitly calculate the generic prequantum operators:
\begin{align*}
Q(z) &= -i\hbar X_z - \theta_{\mathbb{C}}(X_z) + z = -i\hbar\Big(-\frac{i}{\hbar}\partial_{\bar{z}}\Big) - 0 + z = -\partial_{\bar{z}} + z, \\
Q(\bar{z}) &= -i\hbar X_{\bar{z}} - \theta_{\mathbb{C}}(X_{\bar{z}}) + \bar{z} = -i\hbar\Big(\frac{i}{\hbar}\partial_z\Big) - \bar{z} + \bar{z} = \partial_z.
\end{align*}
When evaluating physical states, these operators act strictly on the restricted subspace of holomorphic functions satisfying \(\partial_{\bar{z}}\phi = 0\). As a result, the \(\partial_{\bar{z}}\) derivative effectively vanishes. Thus, this fundamental geometric procedure intrinsically enforces the foundational operator mapping:
\[
Q(z)\phi = z\phi, \qquad Q(\bar{z})\phi = \partial_z\phi.
\]
Substituting these elemental geometric operators back into the inverted phase space coordinates precisely recovers the corresponding prequantum operators for position and momentum, which natively mirror the formal creation and annihilation operators:
\[
Q(q) = \sqrt{\frac{\hbar}{2m\omega}}\,\Big(Q(z)+Q(\bar z)\Big) = \sqrt{\frac{\hbar}{2m\omega}}\,(z+\partial_z),
\]
\[
Q(p) = i\sqrt{\frac{m\hbar\omega}{2}}\,\Big(Q(z)-Q(\bar z)\Big) = i\sqrt{\frac{m\hbar\omega}{2}}\,(z-\partial_z).
\]
Because these linear phase translations generate identically zero intrinsic divergence, i.e., \(L_{X_z}\sqrt{dz} = 0\) and \(L_{X_{\bar{z}}}\sqrt{dz} = 0\), their operators remain trivially unaffected by the metaplectic shift.

Unlike the real polarization, the Hamiltonian \(H\) preserves the complex polarization (\([X_H, \partial_{\bar{z}}] \in P_{\mathbb{C}}\)), allowing it to be quantized directly. Writing \(H = \hbar\omega|z|^2 = \hbar\omega z\bar{z}\), its Hamiltonian vector field evaluated from \(\iota_{X_H}\omega = -dH\) is:
\[
X_H = i\omega (z \partial_z - \bar{z} \partial_{\bar{z}}).
\]
Evaluating the prequantum connection potential on this flow yields
\[
\theta_{\mathbb{C}}(X_H) = (-i\hbar \bar{z} \, dz)(i\omega z \partial_z) = \hbar\omega z\bar{z} = H.
\]
Thus, the uncorrected Kostant--Souriau operator evaluates to:
\[
Q_{\mathrm{pol}}(H) = -i\hbar \nabla_{X_H} + H = -i\hbar X_H - \theta_{\mathbb{C}}(X_H) + H = -i\hbar X_H.
\]
Operating on polarized sections (\(\partial_{\bar{z}}\phi = 0\)), this immediately reduces to \(Q_{\mathrm{pol}}(H) = \hbar\omega z\partial_z\), producing unshifted eigenvalues \(E_n = \hbar\omega n\). Substituting the position and momentum operators symmetrically into the classical formula is not geometrically valid; one must intrinsically quantize \(H\) via its vector field.

To recover the physical zero-point energy, one must introduce the metaplectic correction. The half-form bundle \(\delta_{P_{\mathbb{C}}}\) consists of sections \(\phi(z)\sqrt{dz}\). The correction incorporates the Lie derivative of the half-form along the Hamiltonian flow:
\[
L_{X_H}\sqrt{dz} = \frac{1}{2\sqrt{dz}} L_{X_H}(dz) = \frac{1}{2\sqrt{dz}} d(\iota_{X_H}dz) = \frac{1}{2\sqrt{dz}} d(i\omega z) = \frac{i\omega}{2}\sqrt{dz}.
\]
The augmented metaplectic quantization operator applies this geometric shift:
\begin{align*}
Q_{\mathrm{meta}}(H)(\phi(z)\sqrt{dz}) &= (Q_{\mathrm{pol}}(H)\phi)\sqrt{dz} - i\hbar \phi (L_{X_H}\sqrt{dz}) \\
&= \hbar\omega z\partial_z \phi\sqrt{dz} - i\hbar\Big(\frac{i\omega}{2}\Big)\phi\sqrt{dz} \\
&= \hbar\omega\Big(z\partial_z + \tfrac{1}{2}\Big)\phi\sqrt{dz}.
\end{align*}
This elegantly demonstrates how the metaplectic correction organically secures the zero-point energy shift \(E_0 = \frac{1}{2}\hbar\omega\) through the intrinsic geometry of the complex polarization, generating eigenfunctions \(\phi_n(z)=z^n/\sqrt{n!}\) and the exact spectrum \(E_n = \hbar\omega(n+\tfrac{1}{2})\).

\subsubsection{Equivalence of Polarizations}

A natural question arises: how are these two representations related? The answer is provided by the Blattner--Kostant--Sternberg pairing, which furnishes a unitary isomorphism between \(\mathcal{H}_{\mathrm{pol}}\) (Schr\"odinger) and \(\mathcal{H}_{\mathrm{BF}}\) (Bargmann--Fock). Explicitly, the integral kernel
\[
K(q,z) = \Big(\tfrac{m\omega}{\pi\hbar}\Big){}^{\!1/4}\exp\!\Big(-\tfrac{m\omega}{2\hbar}\,q^2 + \sqrt{\tfrac{2m\omega}{\hbar}}\,q\,z - \tfrac{z^2}{2}\Big)
\]
maps \(\psi(q)\mapsto \phi(z)=\int K(q,z)\,\psi(q)\,dq\), producing a holomorphic \(\phi\) with Gaussian weight. This unitary Segal--Bargmann transform rigorously intertwines the operators of the dual theories. If we denote the transform by the operator \(B\), such that \(\phi(z) = B[\psi(q)]\), then integrating by parts on the integral kernel \(K(q,z)\) shows that multiplying by \(q\) or taking the derivative \(-i\hbar\partial_q\) in the real polarization space is carried into the Bargmann phase space as:
\begin{align*}
B \left[ q \, \psi(q) \right] &= \sqrt{\frac{\hbar}{2m\omega}}(z+\partial_z) \phi(z), \\
B \left[ -i\hbar\partial_q \, \psi(q) \right] &= i\sqrt{\frac{m\hbar\omega}{2}}(z-\partial_z) \phi(z).
\end{align*}
Thus, the seemingly abstract Bargmann operators \(Q(q)\) and \(Q(p)\) derived purely from K\"ahler geometry in the previous section are mathematically identical to the elementary \(\hat{q}\) and \(\hat{p}\) operators of the standard Schr\"odinger representation under this change of basis. This demonstrates that the two starkly different polarizations ultimately yield the exact same quantum theory.

\subsection{The Dirac Magnetic Monopole}

Whereas the harmonic oscillator lives on a topologically trivial phase space, the Dirac monopole provides a paradigmatic example where the non-trivial topology of configuration space enforces prequantum constraints. The requirement that the prequantum line bundle be well-defined leads directly to the celebrated Dirac quantization condition.

Consider a magnetic monopole of magnetic charge \(g\) located at the origin in \(\mathbb{R}^3\). The magnetic field takes the familiar inverse-square form
\[
\mathbf{B} = g\,\frac{\mathbf{r}}{|\mathbf{r}|^3},
\]
where \(\mathbf{r}=(x,y,z)\). A key subtlety is that no globally defined vector potential exists; one must instead define potentials on two overlapping patches. In spherical coordinates \((r,\theta,\phi)\):
\begin{align*}
 A_N &= g\,(1-\cos\theta)\,d\phi\quad (\text{excluding the negative }z\text{-axis}),\\
 A_S &= -g\,(1+\cos\theta)\,d\phi\quad (\text{excluding the positive }z\text{-axis}).
\end{align*}
These potentials are related by a gauge transformation on the overlap: \(A_S - A_N = d\chi\) with the multi-valued function \(\chi(\phi)=-2g\,\phi\). This gauge relation determines the transition function for the prequantum line bundle:
\[
t_{NS}(\phi) = \exp\!\Big( \tfrac{i e}{\hbar}\,\chi(\phi) \Big) = \exp\!\Big( -\,i\,\tfrac{2e g}{\hbar}\,\phi \Big),
\]
which is single-valued under \(\phi\mapsto\phi+2\pi\) if and only if \(\tfrac{2 e g}{\hbar}\in\mathbb{Z}\). This condition is equivalent to the requirement that the total magnetic flux \(\Phi = 4\pi g\) through a sphere enclosing the monopole satisfies the integrality constraint \(e\Phi/h \in \mathbb{Z}\), where \(h = 2\pi\hbar\). (Note that our convention for \(g\) corresponds to the total flux \(\Phi = 4\pi g\); some authors use \(\Phi = g\)).

\medskip

Adopting Gaussian units and setting \(c=1\), the total flux through a surrounding sphere is \(\int_{S^2} \mathbf{B} \cdot d\mathbf{S} = 4\pi g\).

We now derive the symplectic structure on the phase space of a charged particle moving in this background. The Lagrangian for a particle of charge \(e\) and mass \(m\) coupled to an external vector potential \(A_i(\mathbf{x})\) is
\[
L(x,\dot x) \;=\; \tfrac{1}{2}\,m\,\dot x^i\dot x_i \;+\; e\,A_i(x)\,\dot x^i.
\]
The canonical (conjugate) momentum obtained from the Legendre transform is
\[
p_i^{\mathrm{can}} \;=\; \frac{\partial L}{\partial \dot x^i} \;=\; m\,\dot x_i \;+\; e\,A_i(x),
\]
which differs from the kinetic momentum \(\pi_i = m\,\dot x_i\) by the gauge-dependent term \(e\,A_i\). The cotangent bundle \(T^*(\mathbb{R}^3\setminus\{0\})\) carries the canonical symplectic form \(\omega_{\mathrm{can}} = dp_i^{\mathrm{can}}\wedge dx^i\). To obtain a gauge-invariant description, we change variables to the kinetic momenta \(\pi_i = p_i^{\mathrm{can}} - e\,A_i(x)\), so that
\[
dp_i^{\mathrm{can}} \;=\; d\pi_i \;+\; e\,\partial_j A_i\,dx^j.
\]
Substituting into the canonical symplectic form and expanding gives
\[
\omega_{\mathrm{can}}
\;=\; (d\pi_i + e\,\partial_j A_i\,dx^j)\wedge dx^i
\;=\; d\pi_i\wedge dx^i \;+\; e\,\partial_j A_i\,dx^j\wedge dx^i.
\]
The second term is antisymmetrised by the wedge product. Exploiting \(dx^j\wedge dx^i = -dx^i\wedge dx^j\), we write
\[
\partial_j A_i\,dx^j\wedge dx^i
\;=\; \tfrac{1}{2}(\partial_j A_i - \partial_i A_j)\,dx^j\wedge dx^i
\;=\; \tfrac{1}{2}\,F_{ji}\,dx^j\wedge dx^i
\;=\; \tfrac{1}{2}\,F_{ij}\,dx^i\wedge dx^j,
\]
where \(F_{ij} = \partial_i A_j - \partial_j A_i\) is the electromagnetic field-strength tensor and we used its antisymmetry \(F_{ji}=-F_{ij}\) together with the antisymmetry of the wedge product.
Renaming \(\pi_i\to p_i\) (now understood as gauge-invariant kinetic momenta), we arrive at the modified symplectic form
\[
\omega \;=\; d p_i\wedge d x^i \;+\; \frac{e}{2}\,F_{ij}\,dx^i\wedge dx^j.
\]
For the Dirac monopole, the field-strength components are
\[
F_{ij} = \partial_i A_j - \partial_j A_i
= \varepsilon_{ijk}\,B^k
= g\,\varepsilon_{ijk}\,\frac{x^k}{r^3},
\]
so the symplectic form becomes
\[
\omega
= d p_i\wedge d x^i
\;+\; \frac{e\,g}{2\,r^3}\, \varepsilon_{ijk}\,x^k\, d x^i\wedge d x^j.
\]

\medskip

The prequantization of this system requires the construction of a complex line bundle \(L \to T^*(\mathbb{R}^3 \setminus \{0\})\) equipped with a connection \(\nabla\) whose curvature 2-form satisfies
\[
F_\nabla = -\frac{i}{\hbar}\,\omega.
\]
The integrality (prequantization) condition demands that for any closed surface \(\Sigma\) in the zero section projecting to a sphere enclosing the origin,
\[
\frac{1}{2\pi\hbar}\,\int_\Sigma\omega \;\in\;\mathbb{Z}.
\]
Choosing \(\Sigma=S^2_R\) in the zero section:
\[
\int_{S^2_R} \mathbf{B}\cdot d\mathbf{S}
= g \int_0^{2\pi}\!\!\int_0^\pi \sin\theta\,d\theta\,d\phi
= 4\pi\,g,
\]
hence
\[
\int_\Sigma \omega
= e\int_{S^2_R}\mathbf{B}\cdot d\mathbf{S}
= 4\pi\,e\,g.
\]
Thus
\[
\frac{1}{2\pi\hbar}\int_\Sigma\omega
= \frac{4\pi\,e\,g}{2\pi\hbar}
= \frac{2\,e\,g}{\hbar}
\;\in\;\mathbb{Z},
\]
yielding the Dirac quantization condition:
\[
\frac{2 e g}{\hbar}\in\mathbb{Z}.
\]

\subsection{The Quantum Hall Effect}

Our final example combines features of both preceding cases: like the harmonic oscillator, it admits explicit solutions, and like the monopole, it exhibits topological flux quantization. The Landau problem---a charged particle in a uniform magnetic field---underlies the quantum Hall effect~\cite{Klitzing1980} and its topological formulation in terms of Chern numbers~\cite{TKNN1982}, providing a fertile testing ground for the interplay between polarization choices and metaplectic corrections.

We model a charged particle of charge \(e\) and mass \(m\) confined to the plane \(\mathbb{R}^2\) in a uniform magnetic field \(\mathbf{B}=B\,\hat{z}\). For definiteness in the complex-coordinate formulas below, assume \(eB>0\) and define the cyclotron frequency \(\omega_c = eB/m\). The magnetic field modifies the phase-space symplectic form to
\[
\omega_B \;=\; dp_x\wedge dx + dp_y\wedge dy \, + \, eB\,dx\wedge dy.
\]
This is equivalent, via minimal coupling, to the canonical symplectic form with a magnetic term in the Hamiltonian. By absorbing the magnetic field into \(\omega_B\), we emphasize the underlying geometry; this construction is gauge-invariant, and the resulting prequantum operators represent the gauge-covariant (kinetic) momenta. Throughout this section, we use \(\omega_B\) to denote the magnetic symplectic form.

\subsubsection{Prequantum Line Bundle and Curvature}

As in the earlier examples, we begin by constructing the prequantum structure. Choose the symplectic potential
\[
\theta_B = p_x\,dx + p_y\,dy + \frac{eB}{2}\,(x\,dy - y\,dx),\qquad d\theta_B = \omega_B.
\]
The prequantum line bundle \(L\to T^*\mathbb{R}^2\) (trivial here) carries the connection
\[
\nabla = d - \tfrac{i}{\hbar}\,\theta_B,\qquad F_\nabla = -\tfrac{i}{\hbar}\,\omega_B.
\]
The Kostant--Souriau operator for any observable \(f\in C^\infty(M)\) is given by \(Q(f)=-i\hbar\,\nabla_{X_f}+f\), where \(\iota_{X_f}\omega_B=-df\). In the following, we adopt the symmetric gauge.

\subsubsection{Vertical (Real) Polarization and the Landau Hamiltonian}

We first consider the real polarization \(P=\mathrm{span}\{\partial_{p_x},\partial_{p_y}\}\). Polarized sections satisfy \(\nabla_{\partial_{p_i}}s=0\). Since the symplectic potential \(\theta_B\) has no component along the vertical directions (i.e., \(\theta_B(\partial_{p_i})=0\)), the covariant derivative reduces to the partial derivative \(\nabla_{\partial_{p_i}} = \partial_{p_i}\). Thus, polarized sections depend only on the configuration coordinates \((x,y)\), meaning \(\psi(x, y)\). On such polarized sections the position operators are simply
\[
Q(x)=x,\qquad Q(y)=y.
\]

To derive the momentum operators, we first find the Hamiltonian vector fields for \(p_x\) and \(p_y\) from the defining relation \(\iota_{X_f}\omega_B = -df\). For \(f=p_x\), let \(X_{p_x} = A\partial_x + B\partial_y + C\partial_{p_x} + D\partial_{p_y}\). The interior product with the magnetic symplectic form \(\omega_B = dp_x\wedge dx + dp_y\wedge dy + eB\,dx\wedge dy\) is
\begin{align*}
\iota_{X_{p_x}}\omega_B &= -A\,dp_x - B\,dp_y + C\,dx + D\,dy + eB(A\,dy - B\,dx)\\
&= -A\,dp_x - B\,dp_y + (C - eB B)\,dx + (D + eB A)\,dy.
\end{align*}
Equating this to \(-dp_x\) yields \(A=1\), \(B=0\), \(C=0\), and \(D=-eB\). Thus,
\[
X_{p_x} = \partial_x - eB\,\partial_{p_y}.
\]
A similar calculation for \(f=p_y\) yields
\[
X_{p_y} = \partial_y + eB\,\partial_{p_x}.
\]
Next, we evaluate the symplectic potential \(\theta_B = p_x\,dx + p_y\,dy + \frac{eB}{2}(x\,dy - y\,dx)\) on these vector fields:
\begin{align*}
\theta_B(X_{p_x}) &= \theta_B(\partial_x) - eB\,\theta_B(\partial_{p_y}) = p_x - \tfrac{eB}{2}y, \\
\theta_B(X_{p_y}) &= \theta_B(\partial_y) + eB\,\theta_B(\partial_{p_x}) = p_y + \tfrac{eB}{2}x.
\end{align*}
Using the Kostant--Souriau formula \(Q(f) = -i\hbar X_f - \theta_B(X_f) + f\), the prequantum momentum operators are
\begin{align*}
Q(p_x) &= -i\hbar(\partial_x - eB\,\partial_{p_y}) - \Big(p_x - \tfrac{eB}{2}y\Big) + p_x = -i\hbar\partial_x + i\hbar eB\,\partial_{p_y} + \tfrac{eB}{2}y, \\
Q(p_y) &= -i\hbar(\partial_y + eB\,\partial_{p_x}) - \Big(p_y + \tfrac{eB}{2}x\Big) + p_y = -i\hbar\partial_y - i\hbar eB\,\partial_{p_x} - \tfrac{eB}{2}x.
\end{align*}
When these operators are restricted to act on polarized sections \(\psi(x, y)\), the \(\partial_{p_x}\) and \(\partial_{p_y}\) derivatives vanish. Thus, the polarized momentum operators become
\[
Q_{\!\mathrm{pol}}(p_x) = -i\hbar\,\partial_x + \tfrac{eB}{2}\,y, \qquad
Q_{\!\mathrm{pol}}(p_y) = -i\hbar\,\partial_y - \tfrac{eB}{2}\,x.
\]

\subsubsection{Complex (K\"ahler) Polarization: Bargmann Picture and Lowest Landau Level}

As with the harmonic oscillator, a complex polarization offers an alternative viewpoint. To isolate the lowest Landau level rigorously, one should first decompose the full phase space into guiding-center and cyclotron sectors rather than simply set \(dp_x=dp_y=0\). Introduce the guiding-center coordinates
\[
X = x + \frac{p_y}{eB}, \qquad Y = y - \frac{p_x}{eB},
\]
and the cyclotron coordinates
\[
\eta_x = -\frac{p_y}{eB}, \qquad \eta_y = \frac{p_x}{eB},
\]
so that \(x = X + \eta_x\) and \(y = Y + \eta_y\). A direct calculation gives
\[
\omega_B = eB\,dX \wedge dY - eB\,d\eta_x \wedge d\eta_y,
\qquad
H = \frac{p_x^2 + p_y^2}{2m}
= \frac{m\omega_c^2}{2}\,(\eta_x^2+\eta_y^2).
\]
Thus the symplectic form and the Hamiltonian split into independent guiding-center and cyclotron pieces. Quantizing the cyclotron sector produces the usual harmonic oscillator of frequency \(\omega_c\), while the guiding-center sector carries the Landau-level degeneracy.

Projecting to the lowest Landau level means selecting the cyclotron ground state and retaining only operators acting on the guiding-center sector. After this projection, the effective reduced phase space is the guiding-center plane with symplectic form
\[
\omega_{\mathrm{gc}} = eB\,dX\wedge dY.
\]
Introduce the rescaled complex coordinate
\[
z = \frac{X+iY}{\sqrt{2}\,\ell_B}, \qquad \bar{z} = \frac{X-iY}{\sqrt{2}\,\ell_B},
\]
where \(\ell_B=\sqrt{\hbar/(eB)}\). The corresponding differentials satisfy
\[
dz \wedge d\bar{z}
= \frac{1}{2\ell_B^2} (dX + i\,dY) \wedge (dX - i\,dY)
= -\frac{i}{\ell_B^2} dX \wedge dY,
\]
and therefore
\[
\omega_{\mathrm{gc}}
= eB\,dX\wedge dY
= i\hbar\,dz\wedge d\bar z.
\]
This reduced symplectic form shows that, within the LLL, the guiding-center coordinates become a conjugate pair with commutator \([X,Y] = -i\ell_B^2\). With the complex polarization \(P_\mathbb{C}=\mathrm{span}\{\partial_{\bar z}\}\), polarized sections defining the LLL are holomorphic functions with Gaussian weight:
\[
 \mathcal{H}_{\mathrm{LLL}}\cong \Big\{\phi\,\text{holomorphic}:\; \int_{\mathbb{C}} |\phi(z)|^2 e^{-|z|^2}\,\tfrac{i}{2\pi}\,dz\wedge d\bar z < \infty\Big\}.
\]
In this reduced theory, the Hamiltonian is constant on the LLL:
\[
H\big|_{\mathrm{LLL}} = \frac{1}{2}\hbar\omega_c.
\]
The operator \(\hbar\,z\partial_z\) acts on the guiding-center degeneracy in this holomorphic representation, not on the energy.

\subsubsection{Full Phase Space and the Complete Landau Spectrum}

To derive the complete sequence of eigenvalues (all Landau levels) directly from geometric quantization, however, we return to the full phase space \(T^*\mathbb{R}^2 \cong \mathbb{R}^4\). Using the guiding-center coordinates \(X = x + p_y/(eB)\), \(Y = y - p_x/(eB)\) introduced above, we construct complex coordinates for both the cyclotron and guiding-center sectors:
\[
a = \frac{1}{\sqrt{2eB\hbar}}(p_x - i p_y),\qquad b = \sqrt{\frac{eB}{2\hbar}}(X + iY).
\]
A direct calculation reveals that the full magnetic symplectic form \(\omega_B = dp_x\wedge dx + dp_y\wedge dy + eB\,dx\wedge dy\) natively decouples into two independent K\"ahler forms:
\[
\omega_B = eB\, dX \wedge dY - \frac{1}{eB} dp_x \wedge dp_y = i\hbar\, da \wedge d\bar{a} + i\hbar\, db \wedge d\bar{b}.
\]
We adopt the full complex polarization \(P_\mathbb{C} = \mathrm{span}\{\partial_{\bar{a}}, \partial_{\bar{b}}\}\). Thus, polarized sections are holomorphic functions \(\phi(a, b)\) equipped with the global Gaussian weight \(e^{-|a|^2 - |b|^2}\).

Crucially, the classical Hamiltonian \(H = \frac{1}{2m}(p_x^2 + p_y^2) = \hbar\omega_c |a|^2 = \hbar\omega_c a\bar{a}\) depends exclusively on the cyclotron coordinate \(a\) and is remarkably independent of the guiding center \(b\). Because \(H\) is a function purely of \(a\) and \(\bar{a}\), its Hamiltonian vector field \(X_H\) seamlessly preserves the complex polarization. Using the defining geometric relation \(\iota_{X_H}\omega_B = -dH\), we find:
\[
X_H = i\omega_c \left( a\partial_a - \bar{a}\partial_{\bar{a}} \right).
\]
In this K\"ahler polarization, taking the holomorphic-frame primitive \(\theta_{\mathbb{C}} = -i\hbar(\bar{a}\,da + \bar{b}\,db)\) and evaluating it on \(X_H\) yields \(\theta_{\mathbb{C}}(X_H) = \hbar\omega_c a\bar{a} = H\). The bare Kostant--Souriau prequantum operator evaluated on holomorphic polarized sections (\(\partial_{\bar{a}}\phi = \partial_{\bar{b}}\phi = 0\)) is therefore simply:
\[
Q_{\mathrm{pol}}(H) = -i\hbar \left( X_H - \frac{i}{\hbar}\theta_{\mathbb{C}}(X_H) \right) + H = -i\hbar X_H = \hbar\omega_c a\partial_a.
\]
This uncorrected operator yields raw eigenvalues \(\hbar\omega_c n\) with eigenfunctions \(\phi_n(a, b) = f(b) a^n/\sqrt{n!}\), where the arbitrary holomorphic function \(f(b)\) mathematically represents the massive macroscopic degeneracy of each energy level.

To recover the exact physical spectrum and the zero-point energy shift, we incorporate the metaplectic correction. The Lie derivative of the half-form \(\sqrt{da \wedge db}\) along the Hamiltonian flow supplies the geometric phase shift:
\begin{align*}
L_{X_H}\sqrt{da \wedge db}
&= \frac{1}{2\sqrt{da\wedge db}} d\big(\iota_{X_H}(da \wedge db)\big) \\
&= \frac{1}{2\sqrt{da\wedge db}} d(i\omega_c a\,db) \\
&= \frac{i\omega_c}{2}\sqrt{da \wedge db}.
\end{align*}
The completely corrected augmented operator becomes:
\begin{align*}
Q_{\mathrm{meta}}(H)(\phi\sqrt{da\wedge db}) &= (Q_{\mathrm{pol}}(H)\phi)\sqrt{da\wedge db} - i\hbar \phi (L_{X_H}\sqrt{da\wedge db}) \\
&= \hbar\omega_c a\partial_a\phi\sqrt{da\wedge db} - i\hbar\Big(\frac{i\omega_c}{2}\Big)\phi\sqrt{da\wedge db} \\
&= \hbar\omega_c\Big(a\partial_a + \tfrac{1}{2}\Big)\phi\sqrt{da\wedge db}.
\end{align*}
This elegantly secures the exact Landau levels \(E_n = \hbar\omega_c(n + \frac{1}{2})\), \(n\in\mathbb{Z}_{\ge 0}\), while explicitly demonstrating that the full spectrum of eigenvalues precisely emerges from the metaplectically corrected geometric quantization of the full phase space.

\subsubsection{Torus Compactification and Flux Quantization}

On the infinite plane \(\mathbb{R}^2\), the second cohomology \(H^2(\mathbb{R}^2,\mathbb{R})\) vanishes, so the de Rham class \([\omega_B|_{\mathrm{conf}}/(2\pi\hbar)]\) is zero and the integrality condition is satisfied automatically. However, the situation changes dramatically when we compactify the configuration space to a torus \(\mathbb{T}^2 = \mathbb{R}^2 / \Lambda\), where \(\Lambda\) is a lattice generated by two linearly independent vectors. The torus has non-trivial second cohomology \(H^2(\mathbb{T}^2, \mathbb{Z}) \cong \mathbb{Z}\), which imposes a genuine constraint on the magnetic field.

The integrality condition for geometric quantization requires that the de Rham cohomology class of the symplectic form, scaled by \(1/(2\pi\hbar)\), lies in the image of the natural map \(H^2(\mathbb{T}^2, \mathbb{Z}) \to H^2(\mathbb{T}^2, \mathbb{R})\). In other words, there must exist a prequantum line bundle \(L \to \mathbb{T}^2\) with connection \(\nabla\) such that
\[
\frac{i}{2\pi}\, F_\nabla = \frac{\omega_B|_{\mathrm{conf}}}{2\pi\hbar}.
\]
The left-hand side represents the real image of the first Chern class \(c_1(L) \in H^2(\mathbb{T}^2, \mathbb{Z})\). Pairing this class with the fundamental class of \(\mathbb{T}^2\) yields the quantization condition:
\[\begin{aligned}
\left\langle c_1(L),[\mathbb{T}^2]\right\rangle
&= \frac{1}{2\pi\hbar}\int_{\mathbb{T}^2} \omega_B\big|_{\mathrm{conf}} \\
&= \frac{1}{2\pi\hbar}\int_{\mathbb{T}^2} eB\,dx\wedge dy \\
&= \frac{e\Phi}{2\pi\hbar}
= \frac{e\Phi}{h}
=: N_\phi \in \mathbb{Z},\\
\Phi &= \int_{\mathbb{T}^2} B\,dx\wedge dy = B \cdot \mathrm{Area}(\mathbb{T}^2),\qquad h=2\pi\hbar.
\end{aligned}\]
Here \(\Phi\) is the total magnetic flux through the torus, and \(N_\phi\) is the number of flux quanta. This is the celebrated \emph{flux quantization condition}: on a compact surface, the total magnetic flux must be an integer multiple of the flux quantum \(\Phi_0 = h/e\).

The physical interpretation is clear: the prequantum line bundle \(L\) has a well-defined first Chern class, which measures the ``twisting'' of the bundle around the torus. This topological invariant must be an integer, and it equals the number of flux quanta piercing the surface. Unlike the plane, where arbitrary magnetic fields are allowed, the compact geometry of the torus enforces a discrete spectrum of allowed fluxes.

The physical consequences of this topological constraint are most vividly exhibited by the Laughlin gauge invariance argument~\cite{Laughlin1981}. Consider the system on an annular (Corbino disk) geometry, which is topologically a cylinder \(S^1 \times \mathbb{R}\). An Aharonov--Bohm flux \(\Phi_{\mathrm{AB}}\) is threaded through the hole of the cylinder. Because the Hamiltonian depends on \(\Phi_{\mathrm{AB}}\) only through the gauge-invariant combination \(e\Phi_{\mathrm{AB}}/\hbar\), the energy spectrum is periodic: it returns exactly to itself when the flux is advanced by one flux quantum, \(\Phi_{\mathrm{AB}} \to \Phi_{\mathrm{AB}} + \Phi_0\). Laughlin's key insight is that this spectral periodicity, combined with the existence of a spectral gap, constrains the adiabatic response. As the flux is smoothly increased from \(0\) to \(\Phi_0\), each single-particle state evolves into the next one in the spectral ladder---a phenomenon known as \emph{spectral flow}. The net effect is that an integer number \(n\) of electrons is transferred from one edge of the cylinder to the other. Because the electromotive force generated by the time-varying flux is \(\mathcal{E} = -\dot{\Phi}_{\mathrm{AB}}\), the charge transferred \(\Delta Q = n e\) during one period implies a Hall current \(I = \sigma_{xy} \mathcal{E}\) with
\[
\sigma_{xy} = \frac{\Delta Q}{\Delta \Phi_{\mathrm{AB}}} = \frac{n\,e}{\Phi_0} = n\,\frac{e^2}{h},\qquad n\in\mathbb{Z}.
\]
This argument establishes quantized Hall conductivity purely from gauge invariance and the existence of a gap, without reference to band structure or Bloch wavefunctions. The connection to the torus geometry developed above is direct: the periodicity \(\Phi_{\mathrm{AB}} \to \Phi_{\mathrm{AB}} + \Phi_0\) is precisely the statement that the prequantum line bundle over the flux parameter space must have integer Chern class---the same integrality condition enforcing \(N_\phi \in \mathbb{Z}\).

This topological framework extends naturally to the band-theoretic setting. In the presence of a periodic crystalline lattice, the continuous translation symmetry of the plane is broken. The macroscopic Landau levels are split into magnetic Bloch bands. When introducing periodic boundary conditions to model the bulk crystal, the allowed crystal momentum \(\mathbf{k} = (k_x, k_y)\) is restricted to the magnetic Brillouin zone (MBZ). Topologically, because the boundaries of the Brillouin zone are identified with one another \((k_i \sim k_i + G_i)\), the momentum space itself forms a compact torus \(\mathbb{T}^2\).

Within this framework, applying linear response theory via the Kubo formula reveals that the transverse Hall conductivity of a fully occupied band \(n\) is given by the integral of the Berry curvature \(\mathcal{F}_{xy}^{(n)}(\mathbf{k})\) over this momentum-space torus. The Berry connection \(\mathcal{A}_j = i\langle u_n(\mathbf{k}) | \partial_{k_j} | u_n(\mathbf{k}) \rangle\) (where \(|u_n(\mathbf{k})\rangle\) are the periodic parts of the Bloch wavefunctions) plays a mathematically analogous \(U(1)\)-connection role to the prequantum connection explored earlier, except that it is defined over the Brillouin zone rather than over configuration space.

Consequently, the famous Thouless--Kohmoto--Nightingale--den Nijs (TKNN) invariant~\cite{TKNN1982} demonstrates that integrating this Berry curvature over the compact magnetic Brillouin zone mathematically measures the first Chern number of the line bundle formed by the crystalline Bloch states:
\[
\sigma_{xy} = \frac{e^2}{\hbar} \sum_{n\in\mathrm{occ}} \int_{\mathrm{MBZ}} \frac{d^2k}{{(2\pi)}^{2}} \, \mathcal{F}_{xy}^{(n)}(\mathbf{k}) = \frac{e^2}{h} \sum_{n\in\mathrm{occ}} C_n,
\]
where \(C_n \in \mathbb{Z}\) is the Chern number for the \(n\)-th band. While a pure, unperturbed Landau level in a uniform continuum always has a Chern number of exactly 1, a strong periodic lattice potential splits these levels into subbands. The individual Chern numbers \(C_n\) of these subbands can be positive, negative, or zero, with their allowed values constrained by the Bloch geometry, gap-labeling or Diophantine relations, and the total Chern number carried by the parent band. Because the normalized Berry-curvature integral \(\frac{1}{2\pi}\int_{\mathbb{T}^2}\mathcal{F}\) is the first Chern number of the corresponding line bundle, the total combined Chern number \(C = \sum_{n\in\mathrm{occ}} C_n\) is robustly quantized. In this context, the integer \(C\) is physically determined by the position of the Fermi energy: it simply sums the individual geometric Chern numbers of all energy bands that are completely filled by electrons. As long as the Fermi level lies within a spectral gap, the number of occupied bands remains exactly constant, protecting and locking \(C\) to a macroscopic integer. Thus, the striking experimental observation of macroscopic, exact quantization in the Hall conductivity \(\sigma_{xy} = C\frac{e^2}{h}\) is inextricably linked to the very same geometric topology constraint---the integrality of the first Chern class for prequantum bundles---that enforces magnetic flux quantization.

\chapter{Orbifold Theory}

This chapter develops the foundational theory of orbifolds required for geometric quantization in the singular setting. We begin with the basic definitions and examples, then develop the covering theory and orbifold fundamental group. The de Rham cohomology of orbifolds provides the analytical framework for characteristic classes. We then introduce orbifold line bundles and their classification, followed by connections and Chern--Weil theory. The chapter concludes with the special features of two-dimensional orbifolds, including the orbifold Euler characteristic.

\section{Orbifold Definition}\label{sec:orbifold_definition}

Orbifolds generalize smooth manifolds by allowing controlled singularities modeled on quotients of Euclidean space by finite group actions. This additional structure is essential for quantizing phase spaces with discrete symmetries. To formalize this concept, we adapt the standard definition of a manifold atlas, replacing local coordinate charts with local quotient models~\cite{Satake1957,AdemLeidaRuan}.

\begin{definition}[Orbifold]\label{def:orbifold}
An \textbf{orbifold} \( X \) consists of an underlying topological space \( |X| \) equipped with an \emph{orbifold atlas} consisting of local uniformizing systems. The data of the atlas is specified as follows:
\begin{enumerate}
    \item \textbf{Local Charts:} For every point \( p \in |X| \), there exists an open neighborhood \( U_p \subset |X| \), a connected open set \( V_p \subset \mathbb{R}^n \) (the \emph{uniformizing space}), a finite group \( G_p \) (the \emph{local uniformizing group}) acting smoothly and effectively on \( V_p \), and a projection map
    \[
    \pi_p : V_p \to U_p
    \]
    satisfying two conditions:
    \begin{itemize}
        \item \( \pi_p \) is \( G_p \)-invariant, meaning \( \pi_p(g \cdot v) = \pi_p(v) \) for all \( g \in G_p \) and \( v \in V_p \).
        \item \( \pi_p \) induces a homeomorphism \( U_p \cong V_p / G_p \).
    \end{itemize}

    \item \textbf{Compatibility:} The collection of local charts \( \{(V_p, G_p, \pi_p)\} \) covers \( |X| \) and satisfies compatibility conditions on overlaps. Specifically, for any two charts \( (V_p, G_p, \pi_p) \) and \( (V_q, G_q, \pi_q) \) and any point \( x \in U_p \cap U_q \), there must exist a smaller chart \( (V_k, G_k, \pi_k) \) with \( x \in U_k \subset U_p \cap U_q \) that embeds into both. An embedding \( (V_k, G_k) \hookrightarrow (V_p, G_p) \) consists of a smooth open embedding \( \phi_{kp}: V_k \to V_p \) and an injective group homomorphism \( \lambda_{kp}: G_k \hookrightarrow G_p \) such that \( \phi_{kp} \) is \( \lambda_{kp} \)-equivariant (i.e., \( \phi_{kp}(g \cdot v) = \lambda_{kp}(g) \cdot \phi_{kp}(v) \)) and \( \pi_p \circ \phi_{kp} = \pi_k \).
\end{enumerate}
\end{definition}

\subsubsection{Singular Points and Orders.}
The local uniformizing group \( G_p \) is not merely an artifact of the atlas but encodes the local geometry of \( X \). For any point \( x \in |X| \), choose a local chart \( (V_p, G_p, \pi_p) \) covering \( x \) and let \( \tilde{x} \in V_p \) be a preimage of \( x \). The \textbf{isotropy group} (or stabilizer) of \( x \), denoted \( G_x \), is the stabilizer of \( \tilde{x} \) under the \( G_p \)-action: \( G_x = \{ g \in G_p : g \cdot \tilde{x} = \tilde{x} \} \). This abstract group is well-defined up to isomorphism~\cite[Ch.~1]{AdemLeidaRuan}. A point \( x \) is called a \textbf{singular point} if \( G_x \) is nontrivial. The \textbf{order} of the singularity at \( x \) is defined as \( |G_x| \). Consequently, the singular locus---the set of all points with nontrivial isotropy---forms a stratified space whose structure reflects the subgroup lattice of the isotropy groups.

\subsubsection{Geometric Interpretation.}
Geometrically, this definition implies that each point carries a local symmetry group acting on a covering chart. At generic points, this isotropy group is trivial, and the space is locally indistinguishable from a manifold. At singular points, however, a sufficiently small neighborhood of \( x \) is modeled on a quotient \( W / G_x \), where \( W \subset V_p \) is a \( G_x \)-invariant neighborhood of a lift \( \tilde{x} \). The orbifold structure precisely captures how these local quotient models are glued together consistently across the entire space.

\subsubsection{Good and Bad Orbifolds.}
A fundamental distinction in orbifold theory arises when considering the global structure. An orbifold \( X \) is called \textbf{good} (or \textbf{developable}) if it is isomorphic to a global quotient \( [M/\Gamma] \), where \( M \) is a smooth manifold and \( \Gamma \) is a discrete group acting smoothly, effectively, and properly discontinuously on \( M \) with finite stabilizers. In such cases, the orbifold structure is fully captured by the equivariant geometry of the pair \( (M, \Gamma) \), allowing many constructions to be reduced to \(\Gamma\)-equivariant constructions on \(M\)~\cite{Thurston1980,Scott1983}.

Conversely, an orbifold is \textbf{bad} if it cannot be represented as such a global quotient. Bad orbifolds arise when local isotropy requirements are globally incompatible. The canonical example is the \emph{teardrop} orbifold, topologically a sphere \( S^2 \) with a single cone point of order \( n > 1 \). Its orbifold fundamental group is trivial: for a sphere with one cone point, the standard presentation includes both the cone relation \(x^n=1\) and the global relation \(x=1\), forcing \(x\) to be trivial. Hence there are no nontrivial connected orbifold covering spaces, so the universal orbifold cover is the teardrop itself. Since this universal cover is still singular, it is not a manifold; this shows the teardrop is not developable, i.e., not a global quotient of a smooth manifold.

\section{Orbifold Coverings and Fundamental Group}\label{sec:orbifold_fundamental_group}

With the local and global structure of orbifolds defined, we now investigate their topology. The fundamental group of an orbifold is a subtle invariant that detects singular structures invisible to the ordinary fundamental group of the underlying coarse space~\cite{Thurston1980,AdemLeidaRuan}. We adopt the loop-based definition, which is intrinsic and applies equally to both good and bad orbifolds~\cite{AdemLeidaRuan,Scott1983}.

\subsubsection{Orbifold Loops.}
Intuitively, an orbifold loop is a path in the underlying space \( |X| \) decorated with lifting data~\cite{AdemLeidaRuan}. Unlike a loop in a manifold, which is simply a continuous map from \( S^1 \), an orbifold loop must be understood as a path in the local covering charts. Crucially, it remembers not just \emph{where} it goes, but \emph{how} it transitions between charts, allowing it to encode the local group actions~\cite{Thurston1980,MoerdijkPronk1997}.

\begin{definition}[Orbifold Loop]
Let \( X \) be an orbifold and \( x_0 \in |X| \) a basepoint. An \emph{orbifold loop} based at \( x_0 \) consists of the following data:
\begin{enumerate}
    \item A continuous path \( \gamma: [0,1] \to |X| \) such that \( \gamma(0) = \gamma(1) = x_0 \).
    \item A partition \( 0 = t_0 < t_1 < \cdots < t_k = 1 \) of the unit interval.
    \item For each subinterval \( [t_{i-1}, t_i] \), a lift \( \tilde{\gamma}_i: [t_{i-1}, t_i] \to V_i \) to a uniformizing chart \( (V_i, G_i, \pi_i) \) such that \( \pi_i \circ \tilde{\gamma}_i = \gamma|_{[t_{i-1}, t_i]} \).
    \item For each partition point \( t_i \) (\( i=1, \dots, k-1 \)), a transition element \( g_i \) from the relevant chart-transition group embedding that ``glues'' the end of \( \tilde{\gamma}_i \) to the start of \( \tilde{\gamma}_{i+1} \). Specifically, viewing the overlap in a common chart, we require \( \tilde{\gamma}_{i+1}(t_i) = g_i \cdot \tilde{\gamma}_i(t_i) \).
    \item A final ``return'' element \( g \in G_{x_0} \) such that if \( \tilde{x}_0 \) is the starting point of the lift in the chart at \( x_0 \), the endpoint of the lift is \( g \cdot \tilde{x}_0 \). This element \( g \) encodes the monodromy of the loop.
\end{enumerate}
\end{definition}

Two orbifold loops are considered \emph{homotopic} if they can be deformed into each other through a family of such data~\cite{AdemLeidaRuan,Scott1983}. The operations of concatenation and inversion are defined naturally, endowing the set of homotopy classes with a group structure.

\begin{definition}[Orbifold Fundamental Group]
The \emph{orbifold fundamental group} \( \pi_1^{\mathrm{orb}}(X, x_0) \) is the group of homotopy classes of orbifold loops based at \( x_0 \).
\end{definition}

This definition captures the singular nature of the orbifold~\cite{Thurston1980,Scott1983}. For the cone orbifold \( X=[\mathbb{C}/\mathbb{Z}_n] \), consider a small loop in the underlying space \( |X| \) encircling the singular point. As an \emph{orbifold} loop, its lift to the chart \( V \cong \mathbb{C} \) does not close up; instead, it ends at a point related to the starting point by the generator of \( \mathbb{Z}_n \). The ``return element'' \( g \) records this monodromy. Thus, this loop represents a non-trivial element of order \( n \) in \( \pi_1^{\mathrm{orb}}(X) \), even though it is contractible in the underlying space \( |X| \cong \mathbb{C} \).

\subsubsection{Deck Transformations and the Deck Group.}
Let $p \colon \tilde{X} \to X$ be an orbifold covering map.

\begin{definition}[Deck Transformation]
A \emph{deck transformation} (also called a \emph{covering transformation} or \emph{automorphism of the covering}) is a homeomorphism (orbifold isomorphism) $\varphi \colon \tilde{X} \xrightarrow{\;\sim\;} \tilde{X}$ that commutes with the covering map, that is,
\[
  p \circ \varphi = p.
\]
In other words, $\varphi$ permutes the fibres of $p$ while leaving the base $X$ pointwise fixed.
\end{definition}

Deck transformations compose and invert, so they form a group~\cite{Hatcher2002}.

\begin{definition}[Deck Group~\cite{Hatcher2002}]
The \emph{deck group} (or \emph{group of deck transformations}) of the covering $p \colon \tilde{X} \to X$ is
\[
  \Deck(\tilde{X} \to X) \;:=\;  \bigl\{\, \varphi \colon \tilde{X} \xrightarrow{\;\sim\;} \tilde{X} \;\big|\; p \circ \varphi = p \,\bigr\},
\]
with the group operation being composition.
\end{definition}

\subsubsection{Covering Spaces.}
The relationship between the fundamental group and covering spaces generalizes directly from manifold theory~\cite{Hatcher2002,Thurston1980}. An \emph{orbifold covering} \( p: \tilde{X} \to X \) is a map between orbifolds that is locally modeled on quotient maps induced by subgroup inclusions~\cite{Thurston1980,MoerdijkPronk1997}. The orbifold fundamental group classifies connected orbifold coverings~\cite{Thurston1980}.

Every connected orbifold admits a universal covering orbifold~\cite{Thurston1980}. An orbifold \( X \) is called ``good'' if it admits a covering by a smooth manifold; equivalently, its universal orbifold cover \( \tilde{X} \) is a smooth manifold~\cite{Thurston1980,Scott1983}. Since the universal orbifold cover is orbifold simply connected, the orbifold fundamental group is isomorphic to the group of deck transformations~\cite{Thurston1980}:
\[
\pi_1^{\mathrm{orb}}(X, x_0) \cong \Deck(\tilde{X} \to X).
\]
The fundamental group $\pi_1^{\mathrm{orb}}(X)$ acts properly discontinuously on the universal cover \( \tilde{X} \), and \(X\) is recovered as the quotient in the orbifold, equivalently stack, category~\cite{BridsonHaefliger1999,Thurston1980}. Only when \(\tilde{X}\) is a smooth manifold does this become the developable presentation \(X\cong \tilde{X}/\pi_1^{\mathrm{orb}}(X)\) by an ordinary manifold cover.

More generally, for any connected regular orbifold covering $Y \to X$ where $Y$ is a manifold, we have a short exact sequence of groups~\cite{AdemLeidaRuan,Scott1983}:
\[
 1 \longrightarrow \pi_1(Y) \longrightarrow \pi_1^{\mathrm{orb}}(X) \longrightarrow \Deck(Y \to X) \longrightarrow 1.
\]
Because $Y$ is a manifold, its orbifold fundamental group $\pi_1^{\mathrm{orb}}(Y)$ coincides with its ordinary fundamental group $\pi_1(Y)$. For instance, if $X=[M/G]$ is a global quotient of a connected manifold $M$ by an effective, properly discontinuous action of a discrete group $G$, then the quotient map $M \to X$ is a regular \emph{orbifold} covering with deck group $G$~\cite{AdemLeidaRuan,Thurston1980}. This need not be an ordinary topological covering when the action has fixed points. The orbifold covering sequence gives:
\[
 1 \longrightarrow \pi_1(M) \longrightarrow \pi_1^{\mathrm{orb}}([M/G]) \longrightarrow G \longrightarrow 1.
\]
This exact sequence provides a powerful computational tool for good orbifolds, reducing the calculation of the fundamental group to finding the symmetries of a given manifold cover.

\subsubsection{Examples.}
The following examples illustrate how the orbifold fundamental group differs from the topological fundamental group~\cite{Thurston1980,Scott1983}:
\begin{itemize}
    \item \textbf{Cone} \( \mathbb{C}/\mathbb{Z}_n \)\textbf{:} A loop winding once around the cone point lifts to a path in \( \mathbb{C} \) connecting a point \( z \) to \( e^{2\pi i/n}z \). This path is not closed in the covering space but is closed up by the action of the generator of \( \mathbb{Z}_n \). Thus, \( \pi_1^{\mathrm{orb}}(\mathbb{C}/\mathbb{Z}_n) \cong \mathbb{Z}_n \).
    \item \textbf{The ``Football'' (Orbisphere)} \( S^2(n,n) \)\textbf{:} The underlying space is a sphere \( S^2 \) with two singular points of order \( n \) (e.g., at the poles). Let \( \alpha \) be a loop around the north pole and \( \beta \) a loop around the south pole. In the orbifold fundamental group, the local structure imposes the relations \( \alpha^n = 1 \) and \( \beta^n = 1 \). Since the sphere is simply connected, the loop \( \alpha \beta \) (which encloses both poles) is contractible in the underlying space. As an orbifold loop, it lifts to a closed loop in the chart covering the equator (where the action is trivial). Thus \( \alpha \beta = 1 \), implying \( \alpha = \beta^{-1} \). The group is presented as \( \langle \alpha, \beta \mid \alpha^n=1, \beta^n=1, \alpha\beta=1 \rangle \cong \mathbb{Z}_n \).

\end{itemize}

\section{Orbifold de Rham Cohomology}\label{sec:orbifold_de_rham}

While the fundamental group captures discrete topological data, geometric quantization relies heavily on differential forms to describe curvature, polarization, and integration. To extend these tools to orbifolds, we must construct a de Rham complex that accommodates local singularities. The resulting theory mirrors the smooth case, preserving essential features like Stokes' theorem and the de Rham isomorphism, but requires careful handling of local group invariance~\cite{AdemLeidaRuan}.

\begin{definition}[Orbifold Forms]
An \textbf{orbifold \(k\)-form} \(\alpha\) on \(X\) is defined as a collection \(\{\alpha_p\}\) of local forms on the uniformizing charts~\cite{Satake1957}. Specifically:
\begin{enumerate}
	\item For each chart \((V_p, G_p, \pi_p)\), \(\alpha_p\) is a smooth \(k\)-form on \(V_p\) that is \(G_p\)-invariant.
	\item The forms are compatible under chart embeddings: for any embedding \(\phi_{qp}: V_q \to V_p\) between charts, we require \(\phi_{qp}^* \alpha_p = \alpha_q\) (where \(\alpha_q\) is the form on \(V_q\)).
\end{enumerate}
The space of such orbifold \(k\)-forms is denoted \(\Omega^k_{\mathrm{orb}}(X)\).
\end{definition}

\subsubsection{The de Rham Complex.}
The exterior derivative \(d\) extends naturally to this setting~\cite{AdemLeidaRuan}. Since the standard exterior derivative on a manifold commutes with pullbacks and is invariant under diffeomorphisms, it preserves the \(G_p\)-invariance and compatibility conditions of orbifold forms. Thus, we define \(d\alpha\) chartwise by \({(d\alpha)}_{p} = d(\alpha_{p})\). This complex inherits all standard local properties of the smooth de Rham complex: the wedge product of orbifold forms is well-defined and associative, \(d^2 = 0\), and the Leibniz rule \(d(\alpha \wedge \beta) = d\alpha \wedge \beta + {(-1)}^{\deg(\alpha)}\alpha \wedge d\beta\) holds. This yields the \textbf{orbifold de Rham complex}:
\[
0 \longrightarrow \Omega^0_{\mathrm{orb}}(X) \xrightarrow{\,d\,} \Omega^1_{\mathrm{orb}}(X) \xrightarrow{\,d\,} \cdots \xrightarrow{\,d\,} \Omega^n_{\mathrm{orb}}(X) \longrightarrow 0.
\]
The cohomology of this complex is the \textbf{orbifold de Rham cohomology}, denoted \(H^*_{\mathrm{dR,orb}}(X)\).

\subsubsection{Integration on Orbifolds.}
To pair forms with the space, we need a notion of integration. For a compact, oriented orbifold \(X\) (meaning that the charts are oriented and all local group actions and transition maps preserve orientation), the integral of a top-degree form \(\alpha\) is defined by summing local contributions, weighted to account for the multi-valued nature of the charts~\cite{Satake1957}. Using an \emph{orbifold partition of unity} \(\{\rho_i\}\) subordinate to an atlas \(\{(V_i, G_i, \pi_i)\}\), we define:
\[
\int_X \alpha := \sum_i \frac{1}{|G_i|}\int_{V_i} \rho_i \alpha_i.
\]
Here, \(\rho_i\) are \(G_i\)-invariant functions on \(V_i\) that descend to a partition of unity on \(|X|\). The factor \(1/|G_i|\) is crucial: since the projection \(\pi_i: V_i \to U_i\) is generically a \(|G_i|\)-to-1 cover, this weight ensures that the integral correctly measures the ``volume'' of the quotient space without overcounting~\cite{AdemLeidaRuan}.

For a global quotient \([M/G]\) with finite \(G\), this definition simplifies nicely to \(\int_{[M/G]} \alpha = \frac{1}{|G|}\int_M \tilde{\alpha}\), where \(\tilde{\alpha}\) is the \(G\)-invariant representative on \(M\)~\cite{AdemLeidaRuan}.

\subsubsection{Stokes' Theorem.}
This integration theory supports a version of Stokes' theorem~\cite{Satake1957}. For a compact oriented orbifold \(X\) with boundary \(\partial X\) (equipped with the induced orientation), and any \(\beta \in \Omega^{n-1}_{\mathrm{orb}}(X)\):
\[
\int_X d\beta = \int_{\partial X} \beta.
\]
The proof follows by applying the manifold Stokes' theorem on each chart. The partition of unity handles the global patching, while the \(1/|G_i|\) factors ensure that the boundary terms sum correctly, just as the interior terms do.

\subsubsection{Poincar\'e Lemma and the de Rham Theorem.}
The utility of this cohomology theory rests on its comparison with singular cohomology. The link is provided by a local Poincar\'e lemma. On any contractible chart \(V_p\) with linear \(G_p\)-action, any closed invariant form \(\omega\) of positive degree has a primitive \(\eta\) on \(V_p\). By averaging \(\eta\) over the group \(G_p\), we obtain an invariant primitive \(\bar{\eta} = \frac{1}{|G_p|}\sum_{g \in G_p} g^* \eta\)~\cite{AdemLeidaRuan}. In degree \(0\), the closed invariant forms are precisely the locally constant functions.

Consequently, standard sheaf-theoretic arguments (using fine resolutions) lead to the \textbf{orbifold de Rham theorem}~\cite{caramello2019introduction}:
\[
H^*_{\mathrm{dR,orb}}(X) \cong H^*_{\mathrm{coarse}}(X;\mathbb{R})=H^*(|X|;\mathbb{R}).
\]
This isomorphism asserts that, over real coefficients, the orbifold de Rham cohomology is isomorphic to the coarse-space singular cohomology of the underlying topological space \(|X|\)~\cite{AdemLeidaRuan}.

\subsection{Cohomology Notation Convention}\label{sec:cohomology-notation}
Because integral orbifold data and real differential-form data play different roles in quantization, we use separate notation throughout the rest of the manuscript:
\begin{center}
\setlength{\fboxsep}{8pt}
\fbox{%
\begin{minipage}{0.92\textwidth}
\small
\begin{tabularx}{\textwidth}{>{\raggedright\arraybackslash}p{0.26\textwidth}L L}
\textbf{Notation} & \textbf{Definition} & \textbf{Role} \\
\hline
\(H^k_{\mathrm{dR,orb}}(X)\) &
Cohomology of the complex \(\Omega^\bullet_{\mathrm{orb}}(X)\) of invariant orbifold forms. &
Home for curvature forms, symplectic forms, and real Chern--Weil classes. \\
\(H^k_{\mathrm{st}}(X;A)\) &
Stack/classifying-space cohomology \(H^k(B\mathcal{G};A)\cong H^k(\mathcal{X},\underline{A})\), for any orbifold groupoid presentation \(\mathcal{G}\). &
Home for integral orbifold Chern classes and the Picard classification of orbifold line bundles. For \(X=[M/G]\), this is \(H^k_G(M;A)\). \\
\(H^k_{\mathrm{coarse}}(X;A)\) &
Ordinary singular cohomology \(H^k(|X|;A)\) of the underlying coarse topological space. &
Records only the topology of the quotient space; it forgets isotropy representations. \\
\end{tabularx}
\end{minipage}}
\end{center}

The comparison over real coefficients is
\[
H^k_{\mathrm{st}}(X;\mathbb{R})
\cong
H^k_{\mathrm{dR,orb}}(X)
\cong
H^k_{\mathrm{coarse}}(X;\mathbb{R})
=H^k(|X|;\mathbb{R}),
\]
for the finite-isotropy orbifolds considered here. Integrality, however, is not tested in \(H^k_{\mathrm{coarse}}(X;\mathbb{Z})\). The orbifold integral lattice is the image of
\[
\iota_{\mathrm{dR}}\colon
H^k_{\mathrm{st}}(X;\mathbb{Z})
\longrightarrow
H^k_{\mathrm{st}}(X;\mathbb{R})
\cong
H^k_{\mathrm{dR,orb}}(X).
\]
Thus a real de Rham class can be orbifold-integral even when its coarse-space periods are rational rather than integral.

\section{Orbifold Line Bundles}\label{sec:orbifold_line_bundles}

Having established the topological and analytical foundations of orbifold theory, we now turn to the geometric actors that play the central role in prequantization: orbifold line bundles. These bundles carry quantum phases, and their classification weaves together the discrete data of the orbifold fundamental group with the continuous data of cohomology~\cite{AdemLeidaRuan,Satake1957}.

\begin{definition}[Orbifold Vector Bundle]
An \textbf{orbifold vector bundle} \(E \to X\) of rank \(r\) is defined by local data compatible with the orbifold atlas~\cite{AdemLeidaRuan}:
\begin{enumerate}
    \item \textbf{Local Bundles:} For each chart \((V_p, G_p, \pi_p)\), there is a \(G_p\)-equivariant vector bundle \(E_p \to V_p\).
    \item \textbf{Compatibility:} On overlaps, these local bundles are glued via equivariant isomorphisms. Specifically, for any common refinement \((V_k, G_k)\) of charts \((V_p, G_p)\) and \((V_q, G_q)\) with embeddings \(\phi_{kp}: V_k \to V_p\) and \(\phi_{kq}: V_k \to V_q\), there exists a \(G_k\)-equivariant isomorphism \(\psi_{pq}^{(k)}: \phi_{kp}^* E_p \to \phi_{kq}^* E_q\). This collection of isomorphisms must satisfy the standard cocycle condition on triple overlaps: for any common refinement \((V_m, G_m)\) of three charts \((V_p, G_p)\), \((V_q, G_q)\), and \((V_r, G_r)\) with associated embeddings \(\phi_{mp}\), \(\phi_{mq}\), and \(\phi_{mr}\), the isomorphisms compose as
    \[
    \psi_{pr}^{(m)} = \psi_{qr}^{(m)} \circ \psi_{pq}^{(m)}.
    \]
\end{enumerate}
\end{definition}

\begin{definition}[Orbifold Line Bundle]
An \textbf{orbifold line bundle} \(L \to X\) is an orbifold vector bundle of rank one~\cite{AdemLeidaRuan}. Since orbifold charts \((V_p, G_p)\) can typically be chosen such that \(V_p\) is contractible, the local line bundles \(L_p\) are trivializable, allowing us to identify \(L_p \cong V_p \times \mathbb{C}\). The \(G_p\)-equivariant structure on \(L_p\) then reduces to twisting the fibers. This action is determined (up to equivariant isomorphism) by a homomorphism \(\rho_p: G_p \to U(1)\), called the \textbf{isotropy character}~\cite{AdemLeidaRuan}:
\[
\gamma\cdot (z, w) = \bigl(\gamma\cdot z,\, \rho_p(\gamma)w\bigr) \quad \text{for } \gamma\in G_p, (z,w)\in V_p\times\mathbb{C}.
\]
This defines a valid \(G_p\)-equivariant structure because \(\rho_p\) being a homomorphism (\(\rho_p(\gamma_1\gamma_2) = \rho_p(\gamma_1)\rho_p(\gamma_2)\)) strictly enforces the group composition law \(\gamma_1 \cdot (\gamma_2 \cdot (z,w)) = (\gamma_1\gamma_2) \cdot (z,w)\) on the fibers.

The \textbf{compatibility} structure inherited from the general vector bundle definition implies that on any common refinement \((V_k, G_k)\) of \((V_p, G_p)\) and \((V_q, G_q)\) with embeddings \(\phi_{kp}\) and \(\phi_{kq}\), the gluing isomorphism is represented in these trivializations by a smooth transition function \(g_{pq}^{(k)}: V_k \to \mathbb{C}^\times\):
\[
\psi_{pq}^{(k)}(z, w) = \bigl(z,\, g_{pq}^{(k)}(z)w\bigr).
\]
Equivariance imposes an additional constraint on this transition function. If \(h\in G_k\), and if \(\lambda_{kp}:G_k\to G_p\) and \(\lambda_{kq}:G_k\to G_q\) are the group homomorphisms associated with the two chart embeddings, then
\[
g_{pq}^{(k)}(h\cdot z)\,\rho_p(\lambda_{kp}(h))
=
\rho_q(\lambda_{kq}(h))\,g_{pq}^{(k)}(z).
\]
Thus the transition functions do not glue only the underlying ordinary line bundles; they must also intertwine the isotropy characters induced on the common refinement.
The standard cocycle condition on a triple overlap \((V_m, G_m)\) then simplifies to the recognizable multiplicative form:
\[
g_{pr}^{(m)}(z) = g_{qr}^{(m)}(z) \cdot g_{pq}^{(m)}(z) \quad \text{for all } z \in V_m.
\]
\end{definition}

These isotropy characters are the crucial local data distinguishing orbifold line bundles from ordinary line bundles on the coarse space.

\subsection{Connections and Curvature}\label{sec:connections_curvature}

\begin{definition}[Orbifold Connection]
	A \textbf{connection} on an orbifold line bundle \(L \to X\) is specified by a collection of local connections compatible with the atlas~\cite{AdemLeidaRuan}. For each chart \((V_p, G_p, \pi_p)\), choose the canonical equivariant frame \(e_p\) provided by the local trivialization \(L_p \cong V_p \times \mathbb{C}\) and write
	\[
	\nabla_p e_p = -i A_p \otimes e_p,
	\]
	where \(A_p\) is a smooth, real-valued 1-form on \(V_p\). In this canonical frame, where \(G_p\) acts on the fiber by the constant isotropy character \(\rho_p(\gamma) = e^{i\theta_\gamma}\), the equivariance of \(\nabla\) implies that \(A_p\) is \(G_p\)-invariant:
	\[
	\gamma^* A_p = A_p \quad \text{for all } \gamma \in G_p.
	\]
	(For a general, non-equivariant frame \(s = e^{i\chi} e_p\), the connection 1-form instead transforms as \(\gamma^* A' = A' - d(\gamma^*\chi-\chi)\), since the isotropy phase \(\theta_\gamma\) is constant in this local character model. This gauge-like transformation law is discussed further in Section~\ref{sec:orbifold-integrality}.)
	On chart overlaps, these local connections must be strictly compatible with the transition functions gluing the bundle.
\end{definition}

\subsubsection{Leibniz Rule and Covariant Derivative.}
Locally, this definition recovers the standard properties of a connection~\cite{Woodhouse1992}. For a local section \(s\) of \(L|_{V_p}\) and a smooth function \(f\), the Leibniz rule holds:
\[
\nabla_p(fs) = df \otimes s + f\, \nabla_p s.
\]
	If \(e\) is a local frame and \(s = w e\), the covariant derivative is given explicitly by:
\[
\nabla_p s = (dw - i A_p w) \otimes e.
\]

\subsubsection{Curvature.}
In alignment with our conventions, the formal \textbf{curvature} \( F_\nabla \) of the connection is defined locally on each chart as~\cite{AdemLeidaRuan}:
\[
F_\nabla = -i dA_p.
\]
 The \(G_p\)-invariance of the real connection 1-form \( A_p \) ensures that the purely imaginary curvature \( F_\nabla \) is also invariant under the local group actions. Consequently, the closed 2-forms \(\{-i dA_p\}\) descend to a well-defined global orbifold 2-form on \(X\).

\subsubsection{Distributional Curvature.}
A subtle but important feature of orbifold geometry arises when we view the curvature from the perspective of the underlying coarse space \(|X|\)~\cite{AdemLeidaRuan}. The curvature \(F_\nabla\) used in orbifold Chern--Weil theory is a smooth \emph{orbifold} form: on each uniformizing chart it lifts to a smooth invariant form, and these local forms glue in the orbifold sense. A different object appears only after choosing a singular trivialization on the punctured coarse space and pushing the local data down to \(|X|\). In that coarse presentation the connection 1-form may be singular at an orbifold point, and its formal exterior derivative can contain a delta term. This distributional term is not an additional curvature on the orbifold itself; it is a coarse-space encoding of the endpoint isotropy phase in orbifold holonomy.

We use the following conventions throughout the manuscript:
\begin{itemize}
    \item in a unitary local frame, \(\nabla e=-iA\otimes e\), with \(A\) real, so \(F_\nabla=-idA\) and \(c_1^{\mathrm{dR}}(L)=[\frac{i}{2\pi}F_\nabla]=[\frac{dA}{2\pi}]\);
    \item if a cone generator of order \(n\) acts on the fiber by \(e^{2\pi ik/n}\), a positively oriented local orbifold loop has holonomy \(e^{-2\pi ik/n}\), because the endpoint identification uses the inverse group element on the fiber;
    \item when the same cone point is excised from a compact surface, the induced inner-boundary orientation changes the sign, so this local weight contributes \(+k/n\) to the orbifold degree formula.
\end{itemize}

\begin{example}[Fractional Monopole on the Teardrop]
Consider the ``teardrop'' orbifold \(\mathbb{CP}^1(n)\), which is topologically a sphere but has a single cone point of order \(n\) at the North Pole~\cite{Thurston1980,Scott1983}. Let \(U \subset \mathbb{C}\) be a coarse space chart around the cone point and \(V\) its uniformizing cover. The covering structure at this point is given by \(\pi: V \to U\) with \(\pi(z) = z^n\), where \(\mathbb{Z}_n\) acts on \(V\) by \(z \mapsto e^{2\pi i / n}z\).

Consider a line bundle whose local isotropy at the singular point is given by \(\rho: \mathbb{Z}_n \to U(1)\), with generator mapped to \(e^{2\pi i k / n}\). On the local cover \(V\), we can choose a smooth connection 1-form \(\tilde{A}\). However, if we attempt to push this down to a connection 1-form \(A\) on the coarse space, we encounter a problem at the origin: the group action fixes the origin, preventing a smooth pushdown. Away from the origin, the group acts freely, so the punctured quotient \(U \setminus \{0\}\) is a smooth manifold. Choosing an equivariant frame on \(V\setminus\{0\}\) that descends to the punctured quotient gives the singular part
\[
A_{\mathrm{sing}} = -\frac{k}{n} d\theta.
\]
Therefore the singular contribution to the holonomy satisfies
\[
\lim_{\varepsilon\to 0}\exp\!\bigl(i\oint_{\partial U_\varepsilon}A\bigr)=e^{-2\pi i k/n},
\]
which is the inverse of the isotropy character. This inverse appears because parallel transport along the lifted arc ends over \(e^{2\pi i/n}z\), and the quotient identification back to the original fiber uses the action of \(e^{-2\pi i/n}\). Here \(\theta\) is the angular coordinate on \(U\setminus\{0\}\). The real 1-form \(A\) is smooth on \(U \setminus \{0\}\) but singular at the cone point. In distributions, \(d(d\theta)=2\pi\delta_0\), so the formal singular curvature is
\[
F_{\mathrm{sing}} = -i dA_{\mathrm{sing}} = 2\pi i \frac{k}{n}\,\delta_0.
\]
Therefore, in this coarse trivialization,
\[
\frac{i}{2\pi} F_\nabla = -\frac{k}{n} \delta_0 + \text{a smooth 2-form}.
\]
This last display belongs to the singular coarse trivialization, not to the smooth orbifold Chern--Weil representative. The sign records the convention that the holonomy around a positive coarse loop is the inverse of the isotropy character. In the compact surface degree formula, the same local data are evaluated by excising a small cone disk; the induced inner-boundary orientation changes the sign and contributes \(+k/n\) to the orbifold line-bundle degree~\cite{AdemLeidaRuan,Furuta1992SeifertFH,Sakai2017OrbifoldVortices}.
\end{example}

\section{First Chern Class of Orbifold Line Bundles}\label{sec:chern_weil_theory}

\subsection{Curvature as a Cohomology Class}

\begin{theorem}[Chern--Weil homomorphism for orbifold line bundles~\cite{AdemLeidaRuan,MoerdijkPronk1997}]\label{thm:orbifold-chern-weil-line}
Let \(L \to X\) be an orbifold line bundle endowed with a unitary orbifold connection \(\nabla\). Then the following hold:
\begin{enumerate}
    \item The local curvature 2-forms \(\{F_{\nabla,p}\}\) assemble into a global orbifold 2-form \(F_\nabla \in \Omega^2_{\mathrm{orb}}(X;i\mathbb{R})\).
    \item The global curvature form \(F_\nabla\) is closed, i.e., \(dF_\nabla=0\).
    \item Given any two unitary orbifold connections \(\nabla^0\) and \(\nabla^1\) on \(L\), there exists a globally defined orbifold 1-form \(\beta \in \Omega^1_{\mathrm{orb}}(X;i\mathbb{R})\) satisfying
    \[
    F_{\nabla^1}-F_{\nabla^0}=d\beta.
    \]
    Consequently, the de Rham cohomology class
    \[
    \left[\frac{i}{2\pi}F_\nabla\right]\in H^2_{\mathrm{dR,orb}}(X)
    \]
    is a topological invariant of the bundle \(L\), invariant under the choice of connection.
\end{enumerate}
\end{theorem}

\begin{proof}
The first assertion follows directly from a coordinate-patch gluing argument. For the second assertion, the Bianchi identity trivializes in the Abelian case; locally on each chart one obtains
\[
dF_{\nabla,p}=d(-i\,dA_p)=0.
\]
Thus, the global orbifold 2-form \(F_\nabla\) is closed. Regarding the third assertion, consider the local connection 1-forms \(A_p^0\) and \(A_p^1\) associated with \(\nabla^0\) and \(\nabla^1\), respectively. Their difference
\[
\eta_p:=A_p^1-A_p^0
\]
is real-valued. Furthermore, across charting overlaps the non-homogeneous transition terms cancel, enabling the local forms \(\eta_p\) to glue smoothly and define a global orbifold 1-form \(\eta \in \Omega^1_{\mathrm{orb}}(X;\mathbb{R})\). Setting \(\beta:=-i\,\eta\), with local representatives \(\beta_p:=-i\,\eta_p\), one obtains a global orbifold 1-form \(\beta \in \Omega^1_{\mathrm{orb}}(X;i\mathbb{R})\). It then follows that
\[
F_{\nabla^1,p}-F_{\nabla^0,p}=-i\,d(A_p^1-A_p^0)=d(-i\,\eta_p)=d\beta_p
\]
holds uniformly across all charts, establishing the global relation \(F_{\nabla^1}-F_{\nabla^0}=d\beta\). As a result, the normalized curvature yields a rigorous and well-defined de Rham cohomology class.
\end{proof}

\begin{definition}[De Rham first Chern class]\label{def:real-first-chern-class}
For an orbifold line bundle \(L \to X\), the de Rham first Chern class is defined as the cohomology class
\[
c_1^{\mathrm{dR}}(L):=\left[\frac{i}{2\pi}F_\nabla\right]\in H^2_{\mathrm{dR,orb}}(X)\cong H^2_{\mathrm{coarse}}(X;\mathbb{R}),
\]
where the canonical isomorphism stems from the orbifold de Rham theorem elucidated in Section~\ref{sec:orbifold_de_rham}.
\end{definition}

This cohomology class acts as the continuous geometric data complementing the discrete holonomy data explored in Section~\ref{sec:classification}. While the isotropy characters encode the localized topological phases endemic to the singular strata, the class \(c_1^{\mathrm{dR}}(L)\) quantifies the integrated field strength or total curvature sustained by the bundle.

\subsection{Fractional Flux and the Rational Degree}

Over an oriented two-dimensional compact orbifold \(\Sigma\), the integration of the normalized curvature 2-form yields the degree of the line bundle, which corresponds physically to the total magnetic flux piercing the surface:
\[
\deg(L) := \int_\Sigma \frac{i}{2\pi}F_\nabla.
\]
On the smooth locus of \(\Sigma\), this integration behaves according to the standard formulation for smooth manifolds. The distinguishing feature of the orbifold geometry, however, is that a small loop retracting to an orbifold singularity can indefinitely retain nontrivial holonomy. As established in Section~\ref{sec:orbifold_line_bundles}, this asymptotic holonomy is uniquely dictated by the assigned isotropy character. Under the application of Stokes' theorem, these localized holonomic phases manifest as singular boundary contributions to the integrated flux.

To articulate this rigorously, suppose \(\Sigma\) lacks boundary and contains cone points \(p_1, \ldots, p_k\) governed by local uniformizing groups \(G_{p_j} \cong \mathbb{Z}_{m_j}\). If \(k=0\), the following formula reduces to the ordinary Chern--Weil integrality statement \(\deg(L)=b_0\in\mathbb{Z}\). We therefore assume \(k\geq 1\) in the excision argument. Let the respective isotropy representations be specified by \(\rho_{p_j}(\gamma_j) = e^{2\pi i a_j/m_j}\) for \(0 \leq a_j < m_j\). By executing a global excision of sufficiently small orbifold disk neighborhoods \(U_j(\varepsilon)\) centered at each \(p_j\), the orbifold decomposes into a smooth, compact surface with boundary, \(\Sigma_0(\varepsilon) := \Sigma \setminus \bigcup_{j=1}^k \text{int}(U_j(\varepsilon))\). Because the Chern--Weil representative \((i/2\pi)F_\nabla\) is a smooth orbifold \(2\)-form, its integral over \(U_j(\varepsilon)\) tends to zero as \(\varepsilon\to 0\). Thus the full orbifold integral is the limit of the integrals over the truncated surfaces \(\Sigma_0(\varepsilon)\); the fractional contribution appears through the limiting boundary holonomy, not through an additional smooth orbifold curvature term at the cone point.

Since \(\Sigma_0(\varepsilon)\) retracts onto a 1-complex, every line bundle over it is trivializable. One may therefore adopt a global unitary frame and represent the connection as \(\nabla = d - iA\). Invoking Stokes' theorem on this truncated smooth domain, with the boundary orientation \(\partial \Sigma_0(\varepsilon) = -\sum_j \partial U_j(\varepsilon)\), allows the integral to be rigorously reduced to an evaluation of the connection 1-form over the boundary circles:
\[
\deg(L) = \lim_{\varepsilon \to 0} \int_{\Sigma_0(\varepsilon)} \frac{i}{2\pi}F_\nabla = -\frac{1}{2\pi} \sum_{j=1}^k \lim_{\varepsilon \to 0} \oint_{\partial U_j} A.
\]
With the above orientation convention, the boundary integral is determined modulo \(2\pi\) by the local isotropy weight:
\[
\frac{1}{2\pi}\oint_{\partial U_j}A \equiv -\frac{a_j}{m_j}\pmod{\mathbb{Z}}.
\]
Equivalently, for suitable integers \(n_j\in\mathbb{Z}\),
\[
\oint_{\partial U_j}A = 2\pi\left(n_j-\frac{a_j}{m_j}\right).
\]
Substituting this into the Stokes formula shows that the total degree manifests as (cf.\ Theorem~\ref{thm:pic_h2} and the surface classification of Section~\ref{sec:classification})
\[
\deg(L) = b_0 + \sum_{j=1}^k \frac{a_j}{m_j},
\]
for some integer \(b_0 \in \mathbb{Z}\) (namely \(b_0=-\sum_j n_j\)). This integral term \(b_0\) corresponds to the conventional first Chern class evaluated over the smoothed topological surface \(|\Sigma|\), effectively measuring the macroscopic integral twisting of a fully desingularized reference bundle.

These fractional boundary contributions are deeply corroborated by the coarse-space formalism encountered during the treatment of distributional curvature. In the punctured coarse space neighborhood of a cone point, one derives the singular gauge potential \(A_{\mathrm{sing}} = -(a_j/m_j)\,d\theta\), which explicitly yields the holonomy \(e^{-2\pi i a_j/m_j}\). While on the orbifold proper the connection's curvature is everywhere a smooth orbifold 2-form, projecting to the underlying coarse quotient forces the distributional curvature to encode the same fractional holonomy through a Dirac delta singularity.

This synthesis vividly illustrates that the degree of an orbifold line bundle is universally a rational number. Although the real first Chern class is defined as a valid de Rham class, its evaluation over the fundamental class of the orbifold \([\Sigma]\) strictly resides within \(\mathbb{Q}\). The fractional remainder of the degree is thus completely and inextricably determined by the local data---namely, the isotropy representations at the singular loci. Because the bundle is globally constrained, any localized phase modifications must inevitably be balanced by commensurate fractional adjustments to the global topological flux.

In the special case where the surface \(\Sigma\) is a \emph{good orbifold}---meaning it admits a finite manifold cover and can be expressed as a global quotient \([M/G]\) of a smooth compact oriented surface \(M\) by a finite group \(G\)---this rationality is immediately apparent from the covering geometry. An orbifold line bundle \(L \to \Sigma\) corresponds to a \(G\)-equivariant line bundle \(\tilde{L} \to M\). As \(M\) is a smooth manifold, standard Chern--Weil theory dictates that the degree of \(\tilde{L}\) is an integer, \(\deg(\tilde{L}) \in \mathbb{Z}\). Because the natural projection \(\pi: M \to \Sigma\) acts generically as a covering map of degree \(|G|\), integrating the equivariant curvature over \(M\) and passing to the quotient yields the structural relation
\[
\deg(L) = \frac{1}{|G|} \deg(\tilde{L}).
\]
This identity explicitly restricts the orbifold degree to rational values with denominators dividing the discrete group order \(|G|\). The localized excision argument is thus particularly enlightening: it unequivocally secures the strict rationality of \(\deg(L)\) based purely on local asymptotic holonomies, affirming the validity of this rational quantization universally---even for \emph{bad orbifolds} (such as the teardrop or non-developable spindle) which inherently lack any such global uniformizing manifold.

\section{Classification of Orbifold Line Bundles}\label{sec:classification}

The preceding sections isolate the two ingredients that enter the classification problem for orbifold line bundles. On each uniformizing chart, the bundle carries a one-dimensional unitary representation of the local isotropy group. Globally, these local sectors are glued by transition functions exactly as in the manifold case. The orbifold contribution is therefore localized: it appears through the isotropy characters, while the global topology is encoded by the cocycle of transition functions. In this section, we provide a rigorous local-to-global classification of these bundles and prove the central isomorphism relating the smooth/topological orbifold Picard group to stack/classifying-space cohomology.

\subsection{Cohomological Classification}

We begin by formalizing the algebraic structure of the set of orbifold line bundles.

\begin{definition}[Orbifold Picard Group]
Let $X$ be an orbifold. The \textbf{smooth/topological orbifold Picard group} of $X$, denoted $\mathrm{Pic}_{\mathrm{orb}}(X)$, is the abelian group of isomorphism classes of smooth orbifold line bundles over $X$ in the smooth category, or topological orbifold line bundles in the topological category. The group operation is tensor product, and the inverse of a line bundle is its dual.
\end{definition}

This notation is deliberately topological. If \(X\) later carries a complex orbifold structure, we write \(\mathrm{Pic}_{\mathrm{hol}}(X)\) for the group of holomorphic orbifold line bundles. There is then a forgetful map \(\mathrm{Pic}_{\mathrm{hol}}(X)\to\mathrm{Pic}_{\mathrm{orb}}(X)\), but the two groups should not be identified without an additional argument. In purely topological passages where no complex structure is under discussion, the shorter notation \(\mathrm{Pic}(X)\) may still appear as an abbreviation for \(\mathrm{Pic}_{\mathrm{orb}}(X)\).

In the manifold setting, line bundles are classified by the first \v{C}ech cohomology group of the sheaf of non-vanishing continuous complex-valued functions. This result generalizes to orbifolds when the orbifold is treated as a stack or, equivalently, by any Morita-equivalent proper \'etale groupoid presentation~\cite{MoerdijkPronk1997,AdemLeidaRuan,BehrendXu2003,BehrendXu2011}.

\begin{assumption}[Cohomology and effectivity conventions]\label{ass:pic-h2-conventions}
Unless otherwise specified, \(X\) is a paracompact Hausdorff orbifold; in the main smooth part of this manuscript it is an effective smooth orbifold in the sense of Definition~\ref{def:orbifold}. In the smooth category, choose a proper \'etale Lie groupoid presentation
\[
\mathcal{G}_1 \rightrightarrows \mathcal{G}_0
\]
for the associated differentiable stack \(\mathcal{X}\). In the purely topological category, use the analogous proper \'etale topological groupoid presentation. We write
\[
H^k_{\mathrm{st}}(X;A):=H^k(B\mathcal{G};A)
\cong H^k(\mathcal{X},\underline{A}),
\]
where \(B\mathcal{G}\) is the classifying space of the groupoid and \(\underline{A}\) is the corresponding constant sheaf on the stack. This is a Morita-invariant stack cohomology group, not the ordinary cohomology of the coarse space \(|X|\). The notation follows the convention table in Section~\ref{sec:cohomology-notation}. For a finite global quotient \(X=[M/G]\), this convention gives \(H^k_{\mathrm{st}}(X;A)\cong H^k_G(M;A)\).

The same convention applies to non-effective orbifold stacks when they are explicitly used: ineffective stabilizer groups are retained as part of the stack. Replacing a non-effective stack by its effective reduction is a different operation and can change both \(\mathrm{Pic}_{\mathrm{orb}}(X)\) and \(H^2_{\mathrm{st}}(X;\mathbb{Z})\). The proof below does not use effectivity; it uses paracompactness, the proper \'etale groupoid/stack presentation, and the existence of partitions of unity. In the smooth category we use smooth transition functions and the sheaf of smooth functions; in the purely topological category the same argument uses continuous transition functions and continuous functions.
\end{assumption}

\begin{theorem}[Topological and Smooth Classification~\cite{MoerdijkPronk1997,AdemLeidaRuan,BehrendXu2003,BehrendXu2011}]\label{thm:pic_h2}
Let \(X\) satisfy Assumption~\ref{ass:pic-h2-conventions}. Then the orbifold first Chern class defines a canonical isomorphism
\[
c_1^{\mathrm{st}} \colon \mathrm{Pic}_{\mathrm{orb}}(X) \xrightarrow{\ \cong\ } H^2_{\mathrm{st}}(X;\mathbb{Z}).
\]
Here \(\mathrm{Pic}_{\mathrm{orb}}(X)\) is the smooth/topological Picard group fixed above, not the holomorphic Picard group. Equivalently, every smooth/topological complex orbifold line bundle on \(X\) is determined up to isomorphism by its class in the classifying-space/stack cohomology group \(H^2_{\mathrm{st}}(X;\mathbb{Z})\), and every class in \(H^2_{\mathrm{st}}(X;\mathbb{Z})\) arises from a unique isomorphism class of such orbifold line bundles.
\end{theorem}
\begin{proof}
The proof is the same in spirit as the standard manifold proof, except that every construction is carried out on the orbifold stack associated to \(X\). Let \(\mathcal{X}\) denote this stack. Write \(\mathcal{A}_{\mathcal{X}}\) for the sheaf of smooth complex-valued functions in the smooth category, and for the sheaf of continuous complex-valued functions in the topological category; write \(\mathcal{A}_{\mathcal{X}}^\times\) for the subsheaf of nowhere-vanishing functions.

The first step is the standard identification
\[
\mathrm{Pic}_{\mathrm{orb}}(X)\cong H^1(\mathcal{X},\mathcal{A}_{\mathcal{X}}^\times).
\]
Indeed, choose an orbifold groupoid presentation
\[
\mathcal{G}_1 \rightrightarrows \mathcal{G}_0
\]
for \(\mathcal{X}\). An orbifold line bundle on \(X\) is the same as a \(\mathcal{G}\)-equivariant complex line bundle on \(\mathcal{G}_0\). After refining \(\mathcal{G}_0\) by a \(\mathcal{G}\)-invariant open cover \(\{V_i\}\) on which the underlying line bundle is trivial, the bundle is encoded by ordinary transition functions on the overlaps together with compatibility on composable arrows of the groupoid. Equivalently, one obtains a \v{C}ech \(1\)-cocycle on the simplicial nerve of \(\mathcal{G}\) with values in \(\mathcal{A}_{\mathcal{X}}^\times\), and changing the chosen local trivializations alters this cocycle by a coboundary. Thus isomorphism classes are classified by \(H^1(\mathcal{X},\mathcal{A}_{\mathcal{X}}^\times)\).

The next step is to convert this first cohomology group into a second cohomology group with integer coefficients. Consider the exponential sequence of sheaves on $\mathcal{X}$:
\[
0 \to \underline{\mathbb{Z}} \to \mathcal{A}_{\mathcal{X}} \xrightarrow{\exp(2\pi i\,\cdot)} \mathcal{A}_{\mathcal{X}}^\times \to 0,
\]
where \(\underline{\mathbb{Z}}\) is the locally constant sheaf with stalk \(\mathbb{Z}\). This sequence is exact for the same reason as on an ordinary space: locally every nowhere-vanishing complex-valued function admits a logarithm in the chosen category, and the ambiguity in choosing that logarithm is an integer.

By paracompactness, orbifolds admit partitions of unity subordinate to orbifold open covers, so the sheaf \(\mathcal{A}_{\mathcal{X}}\) is fine and hence acyclic. Therefore
\[
H^q(\mathcal{X},\mathcal{A}_{\mathcal{X}})=0
\qquad\text{for all } q>0.
\]
Applying cohomology to the exponential sequence therefore yields the exact segment
\[
H^1(\mathcal{X},\mathcal{A}_{\mathcal{X}})
\to H^1(\mathcal{X},\mathcal{A}_{\mathcal{X}}^\times)
\xrightarrow{\ \delta\ }
H^2(\mathcal{X},\underline{\mathbb{Z}})
\to H^2(\mathcal{X},\mathcal{A}_{\mathcal{X}}).
\]
The outer groups vanish, so the connecting homomorphism
\[
\delta \colon H^1(\mathcal{X},\mathcal{A}_{\mathcal{X}}^\times)\xrightarrow{\ \cong\ } H^2(\mathcal{X},\underline{\mathbb{Z}})
\]
is an isomorphism. By definition,
\[
H^2(\mathcal{X},\underline{\mathbb{Z}})=H^2_{\mathrm{st}}(X;\mathbb{Z}).
\]

It remains to explain why this isomorphism is exactly the orbifold first Chern class. Represent the class of an orbifold line bundle $L$ by a \v{C}ech $1$-cocycle $\{g_{\alpha\beta}\}$ on a cover of the groupoid nerve used to compute \(H^1(\mathcal{X},\mathcal{A}_{\mathcal{X}}^\times)\). Since the exponential map is locally surjective, after refining the cover if necessary, we may choose local logarithms $h_{\alpha\beta}$ such that
\[
g_{\alpha\beta}=\exp(2\pi i\, h_{\alpha\beta}).
\]
On a $2$-simplex of the nerve, the cocycle condition translates to
\[
\exp\!\bigl(2\pi i(h_{\beta\gamma}-h_{\alpha\gamma}+h_{\alpha\beta})\bigr)=1,
\]
so the combination
\[
n_{\alpha\beta\gamma}:=h_{\beta\gamma}-h_{\alpha\gamma}+h_{\alpha\beta}
\]
takes values in $\mathbb{Z}$. The collection $\{n_{\alpha\beta\gamma}\}$ is exactly the \v{C}ech $2$-cocycle $\delta h$ resulting from the coboundary of the local logarithms, and its cohomology class is precisely $\delta([\{g_{\alpha\beta}\}])$. Different choices of local logarithms change $\{n_{\alpha\beta\gamma}\}$ by an integer coboundary, so the class is well-defined.

Geometrically, this integer $2$-cocycle measures the obstruction to choosing the logarithms $h_{\alpha\beta}$ so that they fit together globally, or equivalently the obstruction to flattening all transition data into a globally consistent trivialization. This obstruction is exactly the orbifold analogue of the usual first Chern class. Therefore
\[
c_1^{\mathrm{st}}(L)=\delta([\{g_{\alpha\beta}\}]).
\]

Combining the two steps above gives the canonical isomorphism
\[
c_1^{\mathrm{st}} \colon \mathrm{Pic}_{\mathrm{orb}}(X)
\xrightarrow{\ \cong\ }
H^2_{\mathrm{st}}(X;\mathbb{Z}).
\]
In particular, two orbifold line bundles are isomorphic if and only if they have the same stack first Chern class, and every class in \(H^2_{\mathrm{st}}(X;\mathbb{Z})\) is realized by an orbifold line bundle.
\end{proof}

\begin{remark}[Orbifold vs.\ stack vs.\ coarse-space cohomology~\cite{AdemLeidaRuan}]\label{rem:cohomology_conventions}
The group \(H^2_{\mathrm{st}}(X;\mathbb{Z})\) appearing above is the singular cohomology of the classifying space \(B\mathcal{G}\) of any orbifold groupoid presentation \(\mathcal{G}_1\rightrightarrows\mathcal{G}_0\); equivalently, it is the sheaf cohomology on the associated orbifold stack \(\mathcal{X}\). It is not Chen--Ruan cohomology and not the ordinary integral cohomology of the coarse space. This is the integral cohomology theory that classifies orbifold line bundles. Passing to real coefficients and then applying the real comparison isomorphism gives the de Rham/coarse class discussed in Section~\ref{sec:cohomology-notation}. What changes integrally is not only possible torsion: the orbifold integral lattice inside \(H^2_{\mathrm{dR,orb}}(X)\cong H^2_{\mathrm{coarse}}(X;\mathbb{R})\) can be strictly finer than the ordinary integral lattice on the coarse space, which is why rational degrees can occur even when \(H^2_{\mathrm{st}}(X;\mathbb{Z})\) itself has no torsion. For a finite global quotient \(X=[M/G]\), \(H^2_{\mathrm{st}}(X;\mathbb{Z})\) coincides with the Borel equivariant cohomology \(H^2_G(M;\mathbb{Z})\). For a non-effective quotient stack, the same notation retains the generic stabilizer; passing to the effective reduction may remove line-bundle sectors detected by \(H^2_{\mathrm{st}}(X;\mathbb{Z})\).
\end{remark}

\subsection{Good Orbifolds and Equivariant Bundles}

For orbifolds presented as global quotients of a connected manifold by a finite group acting effectively, $X = [M/G]$, the classification of line bundles can be formulated in terms of equivariant geometry on $M$; this is the standard quotient-orbifold dictionary.

\begin{definition}[$G$-Equivariant Picard Group]
Let $M$ be a smooth connected manifold with a smooth effective action by a finite group $G$. The $G$-equivariant Picard group, denoted $\mathrm{Pic}_G(M)$, is the abelian group of isomorphism classes of $G$-equivariant line bundles over $M$, with group operation given by the tensor product.
\end{definition}

\begin{theorem}[Equivariant Classification~\cite{putman2012picard}]\label{thm:good_orbifold_classification}
Let $M$ be a connected smooth manifold equipped with a smooth effective action of a finite group $G$, and let
\[
X=[M/G]
\]
be the associated global quotient orbifold. Then passing from the quotient orbifold back to the covering manifold identifies orbifold line bundles on $X$ with $G$-equivariant line bundles on $M$. More precisely, there is a canonical isomorphism
\[
\mathrm{Pic}_{\mathrm{orb}}(X) \cong \mathrm{Pic}_G(M).
\]

Equivalently, to give a line bundle on the orbifold $[M/G]$ is the same as to give an ordinary complex line bundle on $M$ together with a compatible lift of the $G$-action to the total space of that bundle.

Assume in addition that $H^1(M,\mathbb{Z}) = 0$. Then the classification of equivariant structures is controlled by the exact sequence
\[
0 \to \mathrm{Hom}(G,U(1)) \to \mathrm{Pic}_G(M) \xrightarrow{\phi} \mathrm{Pic}(M)^G \xrightarrow{\delta} H^2(G, U(1))
\]
where:
\begin{enumerate}
    \item $\phi$ is the forgetful map, which drops the equivariant structure and remembers only the underlying line bundle on $M$;
    \item $\mathrm{Pic}(M)^G$ is the subgroup of isomorphism classes of line bundles $L$ on $M$ satisfying $g^*L \cong L$ for every $g \in G$, so these are exactly the line bundles whose isomorphism class is preserved by the group action;
    \item $\mathrm{Hom}(G,U(1))$ records the possible ways of modifying a given equivariant structure by a character of $G$;
    \item $\delta$ is the obstruction map: for a $G$-invariant line bundle $L$ on $M$, the class $\delta(L)\in H^2(G,U(1))$ vanishes if and only if $L$ admits a $G$-equivariant lift.
\end{enumerate}

In particular, when $\delta(L)=0$, the set of all inequivalent $G$-equivariant structures on the underlying bundle $L$ differs by twisting with characters of $G$, so it is governed by $\mathrm{Hom}(G,U(1))$. Thus the exact sequence separates the problem into two parts familiar from manifold theory: first choose an ordinary line bundle on $M$ whose isomorphism class is fixed by $G$, and then determine whether the $G$-action lifts, with the possible lifts differing by characters.
\end{theorem}
\begin{proof}
The isomorphism $\mathrm{Pic}_{\mathrm{orb}}(X) \cong \mathrm{Pic}_G(M)$ follows directly from the quotient presentation $X=[M/G]$: an orbifold line bundle on $X$ is exactly a line bundle on $M$ together with a compatible $G$-linearization.

For the exact sequence, use the canonical identifications
\[
\mathrm{Pic}_G(M)\cong H^2_G(M,\mathbb{Z}),
\qquad
\mathrm{Pic}(M)\cong H^2(M,\mathbb{Z}).
\]
Apply the Leray--Serre spectral sequence to the Borel fibration
\[
M \longrightarrow EG \times_G M \longrightarrow BG.
\]
Its $E_2$-page is
\[
E_2^{p,q}=H^p\!\bigl(G,H^q(M,\mathbb{Z})\bigr).
\]
Because $H^1(M,\mathbb{Z})=0$, we have $E_2^{1,1}=0$. Since $M$ is connected, the $G$-action on $H^0(M,\mathbb{Z}) \cong \mathbb{Z}$ is trivial, giving $E_2^{2,0} = H^2(G, \mathbb{Z}) = H^2(BG,\mathbb{Z})$. In total degree $2$, the only nonzero terms are
\[
E_2^{2,0}=H^2(BG,\mathbb{Z}),
\qquad
E_2^{0,2}=H^2(M,\mathbb{Z})^G,
\]
and there is no incoming differential to $E_2^{0,2}$. The first potentially nontrivial differential out of $E_2^{0,2}$ is the transgression
\[
d_3 \colon H^2(M,\mathbb{Z})^G \to H^3(BG,\mathbb{Z}),
\]
which is exactly the obstruction map. Therefore the low-degree terms give the exact sequence
\[
0 \to H^2(BG,\mathbb{Z}) \to H^2_G(M,\mathbb{Z}) \xrightarrow{\phi} H^2(M,\mathbb{Z})^G \xrightarrow{\delta} H^3(BG,\mathbb{Z}).
\]
For finite $G$, the short exact sequence of trivial $G$-modules $0 \to \mathbb{Z} \to \mathbb{R} \to U(1) \to 0$ identifies
\[
\begin{aligned}
H^2(BG,\mathbb{Z}) &\cong H^1(G,U(1)) = \mathrm{Hom}(G,U(1)),\\
H^3(BG,\mathbb{Z}) &\cong H^2(G,U(1)).
\end{aligned}
\]
Translating back through the Picard/cohomology identifications yields the stated exact sequence.
\end{proof}

\begin{example}[Fundamental Quotient Orbifolds]
\begin{enumerate}

    \item \textbf{The Cone Orbifold:} For the cone $X=[\mathbb{C}/\mathbb{Z}_n]$, the covering manifold is $M=\mathbb{C}$ with generator acting by $z \mapsto e^{2\pi i/n}z$. Since $\mathbb{C}$ is contractible, $H^1(\mathbb{C},\mathbb{Z}) = 0$ and $\mathrm{Pic}(\mathbb{C}) = 0$. The exact sequence therefore gives $\mathrm{Pic}_{\mathbb{Z}_n}(\mathbb{C}) \cong \mathrm{Hom}(\mathbb{Z}_n, U(1)) \cong \mathbb{Z}_n$. Concretely, every orbifold line bundle on the cone is obtained by lifting the $\mathbb{Z}_n$-action to the trivial bundle $\mathbb{C} \times \mathbb{C}$ via
    \[
    (z,w) \longmapsto \bigl(e^{2\pi i/n}z,\, e^{2\pi i k/n}w\bigr),
    \qquad k \in \{0,\dots,n-1\}.
    \]

    \item \textbf{The Football Orbifold:} For the football orbifold $X=[S^2/\mathbb{Z}_n] \cong S^2(n,n)$, the covering manifold is the Riemann sphere $M=S^2$ with the rotational $\mathbb{Z}_n$-action fixing the two poles. Since $H^1(S^2,\mathbb{Z}) = 0$, the exact sequence applies. Even the trivial bundle $S^2 \times \mathbb{C}$ admits $n$ inequivalent equivariant structures, one for each character of $\mathbb{Z}_n$. More generally, the $\mathbb{Z}_n$-action preserves the degree of a line bundle, so every $\mathcal{O}(d)$ lies in $\mathrm{Pic}(S^2)^{\mathbb{Z}_n}$. For the cyclic group $\mathbb{Z}_n$, the obstruction term vanishes: the coefficient sequence $0\to\mathbb{Z}\to\mathbb{R}\to U(1)\to 0$ gives $H^2(\mathbb{Z}_n,U(1))\cong H^3(\mathbb{Z}_n,\mathbb{Z})=0$, since odd-degree integral cohomology of cyclic groups vanishes. Hence every invariant bundle $\mathcal{O}(d)$ admits an equivariant lift, so the forgetful map $\phi$ is surjective with kernel $\mathrm{Hom}(\mathbb{Z}_n,U(1)) \cong \mathbb{Z}_n$. Therefore
    \[
    0 \to \mathbb{Z}_n \to \mathrm{Pic}_{\mathbb{Z}_n}(S^2) \xrightarrow{\phi} \mathbb{Z} \to 0.
    \]
    Since $\mathbb{Z}$ is free, this extension splits, and we obtain $\mathrm{Pic}_{\mathbb{Z}_n}(S^2) \cong \mathbb{Z} \oplus \mathbb{Z}_n$. The local weights at the two orbifold points are still linked by the global integral class, a point made precise in the surface classification below.
\end{enumerate}
\end{example}

\subsection{Global Decomposition for Orbifold Surfaces}

The interplay between localized isotropy characters and the global topology governs the classification of orbifold line bundles over surfaces. We make this precise with the following structural result from the Seifert-invariant description of orbifold line bundles over orbifold Riemann surfaces~\cite{Furuta1992SeifertFH,MrowkaOzsvathYu1997,Sakai2017OrbifoldVortices}.

\begin{theorem}[Orbifold Surface Classification]
Let $\Sigma$ be a compact oriented $2$-dimensional orbifold of genus $g$ whose singularities are cone points with cyclic isotropy groups $\mathbb{Z}_{m_1}, \dots, \mathbb{Z}_{m_k}$. Then the smooth/topological orbifold Picard group $\mathrm{Pic}_{\mathrm{orb}}(\Sigma)$ fits into the short exact sequence
\[
0 \to H^2(|\Sigma|, \mathbb{Z}) \to \mathrm{Pic}_{\mathrm{orb}}(\Sigma) \to \bigoplus_{i=1}^{k} \mathbb{Z}_{m_i} \to 0.
\]
In general this sequence does \emph{not} split. A convenient description of the resulting group law is given by the Seifert-type invariants and fractional degree discussed after the proof.
\end{theorem}
\begin{proof}
If $k=0$, then $\Sigma$ is an ordinary closed oriented surface and the statement reduces to the classical topological isomorphism $\mathrm{Pic}_{\mathrm{orb}}(\Sigma)=\mathrm{Pic}(\Sigma)\cong H^2(|\Sigma|,\mathbb{Z})$. We therefore assume $k\ge 1$.

We compute $\mathrm{Pic}_{\mathrm{orb}}(\Sigma)$ via a Mayer--Vietoris argument. To build physical intuition, think of this procedure as decomposing a magnetic field configuration into localized fluxes (analogous to magnetic monopoles or Aharonov-Bohm fluxes) at the singular points, and a smooth flux spread over the remainder of the surface.

Choose small disjoint orbifold open disk neighborhoods $U_i$ around each of the $k$ cone points $x_1, \dots, x_k$. Each $U_i$ is isomorphic to the quotient $D^2 / \mathbb{Z}_{m_i}$. Let $V = \Sigma \setminus \{x_1, \dots, x_k\}$ be the complement of the cone points. Since all orbifold points have been removed, $V$ is an ordinary smooth non-compact manifold diffeomorphic to a punctured surface.

Because $U_i$ is contractible to a singular point with isotropy $\mathbb{Z}_{m_i}$, its orbifold Picard group reduces to the possible lifts of the isotropy action, yielding \(\mathrm{Pic}_{\mathrm{orb}}(U_i) \cong H^2_{\mathrm{st}}(U_i;\mathbb{Z}) \cong \mathrm{Hom}(\mathbb{Z}_{m_i}, U(1)) \cong \mathbb{Z}_{m_i}\). Physically, these group elements represent fractional local holonomies trapped at the singularities. For the smooth punctured surface $V$, the second integral cohomology vanishes since $V$ is non-compact, so \(H^2_{\mathrm{st}}(V;\mathbb{Z}) = H^2(V;\mathbb{Z}) = 0\). The geometric interpretation is that a finite punctured surface cannot trap a fully quantized, globally non-trivial flux without a boundary or closure. The intersection \(U_i \cap V\) is a punctured orbifold disk, but the isotropy action is free there, so it is just a smooth annulus homotopy equivalent to \(S^1\). Hence \(H^2_{\mathrm{st}}(U_i \cap V;\mathbb{Z}) = 0\).

Since \(H^*_{\mathrm{st}}\) is the singular cohomology of a classifying space (equivalently, sheaf cohomology on the orbifold stack), it satisfies Mayer--Vietoris for this open cover. Applying Mayer--Vietoris with \(\mathbb{Z}\)-coefficients to the cover \(\{V, \coprod_{i} U_i\}\), the relevant portion reads:
\[
\begin{aligned}
H^1_{\mathrm{st}}(V;\mathbb{Z}) \oplus \bigoplus_{i=1}^k H^1_{\mathrm{st}}(U_i;\mathbb{Z})
&\xrightarrow{r} \bigoplus_{i=1}^k H^1_{\mathrm{st}}(U_i \cap V;\mathbb{Z}) \\
&\xrightarrow{\delta} H^2_{\mathrm{st}}(\Sigma;\mathbb{Z}) \\
&\to H^2_{\mathrm{st}}(V;\mathbb{Z}) \oplus \bigoplus_{i=1}^k H^2_{\mathrm{st}}(U_i;\mathbb{Z}) \to 0.
\end{aligned}
\]
Since the orbifold fundamental group of \(U_i\) is finite (namely \(\mathbb{Z}_{m_i}\)), we have \(H^1_{\mathrm{st}}(U_i;\mathbb{Z}) = \mathrm{Hom}(\mathbb{Z}_{m_i}, \mathbb{Z}) = 0\). The intersections give \(H^1_{\mathrm{st}}(U_i \cap V;\mathbb{Z}) \cong H^1(S^1;\mathbb{Z}) \cong \mathbb{Z}\). The restriction map \(r\) from \(H^1_{\mathrm{st}}(V;\mathbb{Z}) \cong H^1(V;\mathbb{Z}) \cong \mathbb{Z}^{2g + k - 1}\) to \(\bigoplus_{i=1}^k H^1(U_i \cap V;\mathbb{Z}) \cong \mathbb{Z}^k\) evaluates a cohomology class on \(V\) (viewed as a homomorphism \(H_1(V;\mathbb{Z}) \to \mathbb{Z}\)) on each small loop \(\gamma_i\) around the \(i\)-th puncture. Physically, this relates the continuous phase of the wavefunction (or the gauge field's holonomy) on the bulk \(V\) to the loop integrating around a specific cone point.

In homology, the loops $\gamma_1, \ldots, \gamma_k$ satisfy the relation $[\gamma_1] + \cdots + [\gamma_k] = 0$ in $H_1(V, \mathbb{Z})$, since they jointly bound the compact surface $\Sigma \setminus \coprod \mathrm{int}(U_i)$ which is homotopy equivalent to $V$. This is a topological analog of Gauss's law or charge conservation: the net flux flowing out of all punctures must sum to the total flux contained on the closed surface. Dually, since $[\gamma_1],\ldots,[\gamma_{k-1}]$ extend to part of a basis of $H_1(V,\mathbb{Z})$, the map $r$ surjects onto the hyperplane $\{(n_1,\ldots,n_k):\sum n_i = 0\}\cong \mathbb{Z}^{k-1}$, so $\mathrm{coker}(r) \cong \mathbb{Z}$. Since the orbifold structure on $V$ and on each annulus $U_i\cap V$ is trivial, the restriction map $r$ coincides with its coarse-space counterpart. By the ordinary Mayer--Vietoris sequence for the cover $\{|\Sigma|\setminus \{x_1,\dots,x_k\}, \coprod_i D_i\}$, this cokernel therefore identifies canonically with $H^2(|\Sigma|, \mathbb{Z})$, i.e.\ with the global integral class on the underlying surface.

Since \(H^2_{\mathrm{st}}(V;\mathbb{Z})=0\) (as \(V\) is a non-compact surface), and the subsequent term \(\bigoplus_{i=1}^k H^2_{\mathrm{st}}(U_i \cap V;\mathbb{Z})\) vanishes because each intersection is homotopy equivalent to a circle, the tail of the Mayer--Vietoris sequence guarantees surjectivity and reduces to:
\[
0 \to \mathrm{coker}(r) \to H^2_{\mathrm{st}}(\Sigma;\mathbb{Z}) \to \bigoplus_{i=1}^k H^2_{\mathrm{st}}(U_i;\mathbb{Z}) \to 0.
\]
Substituting \(\mathrm{coker}(r) \cong \mathbb{Z}\) and \(H^2_{\mathrm{st}}(U_i;\mathbb{Z}) \cong \mathbb{Z}_{m_i}\), and using the identification \(\mathrm{Pic}_{\mathrm{orb}}(\Sigma) \cong H^2_{\mathrm{st}}(\Sigma;\mathbb{Z})\) from Theorem~\ref{thm:pic_h2}, we obtain the short exact sequence
\[
0 \to H^2(|\Sigma|, \mathbb{Z}) \to \mathrm{Pic}_{\mathrm{orb}}(\Sigma) \to \bigoplus_{i=1}^{k} \mathbb{Z}_{m_i} \to 0.
\]
This is the claimed short exact sequence.
\end{proof}

A concrete nonsplit case is the teardrop orbifold $S^2(m)\cong \mathbb{CP}^1(m)$ with one cone point of order $m$. In that case the sequence becomes
\[
0 \to \mathbb{Z} \to \mathrm{Pic}_{\mathrm{orb}}(S^2(m)) \to \mathbb{Z}_m \to 0,
\]
where the kernel is generated by the ordinary degree-$1$ class on the coarse sphere. Choose a class $F$ mapping to $1\in \mathbb{Z}_m$. Its topological degree must be of the form $\deg(F)=d+1/m$ for some $d\in\mathbb{Z}$. Tensoring $F$ with the inverse of the degree-$d$ coarse generator, we may replace $F$ by a class of degree exactly $1/m$ and the same image in $\mathbb{Z}_m$. Every orbifold line bundle is then uniquely of the form
\[
H^{\otimes r}\otimes F^{\otimes a},
\qquad
r\in \mathbb{Z},\quad 0\le a<m,
\]
where $H$ denotes the ordinary degree-$1$ class on $S^2$. Its degree is
\[
\deg\!\bigl(H^{\otimes r}\otimes F^{\otimes a}\bigr)=r+\frac{a}{m},
\]
so the degree map identifies $\mathrm{Pic}_{\mathrm{orb}}(S^2(m))$ with $(1/m)\mathbb{Z}\cong \mathbb{Z}$. Under this identification the inclusion of the kernel is multiplication by $m$, so the extension is not split.

Although the sequence does not split in general, one can still choose Seifert-type invariants $(b_0; a_1,\dots,a_k)$ with $b_0 \in \mathbb{Z}$ and $0 \leq a_i < m_i$ for each cone point. With this normalization, the associated orbifold line bundle has fractional degree~\cite{Furuta1992SeifertFH,MrowkaOzsvathYu1997,Sakai2017OrbifoldVortices}
\[
\deg(L) = b_0 + \sum_{i=1}^k \frac{a_i}{m_i} \in \mathbb{Q}.
\]
The tensor-product law is therefore not componentwise addition on the tuple $(b_0; a_1,\dots,a_k)$: whenever a sum of local weights crosses a multiple of $m_i$, the excess contributes an additional copy of the integral generator in $H^2(|\Sigma|,\mathbb{Z})$.

This classification shows that orbifold line bundles are not obtained by simply adjoining independent torsion labels to the ordinary degree. The local isotropy weights determine the fractional part of the orbifold degree, and the carrying rule in the tensor product records how the extension by $H^2(|\Sigma|,\mathbb{Z})$ is assembled.

\section[Orbifold Euler Characteristic]{Two-Dimensional Orbifolds: Euler Characteristic}\label{sec:orbifold-euler}
\subsubsection{Gauss--Bonnet (Orbifold Form)~\cite{Satake1957}.}

Throughout this section, \(\chi_{\mathrm{orb}}\) denotes the classical Satake orbifold Euler characteristic, the one normalized by orbifold integration. It should not be confused with the stringy or inertia-orbifold Euler characteristics that also appear in quotient and orbifold literature. With this convention, the definition of \(\chi_{\mathrm{orb}}\) is not arbitrary; it is mandated by the analytical properties of the orbifold. If \(K\) is the Gaussian curvature and \(dA\) the area form, the orbifold Gauss--Bonnet theorem states:
\[
\int_X K\,dA \;=\; 2\pi\,\chi_{\mathrm{orb}}(X).
\]
Here, the integral is the orbifold integral defined in Section~\ref{sec:orbifold_de_rham}: locally one integrates on uniformizing charts and divides by the order of the relevant local uniformizing group. This is a chart-level quotient measure, not a point-mass contribution concentrated at the singular points.

This theorem provides the fundamental link between the topology (Euler characteristic) and the geometry (total curvature) of the orbifold.

While the general theory of orbifolds applies in any dimension, the two-dimensional case (orbifold surfaces) is particularly rich and tractable. Their geometry is governed by a single numerical invariant: the orbifold Euler characteristic.

\subsubsection{Orbifolds with Cone Points~\cite{davis2011lectures}.}
The simplest singularities in dimension two are isolated cone points. For a compact oriented surface \(\Sigma\) of genus \(g\) with \(r\) cone points \(x_1, \dots, x_r\) of orders \(m_1, \dots, m_r\), the \emph{orbifold Euler characteristic} is defined as:
\[
\chi_{\mathrm{orb}}(\Sigma; m_1,\dots,m_r)\;=\;\chi(\Sigma)\; - \;\sum_{i=1}^r\Bigl(1-\frac{1}{m_i}\Bigr),
\]
where \(\chi(\Sigma) = 2 - 2g\) is the standard topological Euler characteristic. Intuitively, each cone point reduces the ``total curvature capacity'' of the surface. The term \((1 - 1/m_i)\) represents the ``curvature deficit'' introduced by the singularity; a cone point of order \(m\) has only \(1/m\)-th the area (and thus curvature capacity) of a smooth disk.

\begin{proof}[Derivation of the cone point formula]
We verify this formula using Satake's orbifold Gauss--Bonnet theorem~\cite{Satake1957}. Remove a small orbifold disk \(D_i\) of geodesic radius \(\varepsilon\) around each cone point \(x_i\), producing a smooth compact surface with boundary \(\Sigma_0 = \Sigma \setminus \bigcup_i \mathrm{int}(D_i)\). On \(\Sigma_0\) the classical Gauss--Bonnet theorem applies:
\[
\int_{\Sigma_0} K\,dA - \sum_{i=1}^{r}\int_{\partial D_i} \kappa_g\,ds = 2\pi\,\chi(\Sigma_0),
\]
where \(\kappa_g\) denotes the geodesic curvature of each boundary circle (with the inward-pointing normal convention). A cone point of order \(m_i\) has a cone angle of \(2\pi/m_i\). As \(\varepsilon \to 0\), the integral of the geodesic curvature over the boundary circle \(\partial D_i\) tends to \(2\pi/m_i\) (the ``turning angle'' at the tip of the cone). Meanwhile, \(\chi(\Sigma_0) = \chi(\Sigma) - r\) since we have removed \(r\) open disks from a closed surface, adding \(r\) boundary components. Substituting:
\[
\int_\Sigma K\,dA = 2\pi\bigl(\chi(\Sigma) - r\bigr) + 2\pi\sum_{i=1}^{r}\frac{1}{m_i} = 2\pi\biggl(\chi(\Sigma) - \sum_{i=1}^{r}\Bigl(1 - \frac{1}{m_i}\Bigr)\biggr).
\]
Comparing with the orbifold Gauss--Bonnet statement \(\int_\Sigma K\,dA = 2\pi\,\chi_{\mathrm{orb}}(\Sigma)\), we obtain the claimed formula.
\end{proof}

\subsubsection{Orbifolds with Mirrors and Corner Reflectors.}
Beyond isolated cone points, 2-orbifolds can have boundaries with reflective symmetry. A \emph{mirror} (or \emph{reflector curve}) is a codimension-1 stratum along which the local isotropy is generated by a reflection. Where two mirrors meet, we find a \emph{corner reflector}. If the mirrors meet at an angle \(\pi/k\), the local isotropy is the dihedral group \(D_k\) of order \(2k\). We refer to \(k\) as the \emph{corner order} (note that the order of the singular point in the sense of Definition~\ref{def:orbifold} is \(|D_k| = 2k\)).

For a 2-orbifold \(X\) whose underlying space is a disk bounded entirely by mirrors with \(r\) corner reflectors of corner orders \(k_1,\ldots,k_r\) (and no interior singularities), the orbifold Euler characteristic is:
\[
  \chi_{\mathrm{orb}}(X) \;=\; 1 \;-\; \frac{1}{2}\sum_{j=1}^r \Bigl(1 - \frac{1}{k_j}\Bigr).
\]

\begin{proof}[Proof via the oriented double]
The oriented double \(\tilde{X}\) is constructed by gluing two copies of \(X\) along the mirror boundary~\cite{Scott1983}. The underlying space of \(\tilde{X}\) is a sphere (since doubling a disk produces a sphere). Each corner reflector of corner order \(k_j\) in \(X\) becomes a cone point of order \(k_j\) in \(\tilde{X}\) (one from each copy, identified at the boundary). Thus \(\tilde{X}\) is a sphere with \(r\) cone points of orders \(k_1,\ldots,k_r\), and the cone point formula gives:
\[
\chi_{\mathrm{orb}}(\tilde{X}) = 2 - \sum_{j=1}^{r}\Bigl(1 - \frac{1}{k_j}\Bigr).
\]
The doubling formula \(\chi_{\mathrm{orb}}(\tilde{X}) = 2\,\chi_{\mathrm{orb}}(X)\) (proved below) then gives:
\[
\chi_{\mathrm{orb}}(X) = \frac{1}{2}\chi_{\mathrm{orb}}(\tilde{X}) = 1 - \frac{1}{2}\sum_{j=1}^{r}\Bigl(1 - \frac{1}{k_j}\Bigr). \qedhere
\]
\end{proof}

\emph{Oriented double.} This relationship can be formalized using the \emph{oriented double}. An orbifold \(X\) with mirror isotropy is not orientable as an orbifold in the usual sense, because the local reflection stabilizers reverse orientation, even when the underlying topological surface is orientable. It admits an oriented double \(\tilde{X}\), obtained by gluing two copies of \(X\) along the mirrored boundary. Each mirror in \(X\) lifts to a regular curve in \(\tilde{X}\), and each corner reflector of corner order \(k\) becomes a cone point of order \(k\) in \(\tilde{X}\). The Euler characteristics are related by the doubling formula:
\[
\chi_{\mathrm{orb}}(\tilde{X}) = 2\,\chi_{\mathrm{orb}}(X).
\]
\begin{proof}[Proof of the doubling formula]
The natural projection \(p: \tilde{X} \to X\) is an orbifold covering of degree 2: the deck group is \(\mathbb{Z}_2\), acting by exchanging the two copies. Along smooth mirror arcs, \(\tilde{X}\) is locally smooth and \(p\) is modeled on the quotient map \(\mathbb{R}^2 \to \mathbb{R}^2/\mathbb{Z}_2\), where \(\mathbb{Z}_2\) acts by \((x,y) \mapsto (x,-y)\). At a corner reflector of corner order \(k\), the local model in \(X\) is \(\mathbb{R}^2/D_k\). The oriented double unfolds the reflection, producing a local model \(\mathbb{R}^2/\mathbb{Z}_k\) in \(\tilde{X}\). The map \(\mathbb{R}^2/\mathbb{Z}_k \to \mathbb{R}^2/D_k\) has degree \(|D_k|/|\mathbb{Z}_k| = 2\), consistent with the global degree. Applying the general multiplicativity of the orbifold Euler characteristic for orbifold coverings of degree \(d\)~\cite{Satake1957,Scott1983}:
\[
\chi_{\mathrm{orb}}(\tilde{X}) = d \cdot \chi_{\mathrm{orb}}(X) = 2\,\chi_{\mathrm{orb}}(X). \qedhere
\]
\end{proof}

\subsubsection{Formula for Global Quotients.}
In the case where the orbifold is a global quotient \([M/G]\) by a finite group \(G\), the Satake, or classical orbifold, Euler characteristic satisfies:
\[
  \chi_{\mathrm{orb}}([M/G]) \;=\; \frac{1}{|G|}\chi(M).
\]
\begin{proof}
The projection \(\pi: M \to [M/G]\) is an orbifold covering of degree \(|G|\). The multiplicativity of the orbifold Euler characteristic under orbifold coverings~\cite{Satake1957,Thurston1980} gives
\[
\chi_{\mathrm{orb}}(M) = |G|\cdot\chi_{\mathrm{orb}}([M/G]).
\]
Since \(M\) is a smooth manifold (having no orbifold singularities), its orbifold Euler characteristic coincides with its ordinary Euler characteristic: \(\chi_{\mathrm{orb}}(M) = \chi(M)\). Dividing by \(|G|\) yields the formula.

\emph{Alternative direct proof.} Using the orbifold integration formula of Section~\ref{sec:orbifold_de_rham}, the orbifold Euler characteristic can be computed as
\[
\chi_{\mathrm{orb}}([M/G]) = \int_{[M/G]} e(T[M/G]) = \frac{1}{|G|}\int_M e(TM) = \frac{1}{|G|}\chi(M),
\]
where \(e\) denotes the Euler class (or equivalently the Gaussian curvature density in dimension two) and the second equality uses the definition of orbifold integration for a global quotient~\cite{Satake1957,AdemLeidaRuan}.
\end{proof}

\chapter{Orbifold Examples}

\section{\texorpdfstring{Planar Cone $\mathbb{C}/\mathbb{Z}_n$}{Planar Cone C/Zn}}\label{sec:config-space-1d}

\subsubsection{Orbifold Definition and Basic Geometry.}
Fix an integer \(n \ge 2\). We work with the standard cyclic quotient orbifold~\cite{AdemLeidaRuan,Kordyukov2011}
\[
 X = [\mathbb{C}/\mathbb{Z}_n], \qquad \mathbb{Z}_n = \{\zeta^k : k = 0,\dots,n-1\}, \qquad \zeta = e^{2\pi i/n},
\]
where $\mathbb{Z}_n$ acts by rotation
\[
 \zeta^k\cdot z  = e^{2\pi i k/n}z.
\]
The coarse space is the usual quotient
\[
 |X| = \mathbb{C}/\mathbb{Z}_n,
\]
which is biholomorphic to $\mathbb{C}$ via the invariant map
\[
 \pi\colon \mathbb{C}/\mathbb{Z}_n\longrightarrow\mathbb{C}, \qquad [z]\longmapsto w = z^n
\]
(well-defined since $(\zeta z)^n = z^n$).
Thus $|X|\cong\mathbb{C}$ is contractible and simply connected. As defined in Section~\ref{sec:orbifold_definition}, the origin $[0] \in |X|$ is the unique \textbf{singular point} of the orbifold, with \textbf{isotropy group} $G_{[0]} \cong \mathbb{Z}_n$ and \textbf{order} $n$. Away from the apex, the action is free and the isotropy groups are trivial.  
\subsubsection{Orbifold Fundamental Group and Coverings.}   
We now determine the orbifold fundamental group of \(X\) and classify its connected orbifold coverings.

The coarse space $|X|$ is simply connected, so $\pi_1(|X|)$ is trivial. As an orbifold, however, the cyclic isotropy at the cone point contributes a nontrivial loop. Following the \textbf{loop-based definition} in Section~\ref{sec:orbifold_fundamental_group}, we find
\[
 \pi_1^{\mathrm{orb}}(X)\;\cong\;\mathbb{Z}_n.
\]
The universal orbifold cover is
\[
 \mathbb{C}\longrightarrow[\mathbb{C}/\mathbb{Z}_n].
\]

Viewing $X$ as a global quotient $[M/G]$ with $M=\mathbb{C}$ and $G=\mathbb{Z}_n$, we apply the standard exact sequence for good orbifolds (see Section~\ref{sec:orbifold_fundamental_group})
\[
 1\longrightarrow\pi_1(M)\longrightarrow\pi_1^{\mathrm{orb}}([M/G])\longrightarrow G\longrightarrow1.
\]

In our case $M=\mathbb{C}$ is simply connected, so $\pi_1(\mathbb{C})=1$ and the exact sequence collapses to an isomorphism
\[
 \pi_1^{\mathrm{orb}}\bigl([\mathbb{C}/\mathbb{Z}_n]\bigr)\;\cong\;\mathbb{Z}_n.
\]
Concretely, fix a basepoint $z_0\in\mathbb{C}^\times$ and let $\bar z_0$ be its image in $X$. Any orbifold loop $\alpha$ based at $\bar z_0$ admits a unique lift $\tilde\alpha$ with
\[
 \tilde\alpha(0)=z_0, \qquad \tilde\alpha(1)=\zeta^k z_0
\]
for some $k\in\{0,\dots,n-1\}$. The assignment $[\alpha]\mapsto \zeta^k$ gives the desired isomorphism
$\pi_1^{\mathrm{orb}}(X)\cong\mathbb{Z}_n$.

A geometric generator of $\pi_1^{\mathrm{orb}}(X)$ is represented in the coarse space by a small loop encircling the apex once. On the cover this corresponds to
\[
 \theta\longmapsto z_0 e^{i\theta}, \qquad \theta\in[0,2\pi/n],
\]
which traces an arc of angle $2\pi/n$ around the origin; its projection winds once around the cone point and lifts from $z_0$ to $\zeta z_0 = e^{2\pi i/n}z_0$. Repeating the loop $n$ times brings the lift back to $z_0$, so the generator has order $n$.

Since connected orbifold coverings correspond to subgroups of $\pi_1^{\mathrm{orb}}(X)\cong\mathbb{Z}_n$, they are classified by the divisors of $n$. For each divisor $d \mid n$, the unique subgroup of order $d$ determines the connected intermediate cover
\[
 [\mathbb{C}/\mathbb{Z}_d]\longrightarrow[\mathbb{C}/\mathbb{Z}_n],
\]
induced by the subgroup inclusion $\mathbb{Z}_d\hookrightarrow\mathbb{Z}_n$; the case $d=1$ is the universal manifold cover $\mathbb{C}\to X$, while $d=n$ gives the identity covering.

\subsubsection{Orbifold de Rham Cohomology.}
We next compute the ordinary (Satake) de Rham cohomology of $X=[\mathbb{C}/\mathbb{Z}_n]$. This reflects the contractibility of the coarse space and serves as a baseline for the richer theories considered later~\cite{Satake1957,Kordyukov2011}.

By definition, orbifold differential forms on $X$ are precisely the $\mathbb{Z}_n$-invariant smooth forms on $\mathbb{C}$, with differential induced by the usual exterior derivative on $\mathbb{C}$. Since $\mathbb{C}\cong\mathbb{R}^2$ is contractible, any closed $k$-form $\alpha$ with $k\ge1$ is exact: we can find a primitive $\beta$ with $\alpha=d\beta$. If $\alpha$ is moreover $\mathbb{Z}_n$-invariant, we may average the primitive over the group,
\[
 \bar\beta \,=\, \frac{1}{n}\sum_{g\in\mathbb{Z}_n} g^*\beta,
\]
which is $\mathbb{Z}_n$-invariant and still satisfies $d\bar\beta = \alpha$. Hence every closed invariant form of positive degree is exact, and
\[
 H^k\bigl(\Omega^*(\mathbb{C})^{\mathbb{Z}_n},d\bigr)=0 \qquad\text{for all }k\ge1.
\]
In degree $0$ the invariant closed forms are exactly the constant functions, so $H^0\cong\mathbb{R}$. This averaging argument is a special case of the general result (see Section~\ref{sec:orbifold_de_rham}) that the orbifold de Rham complex is quasi-isomorphic to the invariant subcomplex $\bigl(\Omega^*(M)^G,d\bigr)$.

A slightly different viewpoint uses the coarse space. The invariant map $\mathbb{C}\to\mathbb{C}$ given by $z\mapsto z^n$ descends to a homeomorphism on the coarse space
\[
 |X|=\mathbb{C}/\mathbb{Z}_n\;\xrightarrow{\ \sim\ }\;\mathbb{C}, \qquad [z]\longmapsto z^n.
\]
Since $\mathbb{C}$ is contractible, its singular cohomology with real coefficients is $\mathbb{R}$ in degree $0$ and $0$ otherwise. The orbifold de Rham theorem identifies this with \(H^*_{\mathrm{dR,orb}}(X)\), so in summary
\[
 H^k_{\mathrm{dR,orb}}(X)\;\cong\;
 \begin{cases}
  \mathbb{R}, & k=0,\\[0.2em]
  0, & k>0.
 \end{cases}
\]
In particular, the cone singularity at the origin does not create new de Rham classes in positive degree.

\subsubsection{Orbifold Line Bundles and Holonomy.}
We now describe a distinguished family of orbifold line bundles on $X=[\mathbb{C}/\mathbb{Z}_n]$ and their holonomy around the cone point.

\emph{Definition of the bundles $L_a$.} Let
\[
 X = [\mathbb{C}/\mathbb{Z}_n], \qquad |X|=\mathbb{C}/\mathbb{Z}_n,
\]
with $\mathbb{Z}_n$ acting by rotations as above. A uniformizing chart is $(V,\mathbb{Z}_n,\pi)$ with $V=\mathbb{C}$. For each integer $a\in\{0,\dots,n-1\}$ we define an orbifold line bundle $L_a$ by letting $\mathbb{Z}_n$ act on the trivial bundle $\mathbb{C}\times\mathbb{C}$ via the weight-$a$ character:
\[
 \zeta\cdot(z,v) = (\zeta z,\,e^{2\pi i a/n}v).
\]
Equivalently, $L_a$ is given locally by $L_a|_{\pi(V)} \cong V\times\mathbb{C}$ with equivariant action
\[
 \zeta\cdot(z,v)
 =\bigl(e^{2\pi i/n}z,\;e^{2\pi i a/n}v\bigr).
\]

A local section $s$ of $L_a$ can be written in a local frame $e$ as $s(z)=f(z)e$; the coefficient function $f:V\to\mathbb{C}$ satisfies the equivariance condition
\[
 f\!\bigl(e^{2\pi i/n}z\bigr) \,=\, e^{2\pi i a/n} f(z).
\]

\emph{Connections and holonomy.} We endow $L_a$ with the trivial connection, which on the uniformizing chart $V=\mathbb{C}$ is simply given by $\nabla = d$ in the global frame $e$. Acting on a local section $s=f e$ this gives $\nabla(s) = df \otimes e$. Since the exterior derivative commutes with the linear $\mathbb{Z}_n$-action, this defines a smooth, globally flat \textbf{orbifold connection} on $X$. The curvature vanishes everywhere in the orbifold sense:
\[
 F_{\nabla} = 0 \quad\text{on }\mathbb{C}.
\]

To compute the holonomy and understand the physics on the coarse space $|X|$, it is instructive to push this connection down to the punctured coarse space $|X|\setminus\{[0]\}$. We pass to an invariant unitary frame $\tilde{e} = e^{i a\theta}e$, where $z=re^{i\theta}$ on $\mathbb{C}^\times$. Here $e^{i a\theta}=(z/|z|)^a$ is single-valued on $\mathbb{C}^\times$ because $a\in\mathbb{Z}$. Since $\zeta$ shifts $\theta \mapsto \theta + 2\pi/n$, the phase factor exactly matches the action on $e$, making $\tilde{e}$ invariant. In this frame, the smooth orbifold connection takes the form
\[
 \nabla(\tilde{e}) = \nabla(e^{i a\theta}e) = i\,a\,d\theta \otimes \tilde{e}.
\]
Thus, while the connection is smooth on the cover, on the punctured coarse space it manifests as a singular Abelian gauge field $\nabla = d - iA$ with real connection 1-form $A = -a\,d\theta$.

On the quotient we use the orbifold angular coordinate $\varphi\in[0,2\pi)$ related to the cover coordinate by $\theta = \varphi/n$ (so that a full $2\pi$ circuit in $\varphi$ corresponds to $2\pi/n$ on the cover). Since $d\theta = d\varphi/n$, the descended connection 1-form reads
\[
 A = -\frac{a}{n}\,d\varphi.
\]
A loop $\varphi\mapsto\varphi+2\pi$ generates $\pi_1^{\mathrm{orb}}(X)$, and its holonomy is
\[
 \exp\!\left(i\oint A\right) = \exp\!\left(-\frac{i a}{n}\int_0^{2\pi} d\varphi\right) 
 = e^{-2\pi i a/n}.
\]
Thus the holonomy around a small loop encircling the apex once is precisely
\[
 e^{-2\pi i a/n},
\]
which acts as the inverse of the \textbf{isotropy character} (see Section~\ref{sec:orbifold_line_bundles}) of $L_a$ viewed as a unitary representation of $\pi_1^{\mathrm{orb}}(X)\cong\mathbb{Z}_n$.

\subsubsection{Classification and Integrality.}
We now turn to the classification of orbifold line bundles on the orbifold cone $X = [\mathbb{C}/\mathbb{Z}_n]$ and the resulting quantization conditions.
Since $X$ is the global quotient $[\mathbb{C}/\mathbb{Z}_n]$, Section~\ref{sec:classification} identifies orbifold line bundles on $X$ with $\mathbb{Z}_n$-equivariant line bundles on $\mathbb{C}$. Because $\mathbb{C}$ is contractible, we have $H^1(\mathbb{C},\mathbb{Z})=0$ and $\mathrm{Pic}(\mathbb{C})=0$, so the exact sequence of Section~\ref{sec:classification} reduces to
\[
 0 \longrightarrow \mathrm{Hom}(\mathbb{Z}_n,U(1)) \longrightarrow \mathrm{Pic}_{\mathrm{orb}}(X) \longrightarrow 0.
\]
Hence
\[
 \mathrm{Pic}_{\mathrm{orb}}(X) \;\cong\; \mathrm{Hom}(\mathbb{Z}_n, U(1)) \;\cong\; \mathbb{Z}_n.
\]
Equivalently, one may argue directly that the contraction $H_t(z)=(1-t)z$ is $\mathbb{Z}_n$-equivariant, so $\mathbb{C}$ is $\mathbb{Z}_n$-equivariantly homotopy equivalent to the fixed point $\{0\}$. Therefore a $\mathbb{Z}_n$-equivariant line bundle on $\mathbb{C}$ is determined up to isomorphism by the induced representation on the fiber over $0$. Thus, isomorphism classes are in one-to-one correspondence with the \textbf{isotropy characters} $\rho: \mathbb{Z}_n \to U(1)$. These are indexed by an integer $a \in \{0, \dots, n-1\}$, corresponding to the weight of the action:
\[
 \zeta \cdot v = e^{2\pi i a/n} v, \qquad v \in L|_{z=0}.
\]
The bundles $L_a$ constructed in the previous paragraph correspond precisely to these classes. In particular, every orbifold line bundle on $X$ admits a flat unitary connection, represented by the explicit flat model $L_a$ for the corresponding character. Since the de Rham cohomology \(H^2_{\mathrm{dR,orb}}(X)=0\), all of these classes have vanishing de Rham first Chern class. The classification discussed in Section~\ref{sec:classification} therefore appears here in local form: for the explicit gauge chosen above, the isotropy character $e^{2\pi i a/n}$ forces the boundary flux around a small circle about the apex to be $-\frac{a}{n}$, hence in particular to lie in $\frac{1}{n}\mathbb{Z}$.

\subsubsection{Chern--Weil Theory and Fractional Flux.}
Finally, we interpret the geometry of $L_a$ using the Chern--Weil theory developed in Section~\ref{sec:chern_weil_theory}. Because the orbifold connection $\nabla = d$ is globally flat on the uniformizing cover, its \emph{orbifold} curvature 2-form $F_\nabla$ vanishes identically. Consequently, the real first Chern class evaluates to zero:
\[
 c_1^{\mathrm{dR}}(L_a) = \left[\frac{i}{2\pi}F_\nabla\right] = 0 \in H^2_{\mathrm{dR,orb}}(X),
\]
in perfect agreement with the vanishing of the orbifold de Rham cohomology.

Nevertheless, when viewed on the coarse space $|X| \cong \mathbb{C}$, the associated singular gauge field $A = -\frac{a}{n} d\varphi$ supports a nontrivial local flux concentrated at the singularity. If $U \subset |X|$ is a small disk centered at the apex, then $\partial U \subset |X|\setminus\{[0]\}$ is an ordinary circle in the smooth locus. Following the boundary-holonomy prescription of Section~\ref{sec:chern_weil_theory}, we obtain the local boundary evaluation
\[
 \frac{1}{2\pi} \oint_{\partial U} A = \frac{1}{2\pi} \int_0^{2\pi} \left(-\frac{a}{n}\right) d\varphi = -\frac{a}{n}.
\]
This is a local boundary contribution on the noncompact coarse space, not a global degree of $X$ itself. Viewed distributionally on the coarse space, the effective curvature $F_{|X|}$ of the singular gauge field $A$ acquires a delta-function singularity $+2\pi i\frac{a}{n}\delta_0$, so that the Chern density $\frac{i}{2\pi}F_{|X|}$ carries the signed term $-\frac{a}{n}\delta_0$ in this coarse trivialization. With the excision orientation used in the compact orbifold surface formula of Section~\ref{sec:chern_weil_theory}, the same local holonomy contributes \(+\frac{a}{n}\) to the orbifold degree.


\section{\texorpdfstring{Orbisphere $S^2(n,m)$}{Orbisphere S2(n,m)}}

\subsubsection{Orbifold Definition and Basic Geometry.}
Consider the oriented $2$--orbifold~\cite{Scott1983,AdemLeidaRuan,LermanTolman1997}
\[
 X = S^2(n,m),
\]
obtained from the topological sphere by marking two points $p_+,p_-\in S^2$ with cone orders
$n\ge2$ and $m\ge2$. The underlying topological space is
\[
 |X|\cong S^2,
\]
so all singular behavior is encoded in the local group actions at the two cone points.

Near the first cone point $p_+$ we choose a uniformizing chart, meaning a local orbifold
model $(\tilde D_+,\mathbb{Z}_n)$ where $\tilde D_+\subset\mathbb{C}$ is a small disk, the
generator $\zeta_n=e^{2\pi i/n}$ acts by $w\mapsto \zeta_n w$, and the quotient map
$q_+(w)=w^n:\tilde D_+\to D_+:=\tilde D_+/\mathbb{Z}_n$ identifies $D_+$ with a neighborhood of
$p_+$. Similarly, near the second cone point $p_-$ we take $(\tilde D_-,\mathbb{Z}_m)$ with
generator $\zeta_m=e^{2\pi i/m}$ acting by $w\mapsto \zeta_m w$ and quotient map $q_-(w)=w^m$
onto $D_-:=\tilde D_-/\mathbb{Z}_m$.
Away from $p_\pm$ the isotropy is trivial. The regular part
\[
X^{\mathrm{reg}} := X\setminus\{p_+,p_-\} \cong S^2\setminus\{\text{two points}\} \cong S^1\times(0,1)
\]
is thus an open annulus.
\subsubsection{Orbifold Fundamental Group and Coverings.}
We now describe the orbifold fundamental group of $X$ using the loop-based definition (see Section~\ref{sec:orbifold_fundamental_group}). Let $x_1$ (resp.
$x_2$) be a loop around $p_+$ (resp. $p_-$), based at a point $p_0$. For a general orbisphere with cone points of orders $m_1,\dots,m_r$ the standard
presentation is
\[
\begin{aligned}
 \pi_1^{\mathrm{orb}}\bigl(S^2(m_1,\dots,m_r)\bigr)
 &=\big\langle x_1,\dots,x_r\ \big|\ x_1^{m_1}=\cdots=x_r^{m_r}=1,\\
 &\qquad x_1x_2\cdots x_r=1\big\rangle.
\end{aligned}
\]
Here each generator $x_i$ is a loop encircling the $i$th cone point; the relations
$x_i^{m_i}=1$ reflect that such loops become trivial after lifting to the order-$m_i$ uniformizing
chart, and the product $x_1x_2\cdots x_r=1$ expresses that on the sphere the concatenation of the
 loops is homotopic to the boundary of a disk containing all cone points, hence contractible.
In our case $r=2$, so $x_1^{\,n}=1$ and $x_2^{\,m}=1$, because these loops become trivial after lifting to the
uniformizing charts $(\tilde D_+,\mathbb{Z}_n)$ and $(\tilde D_-,\mathbb{Z}_m)$.

Since the underlying surface is $S^2$, a loop enclosing both cone points bounds a topological disk,
so one has the sphere relation
\[
 x_1 x_2 = 1.
\]
Thus
\[
 \pi_1^{\mathrm{orb}}\bigl(S^2(n,m)\bigr)
 = \big\langle x_1,x_2\ \big|\ x_1^{\,n}=1,\;x_2^{\,m}=1,\;x_1x_2=1\big\rangle.
\]
Equivalently, this is the free product $\mathbb{Z}_n*\mathbb{Z}_m$ modulo $x_1x_2=1$. Substituting
$x_2=x_1^{-1}$ gives $1 = x_2^m = {(x_1^{-1})}^m = x_1^{-m}$, so $x_1^m=1$. Thus,
\[
\begin{aligned}
 \pi_1^{\mathrm{orb}}\bigl(S^2(n,m)\bigr)
 &\cong\big\langle x_1\mid x_1^{\,n}=1,\;x_1^{\,m}=1\big\rangle\\
 &=\big\langle x_1\mid x_1^{\,g}=1\big\rangle
  \;\cong\;\mathbb{Z}_g,\qquad g=\gcd(n,m).
\end{aligned}
\]
Physically, this group detects the residual symmetry allowed by both cone points. A single loop $x_1$ must simultaneously satisfy the phase constraints of both local cone orders. The two relations $x_1^n=1$ and $x_1^m=1$ are equivalent to $x_1^g=1$, so the group only remembers the
common divisor $g=\gcd(n,m)$.

When $n=m$, the orbisphere is a \textbf{good orbifold} (see Section~\ref{sec:orbifold_definition}) and has a manifold universal orbifold cover; specifically, it is a global quotient $[S^2/\mathbb{Z}_n]$. In this case, the homotopy exact sequence (see Section~\ref{sec:orbifold_fundamental_group}) $1 \to \pi_1(S^2) \to \pi_1^{\mathrm{orb}}(X) \to \mathbb{Z}_n \to 1$ immediately yields $\pi_1^{\mathrm{orb}}(X)\cong\mathbb{Z}_n$.
For $n\neq m$, the orbisphere is a \textbf{bad orbifold} (non-developable), meaning a universal \emph{manifold} covering does not exist (see Section~\ref{sec:orbifold_definition}), even though the orbifold fundamental group is still finite cyclic. This distinction is crucial for quantization: while good orbifolds can be studied via equivariant geometry on a manifold cover, bad orbifolds, like $S^2(n,m)$ for $n \neq m$, require the intrinsic orbifold techniques developed in Chapter 2.
More generally, connected orbifold coverings are classified by subgroups of $\pi_1^{\mathrm{orb}}(X)\cong \mathbb{Z}_g$, hence by the divisors of $g=\gcd(n,m)$. In particular, if $g=1$ then $X$ is orbifold simply connected, so its universal orbifold cover is $X$ itself; when $n\neq m$, this universal cover is still singular, reflecting the bad-orbifold phenomenon.

\subsubsection{Orbifold de Rham Cohomology.}

\emph{Satake's framework and general theorem.} An orbifold $k$--form is a $G_p$--invariant $k$--form
on each uniformizing chart~$V_p$, compatible under change of charts. The exterior differential $d$ is defined
chartwise and preserves invariance, giving a complex $\Omega^*_{\mathrm{orb}}(X)$ (see Section~\ref{sec:orbifold_de_rham}). Satake's theory provides partitions of unity, a Poincar\'e lemma,
 and an orbifold de Rham theorem
\[
 H^{*}_{\mathrm{dR,orb}}(X) \cong H^{*}(|X|;\mathbb{R}),
\]
where the superscript $*$ indicates the full graded cohomology in all degrees and $|X|$ is the
underlying topological space; $H^0$ is simply its degree-zero piece (functions constant on
connected components). Since $|S^2(n,m)|\cong S^2$, this theorem gives
\[
 H^0_{\mathrm{dR,orb}}(X)\cong\mathbb{R},\qquad H^1_{\mathrm{dR,orb}}(X)=0,\qquad H^2_{\mathrm{dR,orb}}(X)\cong\mathbb{R},
\]
with all higher groups vanishing. The cone points do not create extra positive-degree classes:
each orbifold disk $(\tilde B,\mathbb{Z}_q)$ has the de Rham cohomology of a point, so they contribute only $H^0$, leaving the global groups equal to those of
the underlying sphere. Satake's construction (partitions of unity and Poincar\'e lemma on local quotients) applies to effective orbifolds; $S^2(n,m)$ is effective since the local $\mathbb{Z}_n$ and $\mathbb{Z}_m$ actions on the charts are faithful. The bad case $n\neq m$ is therefore covered as well.

We use the coefficient ring $\mathbb{R}$ throughout this section unless otherwise noted; when complex coefficients are needed, we write $H^*(-;\mathbb{C})$ explicitly.

\subsubsection{Orbifold Line Bundles and Holonomy.}
We now provide a comprehensive description of orbifold line bundles on the orbisphere $S^2(n,m)$, detailing their local structure, global classification, and characteristic classes~\cite{Furuta1992SeifertFH,MrowkaOzsvathYu1997,Sakai2017OrbifoldVortices,BaierMouraoNunes2018}.

\emph{Definition and local structure.} An orbifold line bundle $L$ over $S^2(n,m)$ is defined by specifying local line bundles over the uniformizing charts together with compatibility conditions. Following the Seifert-type parametrization of Section~\ref{sec:classification}, it is convenient to encode such a bundle by normalized data
\[
 (d;a,b), \qquad d\in\mathbb{Z},\quad 0\le a<n,\quad 0\le b<m,
\]
and to write the corresponding class as $L_{(d;a,b)}$. Here $a$ and $b$ record the isotropy characters at the two cone points, while $d$ is the integral part appearing in the orbifold degree formula below; equivalently, once the local isotropy weights have been fixed, $d$ is the winding number of a transition function on an equatorial annulus.

\textbf{At the cone points:} Near $p_+$, we have the chart $(\tilde D_+, \mathbb{Z}_n)$. The bundle $L$ is specified by a $\mathbb{Z}_n$-equivariant line bundle over $\tilde D_+$. Since $\tilde D_+$ is contractible, the bundle is trivial, $\tilde D_+ \times \mathbb{C}$, and the $\mathbb{Z}_n$-action is determined by its action on the fiber. The generator $\zeta_n = e^{2\pi i/n}$ acts by
\[
 \zeta_n \cdot (w, \xi) = (\zeta_n w, \, \rho_+(\zeta_n)\xi),
\]
where $\rho_+: \mathbb{Z}_n \to U(1)$ is a representation. We write $\rho_+(\zeta_n) = e^{2\pi i a/n}$ for some integer weight $a \in \{0, \dots, n-1\}$. This integer $a$ is the \emph{local isotropy weight} at $p_+$.
Similarly, near $p_-$, the action is defined by a weight $b \in \{0, \dots, m-1\}$ such that the generator $\zeta_m$ acts on the fiber by multiplication by $e^{2\pi i b/m}$.

\textbf{On the regular part:} Over $X^{\mathrm{reg}} \cong S^1 \times (0,1)$, the bundle is an ordinary complex line bundle and hence topologically trivial. However, when gluing the bundle over $D_+$ to the bundle over $D_-$ along an annular overlap $A \cong S^1 \times (0,1)$, the transition function $h: A \to U(1)$ is classified by its winding number (or degree) $d\in\mathbb{Z}$ around the equator.

\emph{Connections and holonomy.} An orbifold connection $\nabla$ on $L$ is a collection of smooth invariant connections on the local charts that glue together.
In the chart $(\tilde D_+, w)$, a unitary connection can be written as $\nabla = d - iA_+$, where $A_+$ is a smooth, real-valued $1$-form on $\tilde D_+$. The invariance condition under the action $w \mapsto \zeta_n w$ and $\xi \mapsto e^{2\pi i a/n}\xi$ requires
\[
 \gamma^* A_+ = A_+ \qquad \text{for all } \gamma \in \mathbb{Z}_n.
\]
(Since $U(1)$ is abelian, the gauge transformation $g(w) = e^{2\pi i a/n}$ is constant, so the term $g^{-1}dg$ vanishes, and we simply require the $1$-form to be invariant.)

A valid, smooth choice near the origin is simply the trivial connection $A_+ = 0$. In this gauge, parallel transport along a path in $\tilde D_+$ from $w_0$ to $\zeta_n w_0$ gives no phase contribution from the connection $1$-form itself. By definition, the orbifold holonomy of a loop in the quotient $D_+$ around $p_+$ once is computed by lifting it to a path in $\tilde D_+$ and composing the parallel transport with the inverse action of the group element $\zeta_n$ on the fiber over the endpoints. Thus, the fractional holonomy acts as the inverse of the isotropy representation:
\[
 \mathrm{Hol}_{p_+} = e^{-2\pi i a/n}.
\]
Likewise, near $p_-$ one obtains
\[
 \mathrm{Hol}_{p_-} = e^{-2\pi i b/m}.
\]
These local holonomies are the inverses of the isotropy characters and constitute the characteristic orbifold contribution at the two cone points.

\subsubsection{Classification of Orbifold Line Bundles.}
To classify orbifold line bundles globally on $X = S^2(n,m)$, we decompose the orbisphere into two orbifold caps $D_+$ and $D_-$ that overlap on an equatorial annulus $A \cong S^1 \times (0,1)$. Over $D_+$, the bundle is entirely determined by the local isotropy weight $a \in \{0, \dots, n-1\}$. Over $D_-$, it is similarly determined by the weight $b \in \{0, \dots, m-1\}$. Over the intersection $A$, the bundle is patched together via a transition function $h: S^1 \to U(1)$, whose homotopy class is an integer winding number $d \in \mathbb{Z}$.

Consequently, every orbifold line bundle uniquely admits normalized Seifert data $(d;a,b)$, denoted as $L_{(d;a,b)}$. Unlike line bundles on smooth manifolds, the local weights $a$ and $b$ are intrinsic invariants anchored at the singular points. These invariants capture the local obstruction to triviality, allowing the smooth/topological orbifold Picard group \(\mathrm{Pic}_{\mathrm{orb}}(X) \cong H^2_{\mathrm{st}}(X;\mathbb{Z})\) to fit into the exact sequence
\[
 0 \to \mathbb{Z} \to \mathrm{Pic}_{\mathrm{orb}}(X) \to \mathbb{Z}_n \oplus \mathbb{Z}_m \to 0.
\]
The subgroup $\mathbb{Z}$ is tracked by the global integral winding $d$, while the quotient $\mathbb{Z}_n \oplus \mathbb{Z}_m$ records the local isotropy representations. Importantly, the Seifert data does \emph{not} imply a direct product decomposition of the group. As established abstractly in Section~\ref{sec:classification}, the extension is generally nonsplit, governed by a tensor product law that involves algebraic carrying between the local weights and the integer part:
\[
 L_{(d;a,b)}\otimes L_{(d';a',b')} \cong L_{\bigl(d+d'+\epsilon_n+\epsilon_m;\,{[a+a']}_n,\,{[b+b']}_m\bigr)},
\]
where ${[r]}_k$ denotes the standard reduction of $r$ modulo $k$, and the carry variables are
\[
 \epsilon_n=\left\lfloor\frac{a+a'}{n}\right\rfloor,\qquad \epsilon_m=\left\lfloor\frac{b+b'}{m}\right\rfloor.
\]
This algebraic behavior precisely realizes the two-cone-point specialization of the general carry rule.

\emph{Flat Bundles and Torsion.} Flat line bundles on $X$ correspond to unitary representations of the fundamental group, and are thus parameterized by elements in $\mathrm{Hom}(\pi_1^{\mathrm{orb}}(X), U(1))$. Since $\pi_1^{\mathrm{orb}}(X) \cong \mathbb{Z}_g$ where $g = \gcd(n,m)$, there are precisely $g$ flat sectors, comprising the torsion subgroup of $\mathrm{Pic}_{\mathrm{orb}}(X)$. In terms of normalized Seifert data, a convenient set of representatives is the trivial bundle $F_0=L_{(0;0,0)}$ along with
\[
 F_r=L_{\left(-1;\,\frac{rn}{g},\,m-\frac{rm}{g}\right)}, \qquad 1\le r\le g-1.
\]
Each $F_r$ represents a purely flat connection, carrying a total degree properly evaluated to zero:
\[
 -1+\frac{1}{n}\left(\frac{rn}{g}\right)+\frac{1}{m}\left(m-\frac{rm}{g}\right)=0.
\]

\emph{Abstract Group Structure.} By invoking the fractional degree formula $\deg_{\mathrm{orb}}(L) = d + \frac{a}{n} + \frac{b}{m}$ (which we formally derive via Chern\textendash{}Weil theory below), we can unequivocally determine the abstract group structure of $\mathrm{Pic}_{\mathrm{orb}}(X)$. The set of all admissible degrees spans the image of the degree map, $\operatorname{Im}(\deg_{\mathrm{orb}}) \subset \mathbb{Q}$. Since $d \in \mathbb{Z}$, $a \in \mathbb{Z}_n$, and $b \in \mathbb{Z}_m$, the image consists of fractions taking the form
\[
 \frac{dnm + am + bn}{nm}.
\]
Because the integer linear combinations of $nm$, $m$, and $n$ generate the ideal $\gcd(nm, m, n)\mathbb{Z} = \gcd(n,m)\mathbb{Z}$, the values of the numerator exactly sweep out multiples of $g=\gcd(n,m)$. Thus, the image of the degree map is explicitly
\[
 \operatorname{Im}(\deg_{\mathrm{orb}}) = \frac{\gcd(n,m)}{nm}\mathbb{Z} = \frac{1}{\mathrm{lcm}(n,m)}\mathbb{Z} \cong \mathbb{Z}.
\]
The kernel of the degree map is exactly the torsion subgroup of flat bundles $\mathbb{Z}_g$ identified above. This constructs a short exact sequence separating the free topology from the torsion holonomy:
\[
 0 \to \mathbb{Z}_{\gcd(n,m)} \to \mathrm{Pic}_{\mathrm{orb}}\bigl(S^2(n,m)\bigr) \xrightarrow{\deg_{\mathrm{orb}}} \frac{1}{\mathrm{lcm}(n,m)}\mathbb{Z} \to 0.
\]
Because the rightmost group is free abelian, this sequence automatically splits, revealing the abstract algebraic decomposition
\[
 \mathrm{Pic}_{\mathrm{orb}}\bigl(S^2(n,m)\bigr) \cong \mathbb{Z} \oplus \mathbb{Z}_{\gcd(n,m)}.
\]
This structure is compatible with the original exact sequence $0 \to \mathbb{Z} \to \mathrm{Pic}_{\mathrm{orb}}(X) \to \mathbb{Z}_n \oplus \mathbb{Z}_m \to 0$ presented previously. The line bundle $L_{(1;0,0)}$ generating the $\mathbb{Z}$ subgroup has an intrinsic degree of $1$, mapping to an index-$\mathrm{lcm}(n,m)$ sublattice in the free part of $\mathrm{Pic}_{\mathrm{orb}}(X)$. Taking the quotient $\mathrm{Pic}_{\mathrm{orb}}(X)/\mathbb{Z}$ yields the group isomorphism $\mathbb{Z}_{\mathrm{lcm}(n,m)} \oplus \mathbb{Z}_{\gcd(n,m)} \cong \mathbb{Z}_n \oplus \mathbb{Z}_m$. 

\emph{Split vs.\ Nonsplit Exact Sequences.} The Seifert data sequence is nonsplit even though the total degree sequence splits. The exact sequence constructed from Seifert data records the local boundary weights $a$ and $b$. Because multiple copies of a local fractional weight can carry into the global integer part (for example, $L_{(0;1,0)}^{\otimes n} = L_{(1;0,0)}$), the fractional quotients in $\mathbb{Z}_n \oplus \mathbb{Z}_m$ are not represented by independent torsion classes in this presentation. Conversely, the sequence built from the total degree map separates the free degree lattice from the zero-degree flat bundles. The flat torsion subgroup is \(\mathbb{Z}_{\gcd(n,m)}\), and the splitting follows because the degree image \(\frac{1}{\mathrm{lcm}(n,m)}\mathbb{Z}\) is free abelian. This gives the abstract direct sum $\mathbb{Z} \oplus \mathbb{Z}_{\gcd(n,m)}$.

For the symmetric football reduction $m=n$, the formula simplifies to $\mathrm{Pic}_{\mathrm{orb}}\bigl(S^2(n,n)\bigr) \cong \mathbb{Z} \oplus \mathbb{Z}_n$. The possible degrees form $\frac{1}{n}\mathbb{Z}$, and the flat subgroup is $\mathbb{Z}_n$. Since $S^2(n,n)$ is a \emph{good} orbifold admitting the global manifold quotient $[S^2/\mathbb{Z}_n]$, this recovers the equivariant Picard group $\mathrm{Pic}_{\mathbb{Z}_n}(S^2) \cong \mathbb{Z} \oplus \mathbb{Z}_n$, where the free part tracks the topology on the smooth $S^2$ manifold cover, and the torsion part records the discrete $\mathbb{Z}_n$ lift choices.

\subsubsection{Chern\textendash{}Weil Theory and Fractional Flux.}
The continuous invariant of the classification data is the orbifold degree representing total curvature flux, matching the framework of Section~\ref{sec:chern_weil_theory}. Let $\Sigma_0:=X\setminus(\operatorname{int}U_+\cup \operatorname{int}U_-)$ be the complement of small orbifold disks around the cone points. On the smooth annulus $\Sigma_0$ we may choose a smooth, globally defined unitary frame and write the connection as $\nabla=d-iA$, with curvature $F=-idA$. Since the continuous degree is a topological invariant, we are free to choose a connection that is completely flat on the interior of the caps $U_\pm$, localizing all curvature on $\Sigma_0$ (which functionally corresponds to evaluating the integral in the geometric limit where the caps shrink entirely to the singular cone points). With the boundary orientation
\[
 \partial\Sigma_0=-\partial U_+-\partial U_-,
\]
Stokes' theorem gives the evaluated degree
\[
 \deg_{\mathrm{orb}}(L)
 =\frac{i}{2\pi}\int_X F
 = \frac{i}{2\pi}\int_{\Sigma_0} F
 =-\frac{1}{2\pi}\oint_{\partial U_+}A-\frac{1}{2\pi}\oint_{\partial U_-}A.
\]
By the local holonomy formulas above, the fractional boundary fluxes are entirely dictated by the local isotropy actions:
\[
 \frac{1}{2\pi}\oint_{\partial U_+}A \equiv -\frac{a}{n}\pmod{\mathbb{Z}},
 \qquad
 \frac{1}{2\pi}\oint_{\partial U_-}A \equiv -\frac{b}{m}\pmod{\mathbb{Z}}.
\]
Evaluating the globally smooth section on $\Sigma_0$ naturally forces continuous gauge transformations at the intersection boundaries whose residual integer winding numbers collectively add up to the evaluated transition function degree $d$ across the equator. Combining these separated integer residuals over the boundaries mathematically recovers the specialized formal degree without jumps:
\[
 \deg_{\mathrm{orb}}(L) = \frac{i}{2\pi} \int_X F = d + \frac{a}{n} + \frac{b}{m}.
\]
Here the integer $d$ accounts for the continuous integral winding in the gluing data. Applying the general surface framework from Section~\ref{sec:chern_weil_theory}, this is the two-cone-point form of \(\deg(L)=b_0+\sum_j a_j/m_j\) with $b_0=d$. This calculation uses the smooth orbifold Chern--Weil representative. If one instead pushes the connection to a singular coarse-space trivialization, the same local isotropy data can be represented by distributional terms as in Section~\ref{sec:connections_curvature}; those terms are a coarse-space encoding of the orbifold holonomy, not additional curvature on the orbifold itself.

\section{\texorpdfstring{Symmetric Product Orbifold $\mathbb{C}^n/S_n$}{Symmetric Product Orbifold Cn/Sn}}\label{sec:config-space}

\subsubsection{Orbifold Definition and Basic Geometry.}
We study the symmetric-product orbifold~\cite{EmmrichRomer1990,Kordyukov2011}
\[
 X = [\mathbb{C}^n/S_n],
\]
where the symmetric group $S_n$ acts on $\mathbb{C}^n$ by permuting coordinates:
\[
 \sigma\cdot(z_1,\dots,z_n)
 \;=\;(z_{\sigma^{-1}(1)},\dots,z_{\sigma^{-1}(n)}),
 \qquad \sigma\in S_n.
\]
When $n=1$, this is just the manifold $\mathbb{C}$ and all orbifold phenomena below are trivial. We therefore assume $n\ge2$ in the nontrivial discussion that follows.
We regard $X$ as a global quotient with a single global chart
$(V,S_n,\pi)$, where $V=\mathbb{C}^n$ and $\pi\colon V\to V/S_n$ is the quotient map.

The coarse space $|X|=\mathbb{C}^n/S_n$ can be identified with $\mathbb{C}^n$ via the elementary symmetric polynomials.
Define the map $e: \mathbb{C}^n \to \mathbb{C}^n$ by
\[
 e(z) = (e_1(z),\dots,e_n(z)),
\]
where $e_k(z)$ is the $k$-th elementary symmetric polynomial,
\[
 e_k(z) = \sum_{1 \le j_1 < \cdots < j_k \le n} z_{j_1} \cdots z_{j_k}.
\]
These are the coefficients (up to sign) of the polynomial having roots $z_1, \dots, z_n$:
\[
 \prod_{j=1}^n (t - z_j)
 = t^n - e_1(z)\,t^{n-1} + \cdots + (-1)^n e_n(z).
\]
Since each $e_k$ is invariant under permutations, we have $e(\sigma\cdot z) = e(z)$ for all $\sigma\in S_n$. Thus $e$ descends to a well-defined map on the quotient,
\[
 \bar e\colon \mathbb{C}^n/S_n\longrightarrow\mathbb{C}^n,
\]
defined by $\bar{e}([z]) = e(z)$, where $[z]$ denotes the unordered $S_n$-orbit or multiset of coordinates.
By the fundamental theorem of symmetric polynomials,
\[
 \bigl(\mathbb{C}[z_1,\dots,z_n]\bigr)^{S_n}=\mathbb{C}[e_1,\dots,e_n],
\]
so $\bar e$ is precisely the affine quotient map. Equivalently, two unordered
$n$-tuples have the same image precisely when they determine the same monic polynomial, hence the
same multiset of roots, and conversely any point of $\mathbb{C}^n$ arises as the corresponding tuple
of such a polynomial. Thus $\bar e$ is a homeomorphism (indeed an isomorphism of affine varieties), and $|X|\cong\mathbb{C}^n$ is
contractible and simply connected. 

\emph{Stratification by isotropy.} The isotropy group at a point $z=(z_1,\dots,z_n)\in\mathbb{C}^n$ is the stabilizer subgroup $\mathrm{Stab}_{S_n}(z)$, which consists of all permutations $\sigma$ such that $z_{\sigma(i)} = z_i$ for all $i$.
Suppose the coordinates of $z$ take $k$ distinct values $\xi_1, \dots, \xi_k$. Let $n_j$ be the multiplicity of the value $\xi_j$ (i.e., the number of times it appears in the tuple $z$). Then $n_1 + \cdots + n_k = n$, and the stabilizer is the direct product of the symmetric groups acting on each block of identical coordinates:
\[
  \mathrm{Stab}_{S_n}(z) \;\cong\; S_{n_1}\times\cdots\times S_{n_k}.
\]
The orbifold admits a natural stratification by isotropy type. The \emph{regular stratum} $X_{\mathrm{reg}}$ consists of points with all $z_j$ distinct, where $\mathrm{Stab}_{S_n}(z)=\{\mathrm{id}\}$ and the orbifold is a manifold. The \emph{discriminant locus} $\Delta\subset X$ is the complement of $X_{\mathrm{reg}}$, consisting of points where at least two coordinates coincide:
\[
  \Delta = \{ [z] \in X \mid z_i = z_j \text{ for some } i \neq j \}.
\]
Geometrically, this is a complex hypersurface (real codimension $2$). Physically, it represents configurations where particles collide, a standard source of orbifold configuration spaces in systems with quotient symmetries~\cite{EmmrichRomer1990}. At a generic point of $\Delta$ where exactly one pair of particles collides, say $z_i=z_j$ and all other coordinates are distinct, local coordinates
\[
  u=\frac{z_i+z_j}{2}, \qquad w=z_i-z_j,
\]
together with the remaining coordinates show that a neighborhood is modeled on
$\mathbb{C}^{n-1}\times[\mathbb{C}/\mathbb{Z}_2]$, with the transposition $(ij)$ acting by
$w\mapsto -w$. Thus a generic collision is locally the basic cone model $[\mathbb{C}/\mathbb{Z}_2]$
times a smooth factor. The maximal-isotropy stratum is not a single point but the \emph{small diagonal}
\[
  \{[(a,\dots,a)] \mid a\in\mathbb{C}\}\cong\mathbb{C},
\]
on which the full group $S_n$ acts trivially; the origin is one distinguished point on this stratum.

\subsubsection{Orbifold Fundamental Group and Coverings.}
Viewing $X$ as the global quotient $[M/G]$ with $M=\mathbb{C}^n$ and $G=S_n$, we can determine its orbifold fundamental group using the homotopy exact sequence for orbifolds.
For any good orbifold $[M/G]$ (see Section~\ref{sec:orbifold_fundamental_group}), there is a short exact sequence
\[
 1\longrightarrow\pi_1(M)\longrightarrow
 \pi_1^{\mathrm{orb}}\bigl([M/G]\bigr)\longrightarrow G\longrightarrow1.
\]
In our case, the covering space $M=\mathbb{C}^n$ is contractible, so $\pi_1(\mathbb{C}^n)=1$. The sequence collapses to
\[
 1 \longrightarrow 1 \longrightarrow \pi_1^{\mathrm{orb}}\bigl([\mathbb{C}^n/S_n]\bigr) \longrightarrow S_n \longrightarrow 1,
\]
which implies the isomorphism
\[
 \pi_1^{\mathrm{orb}}\bigl([\mathbb{C}^n/S_n]\bigr)\;\cong\;S_n.
\]

\emph{Relation to the Braid Group.} It is important to distinguish the orbifold fundamental group from the fundamental group of the regular part. The regular stratum $X_{\mathrm{reg}}$ is the configuration space of $n$ distinct unordered points in $\mathbb{C}$:
\[
 X_{\mathrm{reg}} = \{(z_1,\dots,z_n) \in \mathbb{C}^n \mid z_i \neq z_j \text{ for all } i \neq j\}/S_n.
\]
Its fundamental group is the braid group on $n$ strands, $\pi_1(X_{\mathrm{reg}}) \cong B_n$.
The orbifold inclusion $X_{\mathrm{reg}} \hookrightarrow X$ induces a surjective homomorphism
\[
 B_n \longrightarrow S_n,
\]
which sends the standard Artin generator $\sigma_i$ to the adjacent transposition $(i\,\,i\!+\!1)$. This is the usual permutation representation of the braid group, whose kernel is the pure braid group $P_n$. Equivalently, $\pi_1^{\mathrm{orb}}(X)$ is the Coxeter quotient of $B_n$ obtained by imposing the additional relations
\[
 \sigma_i^2=1 \qquad (1\le i\le n-1),
\]
so that the braid generators become transpositions. Geometrically, these order-two relations come from the fact that near a generic collision the local model is $[\mathbb{C}/\mathbb{Z}_2]$: a small exchange loop is the cone loop of order $2$, and its square is orbifold-nullhomotopic.

\emph{Universal Cover.} The universal orbifold cover of $X$ is the simply connected space $M=\mathbb{C}^n$ itself:
\[
  \mathbb{C}^n \longrightarrow [\mathbb{C}^n/S_n].
\]
The deck transformation group is exactly $S_n$. Concretely, fix a basepoint $z_0\in\mathbb{C}^n$ with all coordinates distinct and let $\bar z_0$ be its image in $X$. Any orbifold loop $\alpha$ based at $\bar z_0$ lifts to a unique path $\tilde\alpha$ in $\mathbb{C}^n$ starting at $z_0$. The endpoint $\tilde\alpha(1)$ must project to $\bar z_0$, so it lies in the orbit of $z_0$:
\[
  \tilde\alpha(1)=\sigma\cdot z_0
\]
for a unique permutation $\sigma\in S_n$. The assignment $[\alpha]\mapsto\sigma$ defines the isomorphism $\pi_1^{\mathrm{orb}}(X)\cong S_n$.
To see why this is a group isomorphism, we verify the three required properties. First, \textbf{homomorphism}: if $\alpha$ lifts to a path ending at $\sigma z_0$ and $\beta$ lifts to a path ending at $\tau z_0$, then the concatenation $\alpha\cdot\beta$ lifts to the path for $\alpha$ followed by the translated path $\sigma\cdot\tilde\beta$; this combined path ends at $\sigma(\tau z_0) = (\sigma\tau)z_0$, matching the group multiplication in $S_n$. Second, \textbf{surjectivity}: since $\mathbb{C}^n$ is path-connected, there exists a path connecting $z_0$ to any $\sigma z_0$, and its projection is a loop in $X$ with holonomy $\sigma$. Third, \textbf{injectivity}: if $\sigma=\mathrm{id}$, the lift $\tilde\alpha$ is a closed loop in $\mathbb{C}^n$, and since $\mathbb{C}^n$ is simply connected, $\tilde\alpha$ can be contracted to a point, implying $\alpha$ is contractible in the orbifold.

\emph{Geometric generators.} Near a generic point of $\Delta$, the local model
$\mathbb{C}^{n-1}\times[\mathbb{C}/\mathbb{Z}_2]$ shows that a small loop in the coarse space
transverse to $\Delta$ is exactly the cone loop of order $2$ from the basic example
$[\mathbb{C}/\mathbb{Z}_2]$.
Consider a path where two particles $z_i$ and $z_j$ exchange positions by rotating $180^\circ$ around their center of mass (counter-clockwise), while all other particles remain far away and stationary.
In the configuration space $X_{\mathrm{reg}}$, this loop represents a standard braid generator
(up to conjugacy).
In the orbifold $X$, this loop lifts to a path in $\mathbb{C}^n$ connecting $z_0$ to $(ij)\cdot z_0$. Thus, it corresponds to the transposition $(ij) \in S_n$.
Since transpositions generate $S_n$, these simple exchange loops generate the entire orbifold fundamental group.

\subsubsection{Orbifold de Rham Cohomology.}
Because the coarse space $|X|\cong\mathbb{C}^n$ is contractible and simply connected, its ordinary
de Rham cohomology is
\[
 H^k_{\mathrm{dR}}(|X|;\mathbb{R})\;\cong\;
 \begin{cases}
  \mathbb{R}, & k=0,\\[0.2em]
  0, & k>0.
 \end{cases}
\]
For a global quotient orbifold $[M/G]$ with $M$ smooth, the orbifold de Rham complex is identified
with the $G$-invariant de Rham complex on $M$. Here $M=\mathbb{C}^n$ is contractible, so by the
same averaging argument (see Section~\ref{sec:orbifold_de_rham}) as for $[\mathbb{C}/\mathbb{Z}_n]$ we obtain
\[
 H^k_{\mathrm{dR,orb}}(X)
 \;\cong\;
 H^k\bigl((\Omega^*(\mathbb{C}^n))^{S_n},d\bigr)
 \;\cong\;
 \begin{cases}
  \mathbb{R}, & k=0,\\[0.2em]
  0, & k>0.
 \end{cases}
\]
Thus, as in the cone case, the orbifold singularities do not create new
ordinary de Rham classes in positive degree.

\emph{Averaging argument.} Explicitly, if $\alpha\in(\Omega^k(\mathbb{C}^n))^{S_n}$ is a closed $S_n$-invariant $k$-form with $k\ge 1$, then $\alpha=d\beta$ for some $\beta\in\Omega^{k-1}(\mathbb{C}^n)$ by contractibility. Averaging over the group,
\[
  \bar\beta = \frac{1}{n!}\sum_{\sigma\in S_n}\sigma^*\beta,
\]
produces an $S_n$-invariant primitive with $d\bar\beta=\alpha$. Hence every closed invariant form of positive degree is exact.

\subsubsection{Orbifold Line Bundles and Holonomy.}
Following the general classification of orbifold line bundles (Section~\ref{sec:classification}),
we apply Theorem~\ref{thm:good_orbifold_classification} to the global quotient
$X=[\mathbb{C}^n/S_n]$. Since $\mathbb{C}^n$ is contractible, one has
$H^1(\mathbb{C}^n,\mathbb{Z})=0$ and $\mathrm{Pic}(\mathbb{C}^n)=0$, so the exact sequence of
Section~\ref{sec:classification} reduces to
\[
  \mathrm{Pic}_{\mathrm{orb}}(X)\;\cong\;\mathrm{Pic}_{S_n}(\mathbb{C}^n)\;\cong\;\mathrm{Hom}(S_n,U(1)).
\]
The one-dimensional unitary representations (characters) of $S_n$ are exactly the trivial
representation and the sign representation for $n\ge2$. Consequently, for $n \ge 2$, there are
exactly two isomorphism classes of complex orbifold line bundles: the trivial bundle and the sign
bundle. Equivalently,
\[
  \mathrm{Pic}_{\mathrm{orb}}(X)\cong H^2_{\mathrm{st}}(X;\mathbb{Z})\cong \mathbb{Z}_2,
\]
and the sign bundle represents the nontrivial torsion class.

\emph{Definition of the bundles $L_\chi$.} For each one-dimensional character $\chi:S_n\to U(1)$ we define an orbifold line bundle
$L_\chi$ by
\[
 \pi^*L_\chi \;\cong\; V\times\mathbb{C},
\]
with the $S_n$-action on $V\times\mathbb{C}$ given by
\[
 \sigma\cdot(z,v) \;=\; \bigl(\sigma\cdot z,\ \chi(\sigma)\,v\bigr).
\]
The two distinguished examples are the trivial bundle $L_{\mathbf{1}}$ (where $\chi = \mathbf{1}$
is the trivial character), encoding bosonic exchange via symmetric wavefunctions, and the sign
bundle $L_{\mathrm{sgn}}$ (where $\chi=\operatorname{sgn}$ is the sign representation), encoding
fermionic exchange phases via antisymmetric wavefunctions.
On the global chart $V=\mathbb{C}^n$, a section of the pullback bundle can be written in the
standard frame $e$ as $s(z)=f(z)e$; the coefficient function $f:V\to\mathbb{C}$ satisfies the
equivariance condition
\[
  f(\sigma\cdot z) = \chi(\sigma)\,f(z).
\]

\emph{Example: the case $n=2$.} Write
\[
  X=[\mathbb{C}^2/S_2], \qquad u=\frac{z_1+z_2}{2}, \qquad w=z_1-z_2.
\]
The nontrivial element of $S_2$ fixes $u$ and sends $w\mapsto-w$, so the collision locus is
$w=0$. For the sign bundle $L_{\mathrm{sgn}}$, the equivariance condition becomes
\[
  f(u,-w)=-f(u,w).
\]
Setting $w=0$ gives $f(u,0)=0$, so every local section of $L_{\mathrm{sgn}}$ vanishes on the
collision divisor. Equivalently, along the fixed locus $w=0$ the isotropy acts by $-1$ on the fiber, so the invariant part of the fiber is zero. This is the orbifold version of fermionic antisymmetry. By contrast, sections of
$L_{\mathbf{1}}$ satisfy $f(u,-w)=f(u,w)$ and need not vanish on $w=0$.

\emph{Connections and holonomy.} To realize these statistics via
holonomy, equip $L_\chi$ with a flat equivariant connection: start from the trivial connection on
$\mathbb{C}^n\times\mathbb{C}$ and let $S_n$ act on the fiber via $\chi$. This descends to a flat
orbifold connection with holonomy $(\chi(\sigma))^{-1}$ (which simply equals $\chi(\sigma)$ since these representations of $S_n$ take values in $\pm 1$) on any orbifold loop representing $\sigma\in S_n$. 
Concretely, for the trivial bundle $L_{\mathbf{1}}$ the holonomy around any loop is $+1$, while for the sign bundle $L_{\mathrm{sgn}}$ the holonomy around a loop corresponding to an odd permutation (such as a transposition exchanging two particles) is $-1$.

\emph{Integral and real Chern classes.} The integral and real first Chern classes behave
differently here, exactly as in Chapter~2. Integrally, the classification above gives
\[
  c_1^{\mathrm{st}}(L_{\mathbf{1}})=0, \qquad
  c_1^{\mathrm{st}}(L_{\mathrm{sgn}})\neq 0
  \quad\text{in } H^2_{\mathrm{st}}(X;\mathbb{Z})\cong\mathbb{Z}_2.
\]
Thus the sign bundle is topologically nontrivial, but only by a torsion class. After passing to real
coefficients, both bundles have vanishing real first Chern class:
\[
  c_1^{\mathrm{dR}}(L_{\chi}) = 0
  \quad\text{in } H^2_{\mathrm{dR,orb}}(X).
\]

\textbf{Topological reason:} The de Rham target cohomology group \(H^2_{\mathrm{dR,orb}}(X)\) is
zero. This follows from the contractibility of the cover: since $\mathbb{C}^n$ is contractible,
$H^2(\mathbb{C}^n;\mathbb{R})=0$, and the invariant subcomplex has the same cohomology. Thus the
torsion class of $L_{\mathrm{sgn}}$ disappears after tensoring with $\mathbb{R}$.

\textbf{Geometric reason:} The bundles $L_\chi$ are geometrically flat. We equip them with the flat
equivariant connection descended from the trivial connection on $\mathbb{C}^n \times \mathbb{C}$.
As an orbifold connection on the groupoid chart, this connection has identically zero curvature:
$F_{\mathrm{orb}}=0$ everywhere. Hence the Chern--Weil representative of
\(c_1^{\mathrm{dR}}(L_\chi)\) vanishes. Near a generic collision, the transverse local model
$[\mathbb{C}/\mathbb{Z}_2]$ shows that the sign bundle restricts in the transverse direction to the nontrivial weight-one line bundle from the cone example. Thus, on the punctured transverse coarse coordinate one may choose the same invariant frame as in Section~\ref{sec:config-space-1d}, obtaining the local singular gauge potential
\[
 A=-\frac{1}{2}\,d\varphi
\]
and holonomy $-1$ around a small loop linking the discriminant once. In this sense the coarse-space singular connection is exactly the basic cone model, now supported transversely to the discriminant.

\emph{Remark on Braiding Statistics.} One might ask why fractional (anyonic) statistics do not appear, given that the particles move in $\mathbb{C}$ (a $2$-dimensional space). Anyons arise when the configuration space has the braid group $B_n$ as its fundamental group. However, the orbifold $[\mathbb{C}^n/S_n]$ includes the singular strata where particles collide. The ability to pass through these singularities allows the ``braids'' to unknot: an elementary braid generator $\sigma_i$ (a $180^\circ$ half-twist exchanging two adjacent particles) satisfies $\sigma_i^2 = 1$ in the orbifold fundamental group, whereas in $B_n$ the element $\sigma_i^2$ represents a full $360^\circ$ winding and is nontrivial. This collapses $B_n$ to $S_n$, leaving only the two possibilities of bosons and fermions.

\section{The Dihedral Cone \texorpdfstring{$\mathbb{C} / D_n$}{[R2/Dn]}}\label{ex:R2Dn}

In this subsection we study the two-dimensional quotient orbifold~\cite{davis2011lectures,AdemLeidaRuan}
\[
  X_n = [\mathbb{C}/D_n], \qquad n \ge 2,
\]
obtained from the standard linear action of the dihedral group \(D_n\) on the plane. In contrast with the cyclic cone \( [\mathbb{C}/\mathbb{Z}_n] \), the dihedral quotient has mirrors as well as a corner reflector. It is therefore the basic local model for the mirror-and-corner geometry introduced in Section~\ref{sec:orbifold-euler}.

\subsubsection{Orbifold Definition and Basic Geometry.}

Recall that the dihedral group of order \(2n\) admits the presentation
\[
  D_n = \langle r,s \mid r^n = e,\; s^2 = e,\; srs = r^{-1} \rangle.
\]
Geometrically, \(r\) is the rotation by angle \(2\pi/n\) about the origin and \(s\) is reflection in the \(x\)-axis. We realize \(D_n\subset \mathrm{O}(2)\) by
\[
  r(z) = e^{2\pi i/n}z, \qquad s(z) = \overline{z}, \qquad z \in \mathbb{C}.
\]
This is an effective action by Euclidean isometries. Since the acting group is finite, the action is proper and all stabilizers are finite, so the quotient defines a good orbifold in the sense of Section~\ref{sec:orbifold_definition}.

The coarse space is
\[
  |X_n| = \mathbb{C}/D_n,
\]
and is homeomorphic to the closed sector
\[
  \{(\rho,\varphi) : \rho \ge 0,\ 0 \le \varphi \le \pi/n\}.
\]
Thus \(|X_n|\) is contractible. The two boundary rays of this sector are mirror strata: one is fixed by the reflection \(s\), the other by \(s' := rs\). Their common endpoint is the image of the origin and is a \emph{corner reflector} of corner order \(n\); its isotropy group is the full dihedral group \(D_n\), so the order of the singular point in the sense of Definition~\ref{def:orbifold} is \(|D_n| = 2n\). Along the open mirror rays the isotropy group is \(\mathbb{Z}_2\), while away from the mirrors and the apex the action is free.

\begin{remark}[Good Orbifold]
The developable presentation \(X_n = [\mathbb{C}/D_n]\) allows all subsequent constructions to be carried out on the smooth cover \(\mathbb{C}\) with \(D_n\)-equivariance.
\end{remark}

\subsubsection{Orbifold Fundamental Group and Coverings.}
Because \(|X_n|\) is contractible, its ordinary fundamental group is trivial. The orbifold fundamental group is nevertheless non-trivial, and Section~\ref{sec:orbifold_fundamental_group} gives the exact sequence for a global quotient:
\[
  1 \longrightarrow \pi_1(\mathbb{C}) \longrightarrow \pi_1^{\mathrm{orb}}([\mathbb{C}/D_n]) \longrightarrow D_n \longrightarrow 1.
\]
Since \(\mathbb{C}\) is simply connected, this collapses to
\[
  \pi_1^{\mathrm{orb}}(X_n) \cong D_n.
\]
Equivalently, the quotient map
\[
  q\colon \mathbb{C} \longrightarrow [\mathbb{C}/D_n]
\]
is the universal orbifold covering, with deck group \(D_n\).
More generally, based connected orbifold coverings of \(X_n\) are classified by subgroups \(H\le D_n\), while unbased connected coverings are classified up to isomorphism by conjugacy classes of subgroups. The covering corresponding to \(H\) is the quotient map
\[
  [\mathbb{C}/H] \longrightarrow [\mathbb{C}/D_n]
\]
induced by the inclusion \(H\hookrightarrow D_n\).

To see the generators geometrically, fix \(z_0 \in \mathbb{C}\) away from the reflection axes and write \(\bar z_0\) for its image in \(X_n\). Any orbifold loop \(\alpha\) based at \(\bar z_0\) lifts uniquely to a path \(\tilde\alpha\) with
\[
  \tilde\alpha(0)=z_0, \qquad \tilde\alpha(1)=g\cdot z_0
\]
for a unique \(g\in D_n\); the class \([\alpha]\) is sent to this \(g\).

The rotation generator \(r\) is represented by projecting the arc
\[
  \tilde\gamma_r(t) = z_0 e^{2\pi i t/n}, \qquad 0 \le t \le 1.
\]
Its projection is a loop in the quotient because the endpoints \(z_0\) and \(r z_0\) map to the same point of \(X_n\). This is the dihedral analogue of the cone loop from Section~\ref{sec:orbifold_fundamental_group}: it winds once around the corner reflector and has return element \(r\).

The reflection generator \(s\) is represented by an orbifold loop that reaches a mirror and closes there by isotropy. Choose a smooth path \(\eta\) in \(\mathbb{C}\) from \(z_0\) to a point \(x\) on the \(s\)-fixed ray, avoiding the origin. Since \(x=sx\), the projected path \(q\circ\eta\) ends at a mirror point whose local isotropy contains \(s\); closing \(q\circ\eta\) by this isotropy element gives an orbifold loop based at \(\bar z_0\) with return element \(s\). In the covering-space picture, the corresponding lifted representative is the concatenated path
\[
  \tilde\gamma_s = \eta * s(\eta)^{-1}
\]
which starts at \(z_0\) and ends at \(s z_0\). The group relations \(r^n=1\), \(s^2=1\), and \(srs=r^{-1}\) are therefore realized by explicit orbifold loops, not merely abstractly inherited from the quotient presentation.

\subsubsection{Orbifold de Rham Cohomology.}
Section~\ref{sec:orbifold_de_rham} identifies orbifold forms on a global quotient with invariant forms on the cover. Hence
\[
  \Omega^*_{\mathrm{orb}}(X_n) \cong \Omega^*(\mathbb{C})^{D_n}.
\]
Since \(\mathbb{C}\) is contractible, every closed form of positive degree on \(\mathbb{C}\) is exact. If \(\alpha\) is \(D_n\)-invariant and \(d\beta=\alpha\), averaging the primitive gives
\[
  \bar\beta = \frac{1}{|D_n|}\sum_{g\in D_n} g^*\beta,
\]
which is still \(D_n\)-invariant and satisfies \(d\bar\beta=\alpha\). Therefore
\[
  H^k_{\mathrm{dR,orb}}(X_n)
  \cong
  H^k(\Omega^*(\mathbb{C})^{D_n},d)
  \cong
  \begin{cases}
    \mathbb{R}, & k=0,\\[0.2em]
    0, & k>0.
  \end{cases}
\]
This agrees with the orbifold de Rham theorem of Chapter~2, since the coarse space \(|X_n|\) is a contractible sector.

\subsubsection{Orbifold Line Bundles and Holonomy.}
We now specialize the classification theory of Sections~\ref{sec:orbifold_line_bundles} and~\ref{sec:classification}. Because \(X_n\) is a global quotient, orbifold line bundles on \(X_n\) are the same as \(D_n\)-equivariant line bundles on \(\mathbb{C}\). Since \(\mathbb{C}\) is contractible, \(H^1(\mathbb{C},\mathbb{Z})=0\) and \(\mathrm{Pic}(\mathbb{C})=0\), so the exact sequence of Theorem~\ref{thm:good_orbifold_classification} reduces to
\[
  \mathrm{Pic}_{\mathrm{orb}}(X_n)
  \cong
  \mathrm{Pic}_{D_n}(\mathbb{C})
  \cong
  \mathrm{Hom}(D_n,U(1)).
\]
Thus every orbifold line bundle is obtained from the trivial bundle \(\mathbb{C}\times\mathbb{C}\) by choosing a unitary character \(\chi:D_n\to U(1)\) and letting
\[
  g\cdot(z,\xi) = (g\cdot z,\chi(g)\xi).
\]
The abelianization of \(D_n\) is~\cite{AdemMilgram2004}
\[
  D_n^{\mathrm{ab}}
  \cong
  \begin{cases}
    \mathbb{Z}_2, & n \text{ odd},\\[0.3em]
    \mathbb{Z}_2\times\mathbb{Z}_2, & n \text{ even},
  \end{cases}
\]
so there are two characters when \(n\) is odd and four when \(n\) is even.

For \(n\) odd the two line bundles are the trivial one and the sign bundle \(L_{\mathrm{sgn}}\), where
\[
  \chi_{\mathrm{sgn}}(r)=1,\qquad \chi_{\mathrm{sgn}}(s)=-1.
\]
For \(n\) even there are four characters. To match the notation used later in Chapters~4 and~6, we write them as
\begin{enumerate}
  \item \(\chi_{\mathbf{1}}\): \(\chi_{\mathbf{1}}(r)=1\), \(\chi_{\mathbf{1}}(s)=1\);
  \item \(\chi_{\mathrm{sgn}}\): \(\chi_{\mathrm{sgn}}(r)=1\), \(\chi_{\mathrm{sgn}}(s)=-1\);
  \item \(\chi_{\mathrm{det}}\): \(\chi_{\mathrm{det}}(r)=-1\), \(\chi_{\mathrm{det}}(s)=1\);
  \item \(\chi_{\mathrm{sgn}\cdot\mathrm{det}}\): \(\chi_{\mathrm{sgn}\cdot\mathrm{det}}(r)=-1\), \(\chi_{\mathrm{sgn}\cdot\mathrm{det}}(s)=-1\).
\end{enumerate}
Here \(\chi_{\mathrm{sgn}}=\det|_{D_n}\) is the genuine determinant character of the defining \(O(2)\)-representation. The symbol \(\chi_{\mathrm{det}}\) is retained only to match the notation used later in Chapters~4 and~6 for the even-\(n\) character with \(\chi(r)=-1\) and \(\chi(s)=1\).

\emph{Flat equivariant twists.} General \(D_n\)-invariant Hermitian connections on the trivial bundle over \(\mathbb{C}\) need not be flat; one may still choose \(D_n\)-invariant connections with nonzero, exact curvature. For flat orbifold line bundles, however, the equivariant twist is forced to be a character of the finite isotropy group. Indeed, let \((L,\nabla)\) be an orbifold line bundle on \(X_n\) equipped with a flat Hermitian connection, and pull it back to the cover \(\mathbb{C}\). Because \(\mathbb{C}\) is simply connected, the pulled-back flat bundle admits a global unitary parallel frame \(e\), so in this frame
\[
  \nabla = d.
\]
For each \(g\in D_n\), equivariance of \(\nabla\) implies that \(g\cdot e\) is again a global unitary parallel frame. Since \(\mathbb{C}\) is connected, it must differ from \(e\) by a constant phase:
\[
  g\cdot e=\chi(g)e,\qquad \chi(g)\in U(1).
\]
The equivariance law for the group action gives
\[
  \chi(gh)=\chi(g)\chi(h),
\]
so \(\chi\) is a unitary character of \(D_n\). Conversely, every character \(\chi:D_n\to U(1)\) defines a flat orbifold line bundle by equipping the trivial connection \(d\) on \(\mathbb{C}\times\mathbb{C}\) with the equivariant action
\[
  g\cdot(z,\xi)=(g\cdot z,\chi(g)\xi).
\]
Therefore the flat sectors are classified exactly by \(\mathrm{Hom}(D_n,U(1))\), which is finite. This is the precise sense in which the dihedral cone has no continuous Aharonov--Bohm parameter: the flat equivariant twist is discrete, as in the cyclic orbifold cone. The dihedral reflection relation further restricts the rotation character to the values \(\chi(r)=\pm1\).

\emph{Holonomy of the flat sectors.} For a flat character bundle \(L_\chi\), an orbifold loop representing \(g\in\pi_1^{\mathrm{orb}}(X_n)\cong D_n\) has trivial parallel transport on the simply connected cover, and the entire holonomy comes from the equivariant identification of the endpoint fiber with the starting fiber. Thus
\[
  \mathrm{Hol}_{L_\chi}(g)=\chi(g)^{-1}.
\]
(which simply equals \(\chi(g)\) since these characters take values in \(\pm 1\)).
In particular, the loop represented above by \(\tilde\gamma_r\) has holonomy \(\chi(r)^{-1}=\chi(r)\), while a mirror loop represented by \(\tilde\gamma_s\) has holonomy \(\chi(s)^{-1}=\chi(s)\). Since \(H^2_{\mathrm{dR,orb}}(X_n)=0\), all de Rham Chern classes vanish; the non-trivial information is purely torsion and is exactly the discrete character data described above.

\chapter{Quantum Mechanics on Orbifolds}

\section{\texorpdfstring{Free Quotient $S^1/\mathbb{Z}_n$: Flat-Holonomy Baseline}{Free Quotient S1/Zn: Flat-Holonomy Baseline}}\label{sec:quantum-s1-zn}
We first analyze a free quotient of the circle, using \(S^1/\mathbb{Z}_n\) as a flat-holonomy baseline rather than as a singular orbifold test case~\cite{Woodhouse1992,Sniatycki1980}.

\subsection{Notation and Conventions}
We begin by considering a free particle on a circle $S^1$ of circumference $L$. Let $\theta_c = 2\pi x/L$ denote the canonical angular coordinate, so that $\theta_c\in[0,2\pi)$ with the identification $\theta_c\sim\theta_c+2\pi$. In terms of the linear coordinate $x\in[0,L)$, the free Hamiltonian is
\begin{equation}\label{eq:s1-hamiltonian}
 H\;=\;-\frac{\hbar^2}{2M}\,\partial_x^2\;=\;\frac{L_z^2}{2I},
\end{equation}
where $I=MR^2$ is the moment of inertia (with $L=2\pi R$) and $L_z = -i\hbar\partial_{\theta_c}$ is the angular momentum operator. The equality of the two forms of $H$ follows from $\partial_x = (2\pi/L)\partial_{\theta_c}$ and hence
\begin{align*}
 H &= -\frac{\hbar^2}{2M}\left(\frac{2\pi}{L}\right)^2\partial_{\theta_c}^2
   = \frac{1}{2MR^2}(-i\hbar\partial_{\theta_c})^2 = \frac{L_z^2}{2I}.
\end{align*}
The eigenstates are plane waves
\begin{equation}\label{eq:s1-eigenstates}
 \Psi_m(x) = \frac{1}{\sqrt{L}}\,e^{2\pi i m x/L}, \qquad m\in\mathbb{Z},
\end{equation}
with quantized momentum $p_m = \hbar k_m = 2\pi\hbar m/L$.

The quotient $X=S^1/\mathbb{Z}_n$ is defined by the discrete translation $g\colon x\mapsto x+L/n$ (equivalently $\theta_c\mapsto\theta_c+2\pi/n$). A convenient fundamental domain is the half-open interval $[0,L/n)$; when imposing endpoint conditions we use the corresponding cut interval $[0,L/n]$. Since the action of $\mathbb{Z}_n$ is free (no fixed points), the stack quotient has no isotropy and is equivalent to the smooth quotient circle; equipped with the induced metric it has circumference $L/n$ and effective radius $R/n$. Thus the untwisted theory is just the free particle on a smaller circle. More generally, one may also couple to a flat $U(1)$ line bundle on the quotient. Because the quotient is again a circle, such flat bundles are classified by their holonomy $e^{2\pi i\alpha}$ with $\alpha\in\mathbb{R}/\mathbb{Z}$~\cite{Woodhouse1992}. The discrete labels $q\in\{0,\dots,n-1\}$ discussed below correspond only to the special holonomies $\alpha=q/n$ that arise when one keeps track of the chosen $n$-fold cover $S^1\to S^1/\mathbb{Z}_n$ and decomposes the covering Hilbert space into $\mathbb{Z}_n$-isotypic components; they are not fixed-point twisted sectors.

\subsection{Spectrum and Degeneracies}
To describe a general flat-holonomy sector, fix $\alpha\in\mathbb{R}/\mathbb{Z}$ and consider wavefunctions on the cut interval $[0,L/n]$ satisfying quasi-periodic boundary conditions at the endpoints:
\begin{equation}\label{eq:s1-twisted-bc}
 \Psi(L/n)=e^{2\pi i\alpha}\,\Psi(0), \qquad \Psi'(L/n)=e^{2\pi i\alpha}\,\Psi'(0).
\end{equation}
More precisely, the self-adjoint realization of the Hamiltonian on this sector is
\[
 H_\alpha=-\frac{\hbar^2}{2M}\partial_x^2
\]
with domain
\[
 D(H_\alpha)=\left\{\Psi\in H^2([0,L/n]) \;\middle|\; \Psi(L/n)=e^{2\pi i\alpha}\Psi(0),\ \Psi'(L/n)=e^{2\pi i\alpha}\Psi'(0)\right\}.
\]
These are the standard quasi-periodic boundary conditions for the Laplacian on a circle with flat holonomy $e^{2\pi i\alpha}$. Intrinsically, for a flat unitary line bundle one should match both the section and its covariant derivative at the endpoints; in the chosen flat trivialization on the cut interval, the connection $1$-form vanishes and this reduces to~\eqref{eq:s1-twisted-bc}. Equivalently, the wavefunction extends to the universal cover $\mathbb{R}$ and satisfies $\Psi(x+L/n)=e^{2\pi i\alpha}\Psi(x)$ for all $x$.

\begin{remark}[Sector Terminology]\label{rem:sector-terminology}
Throughout this text, when we specialize to the discrete values $\alpha=q/n$, we refer to the corresponding subspaces of the covering Hilbert space transforming under a specific irreducible representation of the deck group $\mathbb{Z}_n$ as \emph{sectors} (or \emph{isotypic sectors}). This highlights that the total Hilbert space on the covering circle decomposes into superselection sectors from the perspective of the quotient's observable algebra. Because the action is free, these are cover-isotypic sectors rather than fixed-point twisted sectors, and they capture only the special holonomies $\alpha=q/n$ inside the full $U(1)$ family.
\end{remark}

The allowed plane waves $\Psi(x) \propto e^{ikx}$ must satisfy this condition. Substituting the ansatz yields $e^{ikL/n} = e^{2\pi i \alpha}$, which implies the quantization condition for the wavenumber:
\begin{equation}\label{eq:s1-wavenumber}
 k_\ell^{(\alpha)} = \frac{2\pi n}{L}(\ell+\alpha), \qquad \ell\in\mathbb{Z}.
\end{equation}
The normalized eigenstates on the fundamental domain are
\begin{equation}\label{eq:s1-orb-eigenstates}
 \Psi_\ell^{(\alpha)}(x) = \sqrt{\frac{n}{L}}\,e^{2\pi i n(\ell+\alpha)x/L}, \qquad x\in[0,L/n].
\end{equation}
The corresponding energy levels are
\begin{equation}\label{eq:s1-energy}
 E_\ell^{(\alpha)} = \frac{\hbar^2}{2M} (k_\ell^{(\alpha)})^2 = \frac{\hbar^2}{2M} \left(\frac{2\pi n}{L}\right)^2 (\ell+\alpha)^2.
\end{equation}

The discrete cover-isotypic sectors are recovered by taking the special values $\alpha=q/n$ with $q\in\{0,\dots,n-1\}$. In that case the quantization condition becomes
\begin{equation}\label{eq:s1-wavenumber-ell}
 k_\ell^{(q)} = \frac{2\pi}{L}(\ell n + q), \qquad \ell \in \mathbb{Z}.
\end{equation}
The corresponding eigenstates and energies are the specializations
\[
 \Psi_\ell^{(q)}(x)=\sqrt{\frac{n}{L}}\,e^{2\pi i(\ell n+q)x/L},
 \qquad
 E_\ell^{(q)}=\frac{\hbar^2}{2M}\left(\frac{2\pi}{L}\right)^2(\ell n+q)^2.
\]
Equivalently, writing $m=\ell n+q$, one has
\[
 k_m^{(q)}=\frac{2\pi}{L}m, \qquad m\equiv q \pmod n,
\]
and the spectrum is precisely the subset of states on the original circle $S^1$ whose integer momentum quantum number satisfies that congruence.

The degeneracy structure is easiest to state in two layers. First, for a fixed holonomy $\alpha$, the energies depend only on $|\ell+\alpha|$. Hence $E_{\ell'}^{(\alpha)}=E_\ell^{(\alpha)}$ implies $\ell'+\alpha=\pm(\ell+\alpha)$. The plus sign gives the same state, while the minus sign requires
\[
 \ell'=-\ell-2\alpha.
\]
Thus a fixed holonomy sector is generically non-degenerate, and an in-sector $\pm k$ degeneracy occurs only when $2\alpha\in\mathbb{Z}$. For $\alpha=0$, the ground state $\ell=0$ is unique while all excited states are doubly degenerate; for $\alpha=\tfrac12$, every level is doubly degenerate.

Second, if one forms the direct sum over the discrete cover-isotypic sectors $\alpha=q/n$, then the familiar $\pm k$ pairing relates different sectors. A state with quantum number $m=\ell n+q$ in sector $q$ has the same energy as the state with quantum number $-m$. Since
\[
 -m=-\ell n-q = (-\ell-1)n + (n-q),
\]
the degenerate partner lies in sector $n-q$ modulo $n$. Thus, for generic $q$ (where $q\neq n-q \pmod n$), the degenerate partner lies in a different sector, so the spectrum within a single generic discrete sector is non-degenerate. For the self-conjugate sectors $q=0$, and $q=n/2$ when $n$ is even, the states with $\pm m$ lie in the same sector. In the $q=0$ sector, the ground state $m=0$ is unique and all excited states are doubly degenerate; in the sector $q=n/2$ (when $n$ is even), every level is doubly degenerate.

\section{\texorpdfstring{Planar Cone $\mathbb{C}/\mathbb{Z}_n$}{Planar Cone R2/Zn}}\label{sec:quantum-cone-c-zn}


We derive the free (time-independent) Schr\"odinger problem on the cone \(\mathbb{C}/\mathbb{Z}_n\) in every twisted sector~\cite{DeserJackiw1988,KayStuder1991,Ford2010FlatCones}. The geometric structure and classification of line bundles for this orbifold were established in Section~\ref{sec:config-space-1d} of Chapter~3. The corresponding geometric quantization of the planar cone, following the general program of Chapter~5, is presented in Section~\ref{subsec:gq-r2-zn} of Chapter~6.

\subsection{Coordinates and Boundary Conditions}
Work in orbifold polar coordinates \((r,\phi)\) with \(r\ge0\) and angular coordinate
\begin{equation}\label{eq:cone-wedge-range}
 \phi \in \Big[0,\tfrac{2\pi}{n}\Big).
\end{equation}
When convenient we pass to the \(2\pi\)-periodic full-angle variable
\begin{equation}\label{eq:cone-cover-angle}
 \varphi := n\,\phi \in [0,2\pi),
\end{equation}
so that the cone metric takes the form
\begin{equation}\label{eq:cone-metric}
 ds^2 = dr^2 + r^2\,d\phi^2 = dr^2 + \frac{r^2}{n^2}\,d\varphi^2.
\end{equation}
Discrete twisted sectors are labeled by \(q\in\{0,\dots,n-1\}\). A fixed quantum theory on the cone chooses one such flat orbifold line-bundle sector; when one wants to keep track of all possible flat twists at once, or equivalently decompose the covering-space Hilbert space into \(\mathbb{Z}_n\)-isotypic components, it is convenient to write a bookkeeping direct sum over all \(q\). In the sector labeled by~$q$, the wavefunction obeys
\begin{equation}\label{eq:cone-twist}
 \Psi\!\left(r,\,\phi+\tfrac{2\pi}{n}\right)
 = e^{2\pi i q/n}\,\Psi(r,\phi).
\end{equation}
With this bookkeeping convention, the sector sum is
\begin{equation}\label{eq:cone-hilbert-decomp}
 \mathcal{H} = \bigoplus_{q=0}^{n-1} \mathcal{H}_q.
\end{equation}
Equivalently, on the fundamental wedge \(0\le \phi < 2\pi/n\) we work in a local gauge in which the flat orbifold connection is gauged away and the sector label is encoded entirely in the seam conditions. For the Laplacian, the seam part of the domain in sector~\(\mathcal H_q\) is therefore specified by functions that are \(H^2\) in the interior and satisfy
\begin{equation}\label{eq:cone-seam-domain}
 \Psi(r,2\pi/n)=e^{2\pi iq/n}\Psi(r,0),
 \qquad
 \partial_\phi\Psi(r,2\pi/n)=e^{2\pi iq/n}\partial_\phi\Psi(r,0).
\end{equation}
These matching conditions make the wedge realization of the Laplacian symmetric and are the differential form of the twisted orbifold boundary condition; the additional choice of self-adjoint behavior at the apex is discussed below.

\subsection{Laplacian and Radial Equation}
The Laplacian in wedge coordinates reads
\begin{equation}\label{eq:cone-laplacian-wedge}
 \nabla^2 = \partial_r^2 + \frac{1}{r}\partial_r + \frac{1}{r^2}\,\partial_\phi^2.
\end{equation}
In the full-angle variable \(\varphi = n\,\phi\), this becomes
\begin{equation}\label{eq:cone-laplacian-cover}
 \nabla^2 = \partial_r^2 + \frac{1}{r}\partial_r + \frac{n^2}{r^2}\,\partial_\varphi^2.
\end{equation}
The free-particle Hamiltonian is
\begin{equation}\label{eq:cone-hamiltonian}
 H = -\frac{\hbar^2}{2M}\,\nabla^2.
\end{equation}

\subsection{Angular Eigenmodes}
Seek separable solutions \(\Psi(r,\phi) = R(r)\,\Phi(\phi)\) with
\(\Phi(\phi) = e^{im\phi}\). The twisted boundary condition~\eqref{eq:cone-twist} enforces
\begin{equation}\label{eq:cone-angular-quant}
 e^{im(2\pi/n)} = e^{2\pi i q/n}
 \quad\Longrightarrow\quad
 m = q + n\ell,\qquad \ell\in\mathbb{Z}.
\end{equation}
Thus the allowed angular momenta in sector~$q$ are $m \equiv q\pmod{n}$.

In the full-angle variable, writing $\Psi \propto e^{is_\varphi\,\varphi}$ with $\varphi = n\phi$, one finds $s_\varphi = m/n = \ell + q/n$ (so $s_\varphi \in \mathbb{Z} + q/n$). Here $s_\varphi$ is the Fourier mode number in the \(\varphi\)-description; equivalently, \(-i\hbar\partial_\varphi\) has eigenvalue \(\hbar s_\varphi\) while \(-i\hbar\partial_\phi\) has eigenvalue \(\hbar m = \hbar n s_\varphi\). The Bessel order is $|ns_\varphi| = |n\ell + q| = |m|$, and the two descriptions are equivalent under $\varphi = n\phi$.

Acting on \(e^{im\phi}\), the angular part of the Laplacian gives \(\partial_\phi^2\,e^{im\phi} = -m^2\,e^{im\phi}\).
Hence the radial equation for a particle of mass $M$ and potential $V(r)$ is
\begin{equation}\label{eq:cone-radial-general}
 -\frac{\hbar^2}{2M}\left[
   \frac{1}{r}\frac{d}{dr}\!\left(r\frac{dR}{dr}\right)
   - \frac{m^2}{r^2}\,R
 \right]
 + V(r)\,R = E\,R.
\end{equation}

\subsection{Free Particle (\texorpdfstring{$V=0$}{V=0}): Derivation and Normalization}\label{sec:cone-free}\label{sec:domains-extensions}

With \(V = 0\) and \(k = \sqrt{2ME}/\hbar \ge 0\), the radial ODE~\eqref{eq:cone-radial-general} reduces to Bessel's equation of order \(\nu = |m|\):
\begin{equation}\label{eq:cone-bessel}
 r^2 R'' + r R' + (k^2 r^2 - \nu^2)\,R = 0,
 \qquad \nu = |m| = |q + n\ell|.
\end{equation}

For each fixed \(m\), the local solutions are linear combinations of \(J_\nu(kr)\) and \(Y_\nu(kr)\), where \(Y_\nu\) denotes the Bessel function of the second kind\footnote{The historical term ``Neumann function'' for $Y_\nu$ is unrelated to Neumann boundary conditions on wedge mirrors.}. For \(\nu \ge 1\), the \(Y_\nu\)-branch behaves like \(r^{-\nu}\) near \(r=0\) and is not square-integrable with respect to the measure \(r\,dr\). For \(\nu=0\), however, \(Y_0(kr)\sim \frac{2}{\pi}\log r\) is locally square-integrable, so the radial operator admits additional self-adjoint extensions in that channel~\cite{KayStuder1991}. Throughout this chapter we choose the Friedrichs extension at the apex; this excludes the logarithmic branch in the \(\nu=0\) channel and selects the regular solution
\begin{equation}\label{eq:cone-radial-solution}
 R(r) = J_\nu(kr), \qquad E = \frac{\hbar^2 k^2}{2M},\quad k \ge 0.
\end{equation}

\subsubsection{Behavior at the Apex.}
In the untwisted sector ($q = 0$), the allowed angular momenta are multiples of $n$ ($m = n\ell$). The \(\ell = 0\) mode has \(\nu=0\) and its Friedrichs solution is nonvanishing at the origin: \(J_0(kr) \to 1\). This is the only channel in which the self-adjoint extension must be chosen explicitly, because the alternative logarithmic behavior \(Y_0(kr)\sim \frac{2}{\pi}\log r\) is still locally square-integrable. For all other channels one has \(\nu = |m|\ge 1\), so the regular solution behaves as \(r^{|m|}\) and vanishes at the tip. In particular, for twisted sectors ($q \neq 0$) the smallest angular momentum is \(\min(q,\,n{-}q) > 0\), forcing every Friedrichs solution to vanish at the apex.

\subsubsection{Spectrum and Eigenfunctions.}
The energy spectrum is absolutely continuous for \(E>0\), with threshold \(E=0\), and is independent of the sector~$q$. The functions below are therefore delta-normalized generalized eigenfunctions rather than \(L^2\)-eigenvectors.
We normalize them using the inner-product on the cone with measure \(r\,dr\,d\phi\) and angular range~\eqref{eq:cone-wedge-range}. Bessel orthogonality (for fixed order~\(\nu\), understood distributionally for \(k,k'>0\)) reads
\begin{equation}\label{eq:bessel-orthogonality}
 \int_0^{\infty} r\,dr\; J_\nu(kr)\,J_\nu(k'r)
  = \frac{\delta(k - k')}{k}.
\end{equation}
(Equivalently, the closure relation $\int_0^\infty k\,dk\;J_\nu(kr)\,J_\nu(kr') = \delta(r-r')/r$ uses the measure~$k\,dk$; the factor $1/k$ on the right-hand side of~\eqref{eq:bessel-orthogonality} is the Jacobian between the two.)

\subsubsection{Angular and Radial Normalization.}
The angular integration yields
\begin{equation}\label{eq:cone-angular-ortho}
 \int_0^{2\pi/n} d\phi\;e^{i(m - m')\phi}
  = \frac{2\pi}{n}\,\delta_{mm'}
  = \frac{2\pi}{n}\,\delta_{\ell\ell'},
  \qquad m = q + n\ell,\; m' = q + n\ell'.
\end{equation}
Requiring the full inner product to equal $\delta(k - k')\,\delta_{\ell\ell'}$, we determine the normalization constant:
\begin{equation}\label{eq:cone-normalization}
 |\mathcal{N}(k)|^2 \cdot \frac{2\pi}{n} \cdot \frac{1}{k}
 = 1
 \quad\Longrightarrow\quad
 \mathcal{N}(k) = \sqrt{\frac{nk}{2\pi}}.
\end{equation}
Note that $|\mathcal{N}(k)|$ drops any dependence on the angular quantum numbers $m$ and $\ell$, directly matching the flat continuum. The resulting eigenfunctions are
\begin{equation}\label{eq:cone-eigenfunctions}
 \Psi_{k,q,\ell}(r,\phi)
 = \sqrt{\frac{nk}{2\pi}}\;J_{|m|}(kr)\;e^{i(q + n\ell)\,\phi},
 \qquad m = q + n\ell,
\end{equation}
and they satisfy orthonormality in $k$ and $\ell$ within a fixed $q$-sector:
\begin{equation}\label{eq:cone-orthonormality}
 \langle \Psi_{k,q,\ell} \mid \Psi_{k',q,\ell'} \rangle
  = \delta(k - k')\,\delta_{\ell\ell'}.
\end{equation}
To write the angular completeness relation on the twisted sector, define the distributional kernel
\begin{equation}\label{eq:cone-angular-kernel}
 \delta_q(\phi,\phi')
 := \frac{n}{2\pi}\sum_{\ell\in\mathbb{Z}}
 e^{i(q+n\ell)(\phi-\phi')}.
\end{equation}
It acts as the identity on test functions satisfying the \(q\)-twisted seam condition~\eqref{eq:cone-seam-domain}; for interior points of the wedge it agrees with the ordinary Dirac delta in the angular variable. Completeness of the generalized eigenbasis within sector~\(\mathcal{H}_q\) is then expressed distributionally as
\begin{equation}\label{eq:cone-completeness}
 \sum_{\ell\in\mathbb{Z}} \int_0^{\infty}\!dk\;
  \Psi_{k,q,\ell}(r,\phi)\,\Psi_{k,q,\ell}(r',\phi')^*
  = \frac{\delta(r - r')}{r}\,\delta_q(\phi,\phi')
  \quad\text{on }\mathcal{H}_q.
\end{equation}

\subsubsection{Equivalent Formulation in Full-Angle Variables.}
In the full-angle variable $\varphi = n\phi$, the twist condition~\eqref{eq:cone-twist} becomes $\Psi(r,\varphi + 2\pi) = e^{2\pi iq/n}\Psi(r,\varphi)$, and the integration measure transforms as $r\,dr\,d\phi = (r/n)\,dr\,d\varphi$. Separation with $\Psi \propto e^{is_\varphi\,\varphi}$ gives $s_\varphi = \ell + q/n$ ($\ell \in \mathbb{Z}$); the Bessel order is $|ns_\varphi| = |n\ell + q| = |m|$. All normalizations above are consistent under this change of variables.

\section{\texorpdfstring{Harmonic Oscillator on $\mathbb{C}/\mathbb{Z}_n$}{Harmonic Oscillator on C/Zn}}

We now turn to the harmonic oscillator on the planar cone $\mathbb{C}/\mathbb{Z}_n$~\cite{MaratheMartucci1985,KayStuder1991}. The geometric setup---including the orbifold polar coordinates $(r, \phi)$ with $\phi\in[0,2\pi/n)$, the cone metric, and the twisted boundary conditions parameterized by the sector label $q\in\{0,\dots,n-1\}$ with equivariant twist phase $e^{2\pi i q/n}$---is identical to that of the free particle in the previous section. As in the previous section, we work directly in the physical gauge with the standard angular derivative $\partial_\phi$ and angular momentum $L_z = -i\hbar\partial_\phi$, satisfying the twisted boundary conditions directly. With the coarse-space holonomy convention of Chapter~3, this same sector has inverse holonomy $e^{-2\pi i q/n}$ around the puncture.

The essential difference is the addition of the isotropic confining potential $V(r) = \tfrac{1}{2}M\omega^2 r^2$ with $\omega > 0$. The Hamiltonian becomes
\begin{equation}\label{eq:osc-hamiltonian}
 H = -\frac{\hbar^2}{2M}\left(
   \partial_r^2 + \frac{1}{r}\partial_r + \frac{1}{r^2}\,\partial_\phi^2
 \right)
 + \frac{1}{2}M\omega^2 r^2.
\end{equation}
Unlike the continuous spectrum of the free particle, this potential induces a discrete spectrum. The resulting states may still be organized by the usual oscillator excitation number, now constrained by the orbifold sectors.

\subsection{Wavefunctions and Normalization}
Fix a sector $q$ and write the angular quantum number as
\begin{equation}\label{eq:osc-angular-quant}
 m = q + n\ell,\qquad \ell\in\mathbb{Z},
\end{equation}
exactly as in the free-particle analysis. Seeking separated solutions
\(\Psi(r,\phi) = R(r)e^{im\phi}\), the stationary Schr\"odinger equation
\(H\Psi = E\Psi\) reduces to
\begin{equation}\label{eq:osc-radial-equation}
 -\frac{\hbar^2}{2M}\left(
   R'' + \frac{1}{r}R' - \frac{m^2}{r^2}R
 \right)
 + \frac{1}{2}M\omega^2 r^2 R
 = E R.
\end{equation}
Because the oscillator potential $\frac{1}{2}M\omega^2 r^2$ is regular at the origin and subdominant to the centrifugal term, the indicial structure of the radial equation near $r=0$ is identical to that of the free particle. As in Section~\ref{sec:cone-free}, the cone apex requires a choice of self-adjoint domain only in the untwisted $m=0$ channel: near $r=0$ the local solutions behave as $1$ and $\log r$, both square-integrable with respect to $r\,dr$~\cite{KayStuder1991}. Throughout this subsection we choose the Friedrichs extension, which discards the logarithmic branch. For $m\neq 0$, regularity already forces $R(r)\sim r^{|m|}$ at the apex.

Let
\[
 \beta := \frac{M\omega}{\hbar} > 0,
 \qquad
 \rho := \beta r^2,
 \qquad
 R(r) = r^{|m|}e^{-\rho/2}F(\rho).
\]
Substituting this ansatz into~\eqref{eq:osc-radial-equation} gives Kummer's equation
\begin{equation}\label{eq:osc-kummer}
 \rho F'' + (|m|+1-\rho)F'
 + \left(\frac{E}{2\hbar\omega} - \frac{|m|+1}{2}\right)F = 0.
\end{equation}
Regularity at the apex selects the Kummer-$M$ branch, while square-integrability as $\rho\to\infty$ forces that branch to terminate to a polynomial. Hence
\begin{equation}\label{eq:osc-quantization}
 \frac{E}{2\hbar\omega} - \frac{|m|+1}{2} = n_r,
 \qquad n_r \in \{0,1,2,\dots\},
\end{equation}
so \(F(\rho) = L_{n_r}^{(|m|)}(\rho)\). Therefore an orthonormal basis of Friedrichs eigenfunctions in sector~$q$ is
\begin{equation}\label{eq:osc-wavefunction}
 \Psi_{n_r,m}^{(q)}(r,\phi)
  = \mathcal{N}_{n_r,m}\; r^{|m|}\,e^{-\beta r^2/2}\,
  L_{n_r}^{(|m|)}(\beta r^2)\; e^{im\phi},
  \qquad m \equiv q \pmod{n},
\end{equation}
where $L_{n_r}^{(|m|)}$ denotes the generalized Laguerre polynomial. The corresponding energies are
\begin{equation}\label{eq:osc-energy}
 E = \hbar\omega\big(2n_r + |m| + 1\big).
\end{equation}

\subsubsection{Orbifold Degeneracy Breaking and Ground State.}
While this formula matches the standard two-dimensional isotropic oscillator, the restriction $m \equiv q \pmod{n}$ fundamentally alters the spectrum. On the smooth plane ($n=1$), all integer $m$ are allowed, resulting in the standard highly degenerate levels. On the orbifold ($n>1$), the congruence condition distributes many states with the same $2n_r + |m|$ among different sectors, so a fixed sector generally has reduced degeneracy; the $\pm m$ pairing remains internal only for the self-conjugate sectors $q=0$ and, when $n$ is even, $q=n/2$. Moreover, the ground state energy within sector $q$ occurs at $n_r = 0$ and is given by
\[
 E_{\min}^{(q)} = \hbar\omega\big(\min(q, n-q) + 1\big).
\]
For $q \neq 0$, this local ground state energy is strictly greater than the untwisted zero-point energy $\hbar\omega$, reflecting the centrifugal barrier imposed by the twisted boundary condition.

\subsubsection{Derivation of the Normalization Constant.}
The inner product on the cone is $\langle f | g\rangle = \int_0^\infty r\,dr \int_0^{2\pi/n} d\phi\, f^* g$. The angular integral contributes
\begin{equation}\label{eq:osc-angular-ortho}
 \int_0^{2\pi/n} d\phi\; e^{i(m - m')\phi}
 = \frac{2\pi}{n}\,\delta_{\ell\ell'}
 = \frac{2\pi}{n}\,\delta_{mm'},
 \qquad m = q+n\ell,\; m' = q+n\ell',
\end{equation}
(since $m-m' = n(\ell-\ell')$ is an integer multiple of $n$, so the exponential integrates to zero unless $\ell=\ell'$). For the radial part, substitute $u = \beta r^2$ (so $r\,dr = du/(2\beta)$ and $r^{2|m|} = u^{|m|}/\beta^{|m|}$):
\begin{equation}\label{eq:osc-radial-integral}
\begin{split}
 \int_0^\infty r\,dr\; r^{2|m|}e^{-\beta r^2}\,
 & L_{n_r}^{(|m|)}(\beta r^2)L_{n_r'}^{(|m|)}(\beta r^2) \\
 &= \frac{1}{2\beta^{|m|+1}}
 \int_0^\infty u^{|m|}e^{-u}\,
 L_{n_r}^{(|m|)}(u)L_{n_r'}^{(|m|)}(u)\,du \\
 &= \frac{\Gamma(n_r + |m| + 1)}{2\beta^{|m|+1}\, n_r!}\,\delta_{n_r n_r'}.
\end{split}
\end{equation}
Here we used the standard Laguerre orthogonality formula
\[
 \int_0^\infty u^\alpha e^{-u}
 L_p^{(\alpha)}(u)L_{p'}^{(\alpha)}(u)\,du
 = \frac{\Gamma(p+\alpha+1)}{p!}\,\delta_{pp'},
 \qquad \alpha > -1.
\]

Requiring $\langle\Psi_{n_r,m}^{(q)}|\Psi_{n_r,m}^{(q)}\rangle = 1$ gives
\begin{equation}\label{eq:osc-norm-derive}
 |\mathcal{N}_{n_r,m}|^2 \cdot \frac{2\pi}{n}
  \cdot \frac{\Gamma(n_r + |m| + 1)}{2\beta^{|m|+1}\, n_r!} = 1,
\end{equation}
which yields the normalization constant (recalling $\beta = M\omega/\hbar$):
\begin{equation}\label{eq:osc-normalization}
  \mathcal{N}_{n_r,m}
 = \sqrt{\frac{n\,\beta^{|m|+1}}{\pi}\;\frac{n_r!}{\Gamma(n_r + |m| + 1)}}.
\end{equation}
Full orthonormality then reads
\begin{equation}\label{eq:osc-orthonormality}
 \int_0^\infty\!r\,dr\int_0^{2\pi/n}\!d\phi\;
 \Psi_{n_r,m}^{(q)}(r,\phi)^*\,
 \Psi_{n_r',m'}^{(q)}(r,\phi)
  = \delta_{mm'}\,\delta_{n_r n_r'}.
\end{equation}

Furthermore, by combining the radial completeness of the Laguerre polynomials with the angular completeness of the $q$-sector exponentials (using the distributional kernel $\delta_q(\phi,\phi')$ from~\eqref{eq:cone-angular-kernel}), the oscillator eigenstates form a complete orthonormal basis for $\mathcal{H}_q$:
\begin{equation}\label{eq:osc-completeness}
 \sum_{n_r=0}^\infty \sum_{\ell\in\mathbb{Z}}
  \Psi_{n_r,m}^{(q)}(r,\phi)\,\Psi_{n_r,m}^{(q)}(r',\phi')^*
  = \frac{\delta(r - r')}{r}\,\delta_q(\phi,\phi'),
  \qquad m = q + n\ell.
\end{equation}

\section{\texorpdfstring{Orbifold Football $S^2(n,n)$}{Orbifold Football S2(n,n)}}
\label{sec:s2-nn}
We investigate the quantum mechanics of the football orbifold $S^2(n,n)$, focusing on the isotypic decomposition of spherical harmonics~\cite{LermanTolman1997,BaierMouraoNunes2018}. 

\subsection{Notation and Conventions}
To set the stage, we first review the standard quantum mechanics on the sphere $S^2$ of radius $R$. In terms of the polar angles $(\theta,\varphi)$ with $\theta\in[0,\pi]$ and $\varphi\in[0,2\pi)$, the free Hamiltonian is given by
\begin{equation}\label{eq:s2-hamiltonian}
 H\;=\;-\frac{\hbar^2}{2M}\,\Delta_{S^2}\;=\;\frac{L^2}{2I},
\end{equation}
where $I=MR^2$ is the moment of inertia and $L^2 = L_x^2 + L_y^2 + L_z^2$ is the squared angular momentum operator. The eigenfunctions of this Hamiltonian are the spherical harmonics $Y_{\ell m}(\theta,\varphi)$, with eigenvalues
\begin{equation}\label{eq:s2-spectrum}
 E_\ell = \frac{\hbar^2}{2I}\ell(\ell+1), \qquad \ell \in \{0, 1, 2, \dots\}.
\end{equation}
On the full sphere, each level $E_\ell$ has a degeneracy of $2\ell+1$, corresponding to $m \in \{-\ell, \dots, \ell\}$.

The orbifold $X=S^2(n,n)$ (or $S^2/\mathbb{Z}_n$) is the quotient of the sphere by the group $\mathbb{Z}_n$ acting via $\varphi \sim \varphi + 2\pi/n$. Geometrically, this is the \textit{American football orbifold}, with conical singularities of angle $2\pi/n$ at the north and south poles ($\theta=0, \pi$). A fundamental domain is the spherical lune $\varphi \in [0, 2\pi/n)$.

\subsection{Spectrum and Degeneracies}
Since $\pi_1^{\mathrm{orb}}(S^2(n,n)) \cong \mathbb{Z}_n$, there are $n$ flat orbifold line bundles on $X$ indexed by the characters
\[
 \chi_q(g)=e^{2\pi i q/n}, \qquad q\in\{0,\dots,n-1\}.
\]
We may either quantize each bundle separately, or equivalently decompose the covering Hilbert space $L^2(S^2)$ into $\mathbb{Z}_n$-isotypic subspaces. Throughout this subsection we use the latter realization and write
\[
 \mathcal H_q
 :=
 \bigl\{\Psi\in L^2(S^2)\mid \Psi(\theta,\varphi+2\pi/n)=e^{2\pi iq/n}\Psi(\theta,\varphi)\bigr\}.
\]
On the cut lune $0 \le \varphi \le 2\pi/n$, a wavefunction in the $q$-th sector therefore satisfies
\begin{equation}\label{eq:s2-twisted-bc}
 \Psi(\theta,\varphi+2\pi/n)=e^{2\pi i q/n}\,\Psi(\theta,\varphi).
\end{equation}
For the Laplacian, the corresponding self-adjoint domain also requires matching of the $\varphi$-derivative across the seam:
\begin{equation}\label{eq:s2-twisted-derivative-bc}
 \partial_\varphi \Psi(\theta,\varphi+2\pi/n)=e^{2\pi i q/n}\,\partial_\varphi \Psi(\theta,\varphi),
 \qquad 0<\theta<\pi.
\end{equation}
These seam conditions do not by themselves determine the behavior at the conical poles. We therefore define the Hamiltonian in sector $q$ as the restriction of the smooth spherical Laplacian on the cover $S^2$ to the $H^2$-subspace $\mathcal H_q$; equivalently, on the lune we choose the orbifold-smooth (hence Friedrichs) domain obtained from smooth $\chi_q$-equivariant lifts on $S^2$. In particular, the untwisted $m=0$ channel excludes the logarithmic branch familiar from the planar cone, and for $q\neq 0$ the seam condition at $\theta=0,\pi$ forces
\[
 \Psi(0,\varphi)=\Psi(\pi,\varphi)=0,
\]
because the pole values must satisfy $v=e^{2\pi iq/n}v$.

With this domain understood, the operator $H$ commutes with both $L_z$ and the $\mathbb{Z}_n$-action, so each spherical eigenspace
\[
 \mathcal E_\ell=\operatorname{span}\{Y_{\ell m}\mid -\ell\le m\le \ell\}
\]
is $\mathbb{Z}_n$-stable. On a spherical harmonic one has
\[
 Y_{\ell m}(\theta,\varphi+2\pi/n) = e^{2\pi i m/n} Y_{\ell m}(\theta,\varphi)
\]
so $\mathcal E_\ell\cap\mathcal H_q$ is spanned by those $Y_{\ell m}$ with
\begin{equation}\label{eq:s2-selection-rule}
 m \equiv q \pmod n.
\end{equation}
Thus an orthonormal eigenbasis in sector $q$ is given by
\begin{equation}\label{eq:s2-orb-wavefunctions}
 \Psi_{\ell, m}^{(q)}(\theta, \varphi) = \sqrt{n}\, Y_{\ell m}(\theta, \varphi), \qquad m \in \{q + kn : k \in \mathbb{Z}\} \cap [-\ell, \ell].
\end{equation}
The factor $\sqrt{n}$ ensures normalization on the fundamental lune: since $|Y_{\ell m}|^2$ is independent of $\varphi$, integrating against the invariant measure $d\Omega = \sin\theta\,d\theta\,d\varphi$ over the lune $[0,2\pi/n)$ yields exactly $1/n$ of the full-sphere norm, so $|\sqrt{n}\,Y_{\ell m}|^2$ integrates to unity.

The energy levels remain
\[
 E_\ell = \frac{\hbar^2}{2I}\ell(\ell+1),
\]
but the $(2\ell+1)$-dimensional spherical-harmonic multiplet splits into $\mathbb{Z}_n$-isotypic pieces. The degeneracy $g_\ell^{(q)}$ in sector $q$ is the number of integers $k$ such that $-\ell \le q + kn \le \ell$, or equivalently
\begin{equation}\label{eq:s2-degeneracy}
 g_\ell^{(q)}
 = \left\lfloor \frac{\ell-q}{n}\right\rfloor - \left\lceil \frac{-\ell-q}{n}\right\rceil + 1
 = \left\lfloor \frac{\ell-q}{n}\right\rfloor + \left\lfloor \frac{\ell+q}{n}\right\rfloor + 1.
\end{equation}
In particular, the untwisted sector has
\[
 g_\ell^{(0)} = 2\left\lfloor \frac{\ell}{n}\right\rfloor + 1,
\]
which reduces to the usual $2\ell+1$ when $n=1$.

For $q \neq 0$, the lowest angular-momentum level appearing in the $q$-sector is
\[
 \ell_{\min}(q)=\min(q,n-q),
\]
because this is the smallest nonnegative integer congruent to either $q$ or $-q$ modulo $n$. Moreover, the complex-conjugate partner of a state with magnetic quantum number $m$ lies in sector $n-q$, since $-m \equiv n-q \pmod n$. Thus the familiar $\pm m$ pairing is shared between the conjugate sectors $q$ and $n-q$; only the self-conjugate sectors $q=0$ and, when $n$ is even, $q=n/2$ contain both partners internally.

Therefore the orbifold quotient does not change the spherical energy formula for the modes that survive the sector projection, but a fixed sector generally contains only a subset of the full sequence. In particular, for $q\neq0$ the first surviving level occurs at $\ell_{\min}(q)=\min(q,n-q)$ rather than at $\ell=0$. Across the bookkeeping direct sum over all $q$-sectors the full spherical spectrum is recovered, while each angular-momentum multiplet is refined according to the residue class of $m$ modulo $n$.

\section{\texorpdfstring{Orbisphere $S^2(n,m)$}{Orbisphere S2(n,m)}}\label{sec:quantum-s2-mn}


In this section, we analyze the quantum mechanics obtained by Kaluza--Klein reducing a free particle on $S^3$ along the Seifert fibration over the spindle orbifold $S^2(n,m)$~\cite{LermanTolman1997,BaierMouraoNunes2018}. This space generalizes the standard two-sphere and the symmetric football $S^2(n,n)$ of the previous section by allowing for distinct cone angles, $2\pi/n$ and $2\pi/m$, at the two poles. We allow the value $1$ as a cone order, with the convention that order $1$ means a smooth point; thus $S^2(1,1)$ is the ordinary sphere, while $S^2(1,m)$ is a one-cone-point teardrop. Unless stated otherwise, we assume throughout that $n$ and $m$ are coprime ($\gcd(n,m)=1$), because only in that effective case does the weighted $S^3/U(1)$ construction below model the same two-cone-point orbifold $S^2(n,m)$ studied in Chapter~3. When $\gcd(n,m)>1$, the same weighted action produces a non-effective quotient with generic stabilizer, not the effective spindle of Chapter~3; we return to that distinction in a separate remark below, but all explicit spectral formulas derived in this subsection are stated for $\gcd(n,m)=1$.

\subsection{Seifert-Fibration Formulation}

\subsubsection{Seifert Fibration and Weighted Action.}
In the coprime case $\gcd(n,m)=1$, the spindle orbifold $S^2(n,m)$ can be realized as the quotient space of the unit three-sphere $S^3 \subset \mathbb{C}^2$ under the weighted $U(1)$ action:
\begin{equation}\label{eq:snm-u1-action}
 (z_1, z_2) \sim (e^{in\alpha} z_1, e^{im\alpha} z_2), \quad \alpha \in [0, 2\pi).
\end{equation}
Unlike the standard Hopf fibration ($n=m=1$), which yields a principal $U(1)$ bundle over a smooth Riemann sphere, general coprime integers $n, m \ge 1$ make $S^3$ the total space of a non-trivial \emph{Seifert fibration}. While the $U(1)$ action is free at generic points, it has non-trivial isotropy at $z_1=0$ and $z_2=0$. At $z_2=0$ (north) the orbit period is $2\pi/n$, while at $z_1=0$ (south) it is $2\pi/m$. The base space that emerges from this Seifert fibration is the spindle orbifold $S^2(n,m)$. In this setup, the geometric singularities purely correspond to isolated cone points where specific discrete isotropy subgroups stabilize the phase rotations. Explicitly, the north pole ($z_2 = 0$) has a stabilizer of $\mathbb{Z}_n$, while the south pole ($z_1 = 0$) has a stabilizer of $\mathbb{Z}_m$.

\subsubsection{Motivation: Formulating the Problem on $S^3$.}
Before proceeding, it is helpful to explain why we solve the Laplacian on the smooth total space $S^3$ of the Seifert fibration rather than directly on the base orbifold $S^2(n,m)$. Writing the eigenvalue problem directly on the 2D base metric introduces several technical difficulties:
\begin{enumerate}
    \item \textbf{Resolving Geometric Singularities:} The 2D metric contains conical singularities at the poles, meaning a direct 2D Laplacian approach would require manually imposing boundary conditions to control singular wavefunctions. In contrast, the total space $S^3$ is perfectly smooth, guaranteeing that regular solutions to the 3D differential equation are automatically well behaved at the cone points.
    \item \textbf{Geometrizing Magnetic Flux:} On the base $S^2(n,m)$, states in a twisted quantum sector behave as a particle coupled to a singular background magnetic monopole. Formulating the covariant 2D Hamiltonian $H \propto (\nabla - iA)^2$ introduces complex differential cross-terms. By formulating the problem on $S^3$, we geometrize this flux: the singular gauge interaction simplifies into ordinary translational kinetic energy along the extra $U(1)$ fiber.
    \item \textbf{Linearizing Fractional Twists:} Tracking non-trivial holonomies and topological charges directly on the orbifold surface requires complicated patching between local charts. On the total space $S^3$, these fractional boundary conditions simplify into a straightforward linear Diophantine constraint on the integer angular momenta.
\end{enumerate}
By elevating the problem to three dimensions, we transform a singular, magnetically-coupled 2D boundary-value problem into a much simpler canonical model: a free quantum particle on a smooth hypersphere.

\subsubsection{Kaluza--Klein Reduction and Eigenspace Decomposition.}
With this rationale, we construct the quantum mechanics of a simple free particle using a \emph{Kaluza--Klein dimensional reduction}. We begin with the Hamiltonian for a free non-relativistic particle on the unit sphere $S^3$:
\begin{equation}\label{eq:snm-hamiltonian-s3}
 H_{S^3} = -\frac{\hbar^2}{2I_{S^3}}\,\Delta_{S^3},
\end{equation}
where $I_{S^3}$ is the moment of inertia and $\Delta_{S^3}$ is the Laplace--Beltrami operator for the standard round metric.

To extract dynamics localized to the base orbifold $S^2(n,m)$, we restrict this higher-dimensional theory to the eigenspaces of the $U(1)$ symmetry. The associated dimensionless infinitesimal generator is
\begin{equation}\label{eq:snm-u1-generator}
 \hat{L}_\alpha = -i \partial_\alpha = -i \left( n \frac{\partial}{\partial \phi_1} + m \frac{\partial}{\partial \phi_2} \right),
\end{equation}
where $\phi_1$ and $\phi_2$ are the phase coordinates corresponding to $z_1$ and $z_2$, respectively; the physical fiber momentum is $\hbar \hat L_\alpha$.

Because the vector field $\partial_\alpha$ leaves the standard round metric of $S^3$ invariant, it is a Killing vector field. Its directional derivative therefore commutes with the Laplace--Beltrami operator, meaning $[\hat{L}_\alpha, H_{S^3}] = 0$. This ensures that the fiber momentum $\hat{L}_\alpha$ is a constant of motion. Moreover, since $S^3$ is a compact Riemannian manifold, the Laplace--Beltrami operator is essentially self-adjoint with a purely discrete spectrum, and the Peter--Weyl theorem for $SU(2) \cong S^3$ guarantees that the resulting eigenfunction expansion is complete.

Consequently, wavefunctions on $S^3$ can be simultaneously diagonalized with the internal momentum, satisfying the constraint:
\begin{equation}\label{eq:snm-charge-eigenvalue}
 \hat{L}_\alpha \Psi_Q = Q \Psi_Q.
\end{equation}
This eigenvalue equation enforces the necessary twisting condition: upon traversing the $U(1)$ fiber, the wavefunction acquires an equivariance phase $e^{iQ\alpha}$ determined by the topological charge $Q \in \mathbb{Z}$. It implies that the total Hilbert space decomposes into invariant topological sectors:
\begin{equation}\label{eq:snm-hilbert-decomp}
 \mathcal{H}_{S^3} = \bigoplus_{Q \in \mathbb{Z}} \mathcal{H}_Q.
\end{equation}
By restricting the three-dimensional Laplacian to a chosen subspace $\mathcal{H}_Q$, we naturally produce an effective two-dimensional Hamiltonian on the base. More precisely, throughout this subsection the \emph{reduced Hamiltonian in sector $Q$} means the operator induced by $H_{S^3}$ on the charge-$Q$ eigenspace $\mathcal{H}_Q$, after identifying $\mathcal{H}_Q$ with sections of the associated orbifold line bundle over $S^2(n,m)$. We do \emph{not} subtract the vertical kinetic contribution separately. Because the length of the $U(1)$ fiber ($\rho = \sqrt{n^2 \cos^2\eta + m^2 \sin^2\eta}$) depends on the coordinates, this reduced operator is not the bare Laplace--Beltrami operator of the base metric; rather, it is the full descended gauge-coupled/dilaton operator coming from the restriction of the three-dimensional Laplacian. In the ordinary Hopf case $n=m=1$, where $\rho \equiv 1$, this convention differs from the purely horizontal magnetic Laplacian by the constant energy shift $\hbar^2 Q^2/(2I_{S^3})$. Thus the energies computed below should be compared only with base operators using the same convention, or after an explicit vertical-energy subtraction has been specified. The dynamics are fundamentally organized by $Q$:
\begin{itemize}
    \item \textbf{Sector $\boldsymbol{Q = 0}$ (Scalar Modes):} Wavefunctions in the zero-charge sector have no dependence on the internal fiber coordinate. They descend to globally invariant functions on $S^2(n,m)$, governed by the weighted scalar operator obtained from the reduction.
    \item \textbf{Sectors $\boldsymbol{Q \neq 0}$ (Gauge-Coupled Sectors):} Non-zero integer charges determine the associated orbifold line bundle and connection on the base. In the reduced description the wavefunctions are sections of non-trivial orbifold line bundles and the descended operator is gauge-coupled, with the vertical kinetic and fiber-length terms retained by the convention above.
\end{itemize}
In terms of the orbifold line-bundle classification of Chapter~3, the topological charge $Q$ determines the local isotropy phases of the corresponding bundle at the two cone points: the actual Seifert stabilizer subgroup at the north pole acts on sections by $e^{2\pi i Q/n}$, while the one at the south pole acts by $e^{2\pi i Q/m}$. To convert these phases to the normalized cone-chart generators of Chapter~3, note that on a local north-pole chart the residual $\mu_n$-action rotates the transverse coordinate with weight $m$, so the standard cone generator is the $m^{-1}$th power of the literal Seifert stabilizer generator; similarly, on a south-pole chart the residual $\mu_m$-action has weight $n$. Hence the intrinsic isotropy weights are
\[
 a \equiv Q\,m^{-1} \pmod{n},
 \qquad
 b \equiv Q\,n^{-1} \pmod{m},
\]
where $m^{-1} \in \mathbb{Z}_n$ and $n^{-1} \in \mathbb{Z}_m$ denote the multiplicative inverses. Equivalently, in the weighted-projective notation of Chapter~6 the sector $Q$ corresponds to $\mathcal{O}(Q)$, whose intrinsic data satisfy these congruences. In particular, the scalar sector $Q = 0$ corresponds to the trivial bundle $L_{(0;0,0)}$ with vanishing orbifold degree.

\subsection{Coordinates, Kaluza--Klein Metric, and Gauge Field}

\subsubsection{Coordinates on $S^3$ and the $U(1)$ Action.}
We parameterize the unit sphere $S^3 \subset \mathbb{C}^2$ using Hopf-like coordinates $(\eta, \phi_1, \phi_2)$:
\begin{equation}\label{eq:snm-hopf-coords}
 z_1 = \cos\eta\, e^{i\phi_1}, \quad z_2 = \sin\eta\, e^{i\phi_2},
\end{equation}
where $\eta \in [0, \pi/2]$ and $\phi_1, \phi_2 \in [0, 2\pi)$. In these coordinates, the standard round metric on the unit $S^3$ takes the form:
\begin{equation}\label{eq:snm-s3-metric}
 ds^2_{S^3} = d\eta^2 + \cos^2\eta\, d\phi_1^2 + \sin^2\eta\, d\phi_2^2.
\end{equation}
The continuous $U(1)$ action~\eqref{eq:snm-u1-action} is generated by the Killing vector field (cf.\ the quantum generator $\hat{L}_\alpha = -i\partial_\alpha$ of~\eqref{eq:snm-u1-generator}):
\begin{equation}\label{eq:snm-killing-vector}
 \partial_\alpha = n \frac{\partial}{\partial \phi_1} + m \frac{\partial}{\partial \phi_2}.
\end{equation}

\subsubsection{Metric Decomposition and the Connection 1-Form.}
To execute the Kaluza--Klein reduction and identify the geometry of the orbisphere $S^2(n,m)$, we cast the $S^3$ metric as a Riemannian submersion over the base spindle. This is accomplished by decomposing the metric into a component along the $U(1)$ fiber and a component orthogonal to it. The standard Kaluza--Klein ansatz takes the form:
\begin{equation}\label{eq:snm-kk-decomp}
 ds^2_{S^3} = ds^2_{S^2(n,m)} + \rho^2\, \omega^2,
\end{equation}
where $\rho^2$ is the squared norm of the fiber-generating Killing vector, $\omega$ is the connection 1-form defined such that $\omega(\partial_\alpha) = 1$, and $ds^2_{S^2(n,m)}$ is the metric on the base.

The length squared of the Killing vector $\partial_\alpha$ gives the scalar field $\rho^2$:
\begin{equation}\label{eq:snm-rho-squared}
 \rho^2 = g(\partial_\alpha, \partial_\alpha) = n^2 \cos^2\eta + m^2 \sin^2\eta.
\end{equation}
Because $n,m \ge 1$ and $\cos^2\eta + \sin^2\eta = 1$, $\rho^2$ never vanishes on $S^3$. This guarantees that the Kaluza--Klein connection is globally regular over the entire total space.
The normalized metric-dual 1-form to the fiber generator is therefore:
\begin{equation}\label{eq:snm-connection-form}
 \omega = \frac{g(\partial_\alpha, \cdot)}{\rho^2} = \frac{n \cos^2\eta\, d\phi_1 + m \sin^2\eta\, d\phi_2}{n^2 \cos^2\eta + m^2 \sin^2\eta}.
\end{equation}

\subsubsection{Derivation of the Base Metric.}
To explicitly construct the base metric $ds^2_{S^2(n,m)}$, we perform an algebraic completion of the square on the angular part of $ds^2_{S^3}$. For the coprime case under consideration, we first define the primitive invariant angular coordinate
\[
 \phi = n\phi_2 - m\phi_1.
\]
It naturally uncouples from the fiber action because $\partial_\alpha \phi = 0$, and because $\gcd(n,m)=1$ it is the correct $2\pi$-periodic quotient angle. (If $d=\gcd(n,m)>1$, the primitive invariant angle would instead be $\tilde\phi=\frac{n}{d}\phi_2-\frac{m}{d}\phi_1$.) Recognizing how the coefficients factor, the completion of the square yields:
\begin{align*}
 \cos^2\eta\, d\phi_1^2 + \sin^2\eta\, d\phi_2^2 &= \frac{(n \cos^2\eta\, d\phi_1 + m \sin^2\eta\, d\phi_2)^2}{\rho^2} \\
 &\quad + \frac{\cos^2\eta \sin^2\eta}{\rho^2}\, d\phi^2 \\
 &= \rho^2 \omega^2 + \frac{\cos^2\eta \sin^2\eta}{n^2 \cos^2\eta + m^2 \sin^2\eta} \, d\phi^2.
\end{align*}
By directly matching this decomposition against the Kaluza--Klein ansatz, we read off the metric of $S^2(n,m)$:
\begin{equation}\label{eq:snm-metric-base}
 ds^2_{S^2(n,m)} = d\eta^2 + \frac{\cos^2\eta \sin^2\eta}{n^2 \cos^2\eta + m^2 \sin^2\eta} \, d\phi^2.
\end{equation}

\subsection{Angular Eigenmodes}
We seek separable solutions of the form:
\begin{equation}\label{eq:snm-ansatz}
 \Psi(\eta, \phi_1, \phi_2) = f(\eta)\,e^{ik_1\phi_1}\,e^{ik_2\phi_2},
\end{equation}
where $k_1, k_2 \in \mathbb{Z}$ are the angular momenta associated with the two constituent complex planes. The twisted boundary condition~\eqref{eq:snm-charge-eigenvalue} is equivalent to the linear selection rule:
\begin{equation}\label{eq:snm-selection-rule}
 nk_1 + mk_2 = Q.
\end{equation}

\subsubsection{The Coprime Case ($\gcd(n,m) = 1$).}
Because $n$ and $m$ are coprime, B\'ezout's identity states that the linear Diophantine equation~\eqref{eq:snm-selection-rule} always admits integer solutions $(k_1, k_2)$ for any integer topological charge $Q$. In this scenario, every integer charge sector admits physical wavefunctions. The general solution is governed by:
\begin{equation}\label{eq:snm-diophantine-general}
 k_1 = k_1^{(0)} + m\ell, \qquad k_2 = k_2^{(0)} - n\ell, \qquad \ell \in \mathbb{Z},
\end{equation}
where $(k_1^{(0)}, k_2^{(0)})$ is any fundamental solution to $nk_1^{(0)} + mk_2^{(0)} = Q$.

\subsubsection{Remark on the Non-Coprime Case ($\gcd(n,m) > 1$).}
If $d = \gcd(n,m) > 1$, then the weighted $U(1)$ action on $S^3$ is non-effective and has generic stabilizer $\mathbb{Z}_d$. The quotient obtained from~\eqref{eq:snm-u1-action} is therefore the non-effective weighted projective orbifold $\mathbb{P}(n,m)$, not the effective two-cone-point orbifold $S^2(n,m)$ of Chapter~3. Its effective quotient has cone orders $n/d$ and $m/d$, and the primitive invariant angle is $\tilde\phi=\frac{n}{d}\phi_2-\frac{m}{d}\phi_1$ rather than $n\phi_2-m\phi_1$.

Within this non-effective weighted quotient, the selection rule~\eqref{eq:snm-selection-rule} indeed forces $Q \in d\mathbb{Z}$. Writing $n = dn'$, $m = dm'$, and $Q = dQ_{\mathrm{eff}}$, one obtains
\[
 n'k_1 + m'k_2 = Q_{\mathrm{eff}},
\]
so the actual spectral problem reduces to the coprime effective pair $(n',m')$ treated above.

By contrast, the intrinsic effective orbifold $S^2(n,m)$ with $d>1$ still has
\[
 \pi_1^{\mathrm{orb}}\bigl(S^2(n,m)\bigr)\cong \mathbb{Z}_d
\]
and hence $d$ flat sectors classified by $\operatorname{Hom}(\mathbb{Z}_d,U(1))$, exactly as established in Chapter~3. Those additional flat sectors are genuine features of the effective orbifold, but they are not captured by the present weighted $S^3$ reduction. A faithful treatment of that case requires a different Seifert total space (typically a lens space or a more general Seifert manifold, depending on the bundle data), and we do not attempt that construction here.

\subsection{Laplacian and Radial Equation}
The Laplacian on $S^3$ in continuous Hopf coordinates is:
\begin{equation}\label{eq:snm-laplacian-s3}
 \Delta_{S^3} = \frac{1}{\sin(2\eta)}\frac{\partial}{\partial\eta}
 \left(\sin(2\eta)\frac{\partial}{\partial\eta}\right)
 + \frac{1}{\cos^2\eta}\frac{\partial^2}{\partial\phi_1^2}
 + \frac{1}{\sin^2\eta}\frac{\partial^2}{\partial\phi_2^2}.
\end{equation}
Acting on the separable ansatz~\eqref{eq:snm-ansatz}, the angular derivatives evaluate to $-k_1^2$ and $-k_2^2$. This yields the radial differential equation for $f(\eta)$:
\begin{equation}\label{eq:snm-radial-eta}
 \frac{1}{\sin(2\eta)}\frac{d}{d\eta}\left(\sin(2\eta)\frac{df}{d\eta}\right)
 - \left(\frac{k_1^2}{\cos^2\eta} + \frac{k_2^2}{\sin^2\eta}\right)f
 = -\lambda\, f,
\end{equation}
where $\lambda$ is the eigenvalue of $-\Delta_{S^3}$. Since $S^3$ is compact, $\lambda$ takes a purely discrete set of values determined by the regularity conditions below.

\subsection{Free Particle: Derivation of Jacobi Solutions}\label{sec:snm-free}

To solve the radial equation analytically, we perform the substitution $x = \cos(2\eta)$, mapping the domain to $x \in [-1,1]$. Because $\eta \in [0,\pi/2]$, one has $\sin(2\eta)=\sqrt{1-x^2}$. Applying the chain rule ($\frac{dx}{d\eta} = -2\sin(2\eta)$), the derivative operator becomes $\frac{d}{d\eta} = -2\sqrt{1-x^2}\frac{d}{dx}$, transforming the radial equation to:
\begin{equation}\label{eq:snm-radial-x}
    (1-x^2)\frac{d^2f}{dx^2} - 2x\frac{df}{dx} - \frac{1}{4}\left(\frac{2k_1^2}{1+x} + \frac{2k_2^2}{1-x}\right)f = -\frac{\lambda}{4}f.
\end{equation}
The equation has regular singular points at $x = \pm 1$, corresponding to the geometric poles of the spindle. To extract the leading behavior near each singularity, we factor out the appropriate power-law asymptotics via the substitution:
\begin{equation}\label{eq:snm-factorization}
    f(x) = (1-x)^{|k_2|/2}\,(1+x)^{|k_1|/2}\,g(x).
\end{equation}
Substituting~\eqref{eq:snm-factorization} into~\eqref{eq:snm-radial-x}, the first- and second-derivative terms involving the singular factors $(1\pm x)^{-1}$ cancel against the polar terms $k_1^2/(1+x)$ and $k_2^2/(1-x)$, leaving a residual equation for $g(x)$ that takes the standard Jacobi form:
\begin{equation}\label{eq:snm-jacobi-ode}
    (1-x^2)g''(x) + \bigl[|k_1| - |k_2| - (|k_1| + |k_2| + 2)x\bigr]g'(x) + \nu(\nu + |k_1| + |k_2| + 1)g(x) = 0,
\end{equation}
with Jacobi parameters $(\alpha, \beta) = (|k_2|, |k_1|)$. The index $\nu$ is related to $\lambda$ by:
\begin{equation}\label{eq:snm-nu-lambda}
    \nu(\nu + |k_1| + |k_2| + 1) = \frac{\lambda - (|k_1| + |k_2|)(|k_1| + |k_2| + 2)}{4}.
\end{equation}
Here $\alpha,\beta \ge 0$. Although the differential expression has regular singular endpoints at $x=\pm 1$, standard Jacobi theory (equivalently, the hypergeometric analysis of endpoint-regular solutions for $\alpha,\beta>-1$) implies that the smooth $L^2$ solutions regular at both endpoints are precisely the polynomials $P_\nu^{(\alpha,\beta)}(x)$ with $\nu \in \mathbb{Z}_{\ge 0}$. Hence
\[
 g(x)=P_\nu^{(|k_2|,|k_1|)}(x), \qquad \nu \in \mathbb{Z}_{\ge 0}.
\]
Setting $\sigma = |k_1| + |k_2|$ and inverting~\eqref{eq:snm-nu-lambda}, we obtain $\lambda = (2\nu + \sigma)(2\nu + \sigma + 2) = K(K+2)$, where:
\begin{equation}\label{eq:snm-K-formula}
    K = 2\nu + |k_1| + |k_2|, \qquad \nu \in \{0, 1, 2, \ldots\}.
\end{equation}
The quantized energy levels are therefore:
\begin{equation}\label{eq:snm-energy}
 E_K = \frac{\hbar^2}{2I_{S^3}}\,K(K+2).
\end{equation}
The full radial wavefunctions are:
\begin{equation}\label{eq:snm-jacobi-solution}
 f(x) = (1-x)^{|k_2|/2}\,(1+x)^{|k_1|/2}\,P_\nu^{(|k_2|,\,|k_1|)}(x).
\end{equation}
For each pair $(k_1, k_2)$ satisfying the selection rule~\eqref{eq:snm-selection-rule}, the allowed energy levels form the tower $K = |k_1| + |k_2|,\; |k_1| + |k_2| + 2,\; |k_1| + |k_2| + 4, \ldots$, stepped by $2$ in $K$ (equivalently, by unit increments in $\nu$).

\subsubsection{Behavior at the Poles.}
Regularity at the poles imposes boundary conditions on the radial factor. Near the north pole, $1-x \approx 2\eta^2$, and near the south pole, $1+x \approx 2(\pi/2-\eta)^2$, giving:
\begin{itemize}
 \item \textbf{North pole} ($\eta = 0$, $x \to 1$, stabilizer $\mathbb{Z}_n$): The radial amplitude scales as $f \sim (1-x)^{|k_2|/2} \sim \eta^{|k_2|}$. Thus regularity forces every state with $k_2 \neq 0$ to vanish at the pole.
 \item \textbf{South pole} ($\eta = \pi/2$, $x \to -1$, stabilizer $\mathbb{Z}_m$): Symmetrically, $f \sim (1+x)^{|k_1|/2} \sim (\pi/2 - \eta)^{|k_1|}$, so regularity forces every state with $k_1 \neq 0$ to vanish at the pole.
\end{itemize}
The cross-pairing---$k_2$ governing the north-pole behavior and $k_1$ the south---is a direct consequence of the coordinate parametrization: the north pole corresponds to $z_2 = 0$ (i.e.\ $\sin\eta = 0$), so the angular factor $e^{ik_2\phi_2}$ controls the singularity there; conversely, the south pole has $z_1 = 0$.

\subsubsection{Degeneracy.}
The degeneracy $g_K^{(Q)}$---the number of independent eigenstates at energy $E_K$ within sector $Q$---equals the number of integer pairs $(k_1, k_2)$ satisfying both the selection rule~\eqref{eq:snm-selection-rule} and the constraint~\eqref{eq:snm-K-formula} with $\nu \ge 0$. Using the parametrization~\eqref{eq:snm-diophantine-general}, this is:
 \begin{equation}\label{eq:snm-degeneracy}
 \begin{aligned}
  g_K^{(Q)} = \#\bigl\{\ell \in \mathbb{Z} : & \,\,|k_1^{(0)} + m\ell| + |k_2^{(0)} - n\ell| \le K, \\
  & \,\,K - (|k_1^{(0)} + m\ell| + |k_2^{(0)} - n\ell|) \in 2\mathbb{Z}_{\ge 0}\bigr\}.
 \end{aligned}
 \end{equation}
The second condition arises because $\nu = (K - |k_1| - |k_2|)/2$ must be a non-negative integer for the Jacobi polynomial $P_\nu^{(|k_2|,|k_1|)}$ to be well-defined, which requires $K - |k_1| - |k_2|$ to be even and non-negative. Each admissible value of $\ell$ yields one independent eigensolution. (From the $SU(2)_L \times SU(2)_R$ viewpoint, the scalar harmonics of level $K$ form the $(j,j)$ representation with $j = K/2$ of the round-$S^3$ isometry group $SO(4)$, while $k_1$ and $k_2$ are convenient linear combinations of the left and right Cartan quantum numbers; the parity constraint $K \equiv |k_1| + |k_2| \pmod{2}$ is the corresponding integrality condition.)

The ground-state energy within charge sector $Q$ is $E_{K_{\min}}$, where:
\begin{equation}\label{eq:snm-Kmin}
 K_{\min}(Q) = \min_{\substack{k_1, k_2 \in \mathbb{Z} \\ nk_1 + mk_2 = Q}} \bigl(|k_1| + |k_2|\bigr),
\end{equation}
corresponding to the $\nu = 0$ ground state. For $Q \neq 0$, $K_{\min} > 0$: indeed, $|k_1| + |k_2| = 0$ would force $k_1 = k_2 = 0$ and hence $Q = 0$ by~\eqref{eq:snm-selection-rule}. This strict positivity reflects the nonzero fiber charge in the full descended operator, rather than the spectrum of a bare horizontal base Laplacian.

\subsubsection{Comparison with the Symmetric Football $S^2(r,r)$.}
It is important to distinguish the present coprime weighted-Seifert construction from the effective football orbifold of Section~\ref{sec:s2-nn}. The formulas derived above assume $\gcd(n,m)=1$, so the only literal symmetric case covered by the current construction is $n=m=1$, namely the ordinary smooth sphere. In that case, for the scalar sector $Q=0$ one has $k_1=-k_2$, hence
\[
 K = 2\nu + |k_1| + |k_2| = 2\bigl(\nu + |k_1|\bigr).
\]
Writing
\[
 \ell := \nu + |k_1| \in \mathbb{Z}_{\ge 0},
\]
every $Q=0$ state has $K=2\ell$, and conversely every even $K$ occurs. Therefore~\eqref{eq:snm-energy} becomes
\begin{equation}
 E_K \;=\; \frac{\hbar^2}{2I_{S^3}}\, (2\ell)(2\ell+2)
 \;=\; \frac{\hbar^2}{2 \left(I_{S^3}/4\right)}\, \ell(\ell+1).
\end{equation}
Under the substitution $\theta = 2\eta$, the base metric~\eqref{eq:snm-metric-base} with $n = m = 1$ reduces to $\frac{1}{4}(d\theta^2 + \sin^2\theta\,d\phi^2)$, confirming an effective radius $R_{\mathrm{base}} = 1/2$ relative to the unit $S^3$ and hence $I_{S^2} = I_{S^3} R_{\mathrm{base}}^2 = I_{S^3}/4$. Thus the entire $Q=0$ sector reproduces the standard spherical spectrum $E_\ell = \frac{\hbar^2}{2 I_{S^2}} \ell(\ell+1)$ of~\eqref{eq:s2-spectrum}. Moreover, for fixed $\ell$ the states are labeled by $k_1=-k_2=s$ with $|s|\le \ell$ and $\nu=\ell-|s|$, which recovers the usual degeneracy $2\ell+1$.

For $r>1$, however, setting $n=m=r$ in the weighted action~\eqref{eq:snm-u1-action} does \emph{not} produce the effective football $S^2(r,r)$ studied in Section~\ref{sec:s2-nn}: the action then has generic stabilizer $\mathbb{Z}_r$, so the quotient is non-effective. The football of the previous section is instead the effective quotient $S^2/\mathbb{Z}_r$, whose sectors are labeled by $q \in \{0,\dots,r-1\}$ and whose spherical-harmonic magnetic quantum number satisfies $m_{\mathrm{sph}} \equiv q \pmod r$ by~\eqref{eq:s2-selection-rule}. Consequently, the charge sectors $Q$ appearing here should not be identified with the twisted sectors $q$ of Section~\ref{sec:s2-nn}, and the degeneracy formulas of the two constructions are genuinely different.

\section{\texorpdfstring{The Dihedral Cone $\mathbb{C}/D_n$}{The Dihedral Cone C/Dn}}\label{sec:quantum-dn-cone}

This section develops quantum mechanics on the dihedral configuration orbifold $\mathbb{C}/D_n$---a cone whose boundary rays are mirror edges~\cite{davis2011lectures,EmmrichRomer1990}. Section~\ref{ex:R2Dn} established the geometric foundations: the orbifold structure, the fundamental group $\pi_1^{\mathrm{orb}} \cong D_n$, the classification of line bundles by characters of $D_n$, and the key constraint that the flat equivariant twists are discrete. The mirror edges make this a configuration-space boundary-value problem, not an immediate instance of the boundaryless symplectic-orbifold hypotheses used in Chapter~\ref{sec:geometric-quantization-orbifolds}. In Chapter~6 the corresponding phase-space quantization is formulated on the cotangent quotient \([T^*\mathbb{C}/D_n]\), where the lifted reflection action is symplectic and the boundaryless hypotheses apply before polarization. Building on the configuration-space geometry here, we now construct the quantum Hilbert space, classify self-adjoint extensions by boundary conditions, solve the spectral problem, and develop the operator algebra. Throughout, we emphasise how representation theory governs the structure of the quantum theory.

\subsection{Geometry and the Hamiltonian}

We begin by fixing notation and recalling the essential geometric data.

\subsubsection{The Dihedral Group Action.}
The dihedral group $D_n$ admits the presentation
\[
  D_n = \langle r, s \mid r^n = 1,\; s^2 = 1,\; srs^{-1} = r^{-1} \rangle.
\]
It acts on $\mathbb{C}$ via a rotation $r$ by angle $2\pi/n$ about the origin and a reflection $s$ across the $x$-axis. The quotient $\mathbb{C}/D_n$ is a wedge of opening angle
\[
  \alpha := \frac{\pi}{n}
\]
with mirror edges along the two boundary rays.

\subsubsection{Coordinate Systems.}
We shall work in two coordinate systems:
\begin{itemize}
\item \emph{Wedge coordinates} $(r, \phi)$ on the quotient, with $r \in [0,\infty)$ and $\phi \in [0, \alpha]$. The boundary ray $\phi = 0$ is fixed by the reflection $s$, while $\phi = \alpha$ is fixed by $rs$.
\item \emph{Cover coordinates} $(r, \theta_{\mathrm{c}})$ on $\mathbb{C}$, with $\theta_{\mathrm{c}} \in [0, 2\pi)$. On the fundamental wedge, we simply identify $\theta_{\mathrm{c}} = \phi$; the full cover is reconstructed by the $D_n$ action.
\end{itemize}

\subsubsection{The Free-Particle Hamiltonian.}
For a particle of mass $M$, the Cartesian momentum operators are $P_x = -i\hbar\,\partial_x$ and $P_y = -i\hbar\,\partial_y$. The free Hamiltonian takes the polar form
\[
  H = \frac{P_x^2 + P_y^2}{2M} = -\frac{\hbar^2}{2M}\left(\partial_r^2 + \frac{1}{r}\partial_r + \frac{1}{r^2}\partial_\phi^2\right).
\]
The underlying Hilbert space is $L^2([0,\infty) \times [0,\alpha],\, r\,dr\,d\phi)$. We write $k = \frac{1}{\hbar}\sqrt{2ME}$ for the wavenumber, and the angular momentum on the cover is $L_z = -i\hbar\,\partial_{\theta_{\mathrm{c}}}$.

\subsection{Boundary Conditions and Self-Adjointness}\label{sec:dihedral-boundary}

The orbifold structure constrains the admissible wavefunctions through equivariance under the dihedral group. For wavefunctions transforming under a one-dimensional representation (scalar sectors), this equivariance reduces to boundary conditions at the mirror edges.

\subsubsection{Mirror Parities and Boundary Conditions.}
Equivariance under the dihedral reflections amounts to choosing a \emph{parity} $\sigma \in \{\pm 1\}$ at each mirror edge:
\begin{itemize}
\item At $\phi = 0$ (fixed by $s$):
  \begin{itemize}
  \item $\sigma_0 = +1$ (even): Neumann condition, $\partial_\phi \Psi\big|_{\phi=0} = 0$;
  \item $\sigma_0 = -1$ (odd): Dirichlet condition, $\Psi\big|_{\phi=0} = 0$.
  \end{itemize}
\item At $\phi = \alpha$ (fixed by $rs$):
  \begin{itemize}
  \item $\sigma_\alpha = +1$: Neumann condition, $\partial_\phi \Psi\big|_{\phi=\alpha} = 0$;
  \item $\sigma_\alpha = -1$: Dirichlet condition, $\Psi\big|_{\phi=\alpha} = 0$.
  \end{itemize}
\end{itemize}
These choices correspond to the eigenvalues of the reflection operators $\rho(s)$ and $\rho(rs)$ in a one-dimensional unitary representation. The four combinations---NN, DD, ND, and DN---define the four possible scalar sectors (with ND and DN existing only when $n$ is even, as we shall see).

\subsubsection{Connection to Characters and Line Bundles.}
In a one-dimensional sector, the parities satisfy $\sigma_0 = \chi(s)$ and $\sigma_\alpha = \chi(rs) = \chi(r)\chi(s)$, from which we deduce $\chi(r) = \sigma_0 \sigma_\alpha$. This establishes the correspondence between boundary conditions and the one-dimensional characters of $D_n$ that classify orbifold line bundles on $[\mathbb{C}/D_n]$ (cf.\ Section~\ref{ex:R2Dn}).

The relation $srs^{-1} = r^{-1}$ imposes $\chi(r)^2 = 1$, so $\chi(r) = \pm 1$. When $n$ is odd, the additional constraint $\chi(r)^n = 1$ forces $\chi(r) = +1$, leaving only two characters. When $n$ is even, both values $\chi(r) = \pm 1$ are allowed, yielding four characters. The complete classification for $n$ even is:

\begin{center}
\renewcommand{\arraystretch}{1.2}
\begin{tabular}{c|c|c|l}
Parities $(\sigma_0, \sigma_\alpha)$ & $\chi(r)$ & Character & Values on $(r, s)$ \\
\hline
$(+,+)$ (NN) & $+1$ & $\chi_{\mathbf{1}}$ (trivial) & $r \mapsto +1,\; s \mapsto +1$ \\
$(-,-)$ (DD) & $+1$ & $\chi_{\mathrm{sgn}}$ (sgn) & $r \mapsto +1,\; s \mapsto -1$ \\
$(+,-)$ (ND) & $-1$ & $\chi_{\mathrm{det}}$ & $r \mapsto -1,\; s \mapsto +1$ \\
$(-,+)$ (DN) & $-1$ & $\chi_{\mathrm{sgn}\cdot\mathrm{det}}$ & $r \mapsto -1,\; s \mapsto -1$
\end{tabular}
\end{center}

\noindent For $n$ odd, only the NN ($\chi_{\mathbf{1}}$) and DD ($\chi_{\mathrm{sgn}}$) sectors exist. Here $\chi_{\mathrm{sgn}}$ is the actual determinant character of the defining $O(2)$-representation; the notation $\chi_{\mathrm{det}}$ is retained only to match the conventions used later in the text.

\subsection{Scalar Sectors: Angular Quantization}\label{sec:dihedral-scalar}

Having established the boundary conditions, we now solve for the angular eigenfunctions in each scalar sector.

\subsubsection{Separation of Variables.}
Seek separated solutions $\Psi(r,\phi) = R(r)\,\Phi(\phi)$, where the angular factor satisfies
\[
  -\Phi''(\phi) = \nu^2 \Phi(\phi)
\]
with general solution $\Phi(\phi) = A\cos(\nu\phi) + B\sin(\nu\phi)$. The boundary conditions at each mirror determine the allowed values of the separation constant $\nu$:
\begin{itemize}
\item At $\phi = 0$:
  \begin{itemize}
  \item $\sigma_0 = +1$ (Neumann): $\Phi'(0) = 0 \Rightarrow B = 0$;
  \item $\sigma_0 = -1$ (Dirichlet): $\Phi(0) = 0 \Rightarrow A = 0$.
  \end{itemize}
\item At $\phi = \alpha$:
  \begin{itemize}
  \item $\sigma_\alpha = +1$ (Neumann): $\Phi'(\alpha) = 0$;
  \item $\sigma_\alpha = -1$ (Dirichlet): $\Phi(\alpha) = 0$.
  \end{itemize}
\end{itemize}

\subsubsection{Allowed Angular Orders.}
Solving these conditions for each parity combination yields the allowed angular orders $\nu_j$. The results are summarised in the following table:

\begin{center}
\renewcommand{\arraystretch}{1.2}
\begin{tabular}{l|l|l}
\hline
Parities & Allowed orders $\nu$ & Angular factor $\Phi(\phi)$ \\
\hline
NN $(+,+)$ & $\nu_j = nj$, $j \ge 0$ & $\cos(\nu_j\phi)$ \\
DD $(-,-)$ & $\nu_j = nj$, $j \ge 1$ & $\sin(\nu_j\phi)$ \\
ND $(+,-)$; $n$ even & $\nu_j = n(j + \tfrac{1}{2})$, $j \ge 0$ & $\cos(\nu_j\phi)$ \\
DN $(-,+)$; $n$ even & $\nu_j = n(j + \tfrac{1}{2})$, $j \ge 0$ & $\sin(\nu_j\phi)$ \\
\hline
\end{tabular}
\end{center}

For the NN and DD sectors, the allowed orders form the lattice $\nu \in n\mathbb{Z}_{\ge 0}$. The NN sector includes $\nu = 0$ (the constant angular mode), while the DD sector starts at $\nu = n$ since $\sin(0) = 0$ does not produce a nontrivial wavefunction.

For the ND sector ($\sigma_0 = +1$, $\sigma_\alpha = -1$), the angular factor is $\Phi(\phi) = \cos(\nu\phi)$ with Dirichlet condition $\cos(\nu\alpha) = 0$, which requires $\nu\alpha = (j + \tfrac{1}{2})\pi$, i.e., $\nu = n(j + \tfrac{1}{2})$. The DN sector is analogous with $\Phi(\phi) = \sin(\nu\phi)$ and Neumann condition $\nu\cos(\nu\alpha) = 0$ at the second edge, yielding identical $\nu$ values. These half-integer ladders exist only when $n$ is even, because the character constraint $\chi(r) = -1$ requires $(-1)^n = 1$.

\begin{remark}
The lowest mode in the DD family has $j = 1$ (giving $\nu = n$), since $j = 0$ would correspond to $\sin(0) = 0$, which vanishes identically.
\end{remark}

\subsection{Radial Equation and Normalised Basis}

With the angular quantization established, we turn to the radial problem.

\subsubsection{Bessel's Equation.}
Substituting the separated ansatz into the Schr\"odinger equation yields Bessel's equation for the radial factor:
\[
  r^2 R''(r) + r R'(r) + (k^2 r^2 - \nu^2) R(r) = 0.
\]
The general solution involves Bessel functions of the first and second kind. Regularity at the apex (the Friedrichs extension) selects the Bessel function of the first kind:
\[
  R(r) = J_\nu(kr), \qquad E = \frac{\hbar^2 k^2}{2M}, \quad k \ge 0.
\]
The parameter $\nu \ge 0$ plays a dual role: it is both the separation constant from the angular equation and the order of the Bessel function.

\begin{remark}
For $\nu \ge 1$, the radial endpoint $r=0$ is limit-point, so the regular/Friedrichs behavior is forced and no self-adjoint extension parameter remains. For $\nu = 0$ (the constant angular mode in the NN sector), both $J_0$ and $Y_0$ are square-integrable near $r = 0$, so alternative self-adjoint extensions exist; we select the Friedrichs extension throughout~\cite{KayStuder1991}. All nonzero scalar and doublet orders satisfy $\nu \ge 1$, so $\nu = 0$ is the only case requiring a choice of extension.
\end{remark}

\subsubsection{Normalized Scalar Basis.}
On the wedge $[0,\alpha]$, a delta-normalised basis for each scalar sector takes the form
\[
  \Psi_{j,k}(r,\phi) = \sqrt{k}\,C_j\,J_{\nu_j}(kr) \times
  \begin{cases}
    \cos(\nu_j\phi), & \sigma_0 = +1,\\
    \sin(\nu_j\phi), & \sigma_0 = -1,
  \end{cases}
\]
where the angular normalization constants are
\[
  C_0 = \frac{1}{\sqrt{\alpha}} \quad \text{(NN sector, } j=0\text{)}, \qquad
  C_j = \sqrt{\frac{2}{\alpha}} \quad \text{(all other modes)}.
\]
The factor $C_0 = 1/\sqrt{\alpha}$ accounts for the constant ($j=0$) mode in the NN sector, while $C_j = \sqrt{2/\alpha}$ provides the standard normalization for $\cos^2$ or $\sin^2$ integrals over $[0,\alpha]$.

Using the Bessel orthogonality relation
\[
  \int_0^\infty r\,dr\, J_\nu(kr) J_\nu(k'r) = \frac{\delta(k - k')}{k}
\]
together with the standard orthogonality of trigonometric functions on $[0,\alpha]$, one verifies
\[
  \int_0^\infty r\,dr \int_0^\alpha d\phi\; \overline{\Psi_{j,k}}\,\Psi_{j',k'} = \delta_{jj'}\,\delta(k - k').
\]
Completeness of these modes in each scalar sector follows from standard Sturm--Liouville theory for the Laplacian on wedges with Neumann or Dirichlet boundaries.

\subsection{Representation Theory of $D_n$}

Over $\mathbb{C}$, the irreducible representations of $D_n$ are at most two-dimensional~\cite{AdemMilgram2004}:
\begin{itemize}
\item For $n$ odd: $2$ one-dimensional irreps and $(n-1)/2$ two-dimensional irreps.
\item For $n$ even: $4$ one-dimensional irreps and $(n/2 - 1)$ two-dimensional irreps.
\end{itemize}
Consequently, there are no higher-dimensional ($>2$) elementary sectors on the cone $\mathbb{C}/D_n$; any larger multiplet would arise only from additional internal symmetries beyond the geometric $D_n$ action.

The one-dimensional irreps classify orbifold line bundles via $\mathrm{Pic}_{\mathrm{orb}}([\mathbb{C}/D_n]) \cong \mathrm{Hom}(D_n, U(1))$ (cf.\ Section~\ref{ex:R2Dn}), while the two-dimensional irreps govern the vector-valued wavefunctions in doublet sectors.

\subsection{Angular Momentum on the Universal Cover}\label{sec:dihedral-cover}

Before constructing the explicitly vector-valued doublet basis, it is illuminating to observe how the allowed angular momenta for all sectors---both scalar and doublet---can be derived uniformly using the universal cover $\mathbb{C}$. This approach makes the connection between $D_n$ representations, angular momentum, and the Bessel order $\nu$ particularly transparent, avoiding redundant derivations.

\subsubsection{Equivariance and Rotation Twists.}
On the $2\pi$-cover, a generic wavefunction decomposes into partial waves $\Psi \propto e^{im\theta_{\mathrm{c}}}$, which have angular momentum $L_z = \hbar m$ ($m \in \mathbb{Z}$). The $D_n$-equivariance constraints under rotation and reflection are encoded as
\begin{align*}
  r: \quad &\Psi(r, \theta_{\mathrm{c}} + \tfrac{2\pi}{n}) = e^{2\pi i\eta}\,\Psi(r, \theta_{\mathrm{c}}), \\
  s: \quad &\Psi(r, -\theta_{\mathrm{c}}) = \sigma_0\,\Psi(r, \theta_{\mathrm{c}}),
\end{align*}
where $\eta \in [0,1)$ is the \emph{rotation twist} and $\sigma_0 = \pm 1$ dictates the parity across the $\phi = 0$ mirror axis. Conjugation by the reflection symmetry $s$ inverts the rotation generator, $srs^{-1} = r^{-1}$. Spatially, $s$ maps the angular coordinate $\theta_{\mathrm{c}} \mapsto -\theta_{\mathrm{c}}$, enforcing that a state with twist $\eta$ must be inextricably paired with a state of twist $-\eta \pmod 1$.

For one-dimensional (scalar) sectors, the representation maps into itself, requiring $\eta \equiv -\eta \pmod 1$. This constrains the scalar twists to
\[
  \eta \in \{0, \tfrac{1}{2}\} \quad \text{($\eta = \tfrac{1}{2}$ is permitted only if $n$ is even)}.
\]
Any generic twist $\eta \notin \{0, \tfrac{1}{2}\}$ necessarily embeds into a two-dimensional doublet irrep wherein the reflection $s$ swaps the chiral twist components.

\subsubsection{Angular Momentum Quantization.}
A physical rotation by the sector angle $2\pi/n$ phase-shifts a partial wave $e^{im\theta_{\mathrm{c}}}$ by $e^{2\pi i m/n}$. Equivariance requires this spatial phase to match the algebraically assigned symmetry twist:
\[
  e^{2\pi i m/n} = e^{2\pi i\eta}, \qquad \text{i.e.,} \quad m = n\eta + n\ell, \quad \ell \in \mathbb{Z}.
\]
The physical radial structure is dictated by the radial Bessel index through $J_{|m|}(kr)$. Since the separation constant satisfies $\nu^2 = m^2$, the non-negative physical Bessel order is
\[
  \nu = |m| = n|\eta + \ell|.
\]
Mapping over these twist parameters recovers the spectra across all unitary sectors:
\begin{itemize}
\item \textbf{NN sector} ($\eta = 0$): $\nu \in \{0, n, 2n, \ldots\}$.
\item \textbf{DD sector} ($\eta = 0$): $\nu \in \{n, 2n, 3n, \ldots\}$, since the $m=0$ sine mode vanishes identically.
\item \textbf{ND/DN sectors} ($\eta = \tfrac{1}{2}$, $n$ even): $\nu \in \{\tfrac{n}{2}, \tfrac{3n}{2}, \tfrac{5n}{2}, \ldots\}$.
\item \textbf{Doublet sectors} ($\eta = q/n$, labeled uniquely by the integer parameter $q \in \{1, \ldots, \lfloor(n-1)/2\rfloor\}$):
      \[
        \nu_j^+ = q + nj, \qquad \nu_j^- = (n-q) + nj \qquad (j \in \mathbb{Z}_{\ge 0}).
      \]
\end{itemize}

\noindent In the scalar sectors, these spectra coincide with the angular orders derived independently from the wedge boundary conditions in Section~\ref{sec:dihedral-scalar}, providing a consistency check on both approaches.

\subsection{Two-Dimensional (Vector) Sectors}\label{sec:dihedral-doublets}

With the underlying angular frequency spectrum unambiguously recovered for generic fractional representations via the universal cover, we now formulate the explicit equivariant vector wavefunctions defining the $D_n$ multiplets.

\subsubsection{Two-Dimensional Irreducible Representations.}
Each discrete doublet sector corresponds symmetrically to the rational twist $q = 1, 2, \ldots, \lfloor(n-1)/2\rfloor$. Selecting an orthonormal internal basis $\{u_q, v_q\}$, the canonical two-dimensional unitary representation $\rho_q$ stipulates that the elementary spatial rotation generator evaluates as
\[
  \rho_q(r) = \begin{pmatrix} \cos(2\pi q/n) & -\sin(2\pi q/n) \\ \sin(2\pi q/n) & \cos(2\pi q/n) \end{pmatrix},
\]
while the transverse axis reflection symmetry enforces
\[
  \rho_q(s) = \begin{pmatrix} 1 & 0 \\ 0 & -1 \end{pmatrix},
\]
thereby asserting $\rho_q(s)u_q = +u_q$ and $\rho_q(s)v_q = -v_q$. Evaluating the eigenvalues of this rotation generator recovers the conjugate phases $e^{\pm 2\pi i q/n}$, consistent with the paired twist parameter $\eta = \pm q/n$ determined on the universal cover.

\subsubsection{Doublet Basis and Equivariance.}
We construct an explicit basis for the doublet sectors by pairing trigonometric angular eigenfunctions with the internal vectors $u_q$ and $v_q$. The ansatz takes a different form on each of the two frequency ladders, because the angular shift $\nu \cdot 2\pi/n$ differs in sign between them. Throughout, we write $\gamma_q = 2\pi q/n$.

The equivariance conditions to be verified are
\[
  \Psi(r, -\phi) = \rho_q(s)\Psi(r, \phi) \quad \text{and} \quad \Psi(r, \phi + \tfrac{2\pi}{n}) = \rho_q(r)\Psi(r, \phi),
\]
where these relations are understood on the universal cover $\mathbb{C}$, with $\phi$ extending beyond the wedge; the physical wavefunction is recovered by restriction to $\phi \in [0,\alpha]$.

\medskip\noindent\textbf{1. The Primary Ladder $\nu_j^+ \equiv q \pmod{n}$:} \\
For angular orders on the primary ladder, the basis states are:
\[
  \Psi_{k,\nu}^{(q)}(r,\phi) = \sqrt{\frac{k}{\alpha}}\,J_\nu(kr)\,\bigl(\cos(\nu\phi)\,u_q + \sin(\nu\phi)\,v_q\bigr).
\]
\emph{Reflection:} Since $\cos$ is even and $\sin$ is odd, $\Psi_{k,\nu}^{(q)}(r,-\phi) = \cos(\nu\phi)u_q - \sin(\nu\phi)v_q = \rho_q(s)\Psi_{k,\nu}^{(q)}(r,\phi)$. \\
\emph{Rotation:} Because $\nu \equiv q \pmod{n}$, the angular shift satisfies $\nu \cdot 2\pi/n \equiv \gamma_q \pmod{2\pi}$, giving
\begin{align*}
  \Psi_{k,\nu}^{(q)}(r,\phi + \tfrac{2\pi}{n}) &= \sqrt{\frac{k}{\alpha}}\,J_\nu(kr)\,\bigl[\cos(\nu\phi + \gamma_q)\,u_q + \sin(\nu\phi + \gamma_q)\,v_q\bigr] \\
  &= \sqrt{\frac{k}{\alpha}}\,J_\nu(kr)\,\bigl[(\cos\nu\phi\cos\gamma_q - \sin\nu\phi\sin\gamma_q)u_q \\
  &\qquad\qquad\qquad\qquad + (\sin\nu\phi\cos\gamma_q + \cos\nu\phi\sin\gamma_q)v_q\bigr] \\
  &= \sqrt{\frac{k}{\alpha}}\,J_\nu(kr)\,\bigl[\cos(\nu\phi)(\cos\gamma_q\,u_q + \sin\gamma_q\,v_q) \\
  &\qquad\qquad\qquad\qquad + \sin(\nu\phi)(-\sin\gamma_q\,u_q + \cos\gamma_q\,v_q)\bigr] \\
  &= \sqrt{\frac{k}{\alpha}}\,J_\nu(kr)\,\bigl[\cos(\nu\phi)\rho_q(r)u_q + \sin(\nu\phi)\rho_q(r)v_q\bigr] \\
  &= \rho_q(r)\Psi_{k,\nu}^{(q)}(r,\phi),
\end{align*}
confirming rotational equivariance.

\medskip\noindent\textbf{2. The Flipped Ladder $\nu_j^- \equiv -q \pmod{n}$:} \\
For angular orders on the flipped ladder, $\nu \equiv -q \pmod{n}$ implies $\nu \cdot 2\pi/n \equiv -\gamma_q \pmod{2\pi}$. To compensate for this sign change while preserving the required reflection parity, the $v_q$ component carries a relative minus sign:
\[
  \Psi_{k,\nu}^{(q)}(r,\phi) = \sqrt{\frac{k}{\alpha}}\,J_\nu(kr)\,\bigl(\cos(\nu\phi)\,u_q - \sin(\nu\phi)\,v_q\bigr).
\]
\emph{Reflection:} $\Psi_{k,\nu}^{(q)}(r,-\phi) = \cos(\nu\phi)u_q + \sin(\nu\phi)v_q = \rho_q(s)\Psi_{k,\nu}^{(q)}(r,\phi)$. \\
\emph{Rotation:} Since $\nu \cdot 2\pi/n \equiv -\gamma_q \pmod{2\pi}$,
\begin{align*}
  \Psi_{k,\nu}^{(q)}(r,\phi + \tfrac{2\pi}{n}) &= \sqrt{\frac{k}{\alpha}}\,J_\nu(kr)\,\bigl[\cos(\nu\phi - \gamma_q)\,u_q - \sin(\nu\phi - \gamma_q)\,v_q\bigr] \\
  &= \sqrt{\frac{k}{\alpha}}\,J_\nu(kr)\,\bigl[\cos(\nu\phi)(\cos\gamma_q\,u_q + \sin\gamma_q\,v_q) \\
  &\qquad\qquad\qquad\qquad + \sin(\nu\phi)(\sin\gamma_q\,u_q - \cos\gamma_q\,v_q)\bigr] \\
  &= \rho_q(r)\Psi_{k,\nu}^{(q)}(r,\phi),
\end{align*}
confirming equivariance for the flipped ladder as well.

\subsubsection{State Normalization and Completeness.}
The inner product on the doublet Hilbert space includes a trace over the two internal components: $\langle\Psi|\Psi'\rangle = \sum_{i=1}^{2}\int r\,dr\,d\phi\;\Psi_i^*\Psi_i'$. With this convention, the angular normalization $\sqrt{1/\alpha}$ (rather than $\sqrt{2/\alpha}$ as in the scalar case) reflects the identity $\cos^2(\nu\phi) + \sin^2(\nu\phi) = 1$, so the angular integral over $[0,\alpha]$ equals $\alpha$.

Alternatively, in a chiral basis where $r$ acts diagonally as $\mathrm{diag}(e^{2\pi i q/n}, e^{-2\pi i q/n})$ and $s$ acts by the Pauli matrix $\sigma_x$ (swapping components), the angular factor takes the form
\[
  \sqrt{\frac{k}{2\alpha}}\,J_\nu(kr) \begin{pmatrix} e^{\pm i\nu\phi} \\ e^{\mp i\nu\phi} \end{pmatrix},
\]
with the upper signs for $\nu \equiv q \pmod{n}$ and lower signs for $\nu \equiv -q \pmod{n}$. The prefactor $\sqrt{k/(2\alpha)}$ accounts for $|e^{\pm i\nu\phi}|^2 + |e^{\mp i\nu\phi}|^2 = 2$, ensuring correct normalization.

Completeness of the doublet modes in each $q$-sector follows from Fourier--Bessel completeness on the cover together with the decomposition of $L^2(\mathbb{C})$ into $D_n$-irreps.

\chapter{Geometric Quantization on Orbifolds}
\label{sec:geometric-quantization-orbifolds}

Geometric quantization provides a rigorous bridge between the symplectic geometry of classical mechanics and the Hilbert space structures of quantum mechanics. When the classical phase space is an orbifold rather than a smooth manifold, this bridge reveals a landscape rich with nuance. This chapter extends the standard quantization pipeline---prequantization, polarization, and metaplectic correction---to the orbifold setting.

Throughout this chapter, unless a later example explicitly says otherwise, the phase space is assumed to be a boundaryless smooth effective symplectic orbifold. This excludes orbifolds with mirror boundary or corner reflectors from the general self-adjointness and half-form statements below. The dihedral examples in Chapters~4 and~6 should be read with this distinction in mind: the mirror strata belong to the configuration quotient \(\mathbb{C}/D_n\), while the symplectic phase space used for geometric quantization is the cotangent quotient \([T^*\mathbb{C}/D_n]\). The cotangent-lifted reflections are symplectic and orientation preserving on phase space, so they fall under the boundaryless hypotheses; mirror boundary appears only after choosing the vertical polarization and passing to the configuration leaf space, where it is handled by explicit sector boundary conditions.

\section{Symplectic Orbifolds}\label{sec:symplectic-orbifolds}

The starting point for any quantization program is a classical phase space. In standard mechanics, this is a symplectic manifold. In our setting, the phase space is a \textbf{symplectic orbifold}~\cite{LermanTolman1997,Kordyukov2011}.

\subsection*{Standing hypotheses and scope}
Unless a result explicitly states otherwise, every orbifold in this chapter is a paracompact smooth effective orbifold in the sense of Definition~\ref{def:orbifold}, with finite local groups acting effectively on smooth uniformizing charts. A symplectic orbifold is assumed to be boundaryless and to have no reflector, mirror, or corner strata: the local groups preserve the symplectic form and hence preserve orientation. Thus the formal symplectic results below are not stated for orbifolds with boundary, mirror boundary, corner reflectors, or other non-symplectic local quotient models without additional boundary conditions or doubling arguments.

This restriction concerns the \emph{ambient symplectic phase space}, not necessarily an underlying configuration quotient whose cotangent orbifold is being quantized. If a finite group acts on a configuration manifold \(Q\), possibly with reflection fixed sets, its cotangent lift to \(T^*Q\) preserves the canonical Liouville form and the canonical symplectic form. A base reflection that is orientation reversing on \(Q\) acts on \(T^*Q\) by the block map \((x,p)\mapsto (gx,g^{-T}p)\), which is symplectic and orientation preserving in phase space. Its fixed locus has even codimension in \(T^*Q\), rather than producing a codimension-one mirror boundary of the symplectic orbifold. Such cotangent-lifted reflection examples therefore fall within the phase-space hypotheses of this chapter. What may still acquire mirror boundary is the leaf space of a later real polarization, and analytic statements about Schr\"odinger operators on that leaf space then require the corresponding boundary-sector conditions to be specified separately.

The notation \(H^*_{\mathrm{st}}(X;\mathbb{Z})\) refers to the stack/classifying-space cohomology of the effective orbifold stack associated to the atlas of \(X\). Comments about explicitly non-effective quotient stacks are comparisons only, not part of the standing hypotheses. Compactness is not assumed unless it is stated in the relevant theorem or paragraph. Likewise, a good/global quotient presentation is not assumed unless it is stated explicitly.

\begin{remark}[Global quotient model for examples]\label{assum:global-quotient}
Many examples and some concrete computations in this chapter are formulated for \textbf{good orbifolds} presented as finite global quotients $X = [M/G]$. In those explicitly marked cases, the orbifold geometry is captured by the $G$-equivariant geometry of $M$, and the definitions are understood as $G$-invariant objects on $M$. This is an examples-and-computations convention, not a standing hypothesis for the intrinsic theorems.
\end{remark}

Recall from Section~\ref{sec:orbifold_definition} that an orbifold chart is a triple $(V, G, \pi)$, where $V$ is a smooth uniformizing neighborhood, $G$ is a finite group acting effectively on $V$, and $\pi\colon V \to V/G$ is the projection onto the local quotient. An orbifold $k$-form $\alpha \in \Omega^k_{\mathrm{orb}}(X)$ is a compatible family of $G$-invariant smooth $k$-forms on the charts.

\begin{definition}[Symplectic Orbifold]\label{def:symplectic-orbifold}
In this chapter, a \emph{symplectic orbifold} is a boundaryless smooth effective orbifold $X$ equipped with a closed, non-degenerate orbifold 2-form $\omega \in \Omega^2_{\mathrm{orb}}(X)$. Consequently each connected component has even dimension $2n$; in the phase-space examples below we take $n\geq 1$. Explicitly, on any local uniformizing chart $(V, G, \pi)$, the lifted form $\omega_V$ is a $G$-invariant symplectic form on $V$.
\end{definition}

\subsection*{Local Normal Form: The Equivariant Darboux Theorem}

In the manifold theory, the Darboux theorem guarantees that every symplectic manifold is locally isomorphic to $(\mathbb{R}^{2n}, \omega_{\mathrm{std}})$ with the standard symplectic form $\omega_{\mathrm{std}} = \sum_i dp_i \wedge dq^i$. The orbifold analogue requires compatibility with the local group action~\cite{LermanTolman1997,Kordyukov2011}.

\begin{proposition}[Equivariant Darboux Theorem]\label{prop:equivariant-darboux}
Let $p$ be a point of a boundaryless smooth effective symplectic orbifold $(X,\omega)$ with finite isotropy group $G_p$. No compactness or global quotient presentation is assumed. There exists a uniformizing chart $(V,G_p,\pi)$ centered at $p$ and local coordinates $(q^1, \ldots, q^n, p_1, \ldots, p_n)$ on $V$ such that:
\begin{enumerate}
    \item The symplectic form takes the standard Darboux form $\omega_V = \sum_{i=1}^n dp_i \wedge dq^i$.
    \item The action of $G_p$ on $V$ is linear with respect to these coordinates: each $g \in G_p$ acts as a linear symplectomorphism $g \colon (q, p) \mapsto S_g(q, p)$ for some $S_g \in \mathrm{Sp}(2n, \mathbb{R})$.
\end{enumerate}
\end{proposition}

\begin{proof}
The proof relies on the equivariant Moser trick, which generalizes the standard manifold case by explicitly incorporating the finite group action. 

First, by the Bochner linearization theorem (since $G_p$ is a finite group acting smoothly on $V$ with a fixed point $p$), there exists a local coordinate chart centered at $p$ in which the action of $G_p$ is strictly linear. Thus, we may assume $V$ is an open ball in $\mathbb{R}^{2n}$ on which $G_p$ acts linearly, and $p$ corresponds to the origin. Evaluating the symplectic form $\omega_V$ at the origin yields a constant symplectic form $\omega_{\mathrm{lin}}$. Since $G_p$ preserves $\omega_V$, its linear action at the origin must also preserve $\omega_{\mathrm{lin}}$, implying that $G_p \subset \mathrm{Sp}(2n, \mathbb{R})$ relative to $\omega_{\mathrm{lin}}$.

To smoothly deform $\omega_V$ into $\omega_{\mathrm{lin}}$, define a family of closed 2-forms $\omega_t = \omega_{\mathrm{lin}} + t(\omega_V - \omega_{\mathrm{lin}})$ for $t \in [0,1]$. Since $\omega_t(0) = \omega_{\mathrm{lin}}$ is non-degenerate for every $t$, and non-degeneracy is an open condition, compactness of $[0,1]$ allows us to shrink to a sufficiently small neighborhood on which every $\omega_t$ is non-degenerate. Averaging an auxiliary Euclidean metric over the finite group makes the linear $G_p$-action orthogonal, so we may choose this neighborhood to be a small $G_p$-invariant ball $U$. In particular, $U$ is star-shaped and each $\omega_t$ is symplectic on $U$ for all $t \in [0,1]$.

Set $\beta=\omega_V-\omega_{\mathrm{lin}}$. Since $\beta$ is closed and $\beta(0)=0$, the relative Poincar\'e lemma on the star-shaped ball $U$ gives a 1-form $\alpha$ with $d\alpha=\beta$ and $\alpha(0)=0$. Concretely, one may use the radial homotopy formula
\[
\alpha_x=\int_0^1 t\,\iota_{R_x}\beta_{tx}\,dt,
\qquad R_x=x^i\partial_i.
\]
Because the $G_p$-action is linear in the Bochner coordinates and preserves both $\omega_V$ and $\omega_{\mathrm{lin}}$, this radial primitive is $G_p$-invariant. In any case, averaging the chosen primitive gives an invariant primitive:
\[
\bar{\alpha} = \frac{1}{|G_p|} \sum_{g \in G_p} g^*\alpha.
\]
Because both $\omega_V$ and $\omega_{\mathrm{lin}}$ are strictly $G_p$-invariant, $\bar{\alpha}$ is a $G_p$-invariant 1-form satisfying $d\bar{\alpha} = \omega_V - \omega_{\mathrm{lin}}$ and vanishing at $p$. 

Next, define a time-dependent vector field $X_t$ by Cartan's equation:
\[
\iota_{X_t} \omega_t = -\bar{\alpha}.
\]
Crucially, because both $\omega_t$ and $\bar{\alpha}$ are strictly $G_p$-invariant, pulling Cartan's equation back by any $g\in G_p$ shows that the transformed vector field $g^*X_t:={(g^{-1})}_*X_t\circ g$ satisfies the same equation. By the non-degeneracy of $\omega_t$, this forces $g^*X_t=X_t$, equivalently $dg_x(X_t(x))=X_t(gx)$. Thus, the vector field $X_t$ is $G_p$-equivariant. After possibly shrinking $U$, the time-dependent vector field $X_t$ admits a flow $\phi_t$ defined on $U$ for all $t \in [0,1]$. This flow commutes with the group action ($\phi_t \circ g = g \circ \phi_t$), fixes the origin (since $\bar{\alpha}$ vanishes there), and satisfies $\frac{d}{dt} \phi_t^* \omega_t = \phi_t^*(\mathcal{L}_{X_t} \omega_t + \frac{d}{dt}\omega_t)$. Since $\mathcal{L}_{X_t} = d \circ \iota_{X_t} + \iota_{X_t} \circ d$ and $d\omega_t = 0$, this evaluates to $\phi_t^*(-d\bar{\alpha} + \omega_V - \omega_{\mathrm{lin}}) = 0$. Hence, $\phi_1^* \omega_V = \omega_{\mathrm{lin}}$.

The Bochner coordinates already make the $G_p$-action linear, and because $\phi_1 \circ g = g \circ \phi_1$, the coordinates obtained by composing with $\phi_1^{-1}$ still have linear $G_p$-action and carry the constant form $\omega_{\mathrm{lin}}$. Finally, by linear symplectic algebra there exists a linear isomorphism $A \colon \mathbb{R}^{2n} \to \mathbb{R}^{2n}$ such that $A^*\omega_{\mathrm{std}} = \omega_{\mathrm{lin}}$, where $\omega_{\mathrm{std}} = \sum_{i=1}^n dp_i \wedge dq^i$. If $z$ denotes the intermediate Moser coordinates and $w=Az$, then ${\bigl(A^{-1}\bigr)}^*\omega_{\mathrm{lin}}=\omega_{\mathrm{std}}$, so the $w$-coordinates are Darboux coordinates; the group action is $A S_g A^{-1}$, hence still linear and symplectic. This establishes the desired local model.
\end{proof}

\subsection*{Symplectic Potentials on Orbifold Charts}

In the manifold theory, a \emph{symplectic potential} (or Liouville 1-form) is a 1-form $\theta$ satisfying $d\theta = \omega$. While a global symplectic potential exists only when $[\omega] = 0$ in de Rham cohomology, local potentials always exist by the Poincar\'e lemma. On orbifolds, local symplectic potentials play an equally essential role.

\begin{proposition}[Existence of invariant local symplectic potentials]\label{prop:invariant-symplectic-potential}
Let $(X,\omega)$ be a boundaryless smooth effective symplectic orbifold. On each contractible uniformizing chart $(V, G, \pi)$, there exists a $G$-invariant local symplectic potential $\bar{\theta}_V$ such that $d\bar{\theta}_V = \omega_V$. No global quotient presentation is required.
\end{proposition}

\begin{proof}
Let $(V, G, \pi)$ be a local chart. After shrinking $V$ if needed, we may assume $V$ is contractible. By the standard Poincar\'e lemma for manifolds, the closed symplectic 2-form $\omega_V$ (which satisfies $d\omega_V = 0$) admits a local primitive 1-form $\theta_V$ such that $d\theta_V = \omega_V$. However, $\theta_V$ is locally not uniquely determined and may fail to be invariant under the local group action.

Since $G$ acts on $V$ by symplectomorphisms, $g^*\omega_V = \omega_V$ for all $g \in G$. This implies that for any $g \in G$, $d(g^*\theta_V) = g^*(d\theta_V) = g^*\omega_V = \omega_V$. Hence, the pullback $g^*\theta_V$ is also a valid local symplectic potential. Because the exterior derivative is a linear operator that commutes with the pullback by diffeomorphisms, we can directly average these local potentials over the finite group $G$:
\[
\bar{\theta}_V = \frac{1}{|G|} \sum_{g \in G} g^* \theta_V.
\]
Evaluating the exterior derivative of $\bar{\theta}_V$ yields:
\[
d\bar{\theta}_V = \frac{1}{|G|} \sum_{g \in G} d(g^*\theta_V) = \frac{1}{|G|} \sum_{g \in G} \omega_V = \omega_V.
\]
Furthermore, for any $h \in G$, we have:
\[
h^* \bar{\theta}_V = \frac{1}{|G|} \sum_{g \in G} {(g \circ h)}^* \theta_V = \frac{1}{|G|} \sum_{g' \in G} {(g')}^* \theta_V = \bar{\theta}_V,
\]
where the second equality uses the substitution $g' = g \circ h$, which is a bijection on $G$. This demonstrates that $\bar{\theta}_V$ is strictly $G$-invariant. Thus, $\bar{\theta}_V$ defines a local orbifold 1-form on the chart $(V,G,\pi)$; in general it need not descend to an ordinary smooth 1-form on the coarse quotient $V/G$ near singular points.
\end{proof}

Under the global quotient presentation $X = [M/G]$ with $G$ finite, if $M$ admits a global symplectic potential $\theta_M$ with $d\theta_M = \omega_M$, the averaged form $\bar{\theta}_M = \frac{1}{|G|}\sum_{g \in G} g^*\theta_M$ is a $G$-invariant symplectic potential on $M$ and therefore defines an orbifold 1-form on $X$. When a global potential does not exist (i.e., when the orbifold de Rham class $[\omega] \neq 0$ in \(H^2_{\mathrm{dR,orb}}(X)\)), one systematically works with local invariant potentials $\bar{\theta}_V$ on chart patches, glued by gauge transformations.

\subsection*{Orbifold Observables and Hamiltonian Mechanics}

To align with the manifold theory (as discussed for quantization in Chapter~1), we formalize the classical phase space mechanics---observables, Hamiltonian vector fields, the Poisson algebra, and the Liouville volume---for a boundaryless smooth effective symplectic orbifold $X$. These concepts are defined locally on uniformizing charts and glue to form global orbifold objects.

\begin{itemize}
    \item \textbf{Classical Observables:} A classical observable is a smooth function on the orbifold, $f \in C^\infty_{\mathrm{orb}}(X)$. Locally, on any uniformizing chart $(V, G, \pi)$, this corresponds to a $G$-invariant smooth function $f_V \in {C^\infty(V)}^G$.
    \item \textbf{Hamiltonian Vector Fields:} For any observable $f$, its Hamiltonian vector field $X_f$ is the unique orbifold vector field defined by the relation $\iota_{X_f}\omega = -df$. Locally on a chart $(V, G, \pi)$, the vector field $X_{f,V}$ satisfies $\iota_{X_{f,V}}\omega_V = -df_V$. Because both $f_V$ and $\omega_V$ are $G$-invariant, $X_{f,V}$ is a $G$-invariant vector field on $V$, making it a well-defined local piece of an orbifold vector field.
    \item \textbf{Poisson Bracket:} The Poisson bracket of two observables $f,h \in C^\infty_{\mathrm{orb}}(X)$ is $\{f, h\} = \omega(X_h, X_f)$. Since $X_f$ and $X_h$ are local $G$-invariant vector fields, $\{f, h\}$ is a $G$-invariant function on each chart, establishing $C^\infty_{\mathrm{orb}}(X)$ as a well-defined Poisson algebra.
    \item \textbf{Liouville Volume:} The symplectic form naturally equips the orbifold with the Liouville top-form $\mu_L = \frac{\omega^n}{n!}$. Locally, $\mu_L$ is represented by the $G$-invariant volume form $\frac{\omega_V^n}{n!}$ on uniformizing charts. Orbifold integration is computed with a partition of unity subordinate to the atlas and, on a chart $(V_i,G_i,\pi_i)$, the contribution is normalized by $1/|G_i|$:
    \[
    \int_X \mu_L
    =
    \sum_i \frac{1}{|G_i|}\int_{V_i}(\rho_i\circ\pi_i)\,\frac{\omega_i^n}{n!}.
    \]
    This chart-group normalization gives the measure needed to define the prequantum Hilbert space; for a global quotient $[M/G]$ it reduces to the familiar factor $|G|^{-1}\int_M$.
\end{itemize}

\subsubsection{Properties of the Orbifold Poisson Bracket.}
We formally state and prove that the orbifold observable algebra is a well-defined Poisson algebra, highlighting the difference with the underlying coarse space.

\begin{proposition}[Orbifold Poisson Algebra]\label{prop:orbifold-poisson-algebra}
For any boundaryless smooth effective symplectic orbifold $(X,\omega)$, without any compactness or global quotient assumption, the space of classical observables $C^\infty_{\mathrm{orb}}(X)$ equipped with the pairing $\{f, h\} = \omega(X_h, X_f)$ forms a Poisson algebra.
\end{proposition}

\begin{proof}
Let $\{(V_i,G_i,\pi_i)\}$ be an orbifold atlas for $X$. An orbifold symplectic form is given by a compatible family of $G_i$-invariant symplectic forms $\omega_i$ on the $V_i$, and an orbifold function $f \in C^\infty_{\mathrm{orb}}(X)$ is given by a compatible family of $G_i$-invariant functions $f_i \in {C^\infty(V_i)}^{G_i}$. On each chart define the Hamiltonian vector field $X_{f_i}$ by $\iota_{X_{f_i}}\omega_i = -df_i$. For any $g \in G_i$, pullback of this equation by $g$ shows that the transformed vector field $g^*X_{f_i}:={(g^{-1})}_*X_{f_i}\circ g$ satisfies the same defining equation, because $g^*\omega_i=\omega_i$ and $g^*f_i=f_i$. By non-degeneracy of $\omega_i$, the solution is unique, so $g^*X_{f_i}=X_{f_i}$; equivalently $dg_x(X_{f_i}(x))=X_{f_i}(gx)$. Compatibility of the orbifold data implies these local vector fields agree under chart embeddings; hence they define a global orbifold vector field $X_f$.

For $f,h \in C^\infty_{\mathrm{orb}}(X)$, define $\{f,h\}$ chartwise by ${\{f,h\}}_i = \omega_i(X_{h_i},X_{f_i})$. Each ${\{f,h\}}_i$ is $G_i$-invariant and compatible under chart embeddings, so it defines an orbifold function $\{f,h\} \in C^\infty_{\mathrm{orb}}(X)$. On each chart $V_i$ this is the usual Poisson bracket associated to the symplectic manifold $(V_i,\omega_i)$, so antisymmetry, the Jacobi identity, and the Leibniz rule hold chartwise. Compatibility then implies these identities hold globally on $C^\infty_{\mathrm{orb}}(X)$, making it a Poisson algebra.
\end{proof}

Although $C^\infty_{\mathrm{orb}}(X)$ is smooth when defined intrinsically via orbifold charts, its realization as functions on the coarse quotient space $|X|$ can look ``singular'': smoothness across the singular strata forces compatibility conditions that are invisible in ordinary manifold coordinates. This is a primary motivation for working with the orbifold structure directly rather than the underlying topological space.

To illustrate, consider a phase space singularity modeled by $\mathbb{R}^2 / \mathbb{Z}_2$ acting on Darboux coordinates via $(q, p) \mapsto (-q, -p)$. Topologically, the coarse space $|X|$ forms a cone. A continuous function on this cone is not automatically an orbifold-smooth observable: for instance, the radial distance $r=\sqrt{q^2+p^2}$ is well-defined on the quotient, but its pullback to $\mathbb{R}^2$ is not smooth at the origin. A true orbifold observable $f \in C^\infty_{\mathrm{orb}}(X)$ is a smooth $\mathbb{Z}_2$-invariant function upstairs, so its Taylor expansion at the origin contains only terms of even total degree. Linear functions such as $a q+b p$ do not descend to the quotient at all unless $a=b=0$, while functions like $r$ descend continuously but fail the orbifold smoothness test. Working directly with invariant functions on orbifold charts enforces exactly these differentiability conditions and keeps the Poisson algebra well defined.

\section{Prequantization}%
\label{sec:prequantization-orbifolds}

With the geometric foundations established, we now turn to the first stage of the quantization program. Prequantization associates to a symplectic orbifold $(X, \omega)$ a Hilbert space carrying a representation of the Poisson algebra of classical observables~\cite{Kostant1970,Souriau1970,MaratheMartucci1985}. As in the manifold theory of Chapter~1, prequantization provides a map from the Poisson algebra of orbifold observables to an algebra of operators on a prequantum Hilbert space. The construction, adapted to the orbifold setting, formalizes the structural requirements first outlined by Dirac.

\subsection{Prequantization Requirements}%
\label{sec:prequantization-requirements}

As in the manifold theory, prequantization encodes the Dirac correspondence principle into a precise operator map. The map must satisfy the following three structural requirements, which are identical in form to the manifold case but now applied to orbifold observables (formulated locally via $G$-invariant functions on uniformizing charts):

\begin{enumerate}
    \item \textbf{Linearity}: For any two orbifold observables $f, g \in C^\infty_{\mathrm{orb}}(X)$ and constants $a, b \in \mathbb{R}$, the quantization map $Q$ must satisfy:
    \[
    Q(a f + b g) = a Q(f) + b Q(g).
    \]
    This ensures that the quantum theory respects the linear structure of classical observables on the orbifold.

    \item \textbf{Constant Function}: The constant function $f=1$ must map to the identity operator $\hat{I}$:
    \[
    Q(1) = \hat{I}.
    \]

    \item \textbf{Commutation Relations}: The map must preserve the algebraic structure by relating the commutator to the Poisson bracket (with our convention $\{f,g\}=\omega(X_g,X_f)$):
    \[
    [Q(f), Q(g)] = i \hbar Q(\{f, g\}),
    \]
    where $[\cdot, \cdot]$ is the commutator, $\hbar$ is the reduced Planck constant, and $\{f, g\}$ is the orbifold Poisson bracket.

    This requirement preserves the fundamental classical structure: the Poisson bracket on $C^\infty_{\mathrm{orb}}(X)$ encodes the orbifold dynamics, and the commutator mirrors it.
\end{enumerate}

\noindent
To implement these requirements concretely, we realize quantum states as sections of an orbifold line bundle whose curvature reproduces the symplectic form. The orbifold structure ensures that the quantum theory is independent of the choice of charts.

\subsubsection{The Fundamental Lemma of Prequantization}

The next lemma packages the curvature condition and the commutator identity into one precise statement, paralleling the fundamental lemma of prequantization in Chapter~1. In the orbifold setting, the operator is defined on orbifold sections and the computation is local on uniformizing charts. Under the standing effective-orbifold convention of this chapter, the principal stratum is dense in each chart.

\begin{lemma}[Orbifold Prequantization]%
\label{lem:fundamental-commutation}
Let $(X,\omega)$ be a boundaryless smooth effective symplectic orbifold, not necessarily compact and not necessarily a global quotient. Let $L$ be a Hermitian orbifold line bundle over $X$ equipped with a unitary orbifold connection $\nabla$ whose curvature is denoted by
\[
F_\nabla = \operatorname{Curv}(\nabla).
\]
Define, for any orbifold-smooth function $f\in C^\infty_{\mathrm{orb}}(X)$, the prequantum operator
\[
Q(f) = -i\hbar\, \nabla_{X_f} + f,
\]
where $f$ acts by pointwise multiplication and $X_f$ is the Hamiltonian vector field associated with $f$ (defined via $\iota_{X_f}\omega = -df$). Then the commutator of the operators satisfies
\[
[Q(f), Q(g)] = i\hbar\, Q(\{f, g\})
\]
for all $f, g \in C^\infty_{\mathrm{orb}}(X)$ if and only if the curvature of the connection satisfies
\[
F_\nabla = -\frac{i}{\hbar}\,\omega.
\]
\end{lemma}

\begin{proof}
We evaluate the commutator locally on uniformizing charts. Let $(V, G, \pi)$ be a chart for $X$, where the line bundle, connection, and symplectic form are represented by a $G$-equivariant bundle $\tilde{L} \to V$, a $G$-equivariant connection $\tilde{\nabla}$, and a $G$-invariant symplectic form $\tilde{\omega}$. 

For any orbifold-smooth functions $f, g \in C^\infty_{\mathrm{orb}}(X)$, their pullbacks $\tilde{f}, \tilde{g}$ to $V$ are $G$-invariant, and their associated Hamiltonian vector fields $\tilde{X}_f, \tilde{X}_g$ are consequently $G$-invariant vector fields on $V$. Likewise, any orbifold section $s$ lifts to a $G$-invariant section $\tilde{s}$ of $\tilde{L}$.

Direct algebraic computation on the smooth manifold $V$, exactly as in the manifold case, yields the operator identity:
\[
[\tilde{Q}(\tilde{f}), \tilde{Q}(\tilde{g})] - i\hbar\, \tilde{Q}(\{\tilde{f}, \tilde{g}\}) = -\hbar^2 \left( F_{\tilde{\nabla}}(\tilde{X}_f, \tilde{X}_g) + \frac{i}{\hbar}\tilde{\omega}(\tilde{X}_f, \tilde{X}_g)\operatorname{Id} \right)
\]
for any smooth functions $\tilde{f}, \tilde{g}$ on $V$. Therefore, the commutator identity holds for all $G$-invariant functions if and only if
\[
F_{\tilde{\nabla}}(\tilde{X}_f, \tilde{X}_g) = -\frac{i}{\hbar}\,\tilde{\omega}(\tilde{X}_f, \tilde{X}_g)\operatorname{Id}
\]
for all $G$-invariant functions. However, at singular points (where the isotropy group is non-trivial), the Hamiltonian vector fields of $G$-invariant functions do not span the entire tangent space. This is because, at a point $x$, the differentials of $G$-invariant functions lie in the $G_x$-fixed subspace of the cotangent space, where $G_x$ is the stabilizer of $x$, so their Hamiltonian vector fields (obtained via $\tilde{\omega}$-duality) are confined to the corresponding $G_x$-fixed subspace of the tangent space, which is proper whenever the stabilizer acts non-trivially on $T_x V$. Thus, evaluating the relation exclusively on invariant functions does not immediately determine the full curvature tensor everywhere.

To overcome this, we restrict our attention to the principal stratum of regular points where the stabilizer is trivial. At such a point $x$, one may choose a sufficiently small neighborhood $U$ such that the translates $gU$ are pairwise disjoint for $g\neq e$. Any smooth local function on $U$ can be multiplied by a bump function that is identically one near $x$, then extended by zero to $\bigcup_{g\in G} gU$ and averaged over $G$ to produce a $G$-invariant function with the same germ at $x$. Hence the differentials of $G$-invariant functions span $T_x^*V$, and their Hamiltonian vector fields span $T_xV$. This algebraically forces the local curvature to satisfy
\[
F_{\tilde{\nabla}} = -\frac{i}{\hbar}\,\tilde{\omega}\operatorname{Id}
\]
on the principal stratum of $V$. For an effective orbifold, the principal stratum is a dense open subset of $V$. Since both the connection curvature and symplectic form are smooth tensors, this equality uniquely extends to the singular strata by continuity. Because the orbifold structures are defined precisely by patching these invariant local data, the chartwise equivalence descends to the global identity on the orbifold $X$.
\end{proof}

\subsubsection{Orbifold Compatibility.}
The prequantum operator $Q(f)$ is well-posed on orbifold sections. Locally on any uniformizing chart $(V, G, \pi)$, the Hamiltonian vector field $X_f$ is $G$-invariant and the connection $\nabla$ is $G$-equivariant, so $\nabla_{X_f}$ maps invariant sections to invariant sections. Therefore, $Q(f)$ preserves the space of orbifold sections. For a global quotient $X=[M/G]$, this reduces to the requirement that $Q(f)$ preserves the invariant subspace ${L^2(M, L_M)}^G$.

\begin{proposition}[Formal Self-Adjointness]%
\label{prop:self-adjointness}
Let $(X,\omega)$ be a boundaryless smooth effective symplectic orbifold and let $L$ carry a unitary orbifold connection. No global quotient presentation is assumed. For any real-valued orbifold observable $f \in C^\infty_{\mathrm{orb}}(X, \mathbb{R})$, the prequantum operator $Q(f)$ is formally self-adjoint on the domain of smooth, compactly-supported orbifold sections of~$L$:
\[
\langle Q(f) s_1, s_2 \rangle = \langle s_1, Q(f) s_2 \rangle \quad \text{for all } s_1, s_2 \in \Gamma_c^\infty(X, L).
\]
If \(X\) is compact, the compact-support condition may be replaced by smooth orbifold sections. Orbifolds with boundary, mirror boundary, or corner reflectors would require separate boundary conditions and are outside this statement.
\end{proposition}

\begin{proof}
By our convention, the Hermitian inner product on the fibers of $L$ is conjugate-linear in the first argument and linear in the second. For any two smooth, compactly-supported orbifold sections $s_1, s_2$, we have:
\[
\langle Q(f) s_1, s_2 \rangle = \langle -i\hbar\nabla_{X_f} s_1 + f s_1, s_2 \rangle = i\hbar \langle \nabla_{X_f} s_1, s_2 \rangle + \langle f s_1, s_2 \rangle.
\]
Because the metric is Hermitian and $f$ is real-valued, $\langle f s_1, s_2 \rangle = \langle s_1, f s_2 \rangle$. For the derivative term, metric compatibility (unitarity) yields:
\[
X_f \langle s_1, s_2 \rangle = \langle \nabla_{X_f} s_1, s_2 \rangle + \langle s_1, \nabla_{X_f} s_2 \rangle.
\]
Integrating this over the orbifold $X$ gives:
\[
\int_X X_f \langle s_1, s_2 \rangle \,\frac{\omega^n}{n!} = \int_X\Bigl(\langle \nabla_{X_f} s_1, s_2 \rangle + \langle s_1, \nabla_{X_f} s_2 \rangle\Bigr)\frac{\omega^n}{n!}.
\]
Now set $\mu:=\omega^n/(n!)$. Since $X_f$ is Hamiltonian, its flow preserves $\omega$ and hence $\mu$ (Liouville's theorem), meaning $\mathcal{L}_{X_f}\mu=0$. Consequently, the Lie derivative of the top form $\langle s_1, s_2 \rangle \mu$ is exactly $\mathcal{L}_{X_f}\bigl(\langle s_1, s_2 \rangle\,\mu\bigr) = \bigl(X_f \langle s_1, s_2 \rangle\bigr)\mu$. By Cartan's formula, $\mathcal{L}_{X_f}\bigl(\langle s_1, s_2 \rangle\,\mu\bigr) = d\bigl(\iota_{X_f}(\langle s_1, s_2 \rangle\,\mu)\bigr) + \iota_{X_f}\bigl(d(\langle s_1, s_2 \rangle\,\mu)\bigr)$, and the second term vanishes because $\langle s_1, s_2 \rangle\,\mu$ is a top form on the $2n$-dimensional orbifold (hence closed).

Because $s_1, s_2$ have compact support (or $X$ is compact without boundary), the orbifold Stokes' theorem~\cite{Satake1957} guarantees that the integral of this exact form over $X$ vanishes. Crucially, in the boundaryless symplectic-orbifold setting used here, singular strata do not act as boundaries and introduce no boundary terms. Indeed, each local isotropy representation is symplectic, so the fixed subspace of any nontrivial isotropy element has even codimension, hence codimension at least two. Thus singular strata do not create codimension-one boundary faces, and Stokes' theorem applies chartwise with the usual orbifold weights. Thus:
\[
0=\int_X d\bigl(\iota_{X_f}(\langle s_1, s_2 \rangle\,\mu)\bigr) = \int_X \mathcal{L}_{X_f}\bigl(\langle s_1, s_2 \rangle\,\mu\bigr)=\int_X X_f \langle s_1, s_2 \rangle\,\mu.
\]
Therefore,
\[
\int_X \langle \nabla_{X_f} s_1, s_2 \rangle\,\mu \;=\; -\int_X \langle s_1, \nabla_{X_f} s_2 \rangle\,\mu,
\]
and hence
\begin{align*}
\langle -i\hbar\nabla_{X_f} s_1, s_2 \rangle
&= i\hbar \int_X \langle \nabla_{X_f} s_1, s_2 \rangle\,\mu \\
&= -i\hbar \int_X \langle s_1, \nabla_{X_f} s_2 \rangle\,\mu \\
&= \langle s_1, -i\hbar\nabla_{X_f} s_2 \rangle,
\end{align*}
where the first equality uses conjugate-linearity in the first argument. Thus the differential operator $-i\hbar\nabla_{X_f}$ is formally self-adjoint, and hence so is $Q(f)$. Explicitly:
\[
\langle Q(f) s_1, s_2 \rangle = \langle s_1, -i\hbar\nabla_{X_f} s_2 \rangle + \langle s_1, f s_2 \rangle = \langle s_1, Q(f) s_2 \rangle. \qedhere
\]
\end{proof}

\subsubsection{The Prequantum Line Bundle and Hilbert Space}%
\label{sec:prequantum-hilbert-space}%
\label{sec:prequantum-operators}

To realize this map, quantum states are identified with sections of the prequantum orbifold line bundle $L$ over the orbifold phase space $(X, \omega)$, equipped with the Hermitian metric and prequantum connection $\nabla$.

The objective is to construct a pair $(L, \nabla)$ such that the curvature two-form $F_\nabla = \operatorname{Curv}(\nabla)$ is proportional to the symplectic form $\omega$. This geometric condition is necessary to satisfy the commutation relations, as established by the fundamental lemma. The natural prequantum Hilbert space is the $L^2$-completion of the space of smooth orbifold sections of $L$, with inner product defined using the Liouville volume form $\omega^n/(n!)$ and orbifold integration: 
\[
\mathcal{H}_{\mathrm{pre}}(X) = L^2\big(X, L, \mu\big), \qquad \langle s_1, s_2 \rangle \;:=\; \int_X \langle s_1(x), s_2(x) \rangle_L \, \frac{\omega^n}{n!}.
\]
While the integral can be evaluated directly on the regular part of the coarse space (since the singular strata have measure zero), orbifold integration is rigorously defined chartwise with the usual $1/|G|$ normalization on uniformizing charts and globally patched together using an orbifold partition of unity. In the specific case where the orbifold is a global quotient $X=[M/G]$ with $G$ a finite group, this Hilbert space is isometrically isomorphic to the space of $G$-invariant sections on $M$ with the renormalized inner product:
\[
\mathcal{H}_{\mathrm{pre}}([M/G]) \;\cong\; {L^2(M, L_M)}^G, \qquad \langle s_1, s_2 \rangle_X = \frac{1}{|G|} \langle s_1, s_2 \rangle_M.
\]

\subparagraph{Self-Adjoint Extensions on Orbifolds.}
A core theorem from the manifold theory of geometric quantization relates complete Hamiltonian flows to self-adjoint operators. In the orbifold setting, completeness of the Hamiltonian flow still produces a canonical self-adjoint realization of the prequantum operator.

\begin{theorem}[Canonical Self-Adjoint Extension from Complete Flows]%
\label{thm:canonical-self-adjoint-extension}
Let $(X, \omega)$ be a boundaryless smooth effective symplectic orbifold, with no compactness or global quotient assumption, and let $(L, \nabla)$ be a prequantum line bundle over $X$. Let
\[
Q_c(f)=-i\hbar\nabla_{X_f}+f
\]
denote the prequantum differential operator on the dense domain \(\Gamma_c^\infty(X,L)\subset\mathcal{H}_{\mathrm{pre}}(X)\) (or \(\Gamma^\infty(X,L)\) when \(X\) is compact). If \(f \in C^\infty_{\mathrm{orb}}(X,\mathbb{R})\) is real-valued and its Hamiltonian vector field \(X_f\) is complete, then the Kostant--Souriau lift of the Hamiltonian flow induces a strongly continuous unitary group \(U_t\) on \(\mathcal{H}_{\mathrm{pre}}(X)\). Let \(A_f\) be its skew-adjoint Stone generator, with domain
\[
\mathcal{D}(A_f)
=
\left\{s\in \mathcal{H}_{\mathrm{pre}}(X)\ \middle|\ 
\lim_{t\to0}\frac{U_t s-s}{t}\ \text{exists in }\mathcal{H}_{\mathrm{pre}}(X)
\right\}.
\]
Then
\[
Q_{\mathrm{fl}}(f):=i\hbar A_f,\qquad
\mathcal{D}(Q_{\mathrm{fl}}(f))=\mathcal{D}(A_f),
\]
is a self-adjoint operator and \(Q_c(f)\subset Q_{\mathrm{fl}}(f)\). Thus the complete flow gives a distinguished self-adjoint realization of the prequantum operator. Essential self-adjointness of \(Q_c(f)\) is not asserted here; it follows only under an additional core hypothesis, equivalently when \(\Gamma_c^\infty(X,L)\) is a core for \(A_f\).
\end{theorem}

\begin{proof}
Because $X_f$ is complete, it integrates to a global 1-parameter group of orbifold symplectomorphisms $\phi_t: X \to X$. Let $P \subset L$ denote the unit circle bundle, equipped with its connection $1$-form $\alpha$. The prequantization condition $F_\nabla = -\frac{i}{\hbar}\omega$ implies that the standard Kostant--Souriau lift of $X_f$ to $P$ preserves both $\alpha$ and the Hermitian metric. This lifted vector field combines the horizontal lift of $X_f$ with the vertical phase rotation determined by $f$. Since $X_f$ is complete and the vertical part lies in the complete circle action, the Kostant--Souriau lift is complete as well. On each uniformizing chart $(V,G,\pi)$, this construction is $G$-equivariant, so the resulting flow descends to a global 1-parameter group of orbifold bundle automorphisms of $P$.

Passing to the associated line bundle gives a 1-parameter group of unitary operators $U_t$ on $\mathcal{H}_{\mathrm{pre}}(X)$: the lifted bundle flow preserves the Hermitian metric, while $\phi_t$ preserves the Liouville measure because it is symplectic. For \(s\in\Gamma_c^\infty(X,L)\), smooth dependence of the lifted flow in local uniformizing charts and compactness of the support give \(\|U_t s-s\|_{L^2}\to0\) as \(t\to0\). Since \(\Gamma_c^\infty(X,L)\) is dense and the \(U_t\) are unitary, this continuity extends to every \(L^2\)-section; hence \(U_t\) is strongly continuous.

By Stone's theorem, its infinitesimal generator \(A_f\) with domain \(\mathcal{D}(A_f)\) is skew-adjoint. A standard local computation on uniformizing charts shows that for every \(s\in \Gamma_c^\infty(X,L)\),
\[
A_f s=\left.\frac{d}{dt}\right|_{t=0}U_t s = -\frac{i}{\hbar}\,Q_c(f)s.
\]
Thus \(Q_c(f)\subset i\hbar A_f\). Since \(A_f\) is skew-adjoint, \(i\hbar A_f\) is self-adjoint on \(\mathcal{D}(A_f)\). This proves the asserted canonical self-adjoint extension. The final statement is a matter of domains: \(Q_c(f)\) is essentially self-adjoint precisely when its closure equals \(i\hbar A_f\), which is equivalent to \(\Gamma_c^\infty(X,L)\) being a core for \(A_f\).
\end{proof}

\subparagraph{Isotypic Decomposition in the Global Quotient Case.}\label{sec:isotypic-decomposition}
When the orbifold phase space admits a global quotient presentation $X = [M/G]$, the prequantum Hilbert space possesses additional structure~\cite{SilvaGuillemin1999,Kordyukov2011}. The full space of $L^2$-sections on $M$ decomposes under the $G$-action into isotypic components:
\[
L^2(M, L_M) \;=\; \bigoplus_{\rho \in \widehat{G}} \; {L^2(M, L_M)}_\rho,
\]
where ${L^2(M, L_M)}_\rho$ denotes the $\rho$-isotypic summand. For the orbifold line bundle represented by this chosen $G$-equivariant bundle $L_M$, the orbifold Hilbert space is the invariant, or trivial-isotypic, component:
\[
\mathcal{H}_{\mathrm{pre}}(X;L) \;=\; {L^2(M, L_M)}_{\mathbf{1}} \;=\; {L^2(M, L_M)}^G.
\]
Since every orbifold observable $f\in C^\infty_{\mathrm{orb}}(X)$ yields a $G$-equivariant operator $Q(f)$, each isotypic summand is preserved by $Q(f)$. Nontrivial flat or isotropy-twisted sectors are accounted for by changing the equivariant structure, hence by changing the orbifold line bundle being quantized, rather than by taking nontrivial isotypic summands inside a fixed scalar orbifold bundle. For an intrinsic orbifold without a chosen global quotient presentation, however, there is no canonical global isotypic decomposition; intrinsically, one only has local equivariance on uniformizing charts.

\subsection{Orbifold Line Bundles and the Integrality Condition}%
\label{sec:orbifold-integrality}

\subsubsection{Classification of Orbifold Line Bundles}
The curvature condition forces a topological quantization: admissible symplectic forms must satisfy an integrality constraint. The construction of the prequantum line bundle requires understanding the classification of orbifold line bundles, which directly parallels the classification of complex line bundles over manifolds discussed in Chapter~1. In the notation fixed in Section~\ref{sec:cohomology-notation}, this classification uses the stack/classifying-space group \(H^2_{\mathrm{st}}(X;\mathbb{Z})\), while curvature forms live in the de Rham group \(H^2_{\mathrm{dR,orb}}(X)\).

On a smooth manifold $M$, the group $H^2(M, \mathbb{Z})$ classifies complex line bundles via the first Chern class. On a paracompact smooth effective orbifold $X$, the corresponding statement is the stack/classifying-space classification established in Chapter~2, Theorem~\ref{thm:pic_h2}: distinct isomorphism classes of smooth orbifold line bundles are in bijective correspondence with elements of \(H^2_{\mathrm{st}}(X;\mathbb{Z})\). This group is the integral cohomology of the classifying space of a proper \'etale groupoid presenting the effective orbifold stack (equivalently, sheaf cohomology on that stack), not the ordinary integral cohomology of the coarse space. In the standing effective-orbifold convention this keeps the local isotropy data; in explicitly non-effective quotient stacks, which are not part of the theorem statements in this chapter, it would also keep the ineffective stabilizer data. The proof uses smooth partitions of unity: the sheaf of smooth complex-valued orbifold functions is fine, its higher cohomology vanishes, and the long exact sequence associated with the exponential sheaf sequence yields \(\mathrm{Pic}_{\mathrm{orb}}(X) \cong H^2_{\mathrm{st}}(X;\mathbb{Z})\), with \(\mathrm{Pic}_{\mathrm{orb}}\) denoting the smooth/topological Picard group rather than a holomorphic Picard group~\cite{MoerdijkPronk1997,AdemLeidaRuan,BehrendXu2003,BehrendXu2011}. When \(X = [M/G]\) is a finite global quotient, orbifold line bundles correspond to \(G\)-equivariant line bundles on \(M\) and are classified by \(H^2_G(M;\mathbb{Z})\).

The stack integral cohomology group can contain torsion, and torsion Chern classes can be represented by flat unitary orbifold line bundles. In addition, even non-torsion stack Chern classes can map to de Rham classes whose evaluation on coarse fundamental cycles is rational, because the orbifold integer lattice includes the local isotropy weights. Thus the discrete data relevant to quantization have two related manifestations: flat holonomy around genuine orbifold loops, and fractional local contributions to the real Chern class when one evaluates it on cycles meeting singular strata.

\subparagraph{Orbifold Connections and Curvature}\label{sec:orbifold-connections}
\begin{definition}[Equivariant Connection]
\label{def:equivariant-connection}
For a smooth effective orbifold \(X\), an orbifold line bundle $L \to X$ with connection $\nabla$ (often called an \textbf{equivariant connection} or \textbf{orbifold connection}) is characterized locally on each uniformizing chart $(V, G, \pi)$ by a $G$-equivariant line bundle $L_V\to V$ with a $G$-equivariant connection. In a local non-vanishing frame $s$ one writes
\[
\nabla s=-iA\otimes s,
\]
so that a section $u\,s$ satisfies $\nabla(u\,s)=(du-iAu)\otimes s$.
\end{definition}

\begin{definition}[Unitary Connection]
\label{def:unitary-connection}
For a \textbf{unitary connection} on an orbifold line bundle over a smooth effective orbifold, the local frame $s$ is chosen to be unitary.
\end{definition}

\begin{proposition}[Unitary Real Potential]
\label{prop:unitary-real-potential}
For a unitary connection on an orbifold line bundle over a smooth effective orbifold, the local frame $s$ is chosen to be unitary and the \textbf{real gauge potential} $A$ is a real 1-form. If a group element acts on this frame by
\[
g^*s=e^{i\theta_g}s
\]
for a local phase function $\theta_g$, equivariance of $\nabla$ forces
\[
g^*A=A-d\theta_g,
\]
exactly as in a gauge transformation with parameter $\theta_g$. In the canonical equivariant frame of the local trivialization used in Section~\ref{sec:connections_curvature}, the isotropy character is constant, so $d\theta_g=0$ and $A$ is strictly $G$-invariant.
\end{proposition}

Just as in the manifold case, the local curvature computation holds identically on the smooth chart $V$:

\begin{proposition}[Curvature Equivalence]
\label{prop:curvature-equivalence}
For a unitary orbifold connection over a smooth effective orbifold, the real closed 2-form $R = dA$ relates to the $\operatorname{End}(L)$-valued curvature $F_\nabla$ via the identity $F_\nabla = \nabla^2 = -i dA = -i R$, which we refer to as the \textbf{curvature-equivalence}.
\end{proposition}

\begin{proposition}[Equivariant Gauge Transformation]
\label{prop:equivariant-gauge-transform}
For a unitary orbifold connection over a smooth effective orbifold, taking the exterior derivative eliminates the exact gauge term, yielding
\[
g^*R=d(g^*A)=d(A-d\theta_g)=dA=R.
\]
Thus, while the connection 1-form $A$ is only locally defined and transforms non-trivially both under frame changes ($A' = A-d\chi$ for $s'=e^{i\chi}s$) and under the $G$-action ($g^*A=A-d\theta_g$), the real curvature 2-form $R$ is strictly $G$-invariant, global, and gauge-invariant across the orbifold.
\end{proposition}

The topological obstruction to the existence of a prequantum bundle with a given curvature $R$ is encoded in the orbifold holonomy. For a standard closed loop $\gamma$ in $X$, the holonomy is simply the manifold phase factor:
\[
\mathrm{hol}(\gamma) = \exp\left( i \oint_\gamma A \right).
\]
On an orbifold, however, one must also evaluate the holonomy along \emph{orbifold loops}---paths $\tilde{\gamma}$ on a uniformizing chart satisfying $\tilde{\gamma}(1) = g \cdot \tilde{\gamma}(0)$ for some isotropy element $g$. The orbifold holonomy around such a loop combines the parallel transport phase $\exp\left( i \int_{\tilde{\gamma}} A \right)$ with the inverse endpoint identification, equivalently the fiber map $\Phi_{g^{-1}}$. For the bundle to be well-defined globally, these fractional holonomies must be compatible with the transition data, leading to the quantization of flux.

\subparagraph{The Integrality Condition}
On a smooth manifold, the integrality of a cohomology class translates directly to the condition that its integral over any smooth closed 2-cycle is an integer. On an orbifold, the corresponding period statement must include the local isotropy data. Consider a closed oriented sub-orbifold surface $\Sigma \subset X$ containing isolated cone points $\{x_j\}$ with local isotropy groups $G_j$. 

Applying the smooth Stokes theorem only on the coarse space misses the equivariant topology. Instead, integrality on an orbifold is rigorously formulated via orbifold Chern--Weil theory. The connection $\nabla$ defines the real curvature $2$-form $R$. Around each singular point $x_j$, the local isotropy group $G_j$ acts on the 1-dimensional fiber of $L$ via a character (a 1-dimensional unitary representation). If this local action is parametrized by an isotropy weight $a_j \in \{0, 1, \dots, |G_j|-1\}$, the orbifold holonomy records the corresponding fractional phase needed to close compatibly with the discrete group action. 

When the integral of the real Chern class $[R/2\pi]$ is evaluated over the orbifold cycle $\Sigma$, one integrates locally on uniformizing charts with the usual $1/|G|$ normalizations, as in Satake's integration theory for \(V\)-manifolds~\cite{Satake1957}. The fractional correction itself is line-bundle data: orbifold line bundles over an orbifold Riemann surface are described by Seifert invariants \((b;\beta_1,\ldots,\beta_k)\), and their rational degree is \(b+\sum_j \beta_j/m_j\)~\cite{Furuta1992SeifertFH,MrowkaOzsvathYu1997,Sakai2017OrbifoldVortices}. In the notation used here, the local invariant \(\beta_j\) is the isotropy weight \(a_j\), and \(m_j=|G_j|\), up to the sign convention determined by whether the fiber character or its inverse is used in the endpoint identification. With the orientation and holonomy convention of Section~\ref{sec:connections_curvature}, excising a small disk around \(x_j\) and computing the endpoint identification in the corresponding orbifold loop gives the contribution \(a_j/|G_j|\). Thus, for a compact oriented orbifold surface with isolated cone points, the standard manifold holonomy argument generalizes to the surface formula
\[
\frac{1}{2\pi} \int_\Sigma R = k + \sum_j \frac{a_j}{|G_j|}, \qquad k \in \mathbb{Z}.
\]
Because of these singular contributions, the integral over a surface evaluated in the coarse space yields a rational number. However, this fractional value is not a failure of integrality; rather, it is precisely the geometric requirement for the de Rham class \([R/2\pi]\in H^2_{\mathrm{dR,orb}}(X)\) to lift to a well-defined integer class in \(H^2_{\mathrm{st}}(X;\mathbb{Z})\). For higher-dimensional orbifolds, the rigorous formulation is the cohomological one: \([R/2\pi]\) must lie in the image of the orbifold integral lattice map \(\iota_{\mathrm{dR}}\colon H^2_{\mathrm{st}}(X;\mathbb{Z})\to H^2_{\mathrm{dR,orb}}(X)\).

\begin{definition}[Integral Orbifold Cohomology Class]%
\label{def:integral-orbifold-class}
Let \(X\) be a smooth effective orbifold, with the stack/classifying-space cohomology convention of Section~\ref{sec:cohomology-notation}. The change of coefficients induced by the standard inclusion $\mathbb{Z} \hookrightarrow \mathbb{R}$ yields a natural homomorphism in cohomology:
\[
\iota_{\mathrm{dR}}\colon
H^2_{\mathrm{st}}(X;\mathbb{Z})
\longrightarrow
H^2_{\mathrm{st}}(X;\mathbb{R})
\cong
H^2_{\mathrm{dR,orb}}(X).
\]
By the orbifold de Rham theorem reviewed in Chapter~2, \(H^2_{\mathrm{dR,orb}}(X)\) is also canonically identified with \(H^2_{\mathrm{coarse}}(X;\mathbb{R})=H^2(|X|;\mathbb{R})\). It may be computed from invariant forms on charts (or from \(G\)-invariant forms in a finite global quotient presentation), while the integral lattice still comes from the stack/classifying-space group \(H^2_{\mathrm{st}}(X;\mathbb{Z})\). A de Rham cohomology class \([\Omega]\in H^2_{\mathrm{dR,orb}}(X)\) is called \textbf{orbifold-integral} if it lies in the image of \(\iota_{\mathrm{dR}}\), meaning there exists an underlying integer class in \(H^2_{\mathrm{st}}(X;\mathbb{Z})\) that maps to it. 
\end{definition}

Equivalently, integrality means that the de Rham class of \(\Omega\) comes from the orbifold integer lattice in degree two. In concrete terms, whenever such a class is evaluated on a compact oriented orbifold surface via orbifold integration, one obtains the rational values dictated by the isotropy data above. This condition forms the rigid topological backbone of orbifold geometric quantization. Requiring that the scaled class \([\omega / 2\pi\hbar]\in H^2_{\mathrm{dR,orb}}(X)\) be orbifold-integral directly enforces that the symplectic flux over every orbifold \(2\)-cycle is quantized in the orbifold sense. On an orbifold, however, this integrality must be interpreted using \(H^2_{\mathrm{st}}(X;\mathbb{Z})\), not \(H^2_{\mathrm{coarse}}(X;\mathbb{Z})\): when an orbifold-integral class is evaluated on cycles meeting singular strata, rational values may appear on the coarse space, with denominators constrained by the local isotropy orders.

\begin{theorem}[Orbifold Prequantum Line Bundle]%
\label{thm:prequantum-line-bundle}
Let $(X,\omega)$ be a paracompact boundaryless smooth effective symplectic orbifold; no compactness or global quotient presentation is assumed. A Hermitian orbifold line bundle $L \to X$ admits a unitary connection $\nabla$ whose curvature satisfies the \emph{prequantization condition}:
\[
F_{\nabla} \;=\; -\frac{i}{\hbar}\,\omega
\]
if and only if its de Rham Chern class satisfies \(c_1^{\mathrm{dR}}(L) = \frac{1}{2\pi\hbar}[\omega]\) in \(H^2_{\mathrm{dR,orb}}(X)\). When this condition holds, such a pair $(L, \nabla)$ is called a \emph{prequantum line bundle}.
\end{theorem}

\begin{proof}
($\Rightarrow$) Assume $(L,\nabla)$ is a bundle satisfying the prequantization condition $F_\nabla=-\frac{i}{\hbar}\omega$. By Chern--Weil theory on orbifolds, the real first Chern class of $L$ is represented by the closed $2$-form $\frac{i}{2\pi}F_\nabla$. Thus, 
\[
c_1^{\mathrm{dR}}(L) = \left[\frac{i}{2\pi}F_\nabla\right] = \left[\frac{i}{2\pi}\left(-\frac{i}{\hbar}\omega\right)\right] = \frac{1}{2\pi\hbar}[\omega] \in H^2_{\mathrm{dR,orb}}(X).
\]

($\Leftarrow$) Suppose \(c_1^{\mathrm{dR}}(L) = \frac{1}{2\pi\hbar}[\omega]\) in \(H^2_{\mathrm{dR,orb}}(X)\). First, we require an initial unitary connection on $L$. Just as in the manifold case, the space of unitary connections is an affine space (and therefore convex). Consequently, we can construct a global unitary connection $\nabla_0$ on $L$ by choosing local connections on uniformizing charts, averaging them over the finite isotropy groups to ensure local equivariance, and patching them together using an orbifold partition of unity.

Let $F_0$ be the curvature of this initial connection $\nabla_0$. By orbifold Chern--Weil theory, 
\[
\left[\frac{i}{2\pi}F_0\right] = c_1^{\mathrm{dR}}(L) = \frac{1}{2\pi\hbar}[\omega].
\]
Hence, the closed $i\mathbb{R}$-valued 2-form
\[
\eta := -\frac{i}{\hbar}\omega - F_0
\]
has trivial real de Rham class after multiplication by $i/(2\pi)$. Equivalently, its class vanishes in the de Rham complex of $i\mathbb{R}$-valued orbifold forms. By the orbifold de Rham theorem, a closed form with trivial cohomology class is exact. Therefore, there exists a globally defined $i\mathbb{R}$-valued orbifold 1-form $\beta$ such that $\eta=d\beta$. 

Defining a new unitary connection $\nabla := \nabla_0 + \beta$, with $\beta$ acting by scalar multiplication, the curvature shifts exactly by $d\beta$, yielding
\[
F_\nabla = F_0 + d\beta = F_0 + \eta = -\frac{i}{\hbar}\omega.
\]
Thus, $(L, \nabla)$ satisfies the prequantization condition.
\end{proof}

\begin{theorem}[Integrality and Existence]%
\label{thm:orbifold-existence}
Let $(X,\omega)$ be a paracompact boundaryless smooth effective symplectic orbifold; no compactness or global quotient presentation is assumed. A prequantum line bundle over \(X\) exists if and only if the symplectic class $[\omega]/(2\pi\hbar)$ lies in the image of the natural map
\[
H^2_{\mathrm{st}}(X;\mathbb{Z}) \xrightarrow{\ \iota_{\mathrm{dR}}\ } H^2_{\mathrm{dR,orb}}(X).
\]
This is the rigorous orbifold generalization of the Weil integrality theorem. For global quotients \(X = [M/G]\), it reduces exactly to the condition of equivariant integrality using \(H^2_G(M;\mathbb{Z})\).
\end{theorem}

\begin{proof}
($\Rightarrow$) Assume a prequantum line bundle $(L, \nabla)$ exists. By Theorem~\ref{thm:prequantum-line-bundle}, its de Rham Chern class satisfies \(c_1^{\mathrm{dR}}(L) = \frac{1}{2\pi\hbar}[\omega]\). Since the de Rham Chern class is precisely the image of the integral stack first Chern class \(c_1^{\mathrm{st}}(L) \in H^2_{\mathrm{st}}(X;\mathbb{Z})\) under \(\iota_{\mathrm{dR}}\), the forward implication follows immediately.

($\Leftarrow$) Assume \(\frac{1}{2\pi\hbar}[\omega]\in H^2_{\mathrm{dR,orb}}(X)\) lies in the image of \(\iota_{\mathrm{dR}}\). Choose an integral stack lift \(\alpha\in H^2_{\mathrm{st}}(X;\mathbb{Z})\) that maps to \(\frac{1}{2\pi\hbar}[\omega]\).

By the classification theorem for orbifold line bundles (Theorem~\ref{thm:pic_h2}), there exists an orbifold line bundle \(L \to X\) whose stack first Chern class is \(c_1^{\mathrm{st}}(L)=\alpha\). We must endow \(L\) with a Hermitian metric. Because the space of Hermitian inner products forms a convex cone, we can choose arbitrary local positive-definite metrics on uniformizing charts, average them over the local finite isotropy groups to guarantee equivariance, and subsequently glue them globally via an orbifold partition of unity. Both averaging and convex combinations preserve positive-definiteness, endowing \(L\) with a global Hermitian geometry.

Since \(c_1^{\mathrm{st}}(L)=\alpha\) maps to \(\frac{1}{2\pi\hbar}[\omega]\), its de Rham Chern class satisfies \(c_1^{\mathrm{dR}}(L)=\frac{1}{2\pi\hbar}[\omega]\). By Theorem~\ref{thm:prequantum-line-bundle}, \(L\) admits a unitary connection \(\nabla\) with \(F_\nabla = -\frac{i}{\hbar}\omega\). Thus, \((L,\nabla)\) is a valid prequantum line bundle, completing the proof.
\end{proof}

\begin{remark}
It is important to emphasize a direct consequence of this proof that is particularly salient for orbifolds. Torsion classes in \(H^2_{\mathrm{st}}(X;\mathbb{Z})\) always lie in the kernel of \(\iota_{\mathrm{dR}}\colon H^2_{\mathrm{st}}(X;\mathbb{Z})\to H^2_{\mathrm{dR,orb}}(X)\); for the finite-type orbifolds considered in the examples of this book, this kernel is exactly the torsion subgroup. Thus the integral lift \(\alpha\) is generally not unique. While the \emph{existence} theorem mirrors the manifold case perfectly, the \emph{uniqueness} of the underlying prequantum line bundle is often broken on orbifolds. In the finite-type cases emphasized here, topological prequantum line bundles for the same symplectic form differ by tensoring with a flat orbifold line bundle whose stack Chern class is torsion~\cite{AdemMilgram2004}. As prequantum bundles with connection, one may also tensor by a topologically trivial flat unitary bundle, so the full flat ambiguity is governed by \(\mathrm{Hom}(\pi_1^{\mathrm{orb}}(X),U(1))\). Since local isotropy groups frequently contribute such flat holonomy data, an orbifold typically admits multiple inequivalent prequantizations even if its underlying coarse space is simply connected.
\end{remark}

\section{Polarizations and the Quantum Hilbert Space}
\label{sec:polarization-orbifolds}

As in the manifold theory of Chapter~1, the resolution is to choose a \emph{polarization}---a specific selection of directions along which wavefunctions must be constant---and restrict to sections satisfying this constraint~\cite{Woodhouse1992,Sniatycki1980,Huebschmann2004}. On orbifolds, the polarization must additionally respect the local orbifold structure: it must be $G$-invariant on each uniformizing chart. This orbifold constraint enriches both the geometry and the resulting quantum theory.

\subsection{Orbifold Polarizations}
\label{sec:orbifold-polarizations}

For orbifolds, the polarization data must be defined chartwise and must be compatible both with the local isotropy actions and with the embeddings between uniformizing charts.

\begin{definition}[Orbifold Polarization]
\label{def:orbifold-polarization}
An \emph{orbifold polarization} on a boundaryless smooth effective symplectic orbifold $(X, \omega)$ of dimension $2n$ is a choice, for each uniformizing chart $(V,G,\pi)$, of a smooth complex subbundle
\[
P_V \subset (TV) \otimes \mathbb{C}
\]
satisfying:
\begin{enumerate}
    \item \textbf{Chart Compatibility:} For every $g \in G$, one has $dg(P_V)=P_V$, and for every chart embedding $\lambda \colon (V',G',\pi') \hookrightarrow (V,G,\pi)$ one has $d\lambda(P_{V'})=P_V|_{\lambda(V')}$.
    \item \textbf{Lagrangian:} Each fiber $(P_V)_x$ has complex dimension $n$, and $\omega_V(\xi, \eta) = 0$ for all $\xi, \eta \in (P_V)_x$.
    \item \textbf{Involutive:} On each chart $V$, one has $[\Gamma(P_V), \Gamma(P_V)] \subset \Gamma(P_V)$.
\end{enumerate}
Equivalently, $P$ is an involutive Lagrangian rank-$n$ subbundle of the complexified tangent orbibundle $T_{\mathbb{C}}X=(TX)\otimes\mathbb{C}$. For the resulting leaf space to be well-behaved, one generally additionally requires that the intersection $P \cap \overline{P}$ has constant rank across $X$.
\end{definition}

\subsubsection{Existence Constraints on Orbifolds.}
Unlike in the smooth manifold theory, where real polarizations always exist locally (e.g., the vertical distribution $\mathrm{span}\{\partial/\partial p_i\}$ in Darboux coordinates), invariant real polarizations might not exist even \emph{locally} on an orbifold. For instance, if an isotropy group $G_p$ acts by generic symplectic rotations mixing base and fiber directions, it need not preserve any real Lagrangian subspace at $p$. Consequently, the existence of a real polarization forces a local representation-theoretic condition: at each point, the symplectic representation of $G_p$ on $T_pX$ must preserve the real Lagrangian subspace $P_{\mathbb{R},p}$. This condition is necessary, though not sufficient for a global real polarization~\cite{PoncinRadouxWolak2010}. By contrast, every compact subgroup of $\mathrm{Sp}(2n, \mathbb{R})$ preserves some compatible complex structure; equivalently, after a symplectic change of basis it lies in $U(n)$. In particular, this applies to all finite isotropy groups of an orbifold. Combined with the equivariant Darboux theorem (Proposition~\ref{prop:equivariant-darboux}), this shows that near any orbifold point one can choose $G_p$-equivariant Darboux coordinates in which $G_p$ acts linearly and symplectically and preserves a compatible complex structure $J$ on the local model. Relative to such a choice, the resulting $(0,1)$-distribution gives a local $G_p$-invariant complex polarization. Finding a global K\"ahler structure remains a substantial topological restriction, but this local algebraic compatibility makes complex/K\"ahler polarizations a robust framework for orbifold geometric quantization.

\subsection{Polarized Sections and the Quantum Hilbert Space}
\label{sec:polarized-hilbert-space}

Once an orbifold polarization $P$ is chosen, the physical Hilbert space $\mathcal{H}_{\mathrm{quant}}(X)$ consists of those prequantum sections $s$ that are covariantly constant along the polarized directions.

\begin{definition}[Polarized Section]
\label{def:polarized-section}
Let \((X,\omega)\) be a boundaryless smooth effective symplectic orbifold equipped with an orbifold polarization \(P\), and let \((L,\nabla)\) be a prequantum line bundle over \(X\). A section $s$ of $L$ is \emph{polarized} if, using the complex-linear extension of $\nabla$, it is covariantly constant along every local orbifold vector field taking values in $P$:
\[
\nabla_\xi s \;=\; 0 \quad \text{for all local sections } \xi \text{ of } P.
\]
Equivalently, on each uniformizing chart $(V,G,\pi)$, the lifted section $s_V$ satisfies
\[
(\nabla_V)_{\tilde{\xi}} s_V = 0
\]
for every local section $\tilde{\xi}$ of $P_V$. Because $P_V$, $\nabla_V$, and $s_V$ are compatible with the chart group action and with chart embeddings, this chartwise condition is independent of the chosen atlas.
\end{definition}

Since $P$ is orbifold-compatible and the connection $\nabla$ is an orbifold connection, the condition of being polarized is well-defined globally. When the polarization yields honest $L^2$ sections (for instance in the K\"ahler case), we form the polarized Hilbert space:
\[
\mathcal{H}_{\mathrm{quant}}(X) \;:=\; \overline{\Gamma_P(X, L)}^{\|\cdot\|_{L^2}}.
\]
The precise inner product and measure entering this completion depend on the type of polarization; in particular, for real polarizations one often must pass from honest $L^2$ sections to distributional sections supported on Bohr--Sommerfeld leaves, and when available one may further incorporate the half-form (metaplectic) correction (see Section~\ref{sec:metaplectic-correction}).

\begin{proposition}[Sufficient Criterion for Quantizability of Orbifold Observables]
\label{prop:orbifold-quantizable}
Let \((X,\omega)\) be a boundaryless smooth effective symplectic orbifold equipped with an orbifold polarization \(P\), and let \((L,\nabla)\) be a prequantum line bundle over \(X\). No compactness or global quotient presentation is assumed. Let $f \in C^\infty_{\mathrm{orb}}(X, \mathbb{R})$. If its Hamiltonian vector field preserves the polarization distribution locally on each uniformizing chart,
\[
[X_{f,V}, \Gamma(P_V)] \;\subset\; \Gamma(P_V) \quad \text{for all charts } V.
\]
then its prequantum operator preserves the polarization constraint on smooth polarized sections.
\end{proposition}

\begin{proof}
The prequantum operator is $Q(f) = -i\hbar\nabla_{X_f} + f$. Whenever $Q(f)$ preserves the polarization constraint, we write $Q_{\mathrm{pol}}(f)$ for its restriction to polarized sections.

By definition, a smooth polarized section $s \in \Gamma_P(X, L)$ satisfies $\nabla_\xi s = 0$ for all local sections $\xi$ of $P$. To preserve this constraint under quantization, $Q(f)s$ must also be polarized, i.e., $\nabla_\xi(Q(f)s) = 0$ for all such local $\xi$. Using $Q(f)s = -i\hbar\nabla_{X_f}s + fs$ and the Leibniz rule:
\[
\nabla_\xi(-i\hbar\nabla_{X_f}s + fs) = -i\hbar\nabla_\xi\nabla_{X_f}s + \xi(f)s + f\nabla_\xi s.
\]
Because $\nabla_\xi s = 0$, the last term drops. By the definition of curvature, $F_\nabla(\xi, X_f) = [\nabla_\xi, \nabla_{X_f}] - \nabla_{[\xi, X_f]}$, so:
\[
\nabla_\xi\nabla_{X_f}s = [\nabla_\xi, \nabla_{X_f}]s + \nabla_{X_f}\underbrace{\nabla_\xi s}_{=\,0} = F_\nabla(\xi, X_f)s + \nabla_{[\xi, X_f]}s.
\]
Inserting the prequantization condition $F_\nabla = -\frac{i}{\hbar}\omega\,\operatorname{Id}$:
\[
\nabla_\xi(Q(f)s) = -i\hbar \left(-\frac{i}{\hbar}\omega(\xi, X_f)s + \nabla_{[\xi, X_f]}s\right) + \xi(f)s.
\]
Using the defining relation for the Hamiltonian vector field, $\iota_{X_f}\omega = -df$, we have $\omega(X_f, \xi) = -df(\xi) = -\xi(f)$. Because the symplectic form is skew-symmetric, this yields $\omega(\xi, X_f) = \xi(f)$. Substituting this into the previous equation yields:
\[
\nabla_\xi(Q(f)s) = -i\hbar \left(-\frac{i}{\hbar}\xi(f)s + \nabla_{[\xi, X_f]}s\right) + \xi(f)s = -i\hbar\nabla_{[\xi, X_f]}s.
\]
If $[X_f,\xi] \in \Gamma(P)$ locally for every polarized vector field $\xi$, then $[\xi,X_f]=-[X_f,\xi]$ is also a local section of $P$, hence $\nabla_{[\xi,X_f]}s=0$ because $s$ is polarized. Therefore $\nabla_\xi(Q(f)s)=0$ for every local section $\xi$ of $P$, so $Q(f)$ preserves the polarization constraint. Chartwise, this is exactly the condition $[X_{f,V},\Gamma(P_V)]\subset\Gamma(P_V)$, and compatibility with the orbifold structure makes these local statements glue to a global one.
\end{proof}

\subsubsection{Physical Obstructions.}
Observable functions whose Hamiltonian vector fields do not preserve the polarization cannot be quantized directly in this framework; they require more advanced techniques such as the BKS pairing or half-form distributions.

\subsection{Real Polarizations and Bohr--Sommerfeld Leaves}
\label{sec:real-polarizations}

A polarization is \emph{real} if $P = \overline{P}$. In this case, $P$ is the complexification of a real rank-$n$ Lagrangian orbifold subbundle $P_\mathbb{R} \subset TX$. On each uniformizing chart $(V,G,\pi)$, the lifted distribution $P_{\mathbb{R},V}\subset TV$ is a smooth involutive distribution on the smooth manifold $V$, so the standard Frobenius theorem integrates it to a foliation of $V$ by Lagrangian submanifolds. Because $P_{\mathbb{R},V}$ is $G$-invariant, this foliation is $G$-invariant, and the local foliations glue to an orbifold foliation of $X$.

\begin{definition}[Orbifold Foliation and Leaves]
\label{def:orbifold-foliation}
An \emph{orbifold foliation} of dimension $n$ on a smooth effective orbifold \(X\) is a rank-$n$ orbifold subbundle $F\subset TX$ such that, on every uniformizing chart $(V, G, \pi)$, the lifted distribution $F_V \subset TV$ integrates to a smooth foliation of $V$ by connected immersed $n$-manifolds and this foliation is preserved by the $G$-action. A \emph{leaf} of the induced foliation on $X$ is the maximal connected immersed orbifold obtained by gluing the local lifted leaves under chart embeddings and local group actions. In this chapter the ambient symplectic orbifold is boundaryless; foliations on orbifolds with mirror or corner boundary would require separate boundary compatibility conditions.
\end{definition}

In the case where the orbifold is a global quotient $X=[M/G]$, the lifted real distribution $P_{\mathbb{R},M}\subset TM$ is $G$-invariant and integrates to a $G$-invariant foliation of $M$ by Lagrangian leaves. If $\tilde{\mathcal{L}}\subset M$ is a leaf and $\pi\colon M\to X$ is the quotient map, then the $G$-saturation $G\cdot\tilde{\mathcal{L}}$ projects to a leaf $\mathcal{L}\subset X$. In general, the chosen lift $\tilde{\mathcal{L}}$ need not be preserved by all of $G$; rather, its translates are also lifted leaves with the same image in $X$. The symmetry group relevant to a chosen lift is the stabilizer $G_{\tilde{\mathcal{L}}}:=\{g\in G\mid g\cdot \tilde{\mathcal{L}}=\tilde{\mathcal{L}}\}$, and the leaf can be represented by the quotient $[\tilde{\mathcal{L}}/G_{\tilde{\mathcal{L}}}]$.

The standard example is the vertical polarization on a cotangent bundle $T^*Q$: the leaves are the fibers $T_q^*Q$, and requiring wavefunctions to be constant along these vertical directions means they depend only on the base coordinates $q$, recovering the Schr\"odinger position representation. If $G$ acts on $Q$ and we form the orbifold quotient $[T^*Q/G]$ via the cotangent-lifted action, then the vertical distribution is automatically $G$-invariant. A leaf over $[q]\in Q/G$ is modeled by the orbifold quotient $[T_q^*Q/G_q]$, where $G_q\subset G$ acts on the fiber by the cotangent lift.

\subsubsection{Flat Connection on Leaves.}
Since each leaf $\mathcal{L}$ is Lagrangian, the symplectic form vanishes on it: $\omega|_\mathcal{L} = 0$. Consequently, the prequantum connection $\nabla$, whose curvature satisfies $F_\nabla = -\frac{i}{\hbar}\omega$, restricts to a \emph{flat connection} on the prequantum line bundle $L|_\mathcal{L}$ over each leaf. Because the connection is flat, locally on any simply connected open subset of a lifted leaf segment $\tilde{\mathcal{L}}\subset V$, a non-zero covariantly constant section always exists and is determined uniquely by its value at a single point via parallel transport. However, extending this section globally over the orbifold leaf requires trivial orbifold holonomy, including compatibility with the isotropy identifications along orbifold loops.

\subsubsection{Holonomy and the Bohr--Sommerfeld Condition.}
To make the Bohr--Sommerfeld condition explicit, choose a uniformizing chart $(V,G,\pi)$ and a connected lifted leaf segment $\tilde{\mathcal{L}}\subset V$. Assuming the chart $V$ is contractible (which can always be arranged, e.g., using a standard ball), Proposition~\ref{prop:invariant-symplectic-potential} provides a $G$-invariant global symplectic potential $\theta$ on $V$ with $d\theta=\omega_V$. Choose any local unitary frame $e$ of $L_V$ over $V$ and write $\nabla e=-iA\otimes e$. By Proposition~\ref{prop:curvature-equivalence} and the prequantization condition,
\[
dA=\frac{1}{\hbar}\omega_V=\frac{1}{\hbar}d\theta.
\]
Hence $A-\theta/\hbar$ is closed on $V$, so because $V$ is contractible there exists $\chi\in C^\infty(V,\mathbb{R})$ with $A-\theta/\hbar=d\chi$. Note that $\chi$ need not be $G$-invariant; any non-invariance is absorbed into the equivariant phase $\alpha_{s_0}(g)$ defined below. After the unitary gauge change $s_0=e^{i\chi}e$ (Proposition~\ref{prop:equivariant-gauge-transform}), the connection takes the form
\[
\nabla = d - \frac{i}{\hbar}\theta
\qquad(\text{equivalently, the real gauge potential is }A=\theta/\hbar).
\]
Writing a lifted section as $\tilde{s} = f\,s_0$, the covariant constancy equation $\nabla \tilde{s} = 0$ along $\tilde{\mathcal{L}}$ becomes
\[
df - \frac{i}{\hbar}\theta\, f = 0 \quad \text{along }\tilde{\mathcal{L}},
\]
so for a chosen basepoint $x_0\in \tilde{\mathcal{L}}$ one has locally $f(x) \propto \exp\bigl(\frac{i}{\hbar}\int_{x_0}^x \theta\bigr)$. For an ordinary closed loop $\gamma \subset \tilde{\mathcal{L}}$, the corresponding holonomy phase is
\[
\mathrm{Hol}_\nabla(\gamma) = \exp\left(\frac{i}{\hbar}\oint_\gamma \theta\right).
\]
For an orbifold loop, one must additionally compose this parallel transport with the inverse equivariant identification on the endpoint fiber. Thus, ordinary loops require ordinary holonomy to be trivial, and a non-zero covariantly constant orbifold section exists globally along $\mathcal{L}$ if and only if the orbifold holonomy is trivial for every orbifold loop in~$\mathcal{L}$.

\begin{definition}[Bohr--Sommerfeld Leaf]
\label{def:bohr-sommerfeld}
Let \((X,\omega)\) be a boundaryless smooth effective symplectic orbifold with a real orbifold polarization \(P\). Let $\mathcal{L}\subset X$ be a leaf of \(P\) and let $(L,\nabla)$ be a prequantum line bundle over \(X\). Since $\omega|_{\mathcal{L}}=0$, the restricted connection on $L|_{\mathcal{L}}$ is flat. We say that $\mathcal{L}$ is \emph{Bohr--Sommerfeld} if the flat orbifold line bundle $L|_{\mathcal{L}}$ has trivial orbifold holonomy; equivalently, if $L|_{\mathcal{L}}$ admits a non-zero covariantly constant orbifold section. In local charts, this means that on each lifted leaf segment $\tilde{\mathcal{L}} \subset V$ there is a parallel section $\tilde{s}$ satisfying the equivariance relation $\Phi_g(\tilde{s}(x))=\tilde{s}(g\cdot x)$ for all $x \in \tilde{\mathcal{L}}$ and all $g \in G_{\tilde{\mathcal{L}}}$.
\end{definition}

The condition can be decomposed into two parts:

\begin{enumerate}
    \item \textbf{Ordinary loops in a lift of the leaf:}
For any closed loop $\gamma$ in a lifted leaf inside a chart (equivalently, an orbifold loop with $g=\mathrm{id}$), the holonomy must be trivial. This yields the usual Bohr--Sommerfeld condition:
\begin{equation}
\label{eq:BS-smooth}
\frac{1}{2\pi\hbar} \oint_\gamma \theta \;\in\; \mathbb{Z}.
\end{equation}

\item \textbf{Orbifold loops and fractional shifts:}
In general, fix a lifted leaf segment $\tilde{\mathcal{L}} \subset V$ in a uniformizing chart, choose a basepoint $x_0\in\tilde{\mathcal{L}}$, and let
\[
G_{\tilde{\mathcal{L}}}:=\{g\in G \mid g\cdot \tilde{\mathcal{L}}=\tilde{\mathcal{L}}\}.
\]
An orbifold loop in the leaf $\mathcal{L}$ based at the class of $x_0\in\tilde{\mathcal{L}}$ is represented by a pair $(\tilde{\gamma},g)$ where $\tilde{\gamma}\colon[0,1]\to\tilde{\mathcal{L}}$, $\tilde{\gamma}(0)=x_0$, and $\tilde{\gamma}(1)=g\cdot x_0$ for some $g\in G_{\tilde{\mathcal{L}}}$. Let $\mathrm{Par}_{\tilde{\gamma}}\colon (L_V)_{x_0}\to (L_V)_{g\cdot x_0}$ denote parallel transport along $\tilde{\gamma}$, and let $\Phi_g$ denote the equivariant lift of $g$ to the prequantum line bundle (see Section~\ref{sec:prequantization-orbifolds}). A local parallel section descends to the orbifold leaf precisely when the \emph{orbifold holonomy}
\[
\mathrm{Hol}^{\mathrm{orb}}_\nabla(\tilde{\gamma},g):=\Phi_{g^{-1}}\circ \mathrm{Par}_{\tilde{\gamma}}\in \mathrm{End}((L_V)_{x_0}) \cong U(1)
\]
is trivial (i.e., equal to the identity).

In a local unitary frame $s_0$ as above, the local connection form is $-i\theta/\hbar$, so the corresponding real gauge potential $A = \theta/\hbar$ is strictly $G$-invariant. By Proposition~\ref{prop:unitary-real-potential}, for each $g\in G$ the equivariant action on $s_0$ is therefore given by a locally constant phase; on the connected chart $V$ this phase is constant. We can thus write
\[
\Phi_g\bigl(s_0(x)\bigr)=e^{2\pi i\,\alpha_{s_0}(g)}\,s_0(g\cdot x),
\qquad \alpha_{s_0}(g)\in\mathbb{R}/\mathbb{Z}.
\]
Then $\mathrm{Hol}^{\mathrm{orb}}_\nabla(\tilde{\gamma},g)=\mathrm{Id}$ is equivalent to:
\begin{equation}
\label{eq:BS-orbifold}
 \frac{1}{2\pi\hbar} \int_{\tilde{\gamma}} \theta \;-\; \alpha_{s_0}(g) \;\in\; \mathbb{Z}.
\end{equation}
Although $\theta$ and $\alpha_{s_0}(g)$ depend on the choices of local symplectic potential and local unitary frame, the condition $\mathrm{Hol}^{\mathrm{orb}}_\nabla(\tilde{\gamma},g)=\mathrm{Id}$ is rigorously intrinsic. If one performs a generic gauge change $s'_0 = e^{i\chi}s_0$, the local connection 1-form changes by an exact term and the new phase $\alpha_{s'_0}(g; x_0)$ shifts by the compensating endpoint contribution determined by $\chi$, leaving their difference modulo $\mathbb{Z}$ unchanged.
\end{enumerate}

\begin{proposition}[Orbifold Bohr--Sommerfeld Condition]
\label{prop:orbifold-BS}
Let \((X,\omega)\) be a boundaryless smooth effective symplectic orbifold with a real orbifold polarization \(P\), and let \((L,\nabla)\) be a prequantum line bundle over \(X\). No global quotient presentation is assumed. A Lagrangian leaf $\mathcal{L}\subset X$ of \(P\) is Bohr--Sommerfeld if and only if the leafwise flat orbifold holonomy of $L|_{\mathcal{L}}$ is trivial for every orbifold loop in $\mathcal{L}$. In a single uniformizing chart, if such a loop is represented by a path $\tilde{\gamma}\subset\tilde{\mathcal{L}}$ with $\tilde{\gamma}(1)=g\cdot \tilde{\gamma}(0)$ for a local group element $g$ preserving the lifted leaf, this condition is
\[
\Phi_{g^{-1}}\circ \mathrm{Par}_{\tilde{\gamma}}=\mathrm{Id}\quad\text{on }(L_V)_{\tilde{\gamma}(0)}.
\]
\end{proposition}

\begin{proof}
Because $\omega|_{\mathcal{L}}=0$, the connection is flat along each leaf. If $\mathcal{L}$ is Bohr--Sommerfeld, then by Definition~\ref{def:bohr-sommerfeld} there exists a non-zero parallel orbifold section. In a local chart this section is represented by a non-zero parallel lift $\tilde{s}$ satisfying $\Phi_g(\tilde{s}(x))=\tilde{s}(g\cdot x)$ whenever $g$ identifies points in the lifted leaf. For any local representative $(\tilde{\gamma},g)$ from $x_0 = \tilde{\gamma}(0)$ to $g\cdot x_0 = \tilde{\gamma}(1)$, parallel transport gives $\mathrm{Par}_{\tilde{\gamma}}\tilde{s}(x_0)=\tilde{s}(g\cdot x_0)$, so applying $\Phi_{g^{-1}}$ yields $\Phi_{g^{-1}}\circ \mathrm{Par}_{\tilde{\gamma}}\bigl(\tilde{s}(x_0)\bigr)=\tilde{s}(x_0)$. Since this vector is non-zero and the fiber is one-dimensional, the holonomy is the identity.
Conversely, assume the leafwise orbifold holonomy is trivial for every orbifold loop. Fix a basepoint on the connected orbifold leaf and a non-zero value in its fiber; flatness defines a parallel section along any lifted path starting at that basepoint. Trivial orbifold holonomy makes the result independent of the chosen representative path and compatible with the endpoint identifications by the local group elements. On overlaps of uniformizing charts, uniqueness of parallel transport for the restricted flat connection shows that these locally defined parallel sections agree under chart embeddings. Therefore they glue to form a global non-zero parallel orbifold section on $\mathcal{L}$.
\end{proof}

\subsubsection{Fractional Spectra and the Quantum Hilbert Space.}
Ignoring half-forms, the quantum space for a real polarization is heuristically supported on the Bohr--Sommerfeld leaves; in common completely integrable situations, and under the usual compactness/properness assumptions ensuring a discrete Bohr--Sommerfeld set, these leaves form a discrete set. The additional orbifold holonomy constraints in \eqref{eq:BS-orbifold} can shift the allowed phases away from strict integers; for example, if an isotropy element of order $r$ acts on the fiber by $e^{2\pi i m/r}$, then the condition reads $\frac{1}{2\pi\hbar}\int_{\tilde{\gamma}}\theta\equiv m/r \pmod{\mathbb{Z}}$ for a representative path $\tilde{\gamma}$ of the corresponding orbifold loop, producing fractional quantum numbers.

\subsubsection{Relation to the Orbifold Integrality Condition.}
The Bohr--Sommerfeld condition still acts as a dynamical selection rule on phase space, linked to the prequantization integrality condition via Stokes' theorem. On an orbifold, however, this relationship must be interpreted using the stack integral lattice \(H^2_{\mathrm{st}}(X;\mathbb{Z})\) and must account for isotropy phases, flat torsion data, and fractional fluxes on the coarse space.

The orbifold prequantization condition is that the real de Rham class \([\omega]/(2\pi\hbar)\in H^2_{\mathrm{dR,orb}}(X)\cong H^2_{\mathrm{coarse}}(X;\mathbb{R})\) lie in the image of \(\iota_{\mathrm{dR}}\colon H^2_{\mathrm{st}}(X;\mathbb{Z})\to H^2_{\mathrm{dR,orb}}(X)\); see Section~\ref{sec:prequantization-orbifolds} and the cohomology conventions of Chapter~2. When evaluated on cycles meeting singular strata, such a stack-integral class can correspond to rational numbers on the underlying coarse space, with denominators set by isotropy orders. This de Rham class fixes the curvature but not the full prequantum orbifold line bundle: different integral stack lifts can differ by torsion, and flat orbifold line bundles can contribute additional holonomy without changing \([\omega]\) in de Rham cohomology. When a real polarization foliates \(X\) into Lagrangian orbifold leaves, the restriction of the connection is flat on each leaf, and the Bohr--Sommerfeld conditions are precisely the requirement that all leafwise orbifold holonomies \(\Phi_{g^{-1}}\circ \mathrm{Par}_{\tilde{\gamma}}\) are trivial. Thus the phases coming from the equivariant lifts \(\Phi_g\) (encoded locally by the constant \(\alpha_{s_0}(g)\) in \eqref{eq:BS-orbifold}) are part of the full prequantum orbifold line-bundle data; they may either reflect the isotropy contribution to an integral stack Chern class or a flat torsion/holonomy choice.

More precisely, if a compact oriented orbifold surface \(\Sigma\) has boundary \(\partial\Sigma\) lying within a Lagrangian leaf, then the holonomy-flux relation from Section~\ref{sec:prequantization-orbifolds}, applied chartwise and combined with orbifold Stokes' theorem, expresses the orbifold holonomy around \(\partial\Sigma\) in terms of the symplectic flux \(\exp\bigl(\frac{i}{\hbar}\int_\Sigma \omega\bigr)\) together with isotropy phases coming from the singular points of \(\Sigma\). On the underlying coarse space these isotropy contributions appear as rational shifts, with denominators determined by the local isotropy orders. For a closed surface \((\partial\Sigma=\emptyset)\), this recovers the statement that the prequantization class is stack-integral even though its evaluation on the coarse space can be fractional. Ultimately, while prequantization secures the mathematical foundation for the quantum operator algebra over \(X\), the Bohr--Sommerfeld condition acts as a dynamical filter, isolating the physically permitted states onto a discrete lattice of special leaves.

\subsection{Complex and K\"ahler Polarizations}
\label{sec:kahler-polarizations}

Polarizations need not be real. A \emph{complex polarization} is an orbifold subbundle $P \subset (TX) \otimes \mathbb{C}$ satisfying the Lagrangian and involutive conditions but containing complex directions~\cite{Huebschmann2004,PoncinRadouxWolak2009}. The most important class arises from K\"ahler geometry.

\subsubsection{K\"ahler Polarizations.}
A polarization is \emph{K\"ahler} if it is induced by an orbifold K\"ahler structure $(\omega,J)$ on $X$, i.e.\ if $P=T^{0,1}X$ for an \emph{integrable} orbifold complex structure $J$ such that $\omega$ is a positive $(1,1)$-form. In particular, one has $P \cap \overline{P}=\{0\}$ and $P+\overline{P}=(TX)\otimes\mathbb{C}$, and the involutivity of $P$ coincides with the integrability of $J$: chartwise this is the usual Newlander--Nirenberg theorem, and the orbifold complex structure is obtained by requiring the chart changes and finite isotropy actions to be holomorphic with respect to the resulting local complex structures.

The Picard classification in Theorem~\ref{thm:pic_h2} concerns the underlying smooth or topological orbifold line bundle, i.e.\ the group \(\mathrm{Pic}_{\mathrm{orb}}(X)\). Holomorphic quantization uses a different object: after choosing a compatible orbifold complex structure, one works in the holomorphic Picard group \(\mathrm{Pic}_{\mathrm{hol}}(X)\) and then forgets to \(\mathrm{Pic}_{\mathrm{orb}}(X)\) when recording the stack Chern class. A unitary prequantum connection determines such a holomorphic orbifold line bundle precisely when its curvature has no $(0,2)$ component with respect to $J$; then $\nabla^{0,1}$ is an integrable Dolbeault operator. By contrast, the stack Chern class alone usually does not determine a holomorphic line bundle: distinct holomorphic structures can live on the same smooth or topological orbifold line bundle, and not every smooth representative is holomorphic for a fixed complex orbifold structure.

Because $\omega$ is of type $(1,1)$, the prequantum curvature $F_\nabla=-\frac{i}{\hbar}\omega$ has $F_\nabla^{0,2}=0$, so $(\nabla^{0,1})^2=0$. Hence $\bar{\partial}_L:=\nabla^{0,1}$ defines a holomorphic structure on $L$, and the polarization condition $\nabla^{0,1}s=0$ is exactly the Cauchy--Riemann equation for sections. Thus, for a K\"ahler polarization, the physical quantum states are the \emph{holomorphic orbifold sections} of the prequantum line bundle.

In this setting, geometric quantization connects directly with algebraic geometry:

\begin{itemize}
    \item \textbf{Holomorphic Structure:} The operator $\bar{\partial}_L := \nabla^{0,1}$ is integrable and defines a holomorphic structure on $L$.
    \item \textbf{Polarized Sections:} The polarized sections are the holomorphic orbifold sections of $L$, i.e., $H^0(X, L)$.
    \item \textbf{Positivity:} When $\omega$ is the K\"ahler form compatible with $J$, $\omega$ is a positive $(1,1)$-form, and the prequantization condition $F_\nabla = -\frac{i}{\hbar}\omega$ gives $\frac{i}{2\pi}F_\nabla = \frac{1}{2\pi\hbar}\omega > 0$, so $L$ is a positive line bundle. For compact complex orbifolds, $H^0(X,L)$ is finite-dimensional. The Kawasaki--Riemann--Roch theorem~\cite{Kawasaki1979} computes the holomorphic Euler characteristic $\chi(X,L)$, explicitly incorporating the fractional isotropy phases of $L$ at the singular strata; when the higher cohomology vanishes (for example under the relevant orbifold Kodaira/Serre vanishing hypotheses, or for sufficiently positive tensor powers), this Euler characteristic equals the exact dimension of $H^0(X,L)$.
\end{itemize}

\subsubsection{Existence of K\"ahler Polarizations on Orbifolds.}
A K\"ahler polarization exists on a global quotient $X = [M/G]$ whenever $M$ admits a $G$-invariant K\"ahler structure compatible with $\omega$. This is guaranteed, for instance, when $(M, \omega)$ is a K\"ahler manifold and $G$ acts by holomorphic isometries. More generally, many physically relevant phase spaces admit global K\"ahler structures, including orbifold quotients of flag manifolds, weighted projective spaces, and toric orbifolds. By contrast, real polarizations require additional compatibility between the isotropy representations and a real Lagrangian distribution, and their existence is not guaranteed even locally.


\section{The Metaplectic Correction}
\label{sec:metaplectic-correction}

While the metaplectic correction serves the same physical purpose in both settings---accounting for zero-point shifts and restoring formal self-adjointness for the relevant quantizable operators---the transition to orbifolds introduces subtle but significant structural modifications~\cite{Woodhouse1992,Sniatycki1980,MaratheMartucci1985}. On a smooth manifold, the half-form correction for a given polarization $P$ relies on the global existence of a square root of the polarization's canonical bundle. On an orbifold, however, this requirement must be satisfied in an equivariant sense. Locally on each uniformizing chart, one needs a half-form bundle that intertwines correctly with the isotropy action, making the local representation-theoretic constraints visible inside the orbifold topological obstruction. Furthermore, while the manifold half-form correction yields the familiar Maslov phase along closed loops, the equivariant structure on an orbifold gives rise to an additional fractional phase whenever an orbifold loop closes up to an isotropy element. This extra isotropy phase can alter the corrected Bohr--Sommerfeld rule, shifting the quantum spectrum relative to the standard manifold theory. The present section details the rigorous construction of this orbifold metaplectic correction and explores its spectral consequences.

\subsection{Half-Forms and Densities}
\label{sec:half-form-definition}

As explained for manifolds in Chapter~1, the need for half-forms arises from the following physical requirement. In classical mechanics, one integrates functions against a volume form to compute observables. In quantum mechanics, however, the probability density $|\psi|^2$ must \emph{itself} be a density---so that integrating it over an appropriate space gives unity. This implies that the wavefunction $\psi$ should transform as a ``square root'' of a density---a \emph{half-density}. The prequantum line bundle $L$ supplies the complex phase information that encodes the symplectic geometry, but its sections transform merely as scalars. The half-form bundle supplies the additional transformation law needed to ensure that the inner product yields an integrable density.

\subsubsection{The Canonical Bundle and Half-Form Bundle.}
Let $P \subset (TX) \otimes \mathbb{C}$ be an orbifold polarization (Definition~\ref{def:orbifold-polarization}) of complex rank $n = \tfrac{1}{2}\dim X$ on a boundaryless smooth effective symplectic orbifold $X$. The \emph{canonical bundle of the polarization} is the determinant line bundle of the annihilator of $P$:
\[
K_P \;:=\; \det(P^o) \;=\; \bigwedge^n P^o,
\]
where $P^o \subset (T^*X) \otimes \mathbb{C}$ is the annihilator of $P$ (i.e., the covectors that vanish on all vectors in $P$). For a real polarization with leaf distribution $P_{\mathbb{R}}$, the annihilator $P^o$ is the complexified conormal bundle of the foliation, so $K_P$ represents complex-valued \emph{transverse} volume forms; after choosing the usual real structure or passing to absolute values, these determine transverse densities on the leaf space when it is smooth. For a K\"ahler polarization $P = T^{0,1}X$, its annihilator is $P^o = \Omega^{1,0}X$, so $K_P$ is precisely the canonical bundle $K_X = \bigwedge^{n,0}T^*X$ of the K\"ahler orbifold.

\begin{definition}[Half-Form Bundle on a Smooth Effective Orbifold]
\label{def:half-form-bundle}
Let \(X\) be a boundaryless smooth effective symplectic orbifold equipped with an orbifold polarization \(P\). A \emph{half-form bundle} for \(P\) is an orbifold line bundle $\delta_P \to X$ equipped with an isomorphism of orbifold line bundles $\delta_P^{\otimes 2} \cong K_P$. This is an orbifold-stack statement for the effective atlas of \(X\); non-effective quotient stacks would require the same square-root condition with the ineffective stabilizer characters included explicitly.
\end{definition}

\noindent
Taking the tensor product $L \otimes \delta_P$ combines the complex phase geometry of $L$ with the half-volume transformations of $\delta_P$. For a K\"ahler polarization, the canonical K\"ahler metric induces a Hermitian structure on $\delta_P$; combined with the prequantum metric, the inner product of two sections yields a scalar function that is integrated against the symplectic orbifold volume. For a real polarization, $K_P$ is the complexified transverse volume-form line, and a choice of compatible real structure or positive transverse density identifies the product of a half-form with the conjugate half-form as a transverse density. Thus the corrected pairing produces the density that one integrates on the leaf space (when that quotient is sufficiently regular, or after the usual distributional reformulation on Bohr--Sommerfeld leaves). This is the geometric realization of the physical requirement that $|\psi|^2$ define a well-behaved probability density.

For real polarizations there is a closely related, but distinct, fallback: one may work with the square root of the transverse \emph{density} line rather than a square root of the complex line $K_P$. This half-density bundle can exist even when a genuine equivariant half-form is obstructed, because absolute values remove orientation characters. It gives the correct $L^2$ pairing on the leaf space, but it is not a holomorphic/metaplectic half-form bundle and it does not carry the same Maslov and equivariant phase data. Unless explicitly stated otherwise, the Bohr--Sommerfeld--Maslov statements below assume a genuine orbifold half-form bundle $\delta_P$.

\subsubsection{The Corrected Hilbert Space.}
At a formal level, before imposing the analytic completion appropriate to the chosen polarization, the \emph{metaplectically corrected polarized state space} for a boundaryless smooth effective symplectic orbifold $X$ is built from polarized orbifold sections of the corrected tensor bundle $L \otimes \delta_P$:
\begin{equation}
\label{eq:corrected-hilbert-space}
\mathcal{S}_{\mathrm{meta}}^{\mathrm{formal}}(X) \;:=\; \bigl\{ s \in \Gamma(X,\, L \otimes \delta_P) \;\big|\; \nabla^{L \otimes \delta_P}_\xi s = 0 \text{ for all } \xi \in P \bigr\},
\end{equation}
where $\nabla^{L \otimes \delta_P}$ is the tensor product of the prequantum connection on $L$ and a connection on $\delta_P$. For the polarization condition, only a \emph{partial} connection along $P$ is needed: the annihilator bundle $P^o$ (hence $K_P=\det(P^o)$) carries a canonical flat partial connection along $P$ induced by Lie differentiation (the Bott connection), computed on each uniformizing chart where the distribution is smooth and involutive, and therefore flat. Because $P$ is $G$-invariant, this partial connection is $G$-equivariant and descends to an orbifold partial connection. Via the isomorphism $\delta_P^{\otimes 2} \cong K_P$, one obtains locally a compatible flat partial connection along $P$ on $\delta_P$: if $\mu=\nu^{\otimes 2}$ and $\nabla^{K_P}\mu=\eta\otimes\mu$ along $P$, then set $\nabla^{\delta_P}\nu=\tfrac12\eta\otimes\nu$. Local choices differ at most by the square-root transition functions of $\delta_P$, so these prescriptions glue precisely when the orbifold half-form bundle exists. Any extension to a full connection yields the same polarization equation. For the K\"ahler case, this is the space of holomorphic orbifold sections:
\[
\mathcal{H}_{\mathrm{meta}}(X) \;=\; H^0(X,\; L \otimes K_X^{1/2}),
\]
the holomorphic orbifold sections of the corrected line bundle, where $K_X^{1/2}$ denotes a chosen holomorphic orbifold square root of $K_X$.
For real polarizations, the space in~\eqref{eq:corrected-hilbert-space} should be read as a formal polarized solution space; the physical Hilbert space is obtained only after passing to the appropriate leaf-space or distributional completion, usually supported on Bohr--Sommerfeld leaves and paired using the half-form or half-density measure.

\subsection{Obstruction Theory}
\label{sec:obstruction}

\subsubsection{Manifold Obstruction.}
On a smooth manifold $M$, the half-form bundle $\delta_P$ exists if and only if $K_P$ admits a square root as a complex line bundle. By the classification of complex line bundles via $c_1 \in H^2(M, \mathbb{Z})$, a square root exists if and only if:
\begin{equation}
\label{eq:metalinear-obstruction}
c_1(K_P) \in 2H^2(M,\mathbb{Z})
\qquad\text{equivalently}\qquad
c_1(K_P)\bmod 2=0 \text{ in } H^2(M,\mathbb{Z}/2).
\end{equation}
Equivalently, the structure group of the relevant frame bundle can be lifted from $GL(n, \mathbb{C})$ to the \emph{metalinear group} $ML(n, \mathbb{C})$ (the double cover defined by the squaring map on $\det$). In the symplectic setting one also encounters the (polarization-independent) \emph{metaplectic structure}, a lift of the symplectic frame bundle from $Sp(2n, \mathbb{R})$ to its double cover $Mp(2n, \mathbb{R})$, which exists if and only if $w_2(TM)=0$ (equivalently $c_1(TM, J) \equiv 0 \pmod{2}$ for any compatible almost complex structure $J$). A metaplectic structure implies the existence of compatible half-form bundles for all polarizations, but for a fixed polarization the relevant obstruction is precisely~\eqref{eq:metalinear-obstruction}.

\subsubsection{Orbifold Structure and Local Obstructions.}
\label{sec:obstruction-orbifold}
For the boundaryless smooth effective orbifolds considered in this chapter, the condition $\delta_P^{\otimes 2} \cong K_P$ is an orbifold, not merely coarse-space, statement. It contains two complementary pieces of data, which are not independent: the orbifold cohomological condition~(i) already encodes the local isotropy constraints~(ii), but we state both explicitly because the local description is essential for computations.

\subparagraph{(i) Global orbifold Picard constraint.}
The orbifold line bundle \(K_P\) must be divisible by \(2\) in the smooth/topological orbifold Picard group \(\mathrm{Pic}_{\mathrm{orb}}(X)\). Because topological orbifold line bundles are classified by \(H^2_{\mathrm{st}}(X;\mathbb{Z})\), this is strictly equivalent to the condition that the stack first Chern class
\[
c_1^{\mathrm{st}}(K_P)\in H^2_{\mathrm{st}}(X;\mathbb{Z})
\]
lies in \(2H^2_{\mathrm{st}}(X;\mathbb{Z})\); equivalently, its reduction under the coefficient map to \(H^2_{\mathrm{st}}(X;\mathbb{Z}/2)\) is zero. This is the exact obstruction to the existence of a \emph{topological} orbifold square root. If one further requires a \emph{holomorphic} square root in the compact complex/K\"ahler setting used for holomorphic quantization, one instead uses \(\mathrm{Pic}_{\mathrm{hol}}(X)\): in that setting the connected component \(\mathrm{Pic}_{\mathrm{hol}}^0(X)\) is a complex torus and hence divisible, so the parity of \(c_1^{\mathrm{st}}(K_P)\) remains the only discrete obstruction, and holomorphic square roots, when they exist, differ by \(2\)-torsion holomorphic orbifold line bundles.

\subparagraph{(ii) Local equivariant square roots.}
On a uniformizing chart $(V, G_p, \pi)$ centered at $p$ with $V$ contractible, the ordinary topological part is trivial, so the only local issue is the $G_p$-equivariant structure. Equivalently, one asks whether the isotropy representation of $G_p$ on the fiber $(K_{P,V})_p$ admits a square root. This representation is encoded by a character
\[
\chi_{K_P,p}\colon G_p \to U(1).
\]
A local orbifold half-form exists on this chart if and only if this character is divisible by $2$ in the character group $\widehat{G_p}=\operatorname{Hom}(G_p,U(1))$; equivalently, there exists $\chi_{\delta,p}\colon G_p \to U(1)$ with $\chi_{\delta,p}^2=\chi_{K_P,p}$.

If $g \in G_p$ acts on the complex transverse quotient $((T_pX)\otimes\mathbb{C})/P_p$ with eigenvalues $\{e^{2\pi i \theta_j}\}_{j=1}^n$, then the induced action on the canonical fiber is
\[
\chi_{K_P,p}(g) \;=\; \det\!\bigl(g|_{(((T_pX)\otimes\mathbb{C})/P_p)^*}\bigr)
\;=\; \prod_{j=1}^n e^{-2\pi i \theta_j}
\;=\; e^{-2\pi i \sum_j \theta_j}.
\]
In the special case that $G_p=\langle g\rangle$ is cyclic of order $N$, write
\[
\chi_{K_P,p}(g)=e^{2\pi i m/N}
\qquad (m \in \mathbb{Z}/N\mathbb{Z}).
\]
Then a square-root character exists exactly when the congruence $2\ell\equiv m \pmod N$ is solvable. Hence if $N$ is odd, multiplication by $2$ on $\mathbb{Z}/N\mathbb{Z}$ is invertible and the square root is unique. If $N$ is even, existence is equivalent to $m$ being even; when it exists, there are two local square-root characters, differing by the nontrivial character of order two. In terms of the eigenangles above, if $e^{2\pi i\theta_j}=e^{2\pi i a_j/N}$ with $a_j\in\mathbb{Z}$, then $m\equiv -\sum_j a_j \pmod N$, so for even $N$ this is equivalent to
\[
N\sum_j \theta_j \in 2\mathbb{Z}.
\]

\begin{remark}[Odd-order isotropy is locally unobstructed]
\label{rem:more-quantizable}
If $G_p$ has odd order, then $\widehat{G_p}$ is a finite abelian group of odd order, so multiplication by $2$ is an automorphism. Hence every local equivariant structure on $K_P$ has a unique square root. Genuine local obstructions arise only from the $2$-primary part of the isotropy character group; in particular, they can occur only when the local group has even order.
\end{remark}

\subsection{The Corrected Quantum Operator}
\label{sec:corrected-operator}

Just as established for manifolds in Chapter~1, the quantum operator must explicitly account for the transformation of half-forms under Hamiltonian flows to ensure the resulting observable is formally self-adjoint. For a real-valued quantizable observable $f \in C^\infty_{\mathrm{orb}}(X, \mathbb{R})$ (meaning $[X_f, P] \subset P$, so the flow preserves the polarization), the corrected operator acts on a section $s \otimes \nu \in \Gamma(X, L \otimes \delta_P)$ via:
\begin{equation}
\label{eq:corrected-operator}
Q_{\mathrm{meta}}(f)(s \otimes \nu) \;=\; \bigl(Q_{\mathrm{pre}}(f)s\bigr) \otimes \nu \;-\; i\hbar\, s \otimes \bigl(\mathcal{L}^{1/2}_{X_f}\nu\bigr),
\end{equation}
where $Q_{\mathrm{pre}}(f) = -i\hbar\nabla_{X_f} + f$ is the standard prequantum operator on $L$, and $\mathcal{L}^{1/2}_{X_f}$ denotes the Lie derivative on half-forms.

\subsubsection{Lie Derivative on Half-Forms.}
Because $f$ is quantizable, its Hamiltonian vector field $X_f$ preserves the polarization $P$ and hence preserves the annihilator bundle $P^o$ and its determinant $K_P=\det(P^o)$. The induced Lie derivative on $K_P$ is globally defined, but its scalar coefficient is local: after choosing a non-vanishing local frame $\mu \in \Gamma(K_P)$, write $\mathcal{L}_{X_f}\mu = (\operatorname{div}_{P,\mu} X_f)\mu$. Equivalently, if $\{\alpha^j\}$ is a local frame of $P^o$ with $\mu=\alpha^1\wedge\cdots\wedge\alpha^n$ and $\mathcal{L}_{X_f}\alpha^j = \sum_k A^j_{\ k}\alpha^k$, then $\operatorname{div}_{P,\mu} X_f = \mathrm{tr}(A)$. Under a change of frame this coefficient changes by the expected derivative of the transition function, so the resulting first-order Lie derivative on half-forms is global even though the displayed divergence coefficient is local. In a fixed local square-root frame $\nu_0$ with $\nu_0^{\otimes 2}=\mu$, we write this coefficient simply as $\operatorname{div}_P X_f$. Taking the square root of the Lie derivative on $K_P$ gives
\begin{equation}
\label{eq:half-form-lie-full}
2\bigl(\mathcal{L}^{1/2}_{X_f}\nu\bigr)\otimes\nu = \mathcal{L}_{X_f}(\nu\otimes\nu),
\end{equation}
for every local half-form $\nu$. In particular, for the reference frame $\nu_0$,
\begin{equation}
\label{eq:half-form-lie}
\mathcal{L}^{1/2}_{X_f}\nu_0 = \tfrac{1}{2}\bigl(\operatorname{div}_P X_f\bigr)\nu_0.
\end{equation}
Thus if a general local half-form is written as $a\nu_0$, then
\[
\mathcal{L}^{1/2}_{X_f}(a\nu_0)=\left(X_f(a)+\tfrac12 a\,\operatorname{div}_P X_f\right)\nu_0.
\]
Consequently, after choosing the local reference half-form $\nu_0$ and absorbing all scalar coefficients into the $L$-valued factor, so that the corrected section is written as $s\otimes\nu_0$, the corrected operator is represented on the $L$-coefficient by:
\begin{equation}
\label{eq:corrected-operator-local}
Q_{\mathrm{meta}}(f)(s \otimes \nu_0) \;=\; \left[ Q_{\mathrm{pre}}(f)s \;-\; \frac{i\hbar}{2}\,\operatorname{div}_P X_f\, s \right] \otimes \nu_0.
\end{equation}

\begin{proposition}[Formal Self-Adjointness of the Corrected Operator]
\label{prop:meta-self-adjoint}
Let \((X,\omega)\) be a boundaryless smooth effective symplectic orbifold with an orbifold polarization \(P\), a prequantum line bundle \((L,\nabla)\), and an orbifold half-form bundle \(\delta_P\). Let $f \in C^\infty_{\mathrm{orb}}(X, \mathbb{R})$ be a real-valued quantizable orbifold observable. No global quotient presentation is assumed. On the domain of smooth corrected polarized sections compactly supported in the relevant measure space (or on all smooth corrected polarized sections when that measure space is compact and boundaryless), the metaplectically corrected quantum operator $Q_{\mathrm{meta}}(f)$ is formally self-adjoint with respect to the natural corrected pairing in either of the following settings:
\begin{enumerate}
    \item the K\"ahler case, where the pairing is orbifold $L^2$-integration over $X$;
    \item a real polarization for which the leaf space carries a smooth measure-space structure (for example, a smooth orbifold quotient in a regular Lagrangian fibration) and the Hamiltonian flow of $f$ descends to that quotient with the usual integration-by-parts formula for compactly supported transverse densities.
\end{enumerate}
Without compact support, compactness, or explicit decay conditions, additional boundary terms at infinity may appear. Orbifolds with boundary, mirror boundary, or corner reflectors require separate boundary terms or boundary conditions and are not included in this proposition.
\end{proposition}

\begin{proof}
Let $\psi_1$ and $\psi_2$ be polarized sections of $L \otimes \delta_P$, compactly supported on the appropriate measure space $\mathcal{M}$ (namely $X$ in the K\"ahler case, and the leaf space in the real-polarization case). Their pairing is defined by integrating a canonical density $\Omega(\nu_1, \nu_2)$ over $\mathcal{M}$ when locally $\psi_j=s_j\otimes\nu_j$:
\[
\langle \psi_1, \psi_2 \rangle = \int_{\mathcal{M}} \langle s_1, s_2 \rangle_L \, \Omega(\nu_1, \nu_2).
\]
For a real polarization, after choosing the standard local transverse-density identification associated with the half-form bundle, $\Omega(\nu_1, \nu_2)$ is the transverse density determined by $\nu_1$ and $\nu_2$. For a K\"ahler polarization, $\Omega(\nu_1, \nu_2) = \langle \nu_1, \nu_2 \rangle_{\delta_P} \frac{\omega^n}{n!}$ is the density formed using the Hermitian metric on $\delta_P$ induced by the K\"ahler structure. In both cases, the density depends sesquilinearly on the half-forms, and in case~(2) we use the additional hypothesis that it defines a smooth measure on the quotient to which the relevant Hamiltonian flow descends.

The calculation is local, so choose a non-vanishing reference half-form $\nu_0$ and write $\psi_j=\sigma_j\otimes\nu_0$, absorbing the scalar half-form coefficients into the $L$-valued factors $\sigma_j$. Let $\Omega_0:=\Omega(\nu_0,\nu_0)$ and write $D:=\operatorname{div}_P X_f$ in this frame. By~\eqref{eq:corrected-operator-local},
\[
Q_{\mathrm{meta}}(f)(\sigma\otimes\nu_0)
=
\left(-i\hbar\nabla_{X_f}\sigma+f\sigma-\frac{i\hbar}{2}D\sigma\right)\otimes\nu_0.
\]
In the inner product $\langle Q_{\mathrm{meta}}(f)\psi_1, \psi_2 \rangle$, the scalar term $f$ is trivially self-adjoint. The derivative term gives:
\[
\langle -i\hbar\nabla_{X_f}\sigma_1 \otimes \nu_0, \psi_2 \rangle = i\hbar \int_{\mathcal{M}} \langle \nabla_{X_f} \sigma_1, \sigma_2 \rangle_L \, \Omega_0.
\]
Metric compatibility of $\nabla$ yields $X_f \langle \sigma_1, \sigma_2 \rangle_L = \langle \nabla_{X_f} \sigma_1, \sigma_2 \rangle_L + \langle \sigma_1, \nabla_{X_f} \sigma_2 \rangle_L$. Rearranging and integrating over $\mathcal{M}$, the key step is the integration-by-parts identity obtained from the vanishing of $\int_{\mathcal{M}}\mathcal{L}_{X_f}\bigl(\langle \sigma_1, \sigma_2\rangle_L \Omega_0\bigr)$. In the K\"ahler case, $\langle \sigma_1, \sigma_2\rangle_L\,\Omega_0$ is a top-form on the orbifold $X$, so Cartan's formula gives $\mathcal{L}_{X_f}(\langle \sigma_1, \sigma_2\rangle_L\,\Omega_0) = d(\iota_{X_f}(\langle \sigma_1, \sigma_2\rangle_L\,\Omega_0))$, whose integral vanishes by the orbifold Stokes' theorem (the singular strata have codimension $\geq 2$ and contribute no boundary terms). In the regular real-polarization case, the same vanishing is the assumed integration-by-parts property for the descended vector field acting on compactly supported transverse densities. Hence:
\[
\int_{\mathcal{M}} \langle \nabla_{X_f} \sigma_1, \sigma_2 \rangle_L \, \Omega_0 = -\int_{\mathcal{M}} \langle \sigma_1, \nabla_{X_f} \sigma_2 \rangle_L \, \Omega_0 -\int_{\mathcal{M}} \langle \sigma_1, \sigma_2 \rangle_L \mathcal{L}_{X_f}\Omega_0.
\]
Because $X_f$ is Hamiltonian, it preserves the symplectic volume in the K\"ahler case; in the regular real-polarization case, the relevant transverse density is the one supplied by the half-forms. Therefore the nontrivial variation of the pairing density is governed by the half-form Lie derivatives:
\[
\mathcal{L}_{X_f}\Omega_0
=
\Omega\bigl(\mathcal{L}^{1/2}_{X_f}\nu_0,\nu_0\bigr)
+\Omega\bigl(\nu_0,\mathcal{L}^{1/2}_{X_f}\nu_0\bigr)
=\operatorname{Re}(D)\,\Omega_0.
\]
Multiplying by $i\hbar$ and substituting this yields:
\[
\langle (-i\hbar\nabla_{X_f}\sigma_1) \otimes \nu_0, \psi_2 \rangle
=
\langle \psi_1, (-i\hbar\nabla_{X_f}\sigma_2) \otimes \nu_0 \rangle
- i\hbar \int_{\mathcal{M}} \langle \sigma_1, \sigma_2 \rangle_L \operatorname{Re}(D)\Omega_0.
\]
The auxiliary half-form correction terms evaluate to:
\begin{align*}
\langle -\tfrac{i\hbar}{2}D\psi_1, \psi_2 \rangle
&- \langle \psi_1, -\tfrac{i\hbar}{2}D\psi_2 \rangle \\
&= +i\hbar \int_{\mathcal{M}} \langle \sigma_1, \sigma_2 \rangle_L \operatorname{Re}(D)\Omega_0.
\end{align*}
(using the sesquilinearity of the metric $\langle \cdot, \cdot \rangle_L$, with the scalar function $\kappa = -\frac{i\hbar}{2}D$, so that locally\footnote{Recall that our convention throughout (see Proposition~\ref{prop:self-adjointness}) is that the inner product is antilinear in the first argument and linear in the second.} $\langle \kappa \sigma_1, \sigma_2\rangle_L - \langle \sigma_1, \kappa \sigma_2\rangle_L = (\bar{\kappa} - \kappa)\langle \sigma_1, \sigma_2\rangle_L = +i\hbar\operatorname{Re}(D)\langle \sigma_1, \sigma_2\rangle_L$).
Summing these equations identically cancels the divergence terms, establishing:
\[
\langle Q_{\mathrm{meta}}(f)\psi_1, \psi_2 \rangle = \langle \psi_1, Q_{\mathrm{meta}}(f)\psi_2 \rangle.
\]
To pass from manifolds to orbifolds, one performs the same calculation on each uniformizing chart $(V,G)$. The lifted Hamiltonian vector field, connection, Hermitian metric, and half-form bundle are all $G$-equivariant, and the Lie derivative on the lifted half-form bundle commutes with the local group action. Thus the chartwise operator commutes with the $G$-action:
\[
g^*\bigl(Q_{\mathrm{meta}}(f)\psi\bigr)=Q_{\mathrm{meta}}(f)(g^*\psi).
\]
Hence $Q_{\mathrm{meta}}(f)$ descends from the charts to an operator on orbifold sections. The local integration-by-parts argument is exactly the manifold one, and orbifold integration is obtained by integrating the corresponding $G$-invariant densities with the usual orbifold weights. Therefore the descended operator is formally self-adjoint with respect to the corrected pairing.
\end{proof}
\subsection{The Maslov Index and Spectral Shifts}
\label{sec:maslov-index}

For an orbifold loop represented by $(\tilde{\gamma},g)$, where $\tilde{\gamma}$ lies in a smooth lifted leaf segment inside a uniformizing chart, the Maslov index is computed from the smooth lifted path $\tilde{\gamma}$ exactly as in the manifold case whenever the real polarization is locally presented by a Lagrangian fibration or cotangent-type model, so that the usual notion of caustic crossing is available~\cite{maslov}. In that setting, orbifold singularities do not themselves create additional smooth caustics on the chart; the orbifold structure contributes instead a separate equivariant phase through the action of $g$ on the endpoint fiber of $\delta_P$.

For a real polarization on a boundaryless smooth effective symplectic orbifold $X$, the metaplectic correction modifies the Bohr--Sommerfeld condition of Section~\ref{sec:real-polarizations}. The additional holonomy contribution from $\delta_P$ around an orbifold loop combines two effects: the Maslov phase from caustic crossings, and the equivariant phase coming from the orbifold identification at the endpoint.

\subsubsection{The Orbifold Bohr--Sommerfeld--Maslov Rule.}\label{sec:corrected-BS}
Under the standing boundaryless smooth effective hypotheses, the corrected condition accounts for both the Maslov phase and the extra equivariant phases coming from the orbifold structure:

\begin{theorem}[Local Chart Form of the Orbifold Bohr--Sommerfeld--Maslov Rule]
\label{thm:orbifold-BS-corrected}
Let \((X,\omega)\) be a boundaryless smooth effective symplectic orbifold with a real orbifold polarization \(P\), a prequantum line bundle \(L\), and an orbifold half-form bundle \(\delta_P\). No compactness or global quotient presentation is assumed. Assume that near the path under consideration the polarization is locally modeled by a Lagrangian fibration (for example, a cotangent-type polarization), so that the Maslov index of a lifted path is defined as in the manifold case. Let \((V,G,\pi)\) be a uniformizing chart, let \(\tilde{\mathcal{L}}\subset V\) be a connected lifted Lagrangian leaf segment, and let \((\tilde{\gamma},g)\) be a \emph{single-chart representative} of an orbifold loop:
\[
\tilde{\gamma}\colon[0,1]\to\tilde{\mathcal{L}},\qquad
\tilde{\gamma}(0)=x_0,\qquad
\tilde{\gamma}(1)=g\cdot x_0,
\]
where \(g\in G\) preserves \(\tilde{\mathcal{L}}\). Choose a local unitary frame \(s_0\) of the lifted prequantum line bundle \(L_V\) on the chart so that the lifted connection is written as \(\nabla=d-\frac{i}{\hbar}\theta\) for a \(G\)-invariant local symplectic potential \(\theta\), and choose a base-aligned reference half-form frame \(\nu_0\) of \(\delta_{P,V}\) along \(\tilde{\gamma}\), parallel on each smooth segment between caustics. Because \(\theta\) is \(G\)-invariant, the lifted action on \(s_0\) preserves the connection \(1\)-form and is therefore given by a constant phase on the connected chart. Write
\[
\Phi^L_g\bigl(s_0(x)\bigr)=e^{2\pi i\,\alpha_{s_0}(g)}\,s_0(g\cdot x),
\qquad
\Phi^\delta_g\bigl(\nu_0(x_0)\bigr)=e^{2\pi i\,\beta_{\nu_0}(g;x_0)}\,\nu_0(g\cdot x_0),
\]
with \(\alpha_{s_0}(g),\beta_{\nu_0}(g;x_0)\in\mathbb{R}/\mathbb{Z}\). For this single-chart representative, the combined corrected holonomy
\[
\bigl(\Phi^L_{g^{-1}}\circ \mathrm{Par}^{L_V}_{\tilde{\gamma}}\bigr)\otimes
\bigl(\Phi^\delta_{g^{-1}}\circ \mathrm{Par}^{\delta_{P,V}}_{\tilde{\gamma}}\bigr)
\in \operatorname{End}\bigl((L_V\otimes\delta_{P,V})_{x_0}\bigr)
\]
is the identity if and only if
\begin{equation}
\label{eq:BS-corrected}
\frac{1}{2\pi\hbar} \int_{\tilde{\gamma}} \theta
\;-\;
\alpha_{s_0}(g)
\;-\;
\beta_{\nu_0}(g;x_0)
\;-\;
\frac{\mu(\tilde{\gamma})}{4}
\;\in\;
\mathbb{Z},
\end{equation}
where \(\mu(\tilde{\gamma})\) is the Maslov index of the smooth lifted path. Consequently, any non-zero polarized orbifold section of \(L\otimes\delta_P\) along a leaf must satisfy \eqref{eq:BS-corrected} for every orbifold loop that admits such a single-chart representative. This theorem is only the local chart expression of the condition; loops crossing several charts must be evaluated by composing the corresponding local parallel transports, transition maps, and endpoint isotropy identifications.
\end{theorem}

\begin{proof}
In the local frame $s_0$, the first factor contributes the prequantum phase
\[
\exp\left(2\pi i\left(\frac{1}{2\pi\hbar}\int_{\tilde{\gamma}}\theta-\alpha_{s_0}(g)\right)\right).
\]
For the half-form bundle, parallel transport along the smooth lifted path contributes the Maslov phase $e^{-i\pi\mu(\tilde{\gamma})/2}$ (as established in the manifold theory of Chapter~1; each caustic crossing rotates the half-form by $e^{-i\pi/2}$), while the equivariant identification contributes $e^{-2\pi i\,\beta_{\nu_0}(g;x_0)}$. Thus the second factor contributes
\[
\exp\left(2\pi i\left(-\beta_{\nu_0}(g;x_0)-\frac{\mu(\tilde{\gamma})}{4}\right)\right).
\]
The tensor-product holonomy is the product of these two phases, so it is trivial exactly when~\eqref{eq:BS-corrected} holds. Although the individual quantities depend on the chosen local frames, the combined orbifold holonomy condition is intrinsic, so the congruence class modulo \(\mathbb{Z}\) is frame-independent. If a global polarized orbifold section exists, its local lift is preserved by leafwise transport and by the endpoint equivariant identifications, so the corrected holonomy is trivial for every such local representative. The converse requires the global holonomy criterion below, not merely a single local equation.
\end{proof}

\begin{corollary}[Global Leafwise Corrected Holonomy Criterion]
\label{cor:global-BS-maslov}
Let \((X,\omega)\) be a boundaryless smooth effective symplectic orbifold with a real orbifold polarization \(P\), a prequantum line bundle \(L\), and an orbifold half-form bundle \(\delta_P\). Let \(\mathcal{L}\) be a connected Lagrangian orbifold leaf of \(P\). Assume that \(L|_{\mathcal{L}}\) is equipped with its flat prequantum leafwise connection, that \(\delta_P|_{\mathcal{L}}\) has the usual Maslov transport along smooth lifted leaf segments, and that these local transports are compatible with chart changes and isotropy actions. No global quotient presentation is assumed. They define a corrected leafwise holonomy representation
\[
\operatorname{Hol}^{\mathrm{corr}}_{\mathcal{L}}
\colon
\pi_1^{\mathrm{orb}}(\mathcal{L},x_0)
\longrightarrow U(1)
\]
for the line \(L\otimes\delta_P\). A non-zero global polarized orbifold section of \((L\otimes\delta_P)|_{\mathcal{L}}\) exists if and only if \(\operatorname{Hol}^{\mathrm{corr}}_{\mathcal{L}}\) is trivial. Equivalently, it is enough to check triviality on any generating set of the leafwise orbifold fundamental group; whenever a generator is represented inside one uniformizing chart, its condition is exactly \eqref{eq:BS-corrected}.
\end{corollary}

\begin{proof}
This is the flat line-bundle holonomy criterion applied to the corrected leafwise transport. If a non-zero polarized orbifold section exists, parallel transport around any leafwise orbifold loop must return its value to itself after applying the endpoint isotropy and half-form identifications, so \(\operatorname{Hol}^{\mathrm{corr}}_{\mathcal{L}}\) is trivial.

Conversely, assume the corrected holonomy representation is trivial. Fix a basepoint and a non-zero vector in \((L\otimes\delta_P)_{x_0}\). Transport this vector along lifted paths in the leaf, using the prequantum flat transport, the half-form/Maslov transport, and the orbifold transition maps. Trivial corrected holonomy makes the result independent of the chosen path and of the chart representative of the same orbifold path. The resulting local parallel sections agree on overlaps by uniqueness of parallel transport and the equivariance of the orbifold bundles, hence glue to a non-zero global polarized orbifold section along \(\mathcal{L}\).
\end{proof}

\subsubsection{Comparison With the Manifold Result.}
When $G$ is trivial (no orbifold structure), we have $\alpha_{s_0}(g)=\beta_{\nu_0}(g;x_0)=0$ and $\tilde{\gamma}=\gamma$, directly recovering the traditional manifold Bohr--Sommerfeld--Maslov rule $\frac{1}{2\pi\hbar}\oint_\gamma \theta - \frac{\mu(\gamma)}{4} \in \mathbb{Z}$. The extra orbifold terms are therefore genuine orbifold phenomena: they record the additional equivariant phases introduced when loops close only after an orbifold identification. Together with the Maslov term, they produce spectra incorporating both the familiar zero-point shift and the fractional phase shifts generated by orbifold singular strata. The examples in Chapter~6 compute these effects, or their absence, in the circle quotient, planar cone, football/orbisphere, and dihedral cone models.

\chapter{Examples of Geometric Quantization on Orbifolds}
\label{sec:examples-gq-orbifolds}

In this chapter, we apply the general formalism developed in the previous chapter to several concrete examples of increasing complexity.

\section{\texorpdfstring{Geometric Quantization of the Free Quotient $S^1/\mathbb{Z}_n$}{Geometric Quantization of the Free Quotient S1/Zn}}
\label{subsec:gq-s1-zn}

\subsection{Symplectic Orbifolds and Observables}
This example is a smooth free-quotient baseline for the geometric-quantization formalism~\cite{Woodhouse1992,Sniatycki1980}. It has no singular strata or fixed-point twisted sectors; the only nontrivial quantum datum left by the quotient is flat holonomy, equivalently the cover-isotypic bookkeeping used in Chapter~4.

Let $S^1$ be parameterized by the angular covering coordinate $\theta_c\in[0,2\pi)$. The group $\mathbb{Z}_n=\langle g\mid g^n=1\rangle$ acts freely by
\[
g\cdot\theta_c=\theta_c+\tfrac{2\pi}{n}.
\]
The quotient stack is $[S^1/\mathbb{Z}_n]$, which is equivalent to the smooth quotient circle. Because the action is free, its orbifold fundamental group is the ordinary fundamental group of that quotient:
\[
\pi_1^{\mathrm{orb}}([S^1/\mathbb{Z}_n])\cong \mathbb{Z}.
\]

\subsubsection{Phase Space and Symplectic Form.}
The classical phase space is the cotangent cylinder
\[
Y:=T^*S^1\cong S^1\times\mathbb{R},
\]
with coordinates $(\theta_c,p)$. The Liouville $1$-form is $\lambda=p\,d\theta_c$, so
\[
\omega=d\lambda=dp\wedge d\theta_c.
\]
The $\mathbb{Z}_n$-action lifts to $Y$ by
\[
g\cdot(\theta_c,p)=\bigl(\theta_c+\tfrac{2\pi}{n},p\bigr),
\]
and preserves $\omega$. Hence
\[
X:=[Y/\mathbb{Z}_n]
\]
is a symplectic orbifold in the benign sense of a free quotient, hence equivalent to a smooth symplectic cylinder.

\subsubsection{Orbifold Observables and Hamiltonian Flow.}
The smooth orbifold observables are precisely the $\mathbb{Z}_n$-invariant functions on $Y$:
\[
C^\infty_{\mathrm{orb}}(X)=C^\infty(Y)^{\mathbb{Z}_n}.
\]
For the momentum observable $f=p$, the standard Hamiltonian convention,
\[
\iota_{X_p}\omega=-dp,
\]
gives
\[
X_p=\partial_{\theta_c}.
\]
This vector field is $\mathbb{Z}_n$-equivariant, so it descends to a well-defined Hamiltonian flow on $X$.

\subsection{Connections, Curvature, and Prequantization}
A prequantum structure on $(X,\omega)$ is an orbifold Hermitian line bundle with unitary connection satisfying~\cite{Kostant1970,Souriau1970}
\[
F_\nabla=-\frac{i}{\hbar}\omega.
\]
For a global quotient, the general theory identifies this with a $\mathbb{Z}_n$-equivariant Hermitian line bundle on the cover.

\subsubsection{Integrality Condition and Flat Ambiguity.}
Because the $\mathbb{Z}_n$-action on $Y$ is free, the orbifold $X=[Y/\mathbb{Z}_n]$ is equivalent to the smooth quotient manifold $Y/\mathbb{Z}_n$, which is again a cylinder~\cite{AdemLeidaRuan}. Hence
\[
H^2_{\mathrm{st}}(X;\mathbb{Z})\cong H^2(Y/\mathbb{Z}_n;\mathbb{Z})=0.
\]
Moreover, $\omega=dp\wedge d\theta_c=d(p\,d\theta_c)$ descends as an exact $2$-form on the quotient. Therefore the orbifold integrality condition (Theorem~\ref{thm:orbifold-existence}) is automatic: there is no flux quantization constraint. However, the general theory also emphasizes that fixing the curvature does \emph{not} determine the prequantum connection uniquely. In the present example, $H^2_{\mathrm{st}}(X;\mathbb{Z})=0$ means there is no nontrivial topological line-bundle class, but there remains a flat unitary connection ambiguity on the quotient cylinder, equivalently on its circle factor, classified by
\[
\mathrm{Hom}\bigl(\pi_1^{\mathrm{orb}}(X),U(1)\bigr)\cong U(1).
\]
This represents holonomy choices for the prequantum connection rather than distinct smooth line-bundle topologies. We parametrize the flat-connection sector by $\alpha\in\mathbb{R}/\mathbb{Z}$, using the convention that after the flat term is gauged away on a cut interval the endpoint gluing phase is $e^{2\pi i\alpha}$. With the connection convention $\nabla=d-iA$, the parallel-transport holonomy of the flat part itself is the inverse phase $e^{-2\pi i\alpha}$.

\subsubsection{A Convenient Equivariant Model.}
Choose the trivial Hermitian line bundle
\[
L_Y=Y\times\mathbb{C}
\]
with trivial $\mathbb{Z}_n$-action on the fiber:
\[
g\cdot(\theta_c,p,z)=\bigl(\theta_c+\tfrac{2\pi}{n},p,z\bigr).
\]
On $L_Y$ define the unitary connection
\[
\nabla_\alpha
=
d-\frac{i}{\hbar}\bigl(p-\hbar n\alpha\bigr)\,d\theta_c.
\]
Its curvature is
\[
F_{\nabla_\alpha}
=
d\!\left(-\frac{i}{\hbar}(p-\hbar n\alpha)\,d\theta_c\right)
=
-\frac{i}{\hbar}\,dp\wedge d\theta_c
=
-\frac{i}{\hbar}\omega.
\]
Because $d\theta_c$ and $p$ are $\mathbb{Z}_n$-invariant, the connection is equivariant and descends to $X$. Different values of $\alpha$ modulo $\mathbb Z$ give inequivalent descended flat sectors.

\subsubsection{Prequantum Hilbert Space and Operator.}
By Section~\ref{sec:prequantum-hilbert-space}, the orbifold prequantum Hilbert space is the invariant subspace
\[
\mathcal{H}_{\mathrm{pre}}(X;\alpha)=L^2(Y,L_Y)^{\mathbb{Z}_n},
\]
with the usual orbifold normalization factor $1/n$ in the inner product. For $f=p$, the prequantum operator is
\[
Q_{\mathrm{pre}}(p)=-i\hbar\nabla_{X_p}+p.
\]
Since $X_p=\partial_{\theta_c}$,
\[
Q_{\mathrm{pre}}(p)
=
-i\hbar\partial_{\theta_c}+\hbar n\alpha.
\]
On the universal cover $\mathbb{R}\times\mathbb{R}$, or on a cut fundamental domain in $\theta_c$, one may use the local change of unitary frame $s'=e^{-i n\alpha\theta_c}\,s$ (equivalently, $\widetilde{\psi}=e^{in\alpha\theta_c}\psi$) to absorb the flat term. In that description the operator becomes $-i\hbar\partial_{\theta_c}$ and the information is transferred into the boundary condition
\[
\widetilde{\psi}(\theta_c+\tfrac{2\pi}{n})=e^{2\pi i\alpha}\widetilde{\psi}(\theta_c).
\]
This boundary condition is precisely the equivariant phase encoded by the lift $\Phi_g$ in the orbifold connection formalism of Section~\ref{sec:orbifold-connections}. For generic $\alpha$ this gauge factor is not single-valued on the covering circle $S^1$, so this is only a local/fundamental-domain description rather than a global gauge on $Y$. When $\alpha=q/n$, however, the factor $e^{in\alpha\theta_c}=e^{iq\theta_c}$ is single-valued on $S^1$. It is still not $\mathbb{Z}_n$-equivariant unless $q\equiv 0 \pmod n$, so it changes the presentation on the cover rather than trivializing the orbifold holonomy. In that gauge the sections transform by the character $g\mapsto e^{2\pi i q/n}$, reproducing the discrete $\mathbb{Z}_n$-isotypic sectors of Chapter~4 as an alternative cover-based presentation.

\subsection{Polarizations and Quantum Hilbert Space}
To obtain the physical quantum theory, we impose a polarization. In this example, we examine both the vertical polarization (to construct the quantum Hilbert space and quantum operators) and the horizontal polarization (to derive the Bohr--Sommerfeld momentum spectrum).

\subsubsection{Vertical Real Polarization and the Hilbert Space.}
The vertical real polarization is
\[
\mathcal{P}_{\mathrm{vert}}=\langle\partial_p\rangle.
\]
Polarized sections are those satisfying $\nabla_{\partial_p}\psi=0$. Since the connection $\nabla_\alpha$ has no $dp$ component, $\nabla_{\partial_p}=\partial_p$, so these are functions $\psi(\theta_c)$ independent of $p$. They are therefore constant along the noncompact momentum leaves. Consequently, a nonzero polarized section is not an element of the prequantum space $\mathcal{H}_{\mathrm{pre}}(X;\alpha)$: the Liouville measure is infinite in the $p$-direction. As in the usual Schr\"odinger realization for a cotangent bundle, one passes from the raw prequantum $L^2$-space to the leaf space of the polarization, here the quotient circle $S^1/\mathbb{Z}_n$, with its induced transverse measure $d\theta_c$.

Concretely, one may work either with $\mathbb{Z}_n$-invariant sections on the covering circle, represented on a cut fundamental interval by
\[
\mathcal{H}_{\mathrm{vert}}(X;\alpha)
\cong
L^2([0,2\pi/n],d\theta_c),
\]
or, after the local frame change described above, with the equivalent quasi-periodic model
\[
\widetilde{\mathcal{H}}_{\mathrm{vert}}(X;\alpha)
\cong
L^2([0,2\pi/n],d\theta_c).
\]
Pointwise endpoint conditions are instead imposed on the Sobolev domains of the differential operators below. The first presentation keeps the bundle trivial and places the flat connection in the operator; the second presentation moves the same data into the quasi-periodic boundary condition. In the invariant picture a convenient orthonormal basis is
\[
\psi_\ell(\theta_c)=\sqrt{\frac{n}{2\pi}}\,e^{in\ell\theta_c}, \qquad \ell\in\mathbb{Z},
\]
while in the quasi-periodic picture it becomes
\[
\widetilde{\psi}_\ell(\theta_c)=\sqrt{\frac{n}{2\pi}}\,e^{in(\ell+\alpha)\theta_c}, \qquad \ell\in\mathbb{Z}.
\]
Acting on these states, the prequantum momentum operator $Q_{\mathrm{pre}}(p)=-i\hbar\partial_{\theta_c}+\hbar n\alpha$ yields
\[
Q_{\mathrm{pre}}(p)\psi_\ell
=
\hbar n(\ell+\alpha)\psi_\ell.
\]
Equivalently, in the quasi-periodic model one has
\[
-i\hbar\partial_{\theta_c}\widetilde{\psi}_\ell
=
\hbar n(\ell+\alpha)\widetilde{\psi}_\ell.
\]
For the special values $\alpha=q/n$, this cover-based picture reproduces the discrete $\mathbb{Z}_n$-isotypic sectors discussed in Chapter~4; for generic $\alpha$ it gives the full $U(1)$ family of flat sectors on the quotient circle.

\subsubsection{Horizontal Real Polarization and Bohr--Sommerfeld Leaves.}
Take the horizontal real polarization
\[
\mathcal{P}_{\mathrm{hor}}=\langle\partial_{\theta_c}\rangle.
\]
Its leaves are the circles $p=\mathrm{const}$, which descend to quotient circles in $X$. Each such leaf has fundamental group $\mathbb{Z}$, generated by the orbifold loop represented by the path
\[
\tilde{\gamma}(t)=\bigl(\theta_0+\tfrac{2\pi t}{n},p\bigr), \qquad 0\le t\le 1,
\]
together with the deck transformation $g$. Thus it is enough to test the Bohr--Sommerfeld condition on this generator. In the chosen trivializing frame,
\[
\nabla_\alpha=d-\frac{i}{\hbar}\theta_\alpha,
\qquad
\theta_\alpha:=(p-\hbar n\alpha)\,d\theta_c,
\]
and $d\theta_\alpha=\omega$. In the general theory developed previously, the Bohr--Sommerfeld rule \eqref{eq:BS-orbifold} includes a local orbifold phase $\alpha_{s_0}(g)$. To avoid notational confusion with the global flat-sector parameter $\alpha$ of this specific example, we temporarily denote this local orbifold phase by $\widetilde{\alpha}_{s_0}(g)$. 

Geometrically, $\widetilde{\alpha}_{s_0}(g)$ encodes the fractional phase acquired by the chosen local unitary frame $s_0$ under the equivariant lift $\Phi_g$ of the orbifold group action, defined by the transformation rule 
\[
\Phi_g\bigl(s_0(x)\bigr)=e^{2\pi i\,\widetilde{\alpha}_{s_0}(g)}\,s_0(g\cdot x).
\]
In general, this local phase compensates the equivariant action on the frame and, if necessary, any failure of the chosen local symplectic potential to be strictly group-invariant, ensuring that the combined holonomy around a closed orbifold loop is gauge-invariant. In the present model, however, $\theta_\alpha$ is already $\mathbb{Z}_n$-invariant. 

In our current explicit model, the prequantum line bundle $L_Y=Y\times\mathbb{C}$ was defined with a trivial $\mathbb{Z}_n$-action on the fibers. This implies that our chosen trivializing frame $s_0=1$ is strictly invariant under the lift, yielding $\Phi_g(1)=1$. Consequently, the local orbifold phase vanishes:
\[
\widetilde{\alpha}_{s_0}(g)=0.
\]
Instead, the nontrivial topological data of the bundle over the quotient is carried directly by the explicitly invariant shifted potential $\theta_\alpha$. The Bohr--Sommerfeld condition \eqref{eq:BS-orbifold} therefore simplifies to the bare line integral of $\theta_\alpha$:
\[
\frac{1}{2\pi\hbar}\int_{\tilde{\gamma}}\theta_\alpha
\;-\; \widetilde{\alpha}_{s_0}(g)
\;=\;
\frac{1}{2\pi\hbar}\cdot \frac{2\pi}{n}(p-\hbar n\alpha)
\;=\;
\frac{p}{\hbar n}-\alpha
\;\in\;
\mathbb{Z}.
\]
Hence the allowed momenta are
\[
p=\hbar n(\ell+\alpha), \qquad \ell\in\mathbb{Z}.
\]
If we specialize to $\alpha=q/n$, this becomes
\[
p=\hbar(n\ell+q),
\]
which is exactly the momentum quantization found in Chapter~4.

\subsection{Metaplectic Correction}
For the vertical polarization used to realize the Hilbert space, the general theory replaces $L$ by $L\otimes\delta_{\mathcal{P}}$ (Definition~\ref{def:half-form-bundle}), where $\delta_{\mathcal{P}}^{\otimes 2}\cong K_{\mathcal{P}}$. The obstruction is the mod-$2$ class \(c_1(K_{\mathcal{P}})\bmod 2\) from \eqref{eq:metalinear-obstruction} in Section~\ref{sec:obstruction}. In the present free-quotient example this obstruction vanishes because the polarization canonical bundle is explicitly equivariantly trivial, as shown next.

\subsubsection{Half-Forms for the Vertical Polarization.}
For
\[
\mathcal{P}_{\mathrm{vert}}=\langle\partial_p\rangle,
\]
the annihilator is
\[
\mathcal{P}_{\mathrm{vert}}^{o}=\langle d\theta_c\rangle,
\]
so the polarization canonical bundle is
\[
K_{\mathcal{P}_{\mathrm{vert}}}
=
\det\bigl(\mathcal{P}_{\mathrm{vert}}^{o}\bigr)
=
\langle d\theta_c\rangle.
\]
The form $d\theta_c$ is nowhere vanishing and $\mathbb{Z}_n$-invariant, so $K_{\mathcal{P}_{\mathrm{vert}}}$ is equivariantly trivial. Hence a global half-form bundle exists and may be chosen to be the trivial square root
\[
\delta_{\mathcal{P}_{\mathrm{vert}}}
=
\langle \sqrt{d\theta_c}\rangle,
\qquad
\bigl(\sqrt{d\theta_c}\bigr)^{\otimes 2}=d\theta_c,
\]
with trivial $\mathbb{Z}_n$-action.

For polarized half-forms, the coefficient is again independent of $p$, but now the product of a half-form with its conjugate gives the transverse density $|\psi(\theta_c)|^2\,|d\theta_c|$ on the leaf space. Thus the metaplectically corrected Hilbert space is the honest $L^2$-space
\[
\mathcal{H}_{\mathrm{meta}}(X;\alpha)
\cong
L^2(S^1/\mathbb{Z}_n,|d\theta_c|)
\cong
L^2([0,2\pi/n],d\theta_c),
\]
or equivalently the quasi-periodic realization obtained from the local frame change above.

\subsubsection{Corrected Operator and Vanishing of the Extra Phase.}
The corrected operator is
\[
Q_{\mathrm{meta}}(p)(\psi\otimes\sqrt{d\theta_c})
=
\bigl(Q_{\mathrm{pre}}(p)\psi\bigr)\otimes\sqrt{d\theta_c}
-i\hbar\,\psi\otimes\mathcal{L}^{1/2}_{X_p}\sqrt{d\theta_c}.
\]
Since $X_p=\partial_{\theta_c}$ and
\[
\mathcal{L}_{\partial_{\theta_c}}d\theta_c=0,
\]
the polarization divergence is zero, so the half-form Lie derivative vanishes:
\[
\mathcal{L}^{1/2}_{X_p}\sqrt{d\theta_c}=0.
\]
Therefore
\[
Q_{\mathrm{meta}}(p)=Q_{\mathrm{pre}}(p)
\]
on polarized sections. Promoting the invariant functions $\psi_\ell$ to half-form sections, we define the basis states
\[
\Psi_\ell=\psi_\ell(\theta_c)\otimes\sqrt{d\theta_c}
=
\sqrt{\frac{n}{2\pi}}\,e^{in\ell\theta_c}\otimes\sqrt{d\theta_c},
\qquad \ell\in\mathbb{Z}.
\]
On this basis we obtain
\[
Q_{\mathrm{meta}}(p)\Psi_\ell
=
\hbar n(\ell+\alpha)\,\Psi_\ell.
\]

\subsubsection{Corrected Bohr--Sommerfeld Rule.}
For the horizontal polarization, the relevant half-form bundle must be computed separately. Since
\[
\mathcal{P}_{\mathrm{hor}}=\langle\partial_{\theta_c}\rangle,
\qquad
\mathcal{P}_{\mathrm{hor}}^{o}=\langle dp\rangle,
\]
the corresponding polarization canonical bundle is $K_{\mathcal{P}_{\mathrm{hor}}}=\langle dp\rangle$. The form $dp$ is nowhere vanishing and $\mathbb{Z}_n$-invariant, so $K_{\mathcal{P}_{\mathrm{hor}}}$ is also equivariantly trivial and admits the trivial square root $\delta_{\mathcal{P}_{\mathrm{hor}}}=\langle\sqrt{dp}\rangle$ with trivial $\mathbb{Z}_n$-action. For the horizontal leaves used above, the metaplectically corrected Bohr--Sommerfeld rule (Theorem~\ref{thm:orbifold-BS-corrected}) therefore introduces no extra equivariant phase. Moreover, the horizontal leaves are smooth circles with no caustics, and the polarization is trivially preserved by translations, so the Maslov index components also vanish (Section~\ref{sec:maslov-index}):
\[
\beta_{\nu_0}(g;x_0)=0,
\qquad
\mu(\tilde{\gamma})=0.
\]
Thus the corrected Bohr--Sommerfeld condition is the same as the uncorrected one. The corresponding quantum Hilbert space consists of distributional sections supported entirely on these quantized leaves, ensuring that the momentum spectrum remains
\[
p=\hbar n(\ell+\alpha).
\]

\subsection{Energy Spectrum and Consistency}
With the vertical polarization, the momentum observable $p$ is quantizable and yields the self-adjoint operator $Q_{\mathrm{meta}}(p)$ above. In the invariant realization it acts as $-i\hbar\partial_{\theta_c}+\hbar n\alpha$ on the periodic Sobolev domain
\[
H^1_{\mathrm{per}}([0,2\pi/n])
=
\left\{
\psi\in H^1([0,2\pi/n])
\;\middle|\;
\psi(2\pi/n)=\psi(0)
\right\}.
\]
Equivalently, in the quasi-periodic realization it is the standard derivative operator $-i\hbar\partial_{\theta_c}$ on the domain
\[
\left\{
\widetilde{\psi}\in H^1([0,2\pi/n])
\;\middle|\;
\widetilde{\psi}(2\pi/n)=e^{2\pi i\alpha}\widetilde{\psi}(0)
\right\}.
\]
Let $I=MR^2$ denote the inertial constant used in Chapter~4. The classical Hamiltonian is then
\[
H_{\mathrm{cl}}=\frac{p^2}{2I}
\]
and its Hamiltonian vector field does \emph{not} preserve the vertical polarization, since $X_{H_{\mathrm{cl}}}=(p/I)\partial_{\theta_c}$ and $[X_{H_{\mathrm{cl}}},\partial_p]=-(1/I)\partial_{\theta_c}\notin\mathcal{P}_{\mathrm{vert}}$. Thus $H_{\mathrm{cl}}$ is not a directly quantizable first-order Kostant--Souriau observable in this polarization. Since the momentum $p$ itself is quantizable, the physical free Hamiltonian is introduced instead as the second-order functional-calculus operator
\[
\widehat{H}:=\frac{1}{2I}\,Q_{\mathrm{meta}}(p)^2.
\]
Equivalently, this operator is associated with the classical function $p^2/(2I)$ through the spectral theorem for $Q_{\mathrm{meta}}(p)$; in the quasi-periodic realization it is the Laplacian $-\hbar^2\partial_{\theta_c}^2/(2I)$. While the momentum operator is self-adjoint on $H^1$, this second-order Hamiltonian is self-adjoint on the smaller Sobolev domain
\[
\left\{
\widetilde{\psi}\in H^2([0,2\pi/n])
\;\middle|\;
\widetilde{\psi}(2\pi/n)=e^{2\pi i\alpha}\widetilde{\psi}(0),\ 
\widetilde{\psi}'(2\pi/n)=e^{2\pi i\alpha}\widetilde{\psi}'(0)
\right\}.
\]
Acting on the basis $\Psi_\ell$, it gives the quantum energy levels
\[
E_{\ell,\alpha}
=
\frac{1}{2I}\bigl(\hbar n(\ell+\alpha)\bigr)^2,
\qquad
\ell\in\mathbb{Z}.
\]
For the discrete values $\alpha=q/n$, this becomes
\[
E_{\ell}^{(q)}
=
\frac{\hbar^2}{2I}(n\ell+q)^2,
\]
exactly as in Chapter~4. For generic $\alpha$, the flat sector lifts the usual $\pm p$ degeneracy; the degeneracy survives only when $2\alpha\in\mathbb{Z}$. This is fully consistent with the general theory: the curvature condition ensures the operator commutation relations (Lemma~\ref{lem:fundamental-commutation}), the remaining quantum data are carried by the flat connection, the Bohr--Sommerfeld leaves (Proposition~\ref{prop:orbifold-BS}) determine the allowed momenta, and the metaplectic correction is present but contributes no additional shift in this one-dimensional example.

\section{\texorpdfstring{Geometric Quantization of the Planar Cone $\mathbb{C}/\mathbb{Z}_n$}{Geometric Quantization of the Planar Cone R2/Zn}}
\label{subsec:gq-r2-zn}

The planar cone $\mathbb{C}/\mathbb{Z}_n$ is the simplest non-compact orbifold with an isolated singularity~\cite{DeserJackiw1988,KayStuder1991,MaratheMartucci1985}. In contrast with compact examples such as the weighted sphere $S^2(n,m)$, prequantization here is not constrained by a real integrality condition: the relevant orbifold de Rham cohomology of phase space vanishes. The nontrivial data are instead torsion orbifold line bundles, equivalently flat orbifold twists determined by the isotropy character at the apex. Geometric quantization therefore reproduces the same twisted sectors that appeared earlier in the Schr\"odinger treatment, but now from intrinsic symplectic and bundle-theoretic input.

\subsection{Symplectic Orbifold Structure}

Let the covering configuration space be $Q_{\mathrm c}=\mathbb{C}$ with Cartesian coordinates $(x,y)$ or polar coordinates $(r,\phi)$, where $r\ge 0$ and $\phi\in[0,2\pi)$. The group $\mathbb{Z}_n=\langle g\mid g^n=1\rangle$ acts by rotation:
\[
g\cdot(r,\phi)=\bigl(r,\phi+\tfrac{2\pi}{n}\bigr),
\qquad
g\cdot(x+iy)=e^{2\pi i/n}(x+iy).
\]
The orbifold configuration space is
\[
Q=[\mathbb{C}/\mathbb{Z}_n],
\]
topologically a cone whose apex carries isotropy group $\mathbb{Z}_n$.

Its phase space is the cotangent orbifold
\[
X:=T^*Q=[T^*\mathbb{C}/\mathbb{Z}_n].
\]
On the covering space $T^*\mathbb{C}\cong\mathbb{R}^4$, with canonical coordinates $(x,y,p_x,p_y)$, the Liouville form and symplectic form are
\[
\lambda=p_x\,dx+p_y\,dy,
\qquad
\omega=d\lambda=dp_x\wedge dx+dp_y\wedge dy.
\]
Both are invariant under the diagonal rotation on position and momentum, so they descend to the orbifold phase space. On the smooth locus $r>0$, polar cotangent coordinates $(r,\phi,p_r,p_\phi)$ satisfy
\[
\lambda=p_r\,dr+p_\phi\,d\phi,
\qquad
\omega=dp_r\wedge dr+dp_\phi\wedge d\phi.
\]

The cotangent-lifted $\mathbb{Z}_n$-action fixes only the zero covector over the cone point. Thus the unique singular point of phase space is the orbit of
\[
(x,y,p_x,p_y)=(0,0,0,0).
\]
If $r=0$ but $(p_x,p_y)\neq(0,0)$, the isotropy is trivial. This makes the local orbifold model especially simple: near the singular point, $\omega$ is already in Darboux form and the isotropy acts linearly by a symplectic rotation, exactly as in the equivariant Darboux theorem.

Classical observables on the orbifold are the $\mathbb{Z}_n$-invariant smooth functions on the cover:
\[
C^\infty_{\mathrm{orb}}(X)=C^\infty(T^*\mathbb{C})^{\mathbb{Z}_n}.
\]
Examples include the angular momentum
\[
L_z=xp_y-yp_x=p_\phi,
\]
the radial coordinate $r^2=x^2+y^2$, and the kinetic energy $(p_x^2+p_y^2)/(2M)$. By contrast, the individual coordinates $x,y,p_x,p_y$ are not invariant for $n\ge 2$ and therefore do not define orbifold observables.

\subsection{Prequantization and Sector Classification}

We now construct a prequantum line bundle $(L,\nabla)$ with curvature
\[
F_\nabla=-\frac{i}{\hbar}\omega.
\]
Since $T^*\mathbb{C}\cong\mathbb{R}^4$ is contractible,
\[
H^2(\mathbb{R}^4;\mathbb{Z})=0.
\]
For the global quotient presentation \(X=[T^*\mathbb{C}/\mathbb{Z}_n]\cong[\mathbb{R}^4/\mathbb{Z}_n]\), stack cohomology is equivariant cohomology~\cite{AdemMilgram2004,Kordyukov2011}:
\[
H^2_{\mathrm{st}}(X;\mathbb{Z})
\cong
H^2_{\mathbb{Z}_n}(\mathbb{R}^4;\mathbb{Z})
\cong
H^2(B\mathbb{Z}_n;\mathbb{Z})
\cong
\mathbb{Z}_n.
\]
This class is pure torsion, so its image in real cohomology vanishes:
\[
H^2_{\mathrm{dR,orb}}(X)=0.
\]
Hence the usual prequantization condition on $[\omega/(2\pi\hbar)]$ is automatically satisfied for every $\hbar>0$. There is no flux quantization constraint.

On the covering space we take the trivial Hermitian line bundle
\[
L_{\mathrm c}=T^*\mathbb{C}\times\mathbb{C}
\]
with unitary connection
\[
\nabla_0=d-\frac{i}{\hbar}\lambda.
\]
Its curvature is $F_{\nabla_0}=-\frac{i}{\hbar}\omega$. All inequivalent orbifold prequantum bundles arise by choosing different $\mathbb{Z}_n$-equivariant structures on this trivial bundle. These are classified by the characters
\[
\chi_a:\mathbb{Z}_n\to U(1),
\qquad
\chi_a(g)=e^{2\pi ia/n},
\qquad
a\in\{0,1,\dots,n-1\}.
\]
Equivalently,
\[
\mathrm{Pic}_{\mathrm{orb}}(X)
\cong
H^2_{\mathrm{st}}(X;\mathbb{Z})
\cong
\mathbb{Z}_n
\cong
\mathrm{Hom}\bigl(\pi_1^{\mathrm{orb}}(X),U(1)\bigr).
\]
Here the label $a$ records the \emph{equivariant weight} of the lifted $\mathbb{Z}_n$-action on the fiber. When the same flat bundle is rewritten on the punctured coarse cone in an invariant frame, the corresponding holonomy around a positive generator of $\pi_1^{\mathrm{orb}}(X)$ is $e^{-2\pi i a/n}$, as recalled in Section~\ref{sec:config-space-1d}. Thus the isotropy weight and the orbifold holonomy character are inverse conventions for the same sector; throughout this subsection we keep the weight label $a$ because it matches the twisted boundary condition of Chapter~4.

\begin{proposition}[Prequantum sectors]
\label{prop:cone-sectors}
There are exactly $n$ inequivalent prequantum sectors on $X=[T^*\mathbb{C}/\mathbb{Z}_n]$, indexed by $a\in\{0,1,\dots,n-1\}$. Fixing one prequantum bundle $(L_0,\nabla_0)$, the sector $L_a$ is obtained by tensoring with the flat orbifold line bundle $F_a$ determined by the character $\chi_a(g)=e^{2\pi ia/n}$. This changes only the torsion equivariant data and leaves the curvature unchanged.
\end{proposition}

Near the singular point, the bundle $L_a$ is locally modeled by
\[
(\widetilde U\times\mathbb{C})/\mathbb{Z}_n,
\]
where the isotropy acts on the fiber by multiplication with $e^{2\pi ia/n}$. Thus the label $a$ is precisely the isotropy weight at the cone point.

\begin{remark}[Configuration-space interpretation]
\label{rem:pi1-cone}
For the configuration orbifold $Q=[\mathbb{C}/\mathbb{Z}_n]$, the smooth locus $(\mathbb{C}\setminus\{0\})/\mathbb{Z}_n$ has ordinary fundamental group $\mathbb{Z}$, while the orbifold fundamental group is $\pi_1^{\mathrm{orb}}(Q)\cong\mathbb{Z}_n$. The difference is exactly the torsion introduced by the apex isotropy, and that same torsion reappears quantum mechanically as the twisted phase of the wavefunction around the cone point.
\end{remark}

The prequantum Hilbert space in sector $a$ is
\[
\mathcal{H}_{\mathrm{pre},a}(X)=L^2(T^*\mathbb{C},L_a)^{\mathbb{Z}_n},
\]
with orbifold inner product normalized by the factor $1/n$ coming from the quotient volume.

\subsection{Polarization and the Quantum Hilbert Space}

For a cotangent bundle, the natural choice is the vertical real polarization
\[
\mathcal{P}=\langle \partial_{p_x},\partial_{p_y}\rangle=\ker(d\pi),
\qquad
\pi:T^*Q\to Q.
\]
This distribution is Lagrangian, integrable, and $\mathbb{Z}_n$-invariant, since the rotation acts linearly and preserves the momentum subspace. In this example there is therefore no orbifold obstruction to using the usual Schr\"odinger polarization.

The polarization condition
\[
\nabla_\xi s=0
\qquad
\text{for all }\xi\in\mathcal{P}
\]
forces polarized sections to be independent of the momenta. A nonzero polarized section is therefore constant along a noncompact momentum fiber and, for that reason, is not an element of the raw prequantum Hilbert space $\mathcal{H}_{\mathrm{pre},a}(X)$. As is standard for the Schr\"odinger polarization on a cotangent bundle, one passes from the phase-space $L^2$ pairing to the leaf-space pairing on the configuration orbifold $Q=[\mathbb{C}/\mathbb{Z}_n]$~\cite{Woodhouse1992,Sniatycki1980}. The metaplectic discussion below shows that the corrected pairing is precisely the orbifold-normalized configuration-space measure. Concretely, a polarized section in sector $a$ may be represented by a function $\psi$ on the covering plane satisfying
\[
\psi(g\cdot z)=\chi_a(g)\,\psi(z).
\]
In polar coordinates this becomes
\[
\psi\bigl(r,\phi+\tfrac{2\pi}{n}\bigr)=e^{2\pi ia/n}\psi(r,\phi),
\]
which is exactly the twisted boundary condition obtained earlier in the Schr\"odinger treatment. In the equivalent holonomy convention, the same sector is described by the inverse phase $e^{-2\pi i a/n}$ around an orbifold generator.

The resulting orbifold inner product may be written either on the full cover or on a half-open fundamental wedge:
\[
\langle \psi,\varphi\rangle_{\mathrm{orb}}
=
\frac{1}{n}\int_{\mathbb{C}}\overline{\psi(z)}\varphi(z)\,d^2z
=
\int_0^\infty r\,dr\int_0^{2\pi/n}d\phi\,
\overline{\psi(r,\phi)}\varphi(r,\phi).
\]

\begin{definition}[Quantum Hilbert space]
\label{def:cone-hilbert}
After polarization, and with the trivial half-form choice made below, the quantum Hilbert space in sector $a$ is
\[
\mathcal{H}_a(\mathbb{C}/\mathbb{Z}_n)
\cong
\left\{
\psi\in L^2(\mathbb{C})
\;\middle|\;
\psi(e^{2\pi i/n}z)=e^{2\pi ia/n}\psi(z)\ \text{a.e.}
\right\}.
\]
The $L^2$ inner product is the orbifold-normalized one displayed above. Equivalently, a state may be represented by its restriction to the half-open wedge $[0,\infty)\times[0,2\pi/n)$ equipped with the above inner product. The twist relation is then used to glue the boundary rays; when one introduces differential operators, it reappears as a seam condition on the relevant Sobolev domains rather than as an extra condition on arbitrary $L^2$ representatives.
\end{definition}

Not every classical observable is directly quantizable as a first-order Kostant--Souriau operator in this polarization. The criterion for that direct construction is that its Hamiltonian vector field preserve $\mathcal{P}$. Angular momentum does, but the free Hamiltonian does not, so the latter is represented by the second-order Schr\"odinger operator on the cone rather than by direct restriction of the prequantum operator.

On the smooth locus $r>0$, one may also perform an auxiliary Bohr--Sommerfeld computation using the horizontal polarization $\mathcal{P}_{\mathrm{hor}}=\langle\partial_r,\partial_\phi\rangle$. Because the Hamiltonian vector field $X_{L_z}=\partial_\phi$ preserves $\mathcal{P}_{\mathrm{hor}}$, its spectrum can be extracted from the horizontal leaves at fixed $(p_r,p_\phi)$. For the orbifold loop consisting of a $2\pi/n$ angular sweep together with the generator $g\in\mathbb{Z}_n$, the orbifold holonomy is
\[
\exp\!\left(\frac{i}{\hbar}\int_0^{2\pi/n}p_\phi\,d\phi\right)e^{-2\pi ia/n}
=
e^{2\pi i(p_\phi/(n\hbar)-a/n)}.
\]
The inverse phase is the holonomy convention corresponding to the weight-$a$ equivariant sector described above.
The Bohr--Sommerfeld condition gives
\[
p_\phi=\hbar(a+n\ell),
\qquad
\ell\in\mathbb{Z}.
\]
This already foreshadows the angular-momentum spectrum derived below.

\subsection{Angular Momentum and the Free Hamiltonian}

The classical angular momentum is
\[
L_z=xp_y-yp_x=p_\phi.
\]
Its Hamiltonian vector field is
\[
X_{L_z}=\partial_\phi.
\]
The prequantum operator is
\[
Q(L_z)=-i\hbar\,\nabla_{X_{L_z}}+L_z.
\]
On polarized sections, the $p_\phi$ term cancels against the connection term, leaving
\[
\hat L_z\psi=-i\hbar\,\partial_\phi\psi.
\]
Let
\[
W:=(0,\infty)\times(0,2\pi/n).
\]
In the wedge realization, this operator acts on the twisted first-order Sobolev domain
\[
\begin{aligned}
D(\hat L_z):=\bigl\{\psi\in L^2(W;r\,dr\,d\phi)\ \big|\ &
\partial_\phi\psi\in L^2(W;r\,dr\,d\phi),\\
&\psi(r,2\pi/n)=e^{2\pi ia/n}\psi(r,0)\text{ in trace sense}\bigr\}.
\end{aligned}
\]
Equivalently, on the covering plane it is the generator of the unitary rotation group restricted to the sector $\mathcal{H}_a$, so it is self-adjoint by Stone's theorem.

For a separated state
\[
\psi(r,\phi)=f(r)e^{im\phi},
\]
the seam condition requires
\[
e^{2\pi i m/n}=e^{2\pi i a/n},
\]
hence
\[
m\equiv a\pmod n.
\]

\begin{theorem}[Angular momentum spectrum]
\label{thm:cone-spectrum}
In the sector $\mathcal{H}_a$, the allowed angular-momentum quantum numbers are
\[
m=a+n\ell,
\qquad
\ell\in\mathbb{Z},
\]
and therefore
\[
\mathrm{Spec}(\hat L_z|_{\mathcal{H}_a})
=
\hbar(a+n\mathbb{Z}).
\]
Each eigenvalue has infinite radial multiplicity, and the allowed angular quantum numbers agree exactly with both the earlier boundary-condition derivation and the Bohr--Sommerfeld computation above.
\end{theorem}

\subsection{Metaplectic Correction}

For the vertical polarization, the annihilator $\mathcal{P}^\circ$ is spanned by $dx$ and $dy$, so the canonical bundle of the polarization is
\[
K_{\mathcal{P}}=\det(\mathcal{P}^\circ)=\langle dx\wedge dy\rangle.
\]
A half-form bundle $\delta_{\mathcal P}$ satisfies
\[
\delta_{\mathcal P}^{\otimes 2}\cong K_{\mathcal P}.
\]
The pulled-back bundle on $T^*\mathbb{C}$ is trivial. Moreover,
\[
g^*(dx\wedge dy)=g^*(r\,dr\wedge d\phi)=r\,dr\wedge d\phi=dx\wedge dy,
\]
so the isotropy action on $K_{\mathcal P}$ is trivial as well. Hence $K_{\mathcal P}$ is equivariantly trivial, and we choose the trivial equivariant square root
\[
\delta_{\mathcal P}=\langle\sqrt{dx\wedge dy}\rangle.
\]
When $n$ is even there is also a nontrivial flat square root with character $g\mapsto -1$; using it would tensor every sector by that flat character, equivalently relabeling $a$ by $a+n/2$ modulo $n$. In the following we use the trivial metalinear choice, so the half-form bundle contributes no additional orbifold phase.

A polarized corrected section is therefore of the form
\[
\psi(x,y)\otimes\sqrt{dx\wedge dy}.
\]
Its Hermitian pairing with the conjugate half-form produces the transverse density
\[
|\psi(x,y)|^2\,dx\,dy,
\]
so the metaplectically corrected Hilbert space is exactly the configuration-space $L^2$-space of Definition~\ref{def:cone-hilbert}. This is the rigorous replacement for the non-normalizable polarized prequantum sections that were constant along the momentum fibers.

For $L_z$, the relevant vector field is again $X_{L_z}=\partial_\phi$. Its Lie derivative on $dx\wedge dy$ vanishes, so the half-form correction contributes no extra term:
\[
Q_{\mathrm{meta}}(L_z)(\psi\otimes \nu)
=
\bigl(-i\hbar\,\partial_\phi\psi\bigr)\otimes \nu.
\]
Likewise, the smooth-stratum angular orbit has Maslov index zero. Thus, for the trivial half-form choice made above, the metaplectic correction produces no spectral shift in this example.

\subsection{Sector Decomposition and Interpretation}

The covering-space Hilbert space decomposes under the $\mathbb{Z}_n$-action into isotypic components,
\[
L^2(\mathbb{C})=\bigoplus_{a=0}^{n-1}\mathcal{H}_a,
\]
where $\mathcal{H}_a$ is the eigenspace for the unitary action
\[
(U_g\psi)(z):=\psi(g\cdot z)
\]
with character $\chi_a$. Every orbifold observable commutes with this action, so each $\mathcal{H}_a$ is preserved by the observable algebra. The sectors are therefore superselection sectors. If one instead uses the pullback convention $\psi\mapsto\psi\circ g^{-1}$, the same sector would be described as the $\chi_a^{-1}$-isotypic component.

There is no canonical sector-independent physical Hilbert space attached to a fixed choice of orbifold prequantization. Choosing the prequantum bundle $L_a$ selects the corresponding orbifold quantum space $\mathcal{H}_a$. The untwisted invariant space is only the special case $a=0$; the other sectors are equally valid quantizations distinguished by their flat twist around the apex.

\section{\texorpdfstring{Geometric Quantization of the Orbifold Football $S^2(n,n)$}{Geometric Quantization of the Orbifold Football S2(n,n)}}
\label{subsec:gq-s2-nn}

The football orbifold
\[
X=S^2(n,n)
\]
is the equal-order specialization of the general orbisphere treated in Section~\ref{subsec:gq-s2-nm}. In contrast with a generic weighted sphere, it is a good orbifold:
\[
X\cong[\widetilde{S}^2/\mathbb{Z}_n],
\]
where $\widetilde{S}^2\cong S^2\cong\mathbb{P}^1$ and $\mathbb{Z}_n$ acts by rotation about the $z$-axis~\cite{LermanTolman1997,BaierMouraoNunes2018}. This quotient presentation makes the geometric quantization especially transparent: one can formulate everything intrinsically on the orbifold, and then identify the resulting quantum space with the $\mathbb{Z}_n$-invariant part of the corresponding quantization on the smooth cover.

\subsection{Symplectic Orbifold Structure}

Let $(\theta,\widetilde{\phi})$ be spherical coordinates on the covering sphere $\widetilde{S}^2$, with $\theta\in[0,\pi]$ and $\widetilde{\phi}\in[0,2\pi)$. If $\gamma$ denotes the generator of $\mathbb{Z}_n$, then
\[
\gamma\cdot(\theta,\widetilde{\phi})
=
\bigl(\theta,\widetilde{\phi}+\tfrac{2\pi}{n}\bigr).
\]
The quotient has two cone points of order $n$, namely the images of the north and south poles.

Choose the positive $\mathbb{Z}_n$-invariant area form
\[
\widetilde{\omega}
=
B\sin\theta\,d\theta\wedge d\widetilde{\phi},
\qquad
B>0.
\]
Its total area on the cover is
\[
\int_{\widetilde{S}^2}\widetilde{\omega}=4\pi B,
\]
and it descends to an orbifold symplectic form $\omega$ on $X$. Orbifold integration gives
\[
\int_X\omega
=
\frac{1}{n}\int_{\widetilde{S}^2}\widetilde{\omega}
=
\frac{4\pi B}{n}.
\]
Hence the normalized orbifold flux is
\[
\frac{1}{2\pi\hbar}\int_X\omega
=
\frac{2B}{\hbar n}.
\]

As in the general theory, the classical observables on the orbifold are precisely the $\mathbb{Z}_n$-invariant smooth functions on the cover:
\[
C^\infty_{\mathrm{orb}}(X)=C^\infty(\widetilde{S}^2)^{\mathbb{Z}_n}.
\]
Thus the quotient picture already captures the observable algebra.

\subsection{Prequantization and Orbifold Line-Bundle Data}

Because $X$ is a global quotient, a prequantum orbifold line bundle on $X$ is equivalently a $\mathbb{Z}_n$-equivariant Hermitian line bundle on $\widetilde{S}^2$ equipped with a $\mathbb{Z}_n$-equivariant unitary connection of curvature $-\frac{i}{\hbar}\widetilde{\omega}$.

Equivalently, the cover flux must satisfy the usual Dirac condition:
\begin{equation}
\label{eq:football-cover-flux}
N_\phi
:=
\frac{1}{2\pi\hbar}\int_{\widetilde{S}^2}\widetilde{\omega}
=
\frac{2B}{\hbar}
\in
\mathbb{Z}.
\end{equation}
Equivalently,
\[
\frac{1}{2\pi\hbar}\int_X\omega=\frac{N_\phi}{n}.
\]
This is exactly the equal-order specialization of the general orbifold integrality condition: the de Rham class lies in the image of the stack integral lattice even though the coarse-space symplectic area may be fractional.

An orbifold line bundle on $S^2(n,n)$ is described by Seifert data~\cite{Furuta1992SeifertFH,MrowkaOzsvathYu1997,Sakai2017OrbifoldVortices,BaierMouraoNunes2018}
\[
(d_0;a,b),
\qquad
d_0\in\mathbb{Z},
\quad
a,b\in\{0,\dots,n-1\},
\]
with orbifold degree
\[
\deg_{\mathrm{orb}}(L)=d_0+\frac{a}{n}+\frac{b}{n}.
\]
For a prequantum line bundle, the orbifold degree must match the symplectic flux:
\[
\deg_{\mathrm{orb}}(L)
=
\frac{1}{2\pi\hbar}\int_X\omega
=
\frac{N_\phi}{n}.
\]
Hence the Seifert data satisfy
\begin{equation}
\label{eq:football-degree-relation}
N_\phi=nd_0+a+b.
\end{equation}
So fixing the curvature does not fix the bundle uniquely. For a fixed cover degree $N_\phi$, there remain exactly $n$ inequivalent orbifold line bundles, parameterized for example by the north-pole weight $a\in\{0,\dots,n-1\}$; once $a$ is chosen, the south-pole weight $b$ and the integer $d_0$ are determined by \eqref{eq:football-degree-relation} together with the constraint $b\in\{0,\dots,n-1\}$. This count of $n$ sectors matches the order of $\pi_1^{\mathrm{orb}}(S^2(n,n))\cong\mathbb{Z}_n$, which classifies the flat twists available for a given curvature class.

This is the equal-order version of the flat-sector phenomenon already seen for general orbifold line bundles: tensoring by a degree-zero orbifold line bundle shifts the local isotropy data while leaving the curvature unchanged. Concretely, for fixed $N_\phi$, one may move between equivalent-curvature sectors by
\[
(d_0;a,b)\longmapsto(d_0';a+\sigma,b-\sigma),
\]
where the local weights are reduced modulo $n$ and the integer degree $d_0'$ is adjusted by the carries needed to keep \eqref{eq:football-degree-relation} valid.

\subsection{K\"ahler Polarization and Holomorphic Sections}

The standard complex structure on $\widetilde{S}^2\cong\mathbb{P}^1$ is $\mathbb{Z}_n$-invariant, and $\widetilde{\omega}$ is a positive multiple of the Fubini--Study form. Therefore the orbifold inherits a K\"ahler polarization, and geometric quantization gives~\cite{Huebschmann2004,Kawasaki1979}
\[
\mathcal{H}_{\mathrm{quant}}(X,L)=H^0(X,L).
\]
In the quotient model this becomes
\[
H^0(X,L)\cong H^0(\widetilde{S}^2,\widetilde{L})^{\mathbb{Z}_n},
\]
where $\widetilde{L}$ is the equivariant lift of the orbifold line bundle $L$, and the invariants are taken with respect to the lifted action on sections.

The lifted bundle has ordinary degree $N_\phi$, so on the cover
\[
H^0(\widetilde{S}^2,\widetilde{L})
\cong
H^0(\mathbb{P}^1,\mathcal{O}(N_\phi)).
\]
If $N_\phi<0$, this space vanishes. In the symplectic case above $B>0$, so the prequantized value $N_\phi=2B/\hbar$ is a positive integer.

Choose an affine coordinate $z$ on $\mathbb{P}^1$ vanishing at the north pole (given by the stereographic projection $z=\tan(\theta/2)e^{i\widetilde{\phi}}$ from the south pole), so that the rotation generator acts by
\[
\gamma\cdot z=e^{2\pi i/n}z.
\]
Then a basis of $H^0(\mathbb{P}^1,\mathcal{O}(N_\phi))$, represented in the standard affine trivialization, is
\[
1,z,\dots,z^{N_\phi}.
\]
If the isotropy weight at the north pole is $a$, choose a local frame $e$ near $z=0$ with
\[
\gamma\cdot e=e^{2\pi ia/n}e.
\]
For the monomial section $z^C e$, the induced action on sections is
\[
(\gamma\cdot s)(z)=\gamma\bigl(s(\gamma^{-1}\cdot z)\bigr),
\]
and therefore
\[
(\gamma\cdot(z^C e))(z)
=
e^{2\pi i(a-C)/n}z^C e.
\]
Hence $z^C e$ is invariant exactly when
\begin{equation}
\label{eq:football-invariant-condition}
C\equiv a\pmod n.
\end{equation}
So the orbifold quantum Hilbert space consists precisely of the invariant monomials selected by the north-pole isotropy character. Under the cover-isotypic convention used for the Schr\"odinger problem in Chapter~4, this congruence corresponds to the sector label $q=a$; here, however, the K\"ahler polarization selects the holomorphic geometric-quantization subspace rather than the full spherical-harmonic spectrum. The south-pole weight imposes no independent congruence, but verifying this requires care with the local cone-chart generator. At the south pole, the local coordinate is $w=1/z$, on which the global rotation acts by $\gamma\cdot w=e^{-2\pi i/n}w$. Thus the standard positive local cone generator in the $w$-coordinate is represented by $\gamma^{-1}$. Defining the south-pole weight $b$ via $\gamma^{-1}\cdot e_S=e^{2\pi ib/n}e_S$, one can verify this explicitly. In the standard trivialization on the $w$-chart, the section $z^C e$ becomes $w^{N_\phi-C}e_S$, and the equivariant action of $\gamma^{-1}$ gives eigenvalue $e^{2\pi i(b-(N_\phi-C))/n}$. Invariance requires $C\equiv N_\phi-b\pmod n$. From the degree relation $N_\phi=nd_0+a+b$, one has $N_\phi-b=nd_0+a\equiv a\pmod n$, recovering the same congruence~\eqref{eq:football-invariant-condition} already imposed at the north pole.

Using~\eqref{eq:football-degree-relation}, write
\[
N_\phi-a=nd_0+b.
\]
The integers $C\in\{0,\dots,N_\phi\}$ satisfying $C\equiv a\pmod n$ are then
\[
C=a,\ a+n,\ a+2n,\ \dots,\ a+d_0 n,
\]
provided $d_0\ge 0$. Hence
\[
\dim H^0(X,L)=\max(0,d_0+1).
\]
This is the equal-order football analogue of the orbifold Riemann--Roch count derived earlier by intrinsic methods.

\subsection{Metaplectic Correction}

For the K\"ahler polarization, the metaplectically corrected quantum space is
\[
\mathcal{H}_{\mathrm{meta}}(X,L)=H^0(X,L\otimes\delta),
\]
provided there exists an orbifold half-form bundle $\delta$ satisfying
\[
\delta^{\otimes 2}\cong K_X.
\]

For the football $S^2(n,n)$, the orbifold canonical bundle has Seifert data
\[
K_X\cong L_{(-2;n-1,n-1)}.
\]
The coarse canonical degree contributes the $-2$, while each cone point contributes local cotangent weight $n-1$. A half-form bundle must therefore solve both the global divisibility problem in $\operatorname{Pic}_{\mathrm{orb}}(X)$ and the local square-root problem for the isotropy characters.

If $n$ is odd, both conditions are satisfied by
\[
\delta=L_{\left(-1;\frac{n-1}{2},\frac{n-1}{2}\right)},
\]
and one checks immediately that $\delta^{\otimes 2}\cong K_X$. If $n$ is even, then the local weight $n-1$ is odd in $\mathbb{Z}_n$, so the congruence
\[
2\ell\equiv n-1\pmod n
\]
has no solution. Hence no local square root exists, and therefore no orbifold half-form bundle exists. We conclude that
\[
\delta \text{ exists on } S^2(n,n)
\quad\Longleftrightarrow\quad
n \text{ is odd}.
\]

When $n$ is odd and $\delta$ exists, it is unique because $\operatorname{Pic}_{\mathrm{orb}}(S^2(n,n))$ has no $2$-torsion. Its pullback to the cover is the usual half-canonical bundle on the sphere:
\[
\pi^*\delta\cong K_{\widetilde{S}^2}^{1/2}\cong\mathcal{O}(-1).
\]
Indeed, the orbifold degree of $\delta$ is $-1+\tfrac{n-1}{2n}+\tfrac{n-1}{2n}=-\tfrac{1}{n}$, so $\deg(\pi^*\delta)=n\cdot(-\tfrac{1}{n})=-1$.
Consequently,
\[
\pi^*(L\otimes\delta)\cong\mathcal{O}(N_\phi-1).
\]
So the metaplectic correction has the same effect on the cover as in the smooth case: it lowers the covering degree by one, after which one again takes the invariant subspace determined by the equivariant structure. Concretely, the corrected north-pole weight is
\[
a'\equiv a+\frac{n-1}{2}\pmod n,
\qquad
a'\in\{0,\dots,n-1\},
\]
with any carry absorbed into the integer part of the Seifert data. The corrected dimension is
\[
\dim\mathcal{H}_{\mathrm{meta}}(X,L)
=
\#\bigl\{C'\in\{0,\dots,N_\phi-1\}:C'\equiv a'\!\pmod n\bigr\},
\]
which is zero if no such integer $C'$ exists.


\section{\texorpdfstring{Geometric Quantization of the Orbisphere $S^2(n,m)$}{Geometric Quantization of the Orbisphere S2(n,m)}}
\label{subsec:gq-s2-nm}

The two-cone-point orbisphere
\[
X=S^2(n,m)
\]
is the basic compact complex orbifold in which the general machinery of orbifold geometric quantization becomes completely explicit~\cite{LermanTolman1997,BaierMouraoNunes2018,Huebschmann2006}. The symplectic form is encoded by an orbifold K\"ahler class, prequantization is governed by the smooth/topological orbifold Chern class together with the local isotropy characters at the poles, the natural polarization is K\"ahler, and the metaplectic correction is controlled first by divisibility in the underlying orbifold Picard group \(\mathrm{Pic}_{\mathrm{orb}}(X)\), and then, for holomorphic quantization, by the corresponding holomorphic square-root problem in \(\mathrm{Pic}_{\mathrm{hol}}(X)\). When $\gcd(n,m)=1$, the same orbifold can be presented as the weighted projective line
\[
X\cong\mathbb{P}(n,m),
\]
which gives a concrete holomorphic model for the quantum Hilbert space. This weighted-projective presentation is a complex-stack quotient presentation; it does not contradict the Chapter~3 statement that, for unequal cone orders, the effective orbisphere is bad in the sense of not admitting a manifold universal orbifold cover. When $\gcd(n,m)>1$, the weighted quotient $\mathbb{P}(n,m)$ is instead non-effective, so it is not the same effective orbifold as the intrinsic two-cone-point spindle $S^2(n,m)$; moreover the effective spindle still carries $g=\gcd(n,m)$ flat sectors coming from $\pi_1^{\mathrm{orb}}(X)\cong\mathbb{Z}_g$. For that reason, it is conceptually cleaner to begin with the intrinsic orbifold description and only then specialize to the weighted-projective picture in the coprime case.

\subsection{Symplectic Orbifold Structure and Observables}

\subsubsection{Local orbifold charts.}
Let $p_N$ and $p_S$ denote the north and south cone points. Their isotropy groups are
\[
\Gamma_{p_N}\cong\mathbb{Z}_n,
\qquad
\Gamma_{p_S}\cong\mathbb{Z}_m.
\]
Thus $X$ admits standard orbifold charts $(D,\mathbb{Z}_n)$ and $(D,\mathbb{Z}_m)$ around the two poles, while away from these points the orbifold is an ordinary smooth sphere. In local complex coordinates $u_N$ and $u_S$ centered at the poles, the generators act by
\[
u_N\longmapsto e^{2\pi i/n}u_N,
\qquad
u_S\longmapsto e^{2\pi i/m}u_S.
\]
These are the local models that underlie all later bundle and half-form calculations.

\subsubsection{Symplectic form and K\"ahler viewpoint.}
A symplectic form on $X$ is given chartwise by invariant symplectic forms on the local uniformizing disks that agree on overlaps. By the equivariant Darboux theorem, each lifted chart is locally symplectomorphic to the standard symplectic plane, so the cone singularities do not alter the local Hamiltonian formalism. We assume throughout that $\omega$ is a positive orbifold K\"ahler form compatible with the complex orbifold structure on $X$.

When $\gcd(n,m)=1$, the same complex orbifold is the weighted projective line
\[
X\cong\mathbb{P}(n,m)
=
\bigl[(\mathbb{C}^2\setminus\{0\})/\mathbb{C}^*\bigr],
\qquad
\lambda\cdot(z_1,z_2)=(\lambda^n z_1,\lambda^m z_2).
\]
In that presentation the orbifold K\"ahler form may be viewed as the positive weighted-projective analogue of the Fubini--Study form. Nothing in the intrinsic prequantization discussion depends on choosing this presentation, and the unequal-order case remains non-developable in the finite-cover sense of Chapter~3. The presentation becomes useful only once we pass to holomorphic sections.

\subsubsection{Orbifold observables and Hamiltonian flows.}
The classical observables are the orbifold-smooth functions
\[
C^\infty_{\mathrm{orb}}(X),
\]
equivalently compatible families of invariant smooth functions on the local charts. For each observable $f$, the equation
\[
\iota_{X_f}\omega=-df
\]
defines an orbifold Hamiltonian vector field. Since the local lifts are invariant under the chart groups, their flows are equivariant and descend to Hamiltonian symplectomorphisms of the orbifold itself. Thus the ordinary Hamiltonian formalism passes to the orbifold without change at the conceptual level; only the global topology and isotropy data distinguish the orbifold case.

\subsection{Prequantization and Orbifold Line-Bundle Data}

A prequantum structure on $(X,\omega)$ is a Hermitian orbifold line bundle $(L,\nabla)$ with unitary connection satisfying
\[
F_\nabla=-\frac{i}{\hbar}\omega.
\]
By the general orbifold prequantization theorem, such a bundle exists if and only if the de Rham class \([\omega]/(2\pi\hbar)\in H^2_{\mathrm{dR,orb}}(X)\) lies in the image of the stack integral lattice \(H^2_{\mathrm{st}}(X;\mathbb{Z})\).

\subsubsection{Intrinsic orbifold integrality condition.}
For the two-cone-point orbisphere, orbifold line bundles are most naturally described by the intrinsic data~\cite{Furuta1992SeifertFH,MrowkaOzsvathYu1997,Sakai2017OrbifoldVortices}
\[
L=L_{(d_0;a,b)},
\]
where the semicolon notation separates the coarse degree $d_0\in\mathbb{Z}$ from the local data, and
\[
a\in\{0,\dots,n-1\},
\qquad
b\in\{0,\dots,m-1\}
\]
record the isotropy weights at the north and south poles. The orbifold degree is
\[
\deg_{\mathrm{orb}}(L)
=
\frac{i}{2\pi}\int_XF_\nabla
=
d_0+\frac{a}{n}+\frac{b}{m}.
\]
Therefore the prequantization condition becomes
\begin{equation}
\label{eq:s2nm-integrality}
\frac{1}{2\pi\hbar}\int_X\omega
=
d_0+\frac{a}{n}+\frac{b}{m}.
\end{equation}
This is the concrete meaning of orbifold integrality for \(S^2(n,m)\). The normalized orbifold flux need not be an integer when expressed on the underlying topological sphere. Its fractional part is absorbed by the isotropy characters of the prequantum line bundle at the two cone points, i.e.\ by the stack-integral lift of the de Rham class.

\subsubsection{Meaning of the isotropy weights.}
The integers $a$ and $b$ are not auxiliary bookkeeping devices: they are the local representation weights of the isotropy groups on the fibers above the singular points. Concretely, in a local frame $e_N$ above the north-pole chart and $e_S$ above the south-pole chart, the standard cone-chart generators act by
\[
g_N\cdot e_N=e^{2\pi ia/n}e_N,
\qquad
g_S\cdot e_S=e^{2\pi ib/m}e_S.
\]
These local phases are exactly what produce the fractional contributions $a/n$ and $b/m$ in the orbifold degree. They are also the data that later govern the existence of half-forms and the arithmetic of holomorphic sections.

\subsubsection{Flat twists and torsion sectors.}
The curvature determines only the real Chern class. Distinct integral lifts with the same curvature can still differ by torsion. For the orbisphere,
\[
\pi_1^{\mathrm{orb}}\bigl(S^2(n,m)\bigr)\cong\mathbb{Z}_g,
\qquad
g=\gcd(n,m).
\]
Hence, when $g>1$, prequantum bundles with the same curvature form can differ by tensoring with a flat orbifold line bundle, and these flat sectors are classified by
\[
\mathrm{Hom}\bigl(\pi_1^{\mathrm{orb}}(X),U(1)\bigr)
\cong
\mathrm{Hom}(\mathbb{Z}_g,U(1)).
\]
Thus the non-coprime case $\gcd(n,m)>1$ contains genuine discrete superselection data not visible in the real symplectic form alone.

\begin{proposition}[Prequantum sectors for fixed curvature]
\label{prop:s2nm-prequantum-sectors}
Fix a symplectic form $\omega$ on $S^2(n,m)$ for which the integrality condition~\eqref{eq:s2nm-integrality} holds, and fix one prequantum bundle $(L,\nabla)$ with curvature $-\frac{i}{\hbar}\omega$. Then every other prequantum bundle with the same curvature is of the form
\[
(L,\nabla)\otimes F_\chi,
\]
where $F_\chi$ is a flat orbifold line bundle determined by a character
\[
\chi\in\mathrm{Hom}\bigl(\pi_1^{\mathrm{orb}}(X),U(1)\bigr).
\]
In particular, when $\gcd(n,m)=1$ there is no torsion ambiguity, whereas when $\gcd(n,m)>1$ there are precisely $g$ discrete flat sectors.
\end{proposition}
\begin{proof}
If $(L',\nabla')$ is another prequantum bundle with the same curvature, then
\[
M:=L'\otimes L^{-1}
\]
inherits the tensor-product connection
\[
\nabla_M:=\nabla'\otimes\nabla^{-1}.
\]
Its curvature is
\[
F_{\nabla_M}=F_{\nabla'}-F_\nabla=0,
\]
so $M$ is a flat orbifold line bundle. Flat orbifold line bundles are classified by their holonomy characters, equivalently by
\[
\mathrm{Hom}\bigl(\pi_1^{\mathrm{orb}}(X),U(1)\bigr).
\]
Thus $M\cong F_\chi$ for some character $\chi$, and therefore
\[
(L',\nabla')\cong (L,\nabla)\otimes F_\chi.
\]
Conversely, tensoring $(L,\nabla)$ with any flat orbifold line bundle does not change the curvature. Since $\pi_1^{\mathrm{orb}}(S^2(n,m))\cong\mathbb{Z}_g$, the character group has cardinality $g$, which yields the stated count of sectors.
\end{proof}

\subsubsection{Weighted-projective notation in the coprime case.}
When $\gcd(n,m)=1$, the same underlying smooth/topological orbifold line-bundle class can be encoded by a single integer parameter. In the weighted-projective complex model this integer is represented by the holomorphic line bundle:
\[
L\cong\mathcal{O}(q)\quad\text{on }\mathbb{P}(n,m).
\]
The intrinsic and weighted-projective parameters are related by
\[
q=mn\,d_0+ma+nb.
\]
Equivalently,
\[
a\equiv q\,m^{-1}\pmod n,
\qquad
b\equiv q\,n^{-1}\pmod m,
\]
where $m^{-1}\in\mathbb{Z}_n$ and $n^{-1}\in\mathbb{Z}_m$ are the multiplicative inverses of $m$ and $n$. In this notation,
\[
\deg_{\mathrm{orb}}(\mathcal{O}(q))=\frac{q}{nm},
\]
so~\eqref{eq:s2nm-integrality} becomes
\[
\frac{1}{2\pi\hbar}\int_X\omega=\frac{q}{nm}.
\]
Because $m$ and $n$ are coprime, these congruences determine $a$ and $b$ uniquely, and then
\[
d_0=\frac{q-ma-nb}{mn}\in\mathbb{Z}
\]
is forced. Hence $q$ gives a bijective parametrization of smooth/topological orbifold line-bundle classes in the coprime case, equivalently $\operatorname{Pic}_{\mathrm{orb}}(X)\cong\mathbb{Z}$ in this weighted-projective model. Since the weighted projective line has no nontrivial degree-zero holomorphic Picard torus, the same integer also labels the standard holomorphic representatives \(\mathcal{O}(q)\) in \(\mathrm{Pic}_{\mathrm{hol}}(X)\). Thus weighted-projective notation does not introduce a new quantization condition. It simply packages the intrinsic orbifold data into a single integer because the coprime case has no torsion.

\subsection{K\"ahler Polarization and the Quantum Hilbert Space}

\subsubsection{Natural choice of polarization.}
Because $X$ is a compact complex orbifold and $\omega$ is compatible with its complex structure, the natural polarization is the K\"ahler polarization~\cite{Huebschmann2004,Kawasaki1979}
\[
\mathcal{P}=T^{0,1}X.
\]
The $(0,1)$-part of the prequantum connection defines a holomorphic structure on $L$, and the polarized Hilbert space is
\[
\mathcal{H}_{\mathrm{pol}}(X,L)=H^0(X,L).
\]
Compactness of $X$ ensures that this space is finite-dimensional. At the abstract level, this is the orbifold version of the standard K\"ahler quantization rule for compact complex manifolds.

\subsubsection{Holomorphic sections in the coprime weighted-projective model.}
In the coprime case, after the prequantum connection has supplied the holomorphic structure on the underlying class \(L\), writing $L\cong\mathcal{O}(q)$ on $\mathbb{P}(n,m)$ means an isomorphism in \(\mathrm{Pic}_{\mathrm{hol}}(X)\), not merely in \(\mathrm{Pic}_{\mathrm{orb}}(X)\). Global holomorphic sections may then be described directly from the quotient presentation. A section of $\mathcal{O}(q)$ is equivalently a holomorphic function
\[
F\colon \mathbb{C}^2\setminus\{0\}\to\mathbb{C}
\]
satisfying the equivariance condition
\[
F(\lambda^n z_1,\lambda^m z_2)=\lambda^q F(z_1,z_2).
\]
Since the origin is a complex codimension-$2$ subset of $\mathbb{C}^2$, Hartogs' theorem extends $F$ holomorphically across the origin. Expanding the resulting entire function in monomials and imposing the equivariance condition shows that only weighted-homogeneous terms of weight $q$ occur. Therefore $H^0(X,\mathcal{O}(q))$ is spanned by monomials
\[
z_1^A z_2^C,
\qquad
A,C\ge 0,
\qquad
nA+mC=q.
\]
Hence
\begin{equation}
\label{eq:s2nm-dimension}
\dim \mathcal{H}_{\mathrm{pol}}
=
\#\{(A,C)\in\mathbb{Z}_{\ge0}^2\mid nA+mC=q\}.
\end{equation}
This is the explicit weighted-projective realization of orbifold Riemann--Roch: the arithmetic of the quantum Hilbert space reduces to the Diophantine problem of writing $q$ as a nonnegative linear combination of $n$ and $m$.

\subsubsection{Interpretation of the monomial count.}
The counting formula~\eqref{eq:s2nm-dimension} makes several general features transparent.
\begin{itemize}
\item If $q<0$, then there are no nonzero holomorphic sections.
\item Even when $q\ge 0$, there may still be no sections if $q$ is not in the semigroup generated by $n$ and $m$.
\item In the coprime case, the integer $q$ already determines the full orbifold line-bundle class. In the non-coprime intrinsic orbifold, by contrast, fixing the real curvature can still leave additional flat data.
\end{itemize}
For example, on $S^2(2,3)$ the bundle $\mathcal{O}(1)$ has no nonzero holomorphic sections because
\[
2A+3C=1
\]
has no solution with $A,C\ge 0$.

\subsubsection{Relation to the intrinsic orbifold data.}
Although the weighted-projective model is the most efficient way to count sections, the intrinsic orbifold parameters $(d_0;a,b)$ remain conceptually important. They determine the local isotropy action of the bundle at the poles, while the global integer $q$ packages these data through the relation
\[
q=mn\,d_0+ma+nb.
\]
In particular, the arithmetic condition $nA+mC=q$ is the global holomorphic shadow of the local isotropy constraints encoded by $a$ and $b$.

\subsubsection{Remark on dynamics.}
The primary output of geometric quantization at this stage is the holomorphic space $H^0(X,L)$. The formalism directly quantizes only those observables whose Hamiltonian flows preserve the K\"ahler polarization. Consequently, one should not identify $H^0(X,L)$ with the full spectral space of an arbitrary magnetic Schr\"odinger operator without additional analytic input. In favorable physical models, this holomorphic space is interpreted as a lowest-Landau-level sector, but that is a further dynamical statement rather than part of the bare geometric quantization result.

\subsection{Metaplectic Correction}

\subsubsection{Canonical bundle and half-form data.}
For the K\"ahler polarization, the polarization canonical bundle is the orbifold canonical bundle
\[
K_X=\Omega_{\mathrm{orb}}^{1,0}(X).
\]
If there exists an orbifold half-form bundle $\delta$ satisfying
\[
\delta^{\otimes 2}\cong K_X,
\]
then the metaplectically corrected Hilbert space is
\[
\mathcal{H}_{\mathrm{meta}}(X,L)=H^0(X,L\otimes\delta).
\]
Thus the existence question is entirely a square-root problem for the orbifold canonical bundle.

\subsubsection{Intrinsic form of the canonical bundle.}
For the two-cone-point orbisphere,
\[
K_X\cong L_{(-2;n-1,m-1)}.
\]
The coarse degree is $-2$, as on the ordinary sphere, while the local cotangent representation contributes weights $n-1$ and $m-1$ at the north and south poles. A half-form bundle must therefore satisfy two simultaneous requirements:
\begin{enumerate}
\item its square must reproduce the global degree $-2$ in \(\mathrm{Pic}_{\mathrm{orb}}(X)\), and in \(\mathrm{Pic}_{\mathrm{hol}}(X)\) when a holomorphic half-form is required;
\item the corresponding local isotropy characters must square to the cotangent characters, equivalently the additive weights must satisfy $2\ell_N\equiv n-1\pmod n$ and $2\ell_S\equiv m-1\pmod m$.
\end{enumerate}
This is the precise orbifold content of the metaplectic obstruction.

\subsubsection{Coprime weighted-projective criterion.}
When $\gcd(n,m)=1$, one also has
\[
K_X\cong\mathcal{O}(-n-m).
\]
Indeed, inserting the intrinsic canonical data $(-2;n-1,m-1)$ into the relation
\[
q=mn\,d_0+ma+nb
\]
gives
\[
q_K=-2mn+m(n-1)+n(m-1)=-(n+m).
\]
Since the coprime weighted-projective parametrization identifies orbifold line bundles with integers $q$, the global divisibility condition is equivalent to $n+m$ being even. The local square-root conditions are
\[
2\ell_N\equiv n-1\pmod n,
\qquad
2\ell_S\equiv m-1\pmod m.
\]
If $n+m$ is even and $\gcd(n,m)=1$, then $n$ and $m$ must both be odd, so multiplication by $2$ is invertible in $\mathbb{Z}_n$ and $\mathbb{Z}_m$, and both congruences are solvable. Conversely, if $n+m$ is odd, then $\mathcal{O}(-n-m)$ is not divisible by $2$ in either the underlying orbifold Picard group or the holomorphic Picard group. Therefore, in the coprime case,
\[
\delta\text{ exists}
\quad\Longleftrightarrow\quad
n+m\text{ is even}.
\]
When this happens, one may choose
\[
\delta\cong\mathcal{O}\!\left(-\frac{n+m}{2}\right).
\]

\subsubsection{Non-coprime case.}
When $\gcd(n,m)>1$, one must analyze the square-root problem intrinsically because the weighted-projective quotient is non-effective. For the present canonical bundle, however, the intrinsic Seifert calculation is completely explicit. The local congruences
\[
2\ell_N\equiv n-1\pmod n,
\qquad
2\ell_S\equiv m-1\pmod m
\]
are solvable if and only if both $n$ and $m$ are odd. In that case, because the orbifold fundamental group $\mathbb{Z}_{\gcd(n,m)}$ has odd order, there is no $2$-torsion in \(\mathrm{Pic}_{\mathrm{orb}}(X)\). Therefore, the underlying smooth/topological half-form bundle is unique, and is given by
\[
\delta\cong L_{\left(-1;\frac{n-1}{2},\frac{m-1}{2}\right)},
\]
and the carry terms vanish because
\[
2\cdot\frac{n-1}{2}=n-1<n,
\qquad
2\cdot\frac{m-1}{2}=m-1<m.
\]
Therefore
\[
\delta^{\otimes 2}\cong L_{(-2;n-1,m-1)}\cong K_X.
\]
If either $n$ or $m$ is even, the corresponding congruence has no solution, so no half-form bundle exists. Hence, for the effective orbisphere $S^2(n,m)$,
\[
\delta\text{ exists}
\quad\Longleftrightarrow\quad
n\text{ and }m\text{ are both odd}.
\]
This reduces to the coprime parity criterion $n+m$ even when $\gcd(n,m)=1$.

\subsubsection{Effect on the polarized Hilbert space.}
When the half-form bundle exists, one replaces $L=L_{(d_0;a,b)}$ by $L\otimes\delta$. Intrinsically, tensoring with
\[
\delta=L_{\left(-1;\frac{n-1}{2},\frac{m-1}{2}\right)}
\]
gives the normalized Seifert data
\[
L\otimes\delta
\cong
L_{\left(d_0-1+\epsilon_N+\epsilon_S;\,
\left[a+\frac{n-1}{2}\right]_n,\,
\left[b+\frac{m-1}{2}\right]_m\right)},
\]
where
\[
\epsilon_N=\left\lfloor\frac{a+(n-1)/2}{n}\right\rfloor,
\qquad
\epsilon_S=\left\lfloor\frac{b+(m-1)/2}{m}\right\rfloor.
\]
Thus the raw half-form tensor factor shifts the integer part by $-1$, but the normalized integer degree also includes the local carry terms. In the coprime weighted-projective notation this gives
\[
\mathcal{H}_{\mathrm{meta}}(X,\mathcal{O}(q))
\cong
H^0\!\left(X,\mathcal{O}\!\left(q-\frac{n+m}{2}\right)\right).
\]
Thus the metaplectic correction shifts the weighted degree by $-(n+m)/2$. The corrected dimension is therefore
\[
\dim \mathcal{H}_{\mathrm{meta}}
=
\#\left\{(A,C)\in\mathbb{Z}_{\ge0}^2
\;\middle|\;
nA+mC=q-\frac{n+m}{2}\right\},
\]
provided we are in the coprime case with $n+m$ even; intrinsically, for non-coprime effective orbifolds one uses the normalized Seifert data above instead of a single weighted-projective integer.

\subsubsection{Comparison with the Kaluza--Klein picture.}
Before the half-form correction, the weighted-projective bundle $\mathcal{O}(q)$ corresponds to the Chapter~4 Kaluza--Klein charge sector $Q=q$. After the correction, the monomial count identifies the holomorphic subspace inside the charge sector with $Q=q-\frac{n+m}{2}$, not the full Kaluza--Klein spectrum. The monomials $z_1^A z_2^C$ with $A,C\ge 0$ and $nA+mC=q-\frac{n+m}{2}$ form a basis for the corrected holomorphic piece, while the full Chapter~4 model also includes non-holomorphic states and higher Jacobi excitations. Geometric quantization with positive K\"ahler polarization intentionally isolates this holomorphic sector. The metaplectic correction shifts the corresponding line bundle from $\mathcal{O}(q)$ to $\mathcal{O}(q-(n+m)/2)$; that shift is the full geometric content of the half-form correction here.


\section{\texorpdfstring{Geometric Quantization on the Cotangent Dihedral Quotient $T^*(\mathbb{C}/D_n)$}{Geometric Quantization on the Cotangent Dihedral Quotient T*(R2/Dn)}}
\label{subsec:gq-r2-dn}

The dihedral configuration quotient $\mathbb{C}/D_n$ is the first example in this chapter where reflection isotropy, and for $n\ge 3$ genuinely non-abelian isotropy, play essential roles~\cite{davis2011lectures,EmmrichRomer1990}. Because the quotient has mirror strata, it should not be confused with the symplectic phase space to be quantized. In this section the symplectic orbifold is the cotangent quotient
\[
T^*[\mathbb{C}/D_n]\cong [T^*\mathbb{C}/D_n],
\]
with the lifted dihedral action preserving the canonical symplectic form. This places the example inside the standing phase-space hypotheses of Chapter~\ref{sec:geometric-quantization-orbifolds}: the mirrors belong to the configuration orbifold $\mathbb{C}/D_n$, whereas the cotangent-lifted reflections on $T^*\mathbb{C}$ are symplectic, orientation preserving, and have even-codimension fixed loci. Mirror boundary reappears only after choosing the vertical polarization and passing to the configuration leaf space, where it is handled through the sector boundary conditions below. In the cyclic cone $\mathbb{C}/\mathbb{Z}_n$, the residual prequantum ambiguity was encoded by characters of an abelian group, so one obtained twisted scalar sectors with fractional angular momentum. For the dihedral quotient, the reflection relation
\[
srs^{-1}=r^{-1}
\]
forces the scalar equivariant data to collapse to a finite set of discrete signs, while the genuinely non-abelian sectors are more naturally described by flat equivariant vector bundles carrying the irreducible representations of $D_n$. Thus the cotangent quotient over $\mathbb{C}/D_n$ is the basic phase-space model for geometric quantization with reflection strata and, for $n\ge 3$, non-abelian orbifold fundamental group.

\subsection{Cotangent Phase-Space Structure and Observables}

Throughout this subsection $n\ge 2$. The case $n=2$ is the degenerate abelian dihedral case; the formulas below still apply, but the doublet range is empty.

Let the covering configuration space be
\[
Q_{\mathrm c}=\mathbb{C}.
\]
The dihedral group $D_n$ of order $2n$ is generated by a rotation $r$ and a reflection $s$ with relations
\[
r^n=1,
\qquad
s^2=1,
\qquad
srs^{-1}=r^{-1}.
\]
In polar coordinates $(\varrho,\phi)$ on $\mathbb{C}$, the action is
\[
r\cdot(\varrho,\phi)=\bigl(\varrho,\phi+\tfrac{2\pi}{n}\bigr),
\qquad
s\cdot(\varrho,\phi)=(\varrho,-\phi).
\]
The quotient configuration orbifold
\[
Q=[\mathbb{C}/D_n]
\]
may be identified with the wedge
\[
0\le \phi\le \alpha,
\qquad
\alpha:=\frac{\pi}{n},
\]
whose two boundary rays are mirror strata. The ray $\phi=0$ is fixed by $s$, while the ray $\phi=\alpha$ is fixed by $s':=rs$.

\subsubsection{Phase space and symplectic form.}
The classical phase space is the cotangent orbifold
\[
X:=T^*Q=[T^*\mathbb{C}/D_n].
\]
On the cover $T^*\mathbb{C}\cong\mathbb{R}^4$, with canonical coordinates $(x,y,p_x,p_y)$, the Liouville form and symplectic form are
\[
\lambda=p_x\,dx+p_y\,dy,
\qquad
\omega=d\lambda=dp_x\wedge dx+dp_y\wedge dy.
\]
In canonical polar coordinates $(\varrho,\phi,p_\varrho,p_\phi)$, this becomes
\[
\lambda=p_\varrho\,d\varrho+p_\phi\,d\phi,
\qquad
\omega=dp_\varrho\wedge d\varrho+dp_\phi\wedge d\phi.
\]

\subsubsection{$D_n$-invariance and equivariant Darboux form.}
The rotation $r$ acts by the same planar rotation on position and momentum, and the reflection $s$ acts by
\[
(x,y,p_x,p_y)\longmapsto(x,-y,p_x,-p_y).
\]
Hence
\[
s^*(dp_x\wedge dx)=dp_x\wedge dx,
\qquad
s^*(dp_y\wedge dy)=(-dp_y)\wedge(-dy)=dp_y\wedge dy,
\]
so $\omega$ is $D_n$-invariant and descends to the quotient. In the coordinates $(x,y,p_x,p_y)$ every group element acts linearly by
\[
\operatorname{diag}(S_g,S_g),
\qquad
S_g\in O(2),
\]
and this block-diagonal matrix lies in $\mathrm{Sp}(4,\mathbb{R})$. Here the usual cotangent-lift formula is $(x,p)\mapsto(S_gx,S_g^{-T}p)$; it reduces to the displayed form precisely because the dihedral matrices are orthogonal. Thus the origin of phase space already realizes the local equivariant Darboux model explicitly.

The same computation shows that the Liouville form is globally invariant:
\[
s^*\lambda=p_x\,dx+(-p_y)(-dy)=\lambda,
\]
and rotations preserve it as well. Therefore $\lambda$ descends to a global orbifold symplectic potential on $X$.

\subsubsection{Observables and isotropy strata.}
The orbifold observables are exactly the $D_n$-invariant smooth functions on the cover:
\[
C^\infty_{\mathrm{orb}}(X)=C^\infty(T^*\mathbb{C})^{D_n}.
\]
Typical examples are
\[
T=\frac{p_x^2+p_y^2}{2M},
\qquad
\varrho^2=x^2+y^2,
\qquad
xp_x+yp_y,
\qquad
L_z^2=(xp_y-yp_x)^2.
\]
The angular momentum itself,
\[
L_z=xp_y-yp_x,
\]
is not an orbifold observable because
\[
s^*L_z=-L_z.
\]
This is the classical source of the standing-wave, parity-based angular decomposition that replaces ordinary angular-momentum quantization on the quotient.

The singular stratification is likewise read off from the $D_n$-action on phase space:
\begin{itemize}
\item the origin $(0,0,0,0)$ has full isotropy group $D_n$;
\item each reflection $\sigma\in D_n$ fixes the cotangent lift $T^*L_\sigma$ of its mirror axis $L_\sigma\subset\mathbb{C}$;
\item generic points have trivial isotropy.
\end{itemize}
For any $D_n$-invariant observable $f$, the Hamiltonian vector field $X_f$ is equivariant, so its flow preserves these isotropy types.

A basic geometric distinction from the cyclic case is already visible here: the reflection part of $D_n$ reverses the standard complex structure on the configuration plane. Consequently, the most obvious K\"ahler polarization coming from the complex coordinate $z=x+iy$ will not descend to the quotient, even though the cotangent-lifted action remains symplectic and preserves the vertical real polarization used below.

\subsection{Prequantization and Equivariant Bundles}

The first step of geometric quantization is to construct a Hermitian orbifold line bundle with unitary connection satisfying
\[
F_\nabla=-\frac{i}{\hbar}\omega.
\]
Since the cover $T^*\mathbb{C}\cong\mathbb{R}^4$ is contractible, the only nontrivial data come from the $D_n$-equivariant structure.

\subsubsection{Stack and de Rham cohomology and the absence of flux quantization.}
For the global quotient $X=[\mathbb{R}^4/D_n]$,
\[
H^2_{\mathrm{st}}(X;\mathbb{Z})\cong H^2(BD_n;\mathbb{Z}).
\]
Because the covering manifold is contractible, every invariant closed form is exact after averaging a primitive over the finite group. Hence
\[
H^2_{\mathrm{dR,orb}}(X)=0.
\]
So the de Rham prequantization class \([\omega/(2\pi\hbar)]\) vanishes automatically, and the orbifold prequantization condition is always satisfied. There is no flux quantization constraint, exactly as for \(\mathbb{C}/\mathbb{Z}_n\).

The possible integral lifts with zero real image form the torsion group determined by $D_n$ alone~\cite{AdemMilgram2004}:
\[
H^2_{\mathrm{st}}(X;\mathbb{Z})
\cong
\mathrm{Hom}(D_n^{\mathrm{ab}},U(1))
\cong
\begin{cases}
\mathbb{Z}_2, & n \text{ odd},\\
\mathbb{Z}_2\times\mathbb{Z}_2, & n \text{ even}.
\end{cases}
\]
Thus the nontrivial prequantum information is purely torsion.

\subsubsection{A canonical prequantum connection on the cover.}
Take the trivial Hermitian line bundle
\[
L_{\mathrm c}=T^*\mathbb{C}\times\mathbb{C}
\]
with its standard metric. On it define the unitary connection
\[
\nabla=d-\frac{i}{\hbar}\lambda
=
d-\frac{i}{\hbar}(p_x\,dx+p_y\,dy).
\]
Its curvature is
\[
F_\nabla
=
d\!\left(-\frac{i}{\hbar}\lambda\right)
=
-\frac{i}{\hbar}\omega.
\]
Because $\lambda$ is $D_n$-invariant, this connection is equivariant for any choice of unitary $D_n$-action on the fiber.

\subsubsection{Scalar sectors and characters.}
For genuine orbifold line bundles, the equivariant structure is given by a character
\[
\chi:D_n\to U(1).
\]
Since $U(1)$ is abelian, every such character factors through the abelianization of $D_n$. The relation
\[
srs^{-1}=r^{-1}
\]
therefore forces
\[
\chi(r)=\chi(r)^{-1},
\]
so $\chi(r)\in\{\pm1\}$. This is the geometric reason that there is no continuous Aharonov--Bohm parameter in the dihedral case.

The allowed scalar sectors are:
\begin{itemize}
\item the trivial character
\[
\chi_{\mathbf 1}(r)=1,
\qquad
\chi_{\mathbf 1}(s)=1,
\]
corresponding to the NN sector;
\item the orientation sign character
\[
\chi_{\mathrm{sgn}}(r)=1,
\qquad
\chi_{\mathrm{sgn}}(s)=-1,
\]
corresponding to the DD sector;
\item when $n$ is even, two additional characters with $\chi(r)=-1$:
\[
\chi_{\mathrm{det}}(r)=-1,\ \chi_{\mathrm{det}}(s)=1,
\qquad
\chi_{\mathrm{sgn}\cdot\mathrm{det}}(r)=-1,\ \chi_{\mathrm{sgn}\cdot\mathrm{det}}(s)=-1,
\]
corresponding to the ND and DN sectors.
\end{itemize}
When $n$ is odd, the condition $\chi(r)^n=1$ rules out $\chi(r)=-1$, so only the first two sectors survive.
Here $\chi_{\mathrm{sgn}}$ is the determinant character of the defining $O(2)$-representation; the notation $\chi_{\mathrm{det}}$ denotes the even-$n$ character with $\chi(r)=-1$ and $\chi(s)=1$, following the convention of Chapter~4.

\begin{proposition}[Scalar prequantum sectors]
\label{prop:dn-scalar-sectors}
For $X=[T^*\mathbb{C}/D_n]$, inequivalent scalar prequantum line bundles with fixed curvature $-\frac{i}{\hbar}\omega$ are classified by characters of $D_n$, equivalently by characters of $D_n^{\mathrm{ab}}$. There are two such sectors when $n$ is odd and four when $n$ is even.
\end{proposition}
\begin{proof}
Because $T^*\mathbb{C}$ is contractible, every Hermitian line bundle on the cover is trivial. If $(L',\nabla')$ and $(L,\nabla)$ are scalar prequantum line bundles with the same curvature, then
\[
L'\otimes L^{-1}
\]
has flat connection. By pulling back the flat orbifold line bundles classified on the configuration quotient $\mathbb{C}/D_n$ in Section~\ref{ex:R2Dn}, such flat twists on the cotangent phase space are classified by characters of $D_n$. Conversely, tensoring a fixed prequantum line bundle by a flat character bundle does not change the curvature. Hence the fixed-curvature sectors are exactly the elements of
\[
\mathrm{Hom}(D_n,U(1))
\cong
\mathrm{Hom}(D_n^{\mathrm{ab}},U(1)),
\]
whose cardinality is $2$ for $n$ odd and $4$ for $n$ even.
\end{proof}

Thus the line-bundle part of the dihedral example only sees the abelianized group \(D_n^{\mathrm{ab}}\). The non-abelian irreducible representations enter below only after allowing flat vector-bundle coefficients. These doublet sectors should therefore be read as a coefficient-system extension with scalar central curvature, not as additional elements of \(\mathrm{Pic}_{\mathrm{orb}}(X)\) or as new scalar prequantum line-bundle sectors.

\subsubsection{Flat vector bundles and doublet sectors.}
Up to this point, the discussion concerns honest prequantum line bundles. To recover the full $D_n$-representation theory seen in Chapter~4, one may enlarge the coefficient system and allow flat Hermitian orbifold vector bundles carrying the same scalar prequantum curvature. These higher-rank objects are not new line-bundle sectors; rather, they are coefficient-twisted versions of the same prequantum system. On the contractible cover every Hermitian vector bundle is trivial, so an equivariant structure is specified by a unitary representation
\[
\rho:D_n\to U(k),
\]
and the same scalar connection acts componentwise. Because the curvature is central,
\[
F_\nabla=-\frac{i}{\hbar}\omega\cdot \mathbf 1_k,
\]
and because the chosen connection form $\lambda\,\mathbf 1_k$ is $D_n$-invariant while $\rho(g)$ is constant on the cover, the connection commutes with the equivariant structure.

\begin{definition}[Prequantum vector bundle]
\label{def:prequantum-vector-bundle-dn}
A prequantum vector bundle of rank $k$ on $X=[T^*\mathbb{C}/D_n]$ is a Hermitian orbifold vector bundle $E\to X$ with unitary connection $\nabla$ satisfying
\[
F_\nabla=-\frac{i}{\hbar}\omega\cdot\mathbf 1_k.
\]
\end{definition}

The irreducible flat coefficient systems are indexed by $\mathrm{Irr}(D_n)$. Besides the scalar characters above, one has the 2-dimensional irreducible representations
\[
\rho_q,
\qquad
1\le q\le \Bigl\lfloor\frac{n-1}{2}\Bigr\rfloor,
\]
which produce the doublet sectors already seen in Chapter~4.

\subsubsection{Prequantum Hilbert space and operators.}
For an irreducible representation $(\rho,V_\rho)$, write
\[
\boldsymbol{\xi}=(x,y,p_x,p_y)\in T^*\mathbb{C}.
\]
The corresponding coefficient-twisted prequantum Hilbert space is
\[
\mathcal{H}_{\mathrm{pre},\rho}(X)
:=
\left\{
\Psi\in L^2(T^*\mathbb{C},V_\rho)
\;\middle|\;
\Psi(g\cdot\boldsymbol{\xi})=\rho(g)\Psi(\boldsymbol{\xi})
\ \text{for all }g\in D_n
\right\}.
\]
Equivalently,
\[
\mathcal{H}_{\mathrm{pre},\rho}(X)
\cong
\operatorname{Hom}_{D_n}\!\bigl(V_\rho^*,L^2(T^*\mathbb{C})\bigr).
\]
This is the usual multiplicity-space description of the $\rho$-isotypic component. For the dihedral irreducibles considered here, $V_\rho$ is self-dual, so one may harmlessly replace $V_\rho^*$ by an equivalent copy of $V_\rho$. In rank $1$ this reduces to the scalar prequantum line-bundle sectors above.
The orbifold inner product is
\[
\langle \Psi_1,\Psi_2\rangle_X
=
\frac{1}{|D_n|}\int_{T^*\mathbb{C}}
\langle \Psi_1,\Psi_2\rangle_{V_\rho}\,d^4z
=
\frac{1}{2n}\int_{T^*\mathbb{C}}\Psi_1^\dagger\Psi_2\,d^4z.
\]

For any $D_n$-invariant observable $f$, the prequantum operator is
\[
Q(f)\Psi=-i\hbar\,\nabla_{X_f}\Psi+f\Psi
=
\bigl(-i\hbar X_f-\lambda(X_f)+f\bigr)\Psi,
\]
acting componentwise on $V_\rho$-valued sections. Because both $X_f$ and $\nabla$ are equivariant, each isotypic sector is preserved.

\subsection{Vertical Polarization and Quantum Hilbert Spaces}

\subsubsection{The vertical real polarization.}
We choose the vertical polarization
\[
\mathcal P=\ker(d\pi)=\langle \partial_{p_x},\partial_{p_y}\rangle.
\]
It is Lagrangian, integrable, and $D_n$-invariant:
\begin{itemize}
\item $\omega(\partial_{p_x},\partial_{p_y})=0$ and $\dim\mathcal P=2$;
\item $[\partial_{p_x},\partial_{p_y}]=0$;
\item rotations and reflections act linearly on $(p_x,p_y)$ and preserve the span of $\partial_{p_x},\partial_{p_y}$.
\end{itemize}
Thus the vertical polarization descends to the orbifold without obstruction.

\subsubsection{Why the obvious K\"ahler polarization fails.}
The standard complex structure on the configuration plane is preserved by rotations but reversed by reflections:
\[
s:z=x+iy\longmapsto \bar z.
\]
Accordingly, the induced K\"ahler polarization on the cover is not $D_n$-invariant; reflections exchange holomorphic and anti-holomorphic directions. This does not rule out all compatible complex structures on $T^*\mathbb{C}$, but it shows that the most natural configuration-space K\"ahler polarization does not descend. For comparison with the Schr\"odinger picture of Chapter~4, the vertical real polarization is the right choice.

\subsubsection{Polarized sections.}
Write
\[
\mathbf x=(x,y)\in\mathbb C.
\]
The polarization condition
\[
\nabla_\xi\Psi=0
\qquad
\text{for all }\xi\in\mathcal P
\]
forces formal polarized sections to be independent of the momentum variables. Thus, at the level of smooth sections, one obtains pullbacks of functions on the leaf space $Q=[\mathbb{C}/D_n]$ with equivariance determined by the chosen sector:
\[
\psi(g\cdot\mathbf x)=\chi(g)\psi(\mathbf x)
\]
in scalar sectors, and
\[
\Psi(g\cdot\mathbf x)=\rho(g)\Psi(\mathbf x)
\]
in vector sectors.

However, these momentum-independent sections are not square-integrable with respect to the phase-space Liouville measure on $T^*\mathbb{C}$. As emphasized in Chapter~5, a real polarization must be completed on the leaf space rather than on phase space itself. In the present cotangent-bundle example, the leaf space is precisely the configuration orbifold $Q=[\mathbb{C}/D_n]$, so after the standard half-density correction discussed explicitly below, the vertical polarization gives the usual Schr\"odinger realization there. Concretely, the resulting quantum spaces are
\[
\mathcal H_{\mathrm{quant},\chi}(Q)
:=
\left\{
\psi\in L^2(\mathbb C)
\;\middle|\;
\psi(g\cdot\mathbf x)=\chi(g)\psi(\mathbf x)
\right\}
\]
in scalar sectors, and
\[
\mathcal H_{\mathrm{quant},\rho}(Q)
:=
\left\{
\Psi\in L^2(\mathbb C,V_\rho)
\;\middle|\;
\Psi(g\cdot\mathbf x)=\rho(g)\Psi(\mathbf x)
\right\}
\]
in vector sectors. Equivalently, these are the $L^2$ spaces of sections of the corresponding flat line or vector bundle over the configuration orbifold. Their orbifold inner products are computed on the cover with the usual quotient normalization:
\[
\langle \Psi_1,\Psi_2\rangle_Q
=
\frac{1}{|D_n|}\int_{\mathbb C}\langle \Psi_1,\Psi_2\rangle\,dx\,dy
=
\frac{1}{2n}\int_{\mathbb C}\langle \Psi_1,\Psi_2\rangle\,dx\,dy.
\]

\subsubsection{Quantizable observables.}
An observable is directly quantizable in this polarization only if its Hamiltonian vector field preserves $\mathcal P$. Position observables satisfy this automatically: for $f=f(x,y)$,
\[
X_f=-\partial_x f\,\partial_{p_x}-\partial_y f\,\partial_{p_y}
\]
is vertical. In particular,
\[
\varrho^2=x^2+y^2
\]
is quantizable and its prequantum operator reduces on polarized sections to multiplication by $\varrho^2$.

By contrast, the free Hamiltonian
\[
H=\frac{p_x^2+p_y^2}{2M}
\]
has
\[
X_H=\frac{p_x}{M}\partial_x+\frac{p_y}{M}\partial_y,
\]
and
\[
[X_H,\partial_{p_x}]=-\frac{1}{M}\partial_x\notin\mathcal P,
\qquad
[X_H,\partial_{p_y}]=-\frac{1}{M}\partial_y\notin\mathcal P.
\]
Thus $H$ is not directly quantizable as a first-order Kostant--Souriau observable in the vertical polarization. The same first-order criterion excludes direct quantization of $L_z^2$: on the smooth locus $L_z=p_\phi$, hence
\[
X_{L_z^2}=2p_\phi\,\partial_\phi,
\qquad
[X_{L_z^2},\partial_{p_\phi}]=-2\,\partial_\phi\notin\mathcal P.
\]
Although $L_z^2$ is $D_n$-invariant, its Hamiltonian vector field therefore does not preserve the polarization. This does not remove the physical angular or free Hamiltonian operators; it means they are supplied by the second-order Schr\"odinger realization on the configuration orbifold, where the free Hamiltonian is represented by the Laplacian with the appropriate wedge boundary conditions.

\subsubsection{Isotypic decomposition.}
The $D_n$-representation on $L^2(\mathbb C)$ decomposes as
\[
L^2(\mathbb C)\cong
\bigoplus_{\rho\in\mathrm{Irr}(D_n)}
\mathcal H_\rho\otimes V_\rho,
\]
where
\[
\mathcal H_\rho\cong \operatorname{Hom}_{D_n}(V_\rho^*,L^2(\mathbb C))
\]
is the multiplicity space. Since the dihedral irreducibles are self-dual, this can again be written using $V_\rho$ instead of $V_\rho^*$ if desired. After choosing a basis of $V_\rho$, this is exactly the sector decomposition by equivariant scalar or vector-valued Schr\"odinger wavefunctions on the configuration leaf space. The trivial representation gives the untwisted orbifold Hilbert space, and the other summands are realized geometrically by twisting with the corresponding flat line or vector bundle.

\subsubsection{Bohr--Sommerfeld leaves.}
For the vertical polarization, the lifted leaves on the cover are the momentum fibers $T^*_x\mathbb{C}$. Over an orbifold base point $[x]$, the descended leaf is modeled by
\[
[T^*_x\mathbb{C}/(D_n)_x],
\]
where $(D_n)_x$ acts on the fiber by the cotangent lift. Since the Liouville form restricts to zero on each lifted fiber, ordinary leafwise holonomy is trivial. The only nontrivial Bohr--Sommerfeld condition comes from the equivariance of the parallel section under this stabilizer action. Equivalently, for a leafwise orbifold loop represented by an isotropy element, the condition reduces to triviality of the isotropy action on the coefficient fiber:
\begin{itemize}
\item interior leaves are automatically Bohr--Sommerfeld;
\item a mirror leaf fixed by a reflection $\sigma$ is Bohr--Sommerfeld in a scalar sector precisely when $\chi(\sigma)=+1$; in a vector sector it supports values in the invariant subspace $V_\rho^{\langle\sigma\rangle}$, which is one-dimensional for the two-dimensional dihedral irreducibles;
\item the apex leaf supports a nonzero parallel scalar section only in the trivial character sector, and more generally in a vector sector only when the stabilizer representation has a nonzero invariant vector.
\end{itemize}
This is the geometric quantization version of the Chapter~4 distinction between Neumann-type and Dirichlet-type mirror behavior. In particular, among the scalar sectors only the NN sector permits a nonzero invariant value at the apex leaf, equivalently only the NN sector contains the constant angular channel $\nu=0$.

\subsection{Sector Decomposition and Angular Modes}

Because $L_z$ is odd under reflection, one no longer diagonalizes the quantum problem by eigenstates of $\hat L_z$. The correct decomposition is instead by $D_n$-equivariant angular modes.

\subsubsection{Scalar sectors.}
Let $\chi$ be a scalar character, and write
\[
\chi_{\mathrm{rot}}:=\chi(r).
\]
The rotation condition
\[
\psi(\varrho,\phi+\tfrac{2\pi}{n})=\chi_{\mathrm{rot}}\psi(\varrho,\phi)
\]
discretizes the allowed angular frequencies. The reflection condition
\[
\psi(\varrho,-\phi)=\chi(s)\psi(\varrho,\phi)
\]
then forces even or odd standing waves:
\begin{itemize}
\item if $\chi(s)=+1$, the angular dependence is of cosine type;
\item if $\chi(s)=-1$, the angular dependence is of sine type.
\end{itemize}
If $\chi_{\mathrm{rot}}=1$, the allowed frequencies lie in
\[
m\in n\mathbb Z.
\]
If $\chi_{\mathrm{rot}}=-1$, which occurs only when $n$ is even, then
\[
m\in n\bigl(\mathbb Z+\tfrac12\bigr).
\]
Thus the cover Fourier labels are the above congruence classes. Equivalently, the non-negative Bessel orders $\nu=|m|$ are:
\begin{itemize}
\item the NN sector contains $\nu=0,n,2n,\dots$;
\item the DD sector contains $\nu=n,2n,3n,\dots$;
\item when $n$ is even, the ND and DN sectors contain
\[
\nu=\frac n2,\frac{3n}{2},\frac{5n}{2},\dots.
\]
\end{itemize}
This recovers the scalar wedge sectors of Chapter~4.

\subsubsection{Doublet sectors.}
For a 2-dimensional irrep $\rho_q$, choose the chiral basis in which
\[
\rho_q(r)=
\begin{pmatrix}
e^{2\pi iq/n} & 0\\
0 & e^{-2\pi iq/n}
\end{pmatrix},
\qquad
\rho_q(s)=
\begin{pmatrix}
0 & 1\\
1 & 0
\end{pmatrix}.
\]
Then equivariance forces one component to carry angular frequencies
\[
m\equiv q\pmod n
\]
and the other to carry
\[
m\equiv -q\pmod n,
\]
with reflection exchanging the two. The allowed angular set is therefore
\[
\mathcal I_q=(q+n\mathbb Z)\cup(-q+n\mathbb Z).
\]
The corresponding non-negative Bessel orders are
\[
\nu\in\{q+nj,\ n-q+nj\mid j\in\mathbb Z_{\ge0}\}.
\]

\begin{remark}[Exclusion of the case $q=n/2$]
\label{rem:q-exclusion}
When $n$ is even, the value $q=n/2$ does not produce a genuine doublet sector: the corresponding representation is reducible and splits into the two scalar characters with $\chi(r)=-1$, namely the ND and DN sectors.
\end{remark}

\subsubsection{Free-particle spectrum in the Schr\"odinger model.}
Although the free Hamiltonian is not directly quantizable in the vertical polarization, the associated Schr\"odinger operator on the configuration orbifold is
\[
\hat H=-\frac{\hbar^2}{2M}\nabla^2.
\]
After separation of variables, the radial equation is Bessel's equation with order
\[
\nu=|m|.
\]
With the Friedrichs, equivalently regular, apex extension used in Chapter~4, the radial solutions are~\cite{KayStuder1991}
\[
R(\varrho)=J_\nu(k\varrho),
\qquad
E=\frac{\hbar^2k^2}{2M},
\qquad
k\ge 0.
\]
For this extension the continuous energy spectrum is therefore the same as in the cyclic case, while the sector dependence enters only through the admissible Bessel orders listed above. Other self-adjoint extensions in the $\nu=0$ channels can change the spectral problem and are not included in this sector comparison.

\subsection{Metaplectic Correction}
\label{subsec:metaplectic-dn}

\subsubsection{Canonical bundle of the vertical polarization.}
For
\[
\mathcal P=\langle \partial_{p_x},\partial_{p_y}\rangle,
\]
the annihilator is
\[
\mathcal P^\circ=\langle dx,dy\rangle,
\]
so the polarization canonical bundle is
\[
K_{\mathcal P}=\det(\mathcal P^\circ)=\langle dx\wedge dy\rangle.
\]
A genuine half-form bundle would be an orbifold line bundle $\delta_{\mathcal P}$ satisfying
\[
\delta_{\mathcal P}^{\otimes 2}\cong K_{\mathcal P}.
\]

\subsubsection{Topological and equivariant obstruction.}
On the ordinary cover there is no topological square-root obstruction: the space $T^*\mathbb C$ is contractible, so
\[
H^2(T^*\mathbb C;\mathbb Z)=0.
\]
Thus $K_{\mathcal P}$ is topologically trivial as an ordinary line bundle on the cover. The obstruction to a genuine half-form is therefore entirely local and equivariant. Rotations preserve the oriented 2-form:
\[
r^*(dx\wedge dy)=dx\wedge dy.
\]
Reflections reverse it:
\[
s^*(dx\wedge dy)=dx\wedge(-dy)=-(dx\wedge dy).
\]
Thus the isotropy character of $K_{\mathcal P}$ is
\[
\chi_{K_{\mathcal P}}(r)=1,
\qquad
\chi_{K_{\mathcal P}}(s)=-1,
\]
namely the sign character. A genuine half-form would require a character $\chi_\delta$ satisfying
\[
\chi_\delta^2=\chi_{K_{\mathcal P}}.
\]
But on a reflection isotropy group $\langle s\rangle\cong\mathbb Z_2$, every character squares to $1$, never to $-1$. Therefore $K_{\mathcal P}$ admits no $D_n$-equivariant square root.

\subsubsection{Half-densities as the real-polarization substitute.}
For real polarizations the standard replacement is the half-density bundle
\[
|dx\wedge dy|^{1/2}.
\]
Because absolute value removes orientation signs,
\[
s^*|dx\wedge dy|^{1/2}
=
|-(dx\wedge dy)|^{1/2}
=
|dx\wedge dy|^{1/2},
\]
so the half-density bundle carries the trivial $D_n$-equivariant structure. This is the natural corrected object in the present orbifold setting.

\subsubsection{Corrected operators and divergence.}
Since no genuine orbifold half-form exists here, the metaplectic operator formula of Section~\ref{sec:corrected-operator} is not being applied literally. The statement below uses the half-density replacement described in Section~\ref{sec:half-form-definition}.

For a quantizable position observable $f=f(x,y)$,
\[
X_f=-\partial_x f\,\partial_{p_x}-\partial_y f\,\partial_{p_y}.
\]
Since this vector field does not move the base coordinates,
\[
\mathcal L_{X_f}(dx)=0,
\qquad
\mathcal L_{X_f}(dy)=0,
\qquad
\mathcal L_{X_f}(dx\wedge dy)=0.
\]
Hence
\[
\operatorname{div}_{\mathcal P}X_f=0,
\]
so the density-corrected operator agrees with the prequantum operator on such observables. For example, for
\[
\varrho^2=x^2+y^2,
\qquad
X_{\varrho^2}=-2x\,\partial_{p_x}-2y\,\partial_{p_y},
\]
one has
\[
Q_{\mathrm{meta}}(\varrho^2)=Q_{\mathrm{pre}}(\varrho^2).
\]
No metaplectic spectral shift appears.

\subsubsection{Maslov index and Bohr--Sommerfeld correction.}
The lifted vertical leaves are affine planes $T^*_x\mathbb C$, so there are no caustics and the Maslov index is zero:
\[
\mu=0.
\]
Because the half-density bundle is equivariantly trivial, it contributes no extra isotropy phase. Thus the density-corrected Bohr--Sommerfeld condition is the same as the uncorrected one in this example: the only nontrivial leafwise compatibility comes from the prequantum isotropy action discussed above.

\begin{remark}[Why the density replacement is essential]
\label{rem:signed-halfform-dn}
The failure of a genuine half-form is not a matter of choosing a branch of $\sqrt{-1}$. It is a genuine local obstruction: the sign character of $K_{\mathcal P}$ on a reflection isotropy group is not divisible by $2$ in the local character group. Half-densities avoid this obstruction because absolute values erase the reflection sign.
\end{remark}


\backmatter
\bibliographystyle{unsrt}
\bibliography{bibliography}

\end{document}